\documentclass[11pt,a4paper]{article}
\usepackage{jheppub}
\usepackage{longtable}

\usepackage{graphicx}
\usepackage{dcolumn}
\usepackage{bm}
\usepackage{pdflscape}
\usepackage{setspace}
\usepackage{epsfig}
\usepackage[normalem]{ulem}
\usepackage{color}
\usepackage{multirow}
\usepackage{array}
\usepackage{hyperref}
\usepackage{color}

\newcommand*{\La}{\cal{L}}
\newcommand*{\no}{\noindent}
\newcommand*{\bea}{\begin{eqnarray}}
\newcommand*{\eea}{\end{eqnarray}}
\newcommand*{\be}{\begin{equation}}
\newcommand*{\ee}{\end{equation}}
\newcommand*{\pd}{\partial}
\newcommand*{\pdm}{\pd_{\mu}}
\newcommand*{\pdn}{\pd_{\nu}}

\newcommand*{\nn}{}

\title{A study of the influence of the gauge group on the Dyson-Schwinger equations for scalar-Yang-Mills systems}

\author[1]{Veronika Macher}
\author[2]{Axel Maas}\emailAdd{axelmaas@web.de}
\author[1]{Reinhard Alkofer}
\affiliation[1]{Institute of Physics, Karl-Franzens-University Graz, Universit\"atsplatz 5, A-8010 Graz, Austria}
\affiliation[2]{Theoretical-Physical Institute, Friedrich-Schiller-University Jena, Max-Wien-Platz 1, D-07743 Jena, Germany}

\abstract{
The particular choice of the gauge group for Yang-Mills theory plays an important role when it comes to the influence of matter fields. In particular, both the chosen gauge group and the representation of the matter fields yield structural differences in the quenched case. Especially, the qualitative behavior of the Wilson potential is strongly dependent on this selection. Though the algebraic reasons for this observation is clear, it is far from obvious how this behavior can be described besides using numerical simulations.

Herein, it is investigated how the group structure appears in the Dyson-Schwinger equations, which as a hierarchy of equations for the correlation functions have to be satisfied. It is found that there are differences depending on both the gauge group and the representation of the matter fields. This provides insight into possible truncation schemes for practical calculations using these equations.}


\begin{document}

\maketitle

\section{Introduction}

When investigating Yang-Mills theory, it becomes quickly apparent that there exists little qualitative difference when selecting any kind of (semi-)simple Lie algebra as the underlying gauge algebra of the gauge group \cite{Bohm:2001yx,Holland:2003kg,Pepe:2006er,Braun:2010cy,Maas:2010qw,Wellegehausen:2010ai,Lucini:2001ej}. In particular, up to trivial effects the bound-state spectrum and its dynamics is quite similar \cite{Wellegehausen:2010ai,Lucini:2001ej}. The same applies to the elementary correlation functions, as far as they have been investigated \cite{Maas:2010qw,Maas:2007af,vonSmekal:1997vx,Maas:2005ym,Cucchieri:2007zm,Oliveira:2008uf}. Also the finite-temperature behavior, except for the order of the always appearing finite-temperature phase transition, is essentially identical \cite{Braun:2010cy,Cossu:2007dk,Holland:2003kg,Bringoltz:2005rr}. Thus, for Yang-Mills theory only the generic structure of a Lie algebra seems to be of importance.

The situation drastically changes when static matter in a given representation of the gauge group is investigated in the background of the gauge fields. Group-theoretical aspects play now a crucial role. In particular, whether it is possible to combine the matter fields with the gauge boson fields to form a trivial representation, i.\ e., a bound state which is neutral w.\ r.\ t.\ the gauge group. If this is the case, the gauge bosons will screen any such charge. This applies, e.\ g., to the case of matter fields in the same representation as the gauge bosons, i.\ e., the adjoint. An alternative are gauge algebras which have representations which are equivalent to the adjoint one in a certain sense, like the fundamental representation of G$_2$. Otherwise, it is not possible to combine any finite number of gauge bosons with such a static matter field. In that case, a so-called confining Wilson potential \cite{Greensite:2003bk} is observed, and such charges require infinite energy to be produced individually. This applies, e.\ g., to matter fields in the fundamental representation of SU($N$). This latter fact can be understood from the perspective that a single charge in this case cannot be made gauge-invariant, and such a state can therefore not be physical.

The situation changes once more when the matter fields become dynamical, and the number of flavors is not too large such as not to enter either a quasi-conformal or conformal phase or losing asymptotic freedom \cite{Sannino:2009za}. In this case, they can screen each other, and the dynamics become in all cases qualitatively similar, though significant quantitative differences are possible.

Though the qualitative prediction of the behavior is a mere exercise in group theory, and can be accessed even quantitatively using numerical simulations, the underlying dynamics are rather less simple. In particular, the screening proceeds by the dynamical formation of bound states. Also, the formation of the confinement potential is not yet entirely clear, though many proposals for its explanation exists \cite{Greensite:2003bk,Alkofer:2006fu}.

One possible tool for attempting to describe such phenomena are functional methods, here in the form of Dyson-Schwinger equations (DSEs). In general, approximations are required to solve these equations. It is therefore important to capture the essential properties of the gauge group in such approximations to be able to reproduce the behavior described above. The aim here is to study how the gauge group appears in the DSEs for the various possible matter representations in both the quenched and unquenched cases. These results will then provide a basis to construct adequate approximation schemes in future calculations.

To reduce the complexity, in the following only scalar matter will be considered, since the aforementioned patterns are independent of the Lorentz structure \cite{Greensite:2003bk}. The corresponding DSEs will be briefly introduced and discussed in section \ref{sec:dse}. Then it will be outlined that the group plays an important role to distinguish the behavior of the matter representation in section \ref{sec:power}, using power-counting arguments only. The main results on the algebraic structure will be given in section \ref{sec:res}, though many results will be relegated to the appendices \ref{app:res} and \ref{app:res2}, if they do not directly influence the behavior of the system. It should be noted throughout that the results depend only on the gauge algebra, irrespective of which of the related groups are considered. In particular, the results are the same for the su($N$) gauge algebra, independently of whether the gauge group SU($N$) or SU($N$)/Z$_N$ is used. Thus, the distinction of algebra and group will not be made here. However, it should be noted that this distinction is crucial when it comes to the question of anomalies, where the related homotopy groups, single-valuedness and faithful representation are central issues \cite{O'Raifeartaigh:1986vq,Witten:1982fp}.

Some preliminary results can be found in \cite{Macher:2010ad} and some more details in \cite{Macher:2010di}.

\section{The Dyson-Schwinger equations of a scalar-Yang-Mills system}\label{sec:dse}

\subsection{Setup}

The (Euclidean) Lagrangian governing a scalar field $\phi^a$ in representation $R$ of the gauge algebra $G$ coupled to a Yang-Mills field $A_\mu^a$ implementing the gauge group $G$ is given by
\bea
\La&=&\frac{1}{4}F_{\mu\nu}^aF^{\mu\nu}_a+\frac{1}{2}((D_\mu^{ij}\phi_j)^+ D^\mu_{ik}\phi^k+m_h^2\phi^{+}_i\phi^i)+\frac{h}{4!}\phi^{+}_i\phi^i\phi^{+}_j\phi^j\nn\\
F^a_{\mu\nu}&=&\pdm A_\nu^a-\pdn A_\mu^a-gf^a_{\;bc}A_\mu^bA_\nu^c\nn\\
D_\mu^{ij}&=&\delta^{ij}\pdm+gt_{Rc}^{ij}A_\mu^c\nn
\eea
\no where $g$ is the gauge coupling, $f^{abc}$ the totally anti-symmetric structure constants with the indices $a,...$ in the adjoint representation, and $t_R^{ija}$ are the generators of the gauge algebra in the representation $R$ with indices $i,...$ in this representation. The explicit label $R$ will be dropped wherever the context suffices. Here, eventually only the fundamental and the adjoint representation will be used.

The gauge chosen here will be Landau gauge, such that additional ghost fields $c^a$ and anti-ghost fields $\bar{c}^a$ in the adjoint representation appear, alongside with a gauge-fixing Lagrangian. Concerning the definition of the Landau gauge beyond perturbation theory, see \cite{Maas:2010wb} and references therein. However, this plays little role for the present investigation.

\begin{table}
 \caption{\label{tab:cons}Constants for the various gauge algebras, where $N_F$ is the dimensionality of the (chosen) fundamental representation, $N_A$ of the adjoint representation, $T_R$ is the fundamental index constant, $C_A$ and $C_F$ are the adjoint and fundamental Casimirs, respectively. Explicit results for certain contractions of products of symmetric $d$-tensors are also given.}
\vspace{0.2cm}
 \begin{tabular}{|c|c|c|c|c|c|c|c|c|}
 \hline
 Group & $N_F$ & $N_A$ & $T_R$ & $C_A$ & $C_F$ & $d_{33}=(d^{abc})^2$ & $d_{44}=(d^{abcd})^2$ & $d_{444}=d^{abcd}d_{cd}^{\;\;ef}d_{efab}$ \cr
 \hline
 SU(2) & 2 & 3 & $\frac{1}{2}$ & 2 & $\frac{3}{4}$ & 0 & $\frac{45}{16\times 24^2}$ & $\frac{165}{64\times 24^3}$ \cr
 \hline
 SU(3) & 3 & 8 & $\frac{1}{2}$ & 3 & $\frac{4}{3}$ & $\frac{10}{3}$ & $\frac{15}{24 ^2}$ & $\frac{4}{24^3}$ \cr
 \hline
 G$_2$ & 7 & 14 & $\frac{1}{2}$ & 2 & 1 & 0 & $\frac{21}{2\times 24^2}$ & $\frac{1}{8\times 24^3}$ \cr
 \hline
\end{tabular}
\end{table}

In the following also the symmetric structure constants
\be
d^{abc}=t^a\{t^b,t^c\}\nn
\ee 
\no and 
\be
d^{abcd}=\frac{1}{4!}t^a\left(t^b\{t^c,t^d\}+t^c\{t^b,t^d\}+t^d\{t^b,t^c\}\right)\nn
\ee
\no will appear. Note that $d^{abc}$ is only non-zero for the Lie groups SU($N>2$) \cite{Cvitanovic:2008}. Furthermore, most of the results will be kept general using only the generic expressions. However, the most interesting results will also be given explicitly for SU(2), SU(3), and G$_2$. The corresponding constants, which also define our conventions, are listed in table \ref{tab:cons}.

With these definitions, it is sufficient to derive explicitly the DSEs. Since this is an algorithmic task, this can be automatized, and has been performed using the DoDSE package \cite{Alkofer:2008nt}, a specialized version of the package DoFun \cite{Huber:2011qr} for DSEs and functional renormalization group equations. In the present context, only the DSEs for correlation functions up to order four will be used. Due to the number of terms, they will not be displayed explicitly here until truncated in section \ref{sec:trunc}.

The DSEs contain in each diagram besides the full vertices $\Gamma$ also the tree-level vertices. These are listed in our conventions in appendix \ref{app:tl}.

Unfortunately, except for very particular cases \cite{Fischer:2009tn,Fister:2010yw}, it is not possible to obtain information directly from this infinite tower of equation. Therefore, in the current explorative setting, we truncate this system, as described below. It is interesting to first evaluate infrared power-counting rules for the equations.

\subsection{Results at power-counting level}\label{sec:power}

Under certain conditions on the solutions of the DSEs, in particular that the ghost propagator is more divergent than the one of a massless particle, it has been found that all correlation functions of the Yang-Mills system at the symmetric point in the far infrared can be characterized by certain critical exponents \cite{Alkofer:2004it,Fischer:2009tn,Huber:2007kc}. Furthermore, relations between all of these exponents can be determined without truncating the system of DSEs, if in addition the corresponding functional renormalization group equations (FRGEs) are taken into account \cite{Fischer:2006vf,Fischer:2009tn}. If furthermore one of the exponents can be specified by some means, the value of all others are fixed uniquely \cite{Fischer:2006vf,Fischer:2009tn}.

If matter is added to the system, which has so far been done for fundamental quarks and scalars \cite{Fister:2010yw,Alkofer:2008jy,Alkofer:2008tt,Schwenzer:2008vt}, it is no longer possible to find only one solution \cite{Fister:2010yw,Fischer:2009tn}. In particular, in case of the fundamental scalars, there exist at least two different ones under the same assumptions as before, which differ qualitatively in the scalar sector. Furthermore, in addition there may exist additional kinematic singularities beyond the symmetric point, but the constraints for these are even weaker. It should be noted that this analysis will break down at the latest when asymptotic freedom is lost.

Under the assumptions made, it is possible to map a given diagram on an expression which is at most linear in the critical exponents \cite{Alkofer:2004it,Huber:2009wh}. Thus, the determination of the exponents is reduced to the solution of a hierarchy of linear inequalities, which is in principle straightforward, though challenging in detail. In particular, it follows immediately that for the same set of graphs the same set of equations for the exponents is obtained, irrespective of the underlying field types or interactions. As a consequence, this approach yields identically the same solution manifold for scalars (or quarks) in any representation of any gauge algebra. Thus, in the adjoint case the same results are obtained as in the fundamental case. Still, there are multiple solutions in the matter case for the hierarchy of equations, so the different behavior of adjoint and fundamental scalars is not necessarily implying that the analysis is incorrect. However, it implies that at least further results are needed to explain the difference between adjoint and fundamental matter.

It should be noted that in addition other solutions to the DSEs have been found \cite{Fischer:2008uz,Boucaud:2008ji,Binosi:2009qm}, which do not fulfill the assumptions. In this case, a similar analysis cannot (yet) be performed. However, the case investigated here suffices to motivate the importance of the gauge group and algebraic pre-factors for the DSEs (and FRGs).

\subsection{Truncation}\label{sec:trunc}

To make the system of DSEs tractable, it will be truncated. To include the lowest gauge-invariant state, which supposedly has an overlap with the Wilson line, this will be done up to order four, i.\ e., including four-point correlation functions. Furthermore, all genuine two-loop expressions will be dropped. These are sub-leading perturbatively, and in Landau gauge seem to be sub-leading as well at lower momenta under the assumption of an infrared divergent ghost dressing function \cite{Fischer:2009tn}, except in very special circumstances \cite{Bloch:2003yu,Maas:2010wb,Watson:2001yv}.

\begin{figure}
\includegraphics[width=0.55\textwidth]{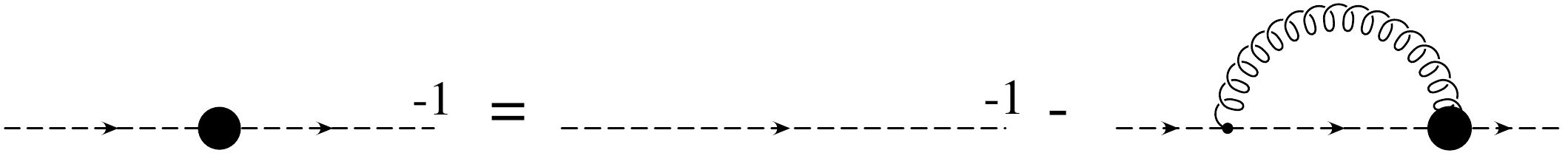}\\
\includegraphics[width=1.00\textwidth]{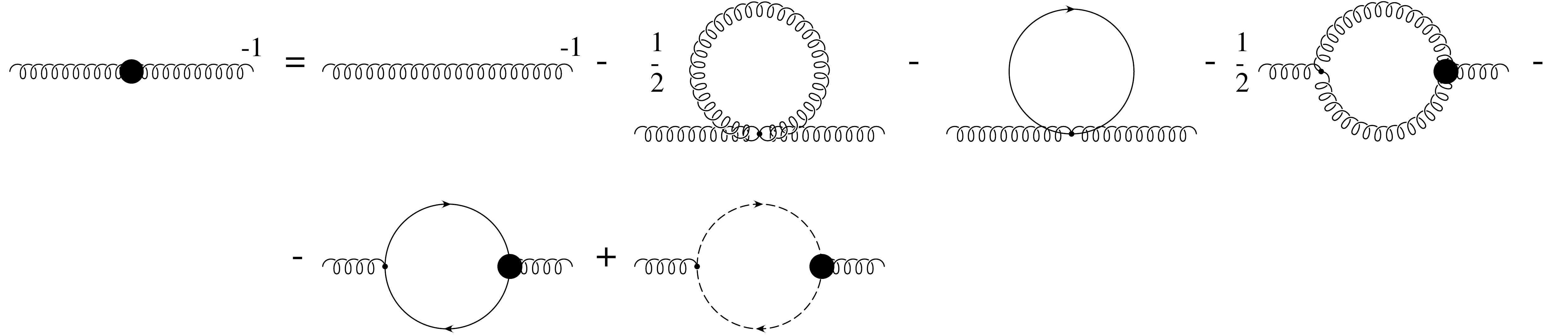}\\
\includegraphics[width=1.00\textwidth]{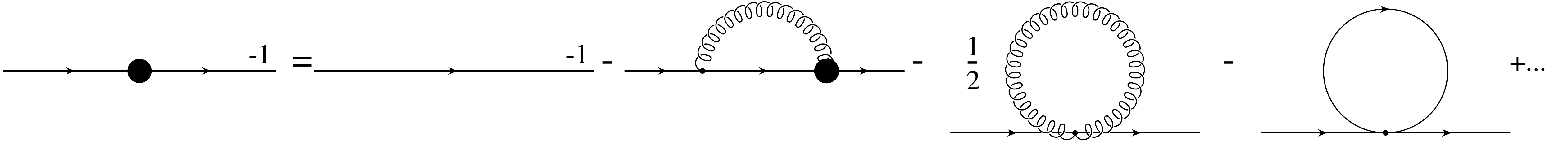}
\caption{\label{dse-prop}The truncated DSEs for the propagators. Curly lines denote gluons, dashed lines denote ghosts, and full lines denotes the scalars of either type. The dots symbolize that these equations are truncated.}
\end{figure}

\begin{figure}
\includegraphics[width=1.00\textwidth]{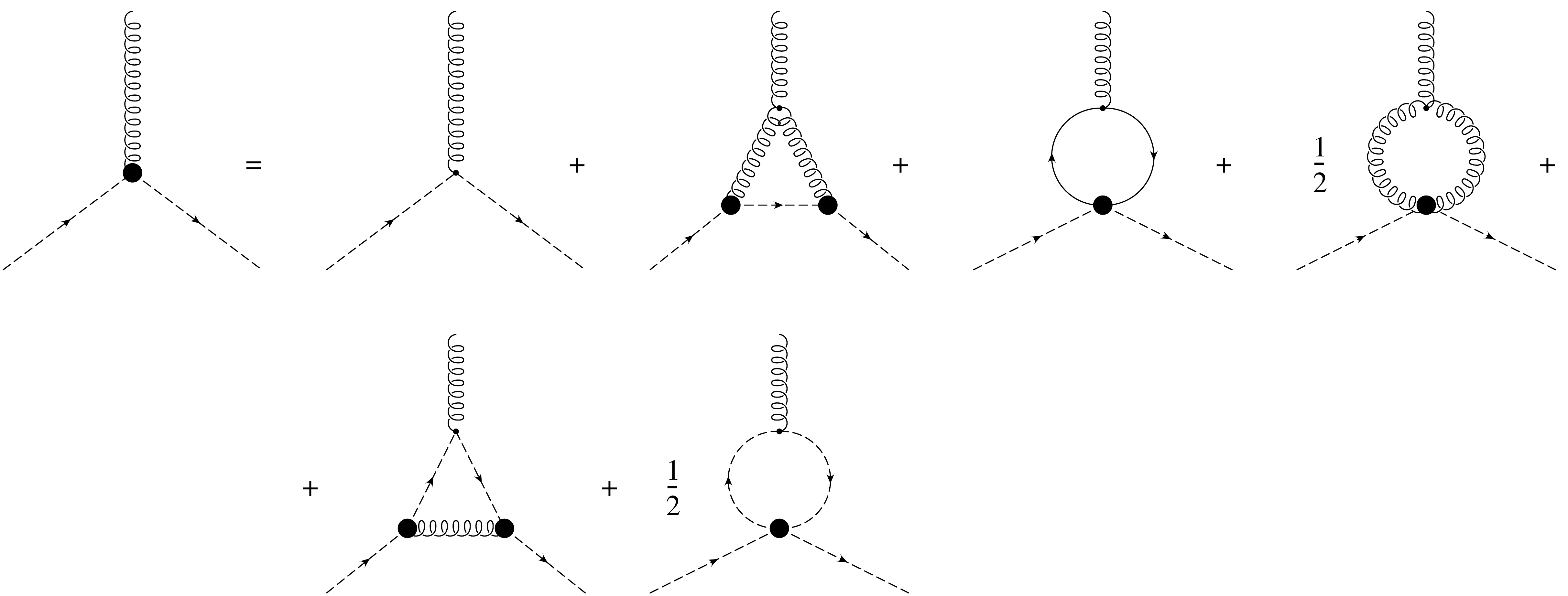}\\
\includegraphics[width=1.00\textwidth]{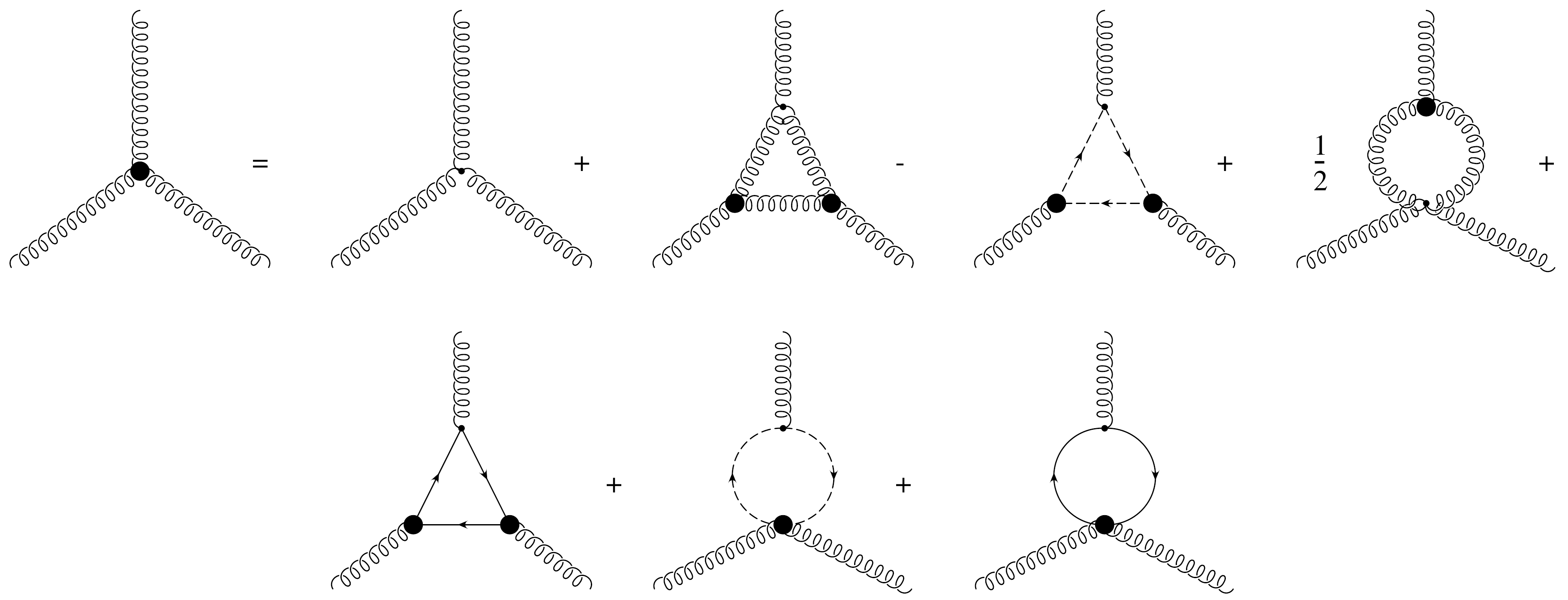}\\
\includegraphics[width=1.00\textwidth]{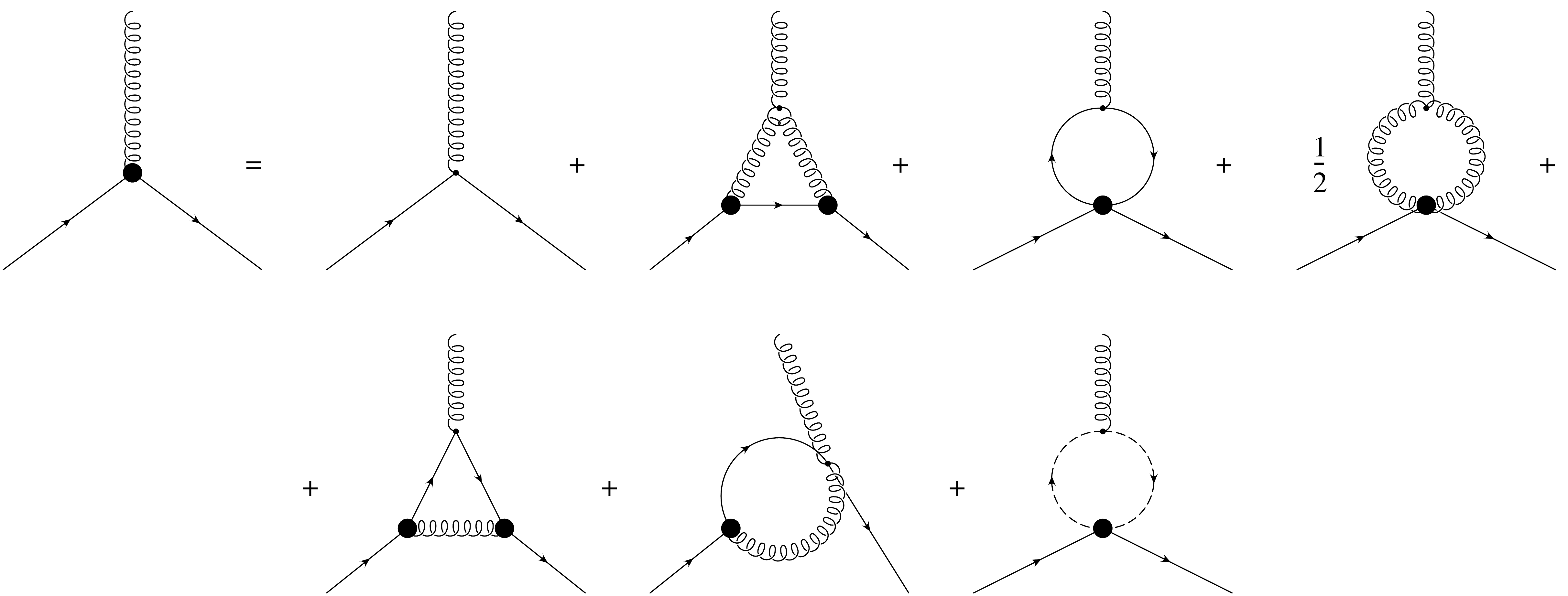}
\caption{\label{dse-3v}The truncated DSEs for the three-point vertices. Curly lines denote gluons, dashed lines denote ghosts, and full lines denotes the scalars of either type.}
\end{figure}

\begin{figure}
\includegraphics[width=1.00\textwidth]{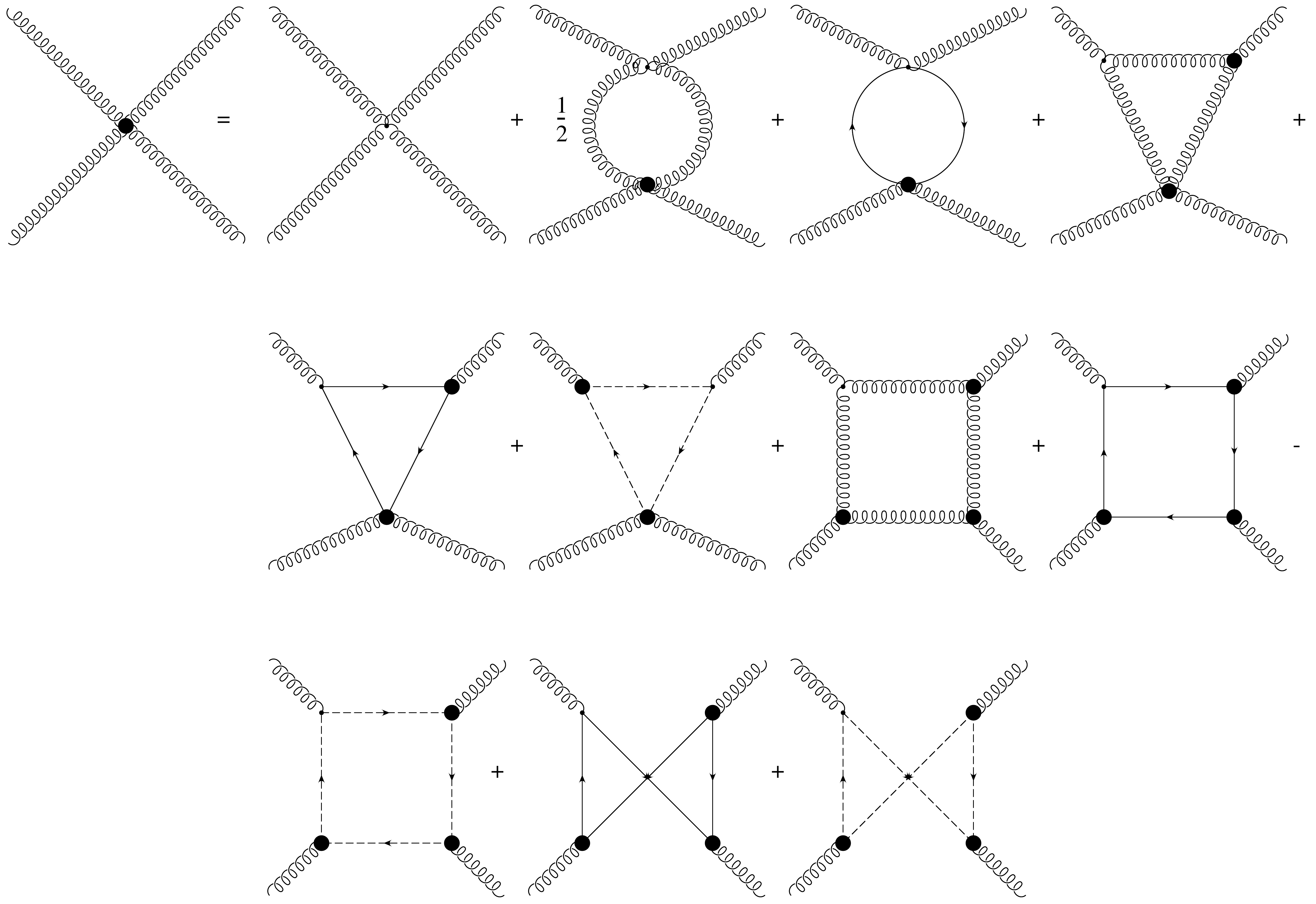}
\caption{\label{dse-4vg}The truncated DSEs for the four-point vertex in the gluonic sector. Curly lines denote gluons, dashed lines denote ghosts, and full lines denotes the scalars of either type.}
\end{figure}

\begin{figure}
\includegraphics[width=1.00\textwidth]{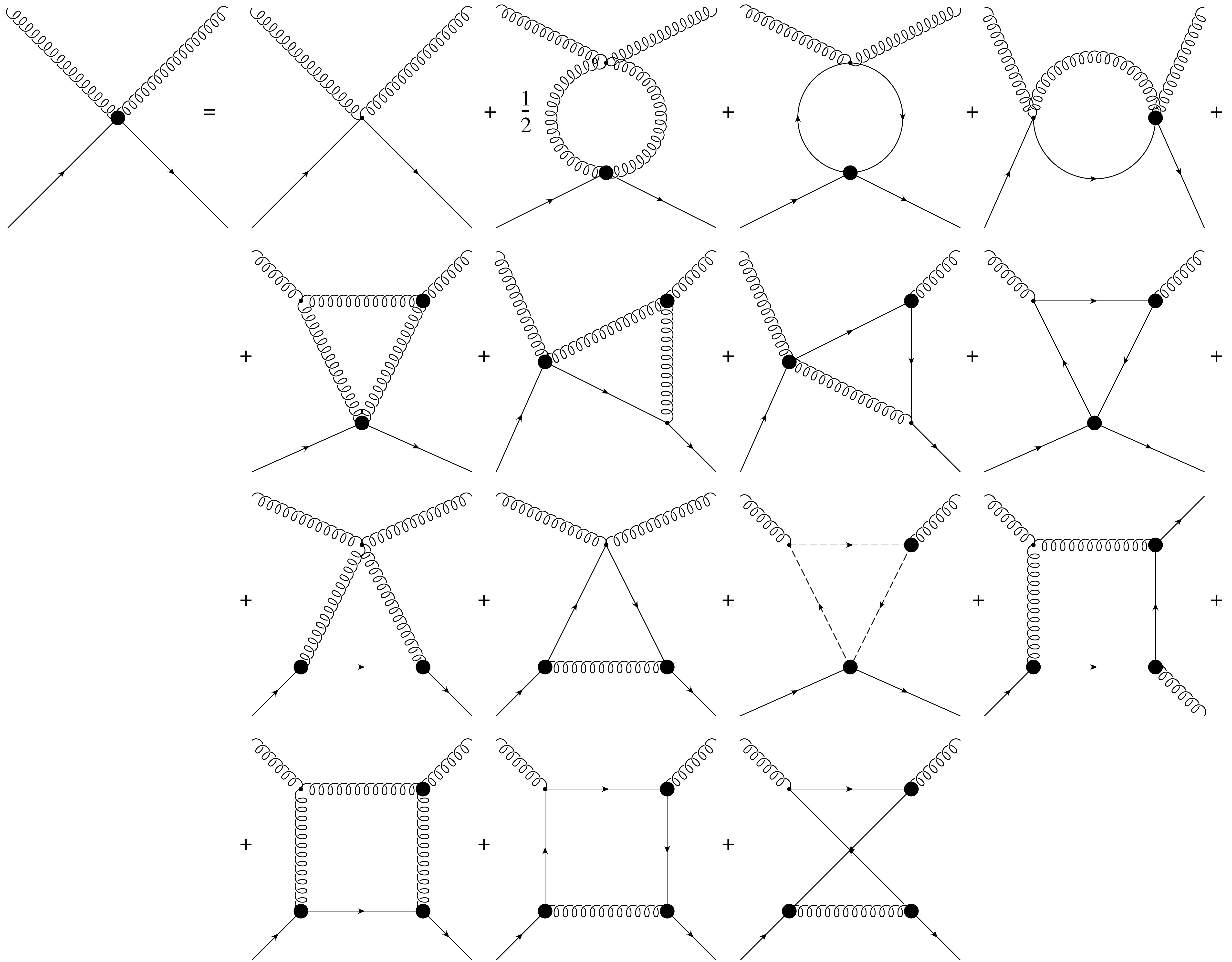}\\
\includegraphics[width=1.00\textwidth]{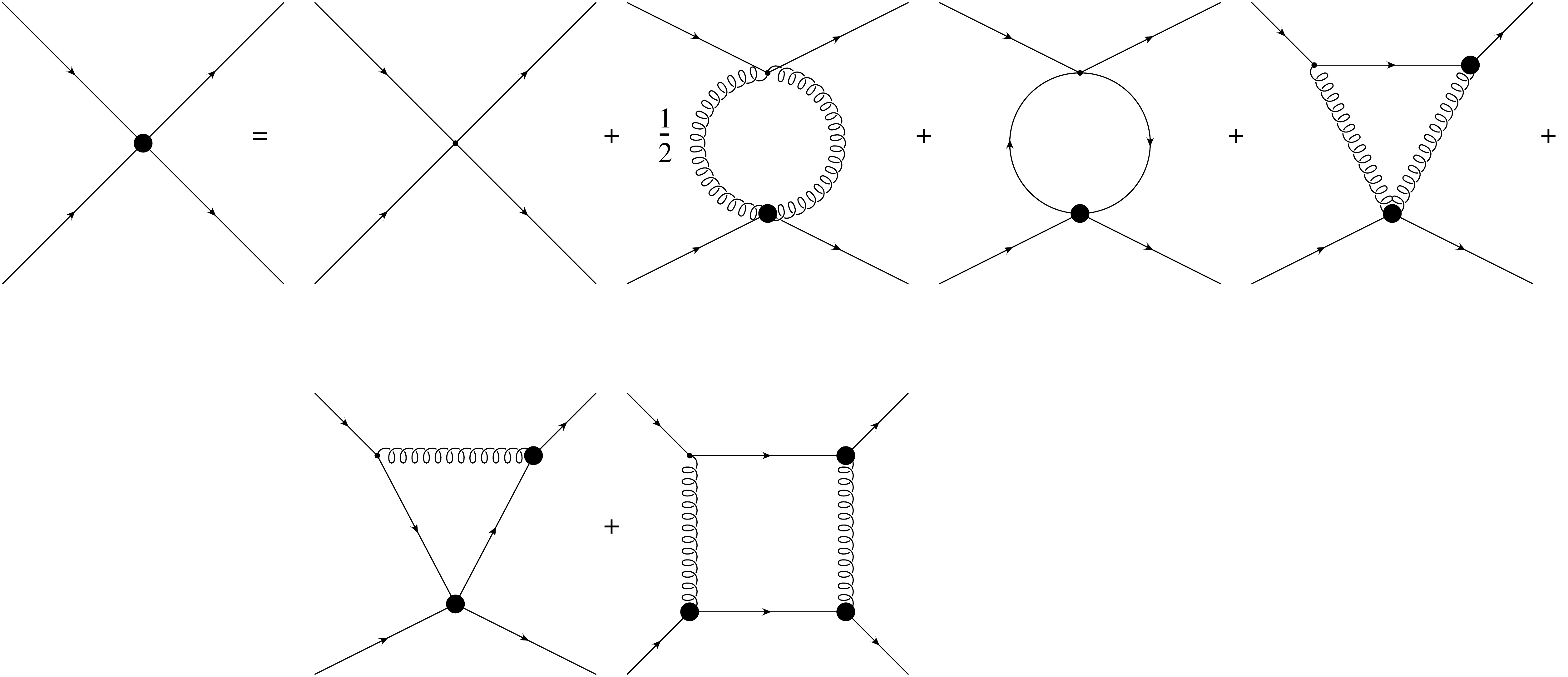}
\caption{\label{dse-4vm}The truncated DSEs for the four-point vertices in the matter sector. Curly lines denote gluons, dashed lines denote ghosts, and full lines denotes the scalars of either type.}
\end{figure}

Here, the main aim is a statement about the color structure which is generated in a first iteration step. For that purpose, all appearing vertices in the loops will be taken to be tree-level. This reduces the amount of algebra at the current state to a manageable amount. The resulting set of DSEs are shown in figure \ref{dse-prop}-\ref{dse-4vm}. These are only the equations for the primitively divergent vertices, which play possibly a special role \cite{Huber:2009wh}. Nonetheless, we also investigated the equations for the non-primitive four-point vertices, the two-ghost-two-scalar vertex, the four-ghost vertex, and the two-gluon-two-ghost vertex. As discussed below, they turn out to play no particular role, and therefore we just list our results for them in appendix \ref{app:res2}, for the sake of brevity, and concentrate in the main text on the primitive divergent vertices.

Still, it is not possible to give a closed expression for all appearing color structures. This is mostly due to the appearance of contractions of the generators $t$ in various representations, where the index contraction is such that the result cannot be expressed in terms of fewer generators. Thus, the resulting color structures are not evident. To make the emerging structure more transparent, the results are therefore deconstructed into an appropriate base structure for the color algebra. This base system is constructed using the primitiveness assumption for simple Lie algebras \cite{Cvitanovic:2008}. The resulting expressions are then projected in this base-system, yielding a coefficient vector, which can then be analyzed. Of course, this coordinate choice is by no means unique. This construction yields the following base systems:

For two-point functions, the only remaining base-vector is $\delta_{ab}$ and $\delta_{ij}$ for adjoint and fundamental quantities, respectively\footnote{Note that in case of a (genuine or analytically connected \cite{Caudy:2007sf}) Higgs phase we use implicitly a non-aligned Landau gauge, which permits to use the same color tensors as in a confinement phase \cite{Maas:unpublished}.}.

For three-point vertices, the case of three adjoint indices encompass $f_{abc}$ and $d_{abc}$. In the case of two fundamental and one adjoint index there remains only the structure $t^a_{ij}$.

The situation becomes then quickly more complicated for four indices. In general, the construction principle was to select an orthogonal system of base tensors. One of them was always the tree-level expression, as the contraction with the tree-level expression is often the only contribution playing a role if the number of external indices is less than the number of internal indices, due to the explicit appearance of bare vertices in the DSEs. The remaining base vectors are then constructed such that they have as good as possible totally symmetric or anti-symmetric structures in the remaining indices. This leads for four adjoint indices to the six base vectors
\bea
&&f^{eac}f^{ebd}+f^{ead}f^{ebc}\nn\\
&&f^{eac}f^{ebd}-f^{ead}f^{ebc}\nn\\
&&\delta_{ab}\delta_{cd}+\delta_{ad}\delta_{bc}+\delta_{ac}\delta_{bd}\nn\\
&&f^{eac}f^{ebd}-f^{ead}f^{ebc}+C_{A}(N_{A}-1)\delta_{ac}\delta_{bd}+C_{A}(1-N_{A})\delta_{ad}\delta_{bc}\nn\\
&&f^{eac}f^{ebd}+f^{ead}f^{ebc}-\frac{2}{3}C_{A}(N_{A}+1)\delta_{ab}\delta_{cd}+\frac{1}{3}C_{A}(N_{A}+1)\delta_{ad}\delta_{bc}+\frac{1}{3}C_{A}(N_{A}+1)\delta_{ac}\delta_{bd}\nonumber\\\nn\\
&&\delta_{ab}\delta_{cd}+\delta_{ad}\delta_{bc}+\delta_{ac}\delta_{bd}+\frac{24(N_{A}+2)}{T_{R}(C_{A}-6C_{F})} d^{abcd}\nn
\eea
\no which includes both the tree-level adjoint four-scalar and the tree-level two-gluon-two-scalar vertices. For convenience, however, these will be reordered by exchange such that always the tree-level vertices are component zero. The four-gluon vertex has at tree-level no simple color structure, but consists of different color tensors for different Lorentz structures, though all with the same symmetry properties. Therefore, in this case the same basis will be used as for the two-adjoint-scalar-two-gluon vertex. Note that these tensors are orthogonal, but not orthonormal, such that no excessive divisions appear at this stage. The same applies to the four-ghost vertex, the two-gluon-two-ghost vertex, and the two-ghost-two-scalar vertex for adjoint matter.

For the two-fundamental-scalar-two-gluon case the base tensors are
\bea
&&t^{a}_{ik}t^{b}_{kj}+t^{b}_{ik}t^{a}_{kj}\nn\\
&&t^{a}_{ik}t^{b}_{kj}-t^{b}_{ik}t^{a}_{kj}\nn\\
&&t^{a}_{ik}t^{b}_{kj}+t^{b}_{ik}t^{a}_{kj}-T_{R}(1+N_{A})\delta_{ab}\delta_{cd}\nn\\
&&C_{A}t^{a}_{ik}t^{b}_{kj}-C_{A}t^{b}_{ik}t^{a}_{kj}-2iT_{R}(N_{A}-1)f^{abc}t^{c}_{ij}\nn\\
&&t^{a}_{ik}t^{b}_{kj}+t^{b}_{ik}t^{a}_{kj}-\frac{T_{R}N_{A}\left(-C_{A}+4C_{F}+T_{R}(1+N_{A})(-4+N_{F}+N_{A}N_{F})\right)}{2N_{F}d33(1+N_{A})}d^{abc}t^{c}_{ij}\nonumber\\
&&\quad-\frac{C_{A}-4C_{F}-4T_{R}(N_{A}+1)}{N_{F}(N_{A}+1)}\delta_{ab}\delta_{cd}\nn
\eea
\no and thus one tensor less. This base system is also used for the fundamental representation of the two-ghost-two-scalar vertex.

Finally, in case of the four-fundamental-scalar vertex the base system is
\bea
&&\delta_{ji}\delta_{lk}+\delta_{jk}\delta_{li}\nn\\
&&\delta_{ji}\delta_{lk}-\delta_{jk}\delta_{li}\nn\\
&&-N_{A}T_{R}\delta_{ji}\delta_{lk}-N_{A}T_{R}\delta_{jk}\delta_{li}+N_{F}(N_{F}+1)t^{a}_{ji}t^{a}_{lk}+N_{F}(N_{F}+1)t^{b}_{jk}t^{b}_{li}\nn\\
&&-N_{A}T_{R}\delta_{ji}\delta_{lk}+N_{A}T_{R}\delta_{jk}\delta_{li}-N_{F}(N_{F}-1)t^{a}_{ji}t^{a}_{lk}+N_{F}(N_{F}-1)t^{b}_{jk}t^{b}_{li}\nn,
\eea
\no leaving only four tensors to ponder. 

There is a second problem. Some of the diagrams in the vertex equations include vertices which do not have a tree-level analogue, and would therefore vanish. There are two different possibilities to treat these. Either they are set to zero as well, or some assumption is made on their color structure. Here, the latter option is employed, and the assumption for the color structure is made using the lowest-order tree-level diagram. Note that beyond the current one-step iteration, these would be obtained from the corresponding vertex equations.

Finally, even using such a base system, the resulting coefficient vector is not always a pure number, but can contain different Lorentz structures. This particularly occurs when a four-gluon vertex appears, but when going beyond the tree-level approximation for the appearing vertices this will happen generically. Since the aim here are patterns, these Lorentz structures will not be detailed, but just kept in the form of not explicitly given functions on which the color coefficients depend.

There are in total four such vertices. Two have under these assumption the same color structure, the four-ghost vertex and the two-ghost-two-gluon vertex, which will be approximated by
\be
f^{eab}f^{ecd}T_1+f^{ead}f^{ebc}T_2\nn,
\ee
\no where $T_1$ and $T_2$ denote some Lorentz tensors.

The second set of vertices are the two-ghost-two-scalar vertices, which will be approximated by a color structure of $f^{eab}t^e_{ij}$ and $f^{eab}f^{ecd}$ for fundamental and adjoint scalars, respectively, multiplying some tensor structure.

This completely fixes the set of truncations made in the present study. Though these approximations are partly rather drastic, the problem still remains of significant complexity, and was partly solved using the algebraic program FORM for color factors \cite{vanRitbergen:1998pn}. However, the results still admitted the observation of certain patterns.

\section{Results at the algebra level}\label{sec:res}

For the remaining presentation of results, it is useful to summarize first a number of observations. These are made with the primary aim of this study in mind: To identify those diagrams which may give rise to the qualitative non-perturbative differences observed for the different gauge groups.

First, there are certain coefficients which turn out to be always zero, irrespective of the group. This can be usually attributed to the symmetry properties of the graph and color structure under scrutiny. The second type of pattern is that a coefficient vanish only for $G_2$, which is not due to a particular expression in terms of the gauge group constants, but occurs due to several possible combinations. Then there are cases where only the contribution for SU(2) is vanishing, which again occurs for several different combinations of the constants. Finally, some situations appear, where only the contribution from SU(3) is non-vanishing, what is usually due to the involvement of the symmetric tensor $d^{abc}$.

In the following, only those results will be presented which are of particular relevance from this point of view. The remaining results can be found in the appendix \ref{app:res}.

The simplest entities are the propagator equations. However, there is no group invariant with two indices, except for the Kronecker-$\delta$, which should be the only one appearing. This is confirmed here. Explicit expressions can be found in the appendix \ref{app:res:prop}. This changes for the three-point vertices.

\subsection{3-point function DSEs in the quenched approximation}

To compare the quenched and unquenched results, it is useful to first study the quenched approximation separately. There are then three tree-level vertices containing three external legs: The ghost-gluon, the scalar-gluon and the 3-gluon vertex. The most interesting diagrams based on the power-counting analysis for the ghost-gluon vertex are given by

\vspace{0.2cm}

\no\begin{minipage}{0.3\linewidth}
    \includegraphics[width=0.70\linewidth]{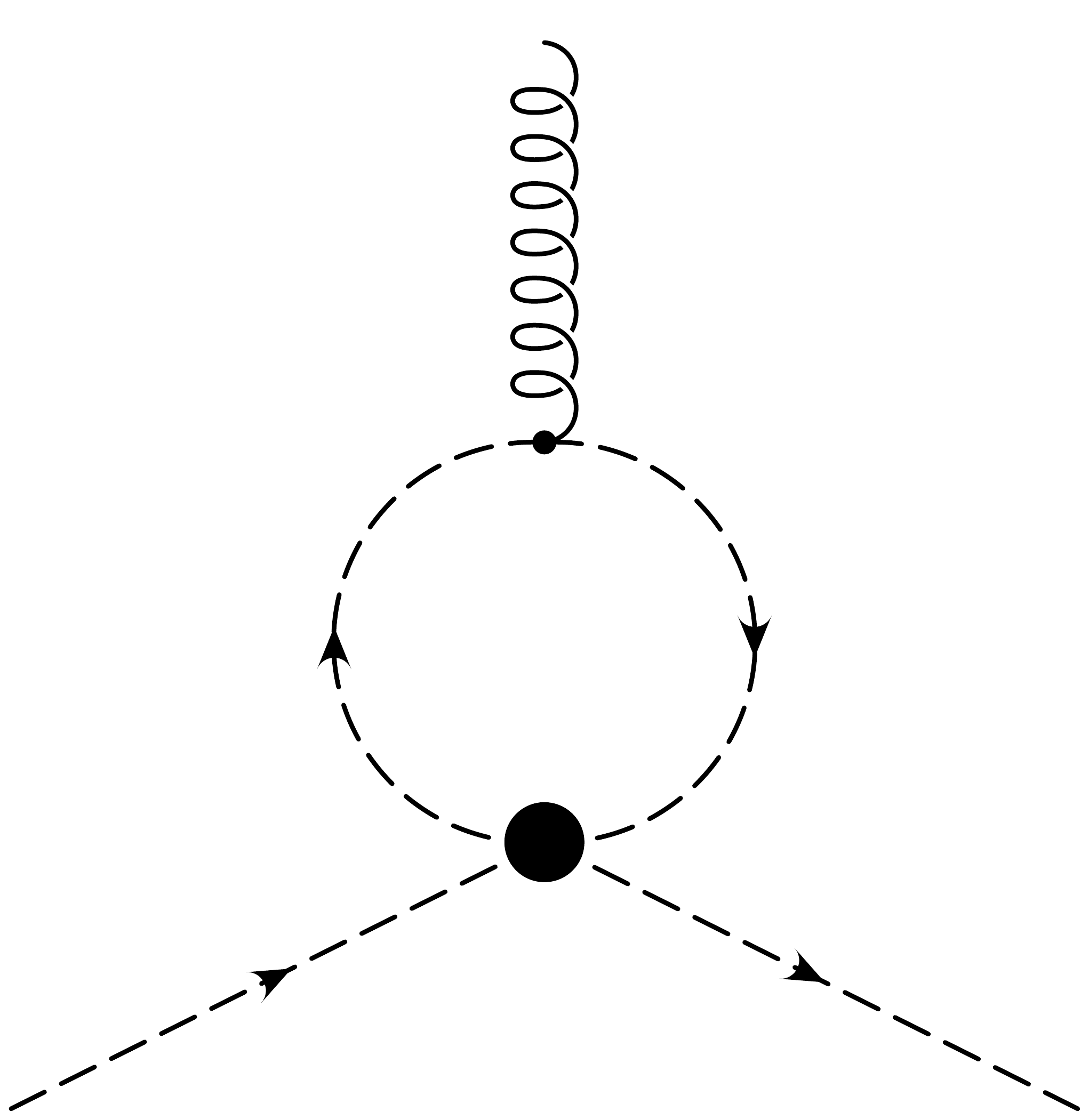}
\end{minipage}
\begin{minipage}{0.6\linewidth}
  \begin{tabular}{| l | l | l | l | l |}
    \hline
	      & group invariant & SU(2) & SU(3) & G$_2$ \\ \hline
    adjoint  & $C_{A}f^{abc}$ & $2f^{abc}$ & $3f^{abc}$ & $2f^{abc}$ \\ 
\hline
  \end{tabular}
\end{minipage}

\no\begin{minipage}{0.3\linewidth}
  \includegraphics[width=0.70\linewidth]{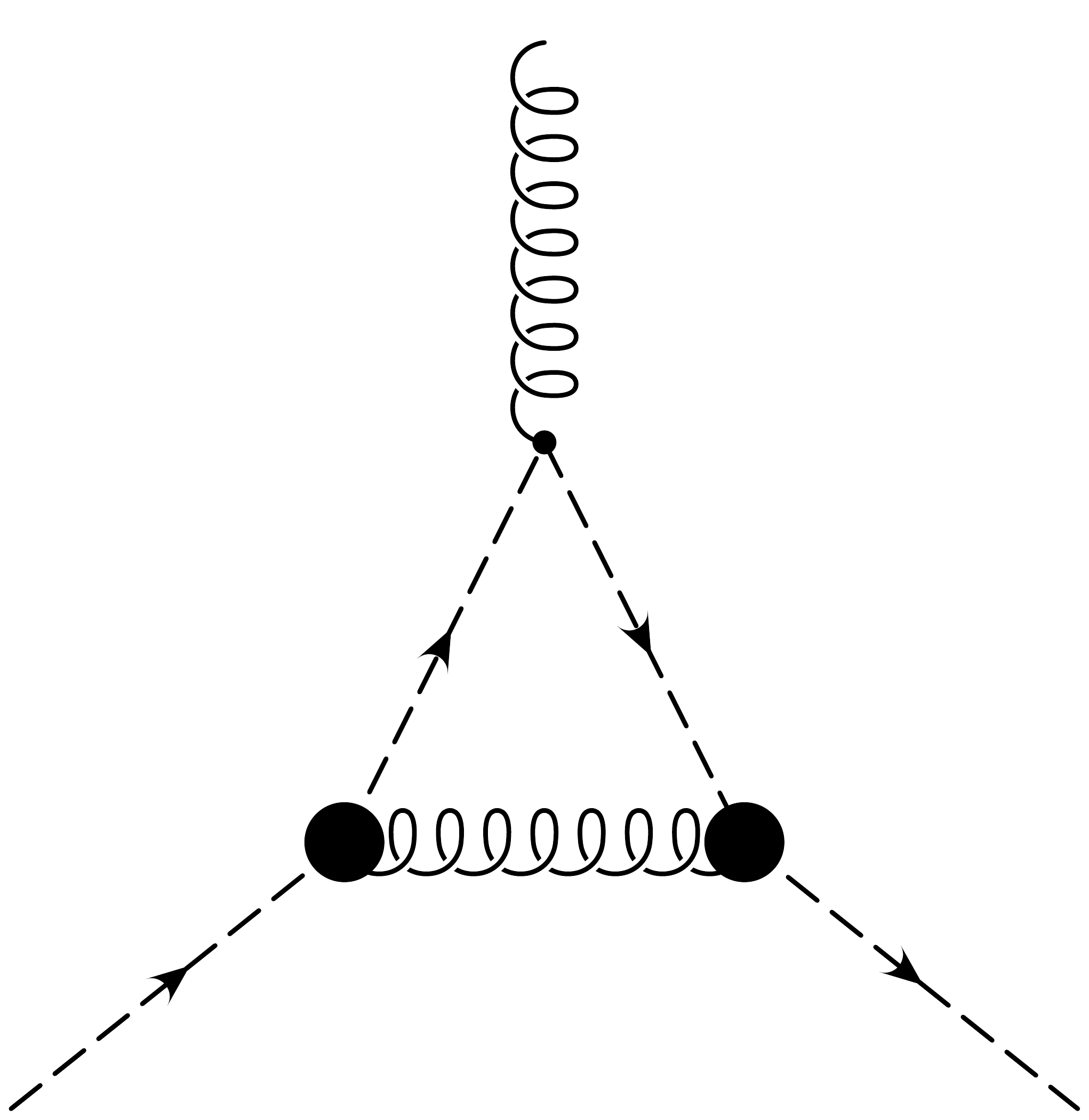}
\end{minipage}
\begin{minipage}{0.6\linewidth}
  \begin{tabular}{| l | l | l | l | l |}
    \hline
	      & group invariant & SU(2) & SU(3) & G$_2$ \\[1mm] \hline
    adjoint  & $-\frac{1}{2}C_{A}f^{abc}$ & $-f^{abc}$ & $-\frac{3}{2}f^{abc}$ & $-f^{abc}$ \\[1mm]
\hline
  \end{tabular}
\end{minipage}

\vspace{0.2cm}

\no which, however, do not show any non-trivial dependence on the group.

For the three-gluon vertex, the results depend already on the approximations for the gluon-ghost scattering kernel. With the employed assumptions, the results read

\vspace{0.2cm}

\no\begin{minipage}{0.3\linewidth}
    \includegraphics[width=0.70\linewidth]{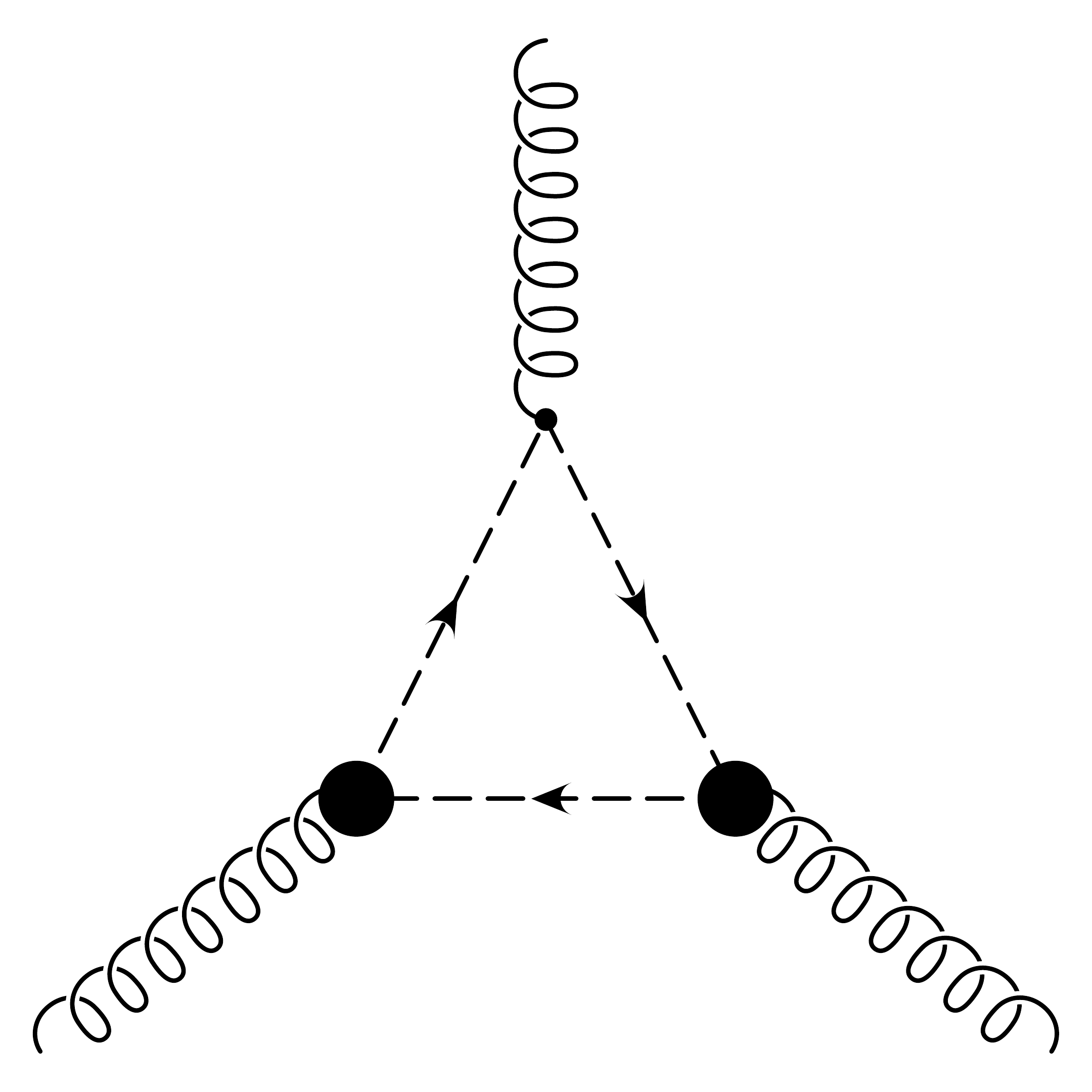}
\end{minipage}
\begin{minipage}{0.6\linewidth}
  \begin{tabular}{| l | l | l | l | l |}
    \hline
	      & group invariant & SU(2) & SU(3) & G$_2$ \\ \hline
	   adjoint   & $\frac{1}{2}C_{A}f^{abc}$ & $f^{abc}$ & $\frac{3}{2}f^{abc}$ & $f^{abc}$ \\[1mm] 
\hline
  \end{tabular}
\end{minipage}

\no\begin{minipage}{0.3\linewidth}
  \includegraphics[width=0.70\linewidth]{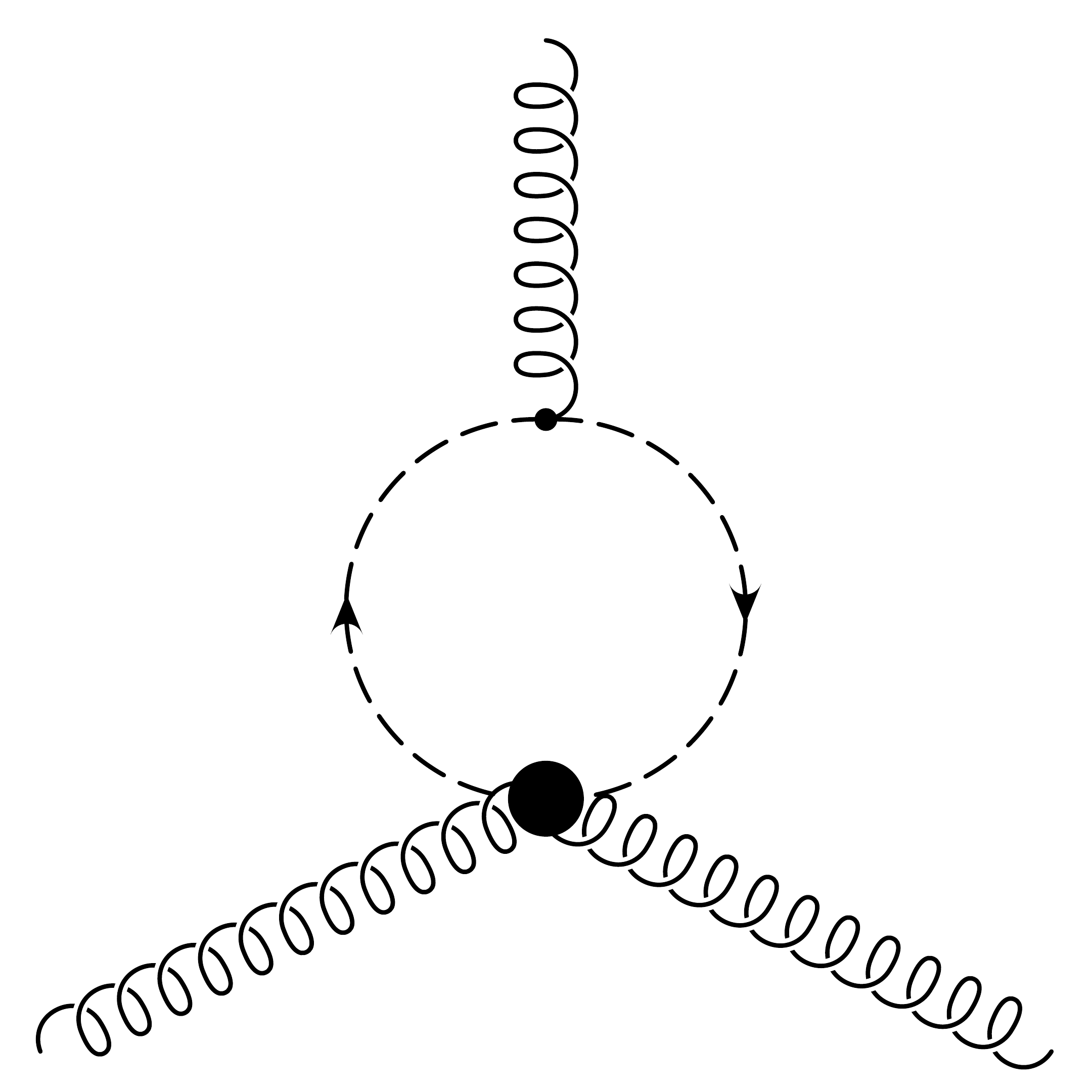}
\end{minipage}
\begin{minipage}{0.6\linewidth}
  \begin{tabular}{| l | l | l | l | l |}
    \hline
	      & group invariant & SU(2) & SU(3) & G$_2$ \\ \hline
	    $T_1$  & $\frac{1}{2}C_{A}f^{abc}$ & $f^{abc}$ & $\frac{3}{2}f^{abc}$ & $f^{abc}$ \\ 
	    $T_2$  & $-C_{A}f^{abc}$ & $-2f^{abc}$ & $-3f^{abc}$ & $-2f^{abc}$ \\ 
\hline
  \end{tabular}
\end{minipage}

\vspace{0.2cm}

\no where $T_i$ denotes the corresponding Lorentz tensor structures. Again, no difference appear. Thus, in agreement with lattice results \cite{Maas:2010qw,Maas:2007af,vonSmekal:1997vx,Maas:2005ym,Cucchieri:2007zm,Oliveira:2008uf}, the pure gluonic sector is at this level of truncation insensitive to the gauge algebra. The results for the quenched two-scalar-gluon vertex are finally given as

\vspace{0.2cm}

\begin{minipage}{0.3\linewidth}
    \includegraphics[width=0.70\linewidth]{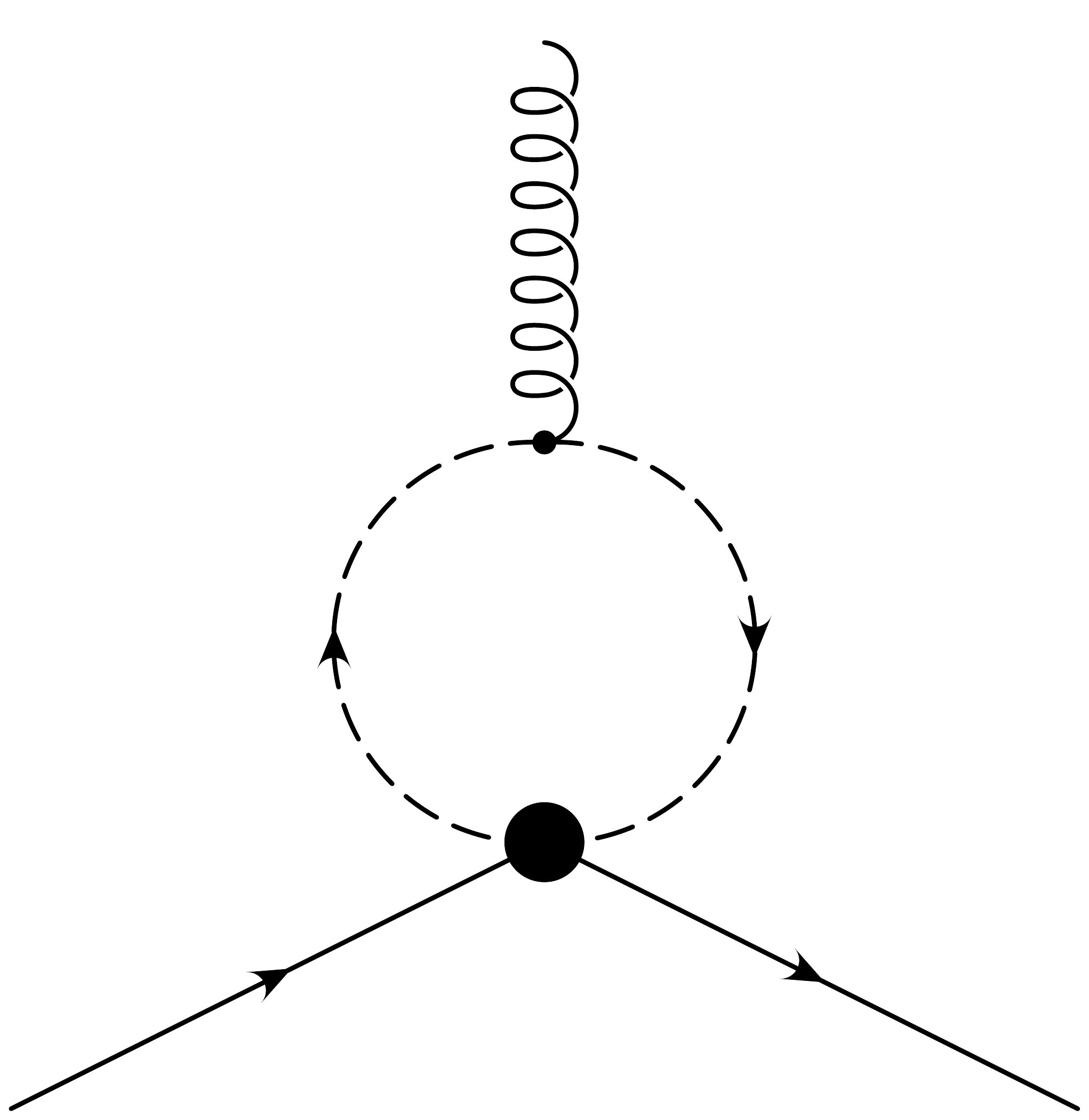}
\end{minipage}
\begin{minipage}{0.6\linewidth}
  \begin{tabular}{| l | l | l | l | l |}
    \hline
	      & group invariant & SU(2) & SU(3) & G$_2$ \\ \hline
    adjoint  & $-C_{A}f^{abc}$ & $-2f^{abc}$ & $-3f^{abc}$ & $-2f^{abc}$ \\[1mm]
    fund.  & $\frac{1}{4}C_{A}t^{a}_{ij}$ & $\frac{1}{2}t^{a}_{ij}$ & $\frac{3}{4}t^{a}_{ij}$ 
	    & $\frac{1}{2}t^{a}_{ij}$ \\[1mm]
\hline
  \end{tabular}
\end{minipage}

\vspace{0.2cm}

\no Here, too, no structural difference between the fundamental and the adjoint case can be observed. This is however important, as based on the scenario discussed in \cite{Fister:2010yw}, a difference between the adjoint and the fundamental case would have been desirable. This does not seem to be the case. In particular, no further Lorentz structures can emerge beyond those at tree level for this vertex. Such an insensitivity is compatible with the results from lattice calculations \cite{Maas:2011yx}, which show so far no pronounced difference between fundamental and adjoint scalars for the two-scalar-gluon vertex.

However, in the contributions

\vspace{0.2cm}

\no\begin{minipage}{0.18\linewidth}
     \includegraphics[width=1.0\linewidth]{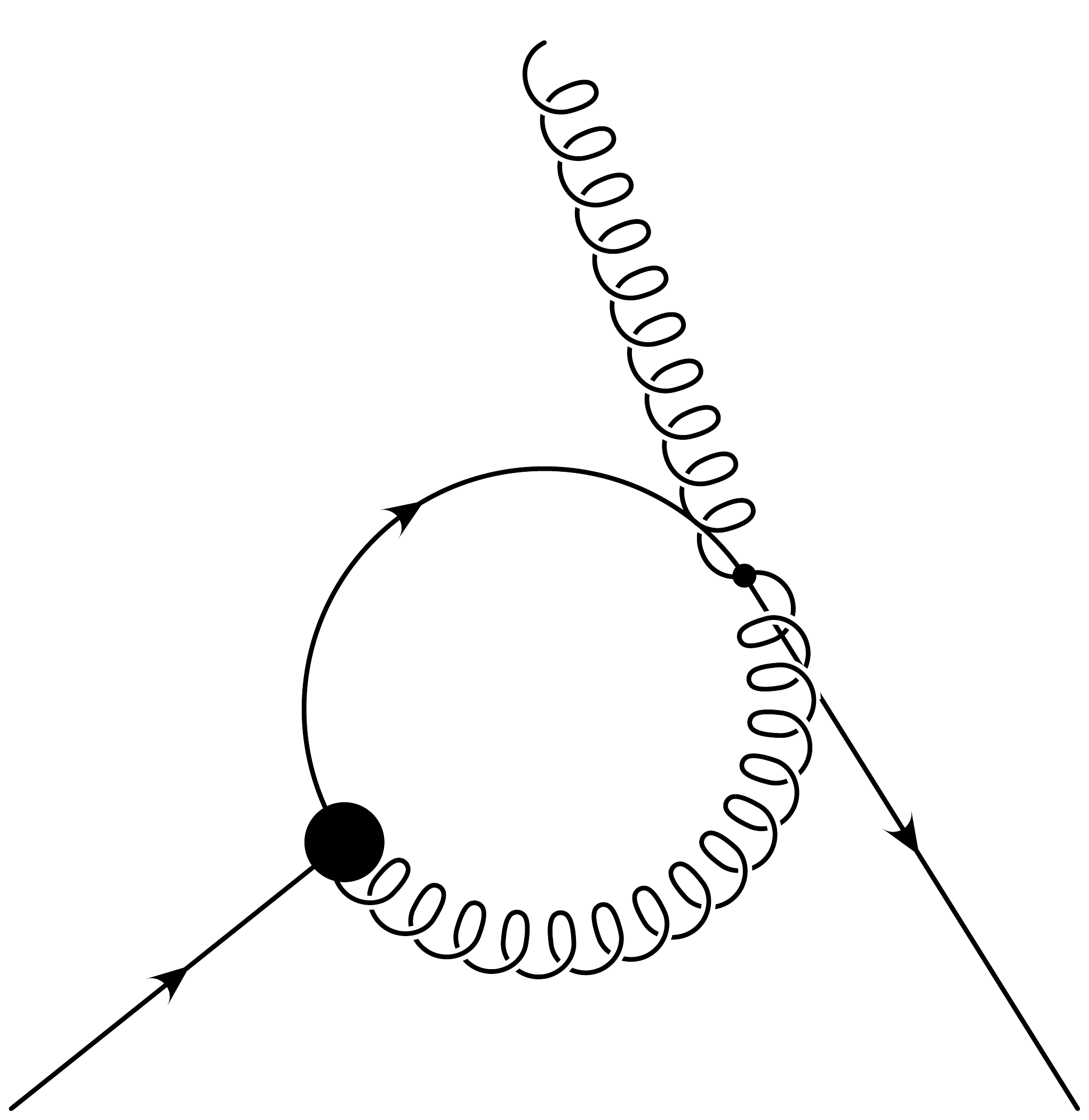}
\end{minipage}
\begin{minipage}{0.72\linewidth}
     \begin{center}
  \begin{tabular}{| l | l | l | l | l |}
    \hline
	      & group invariant & SU(2) & SU(3) & G$_2$ \\ \hline
    adjoint  & $\frac{3}{2}C_{A}f^{abc}$ & $3f^{abc}$ & $\frac{9}{2}f^{abc}$ & $3f^{abc}$ \\[1mm] 
    fund.  & $\frac{1}{2}\left(C_{F}-\frac{1}{2}C_{A}\right)t^a_{ij}$ & $-\frac{1}{8}t^a_{ij}$ & $-\frac{1}{12}t^a_{ij}$ & 0 \\[1mm]
\hline
  \end{tabular}
\end{center}
\end{minipage}

\no\begin{minipage}{0.18\linewidth}
     \includegraphics[width=1.0\linewidth]{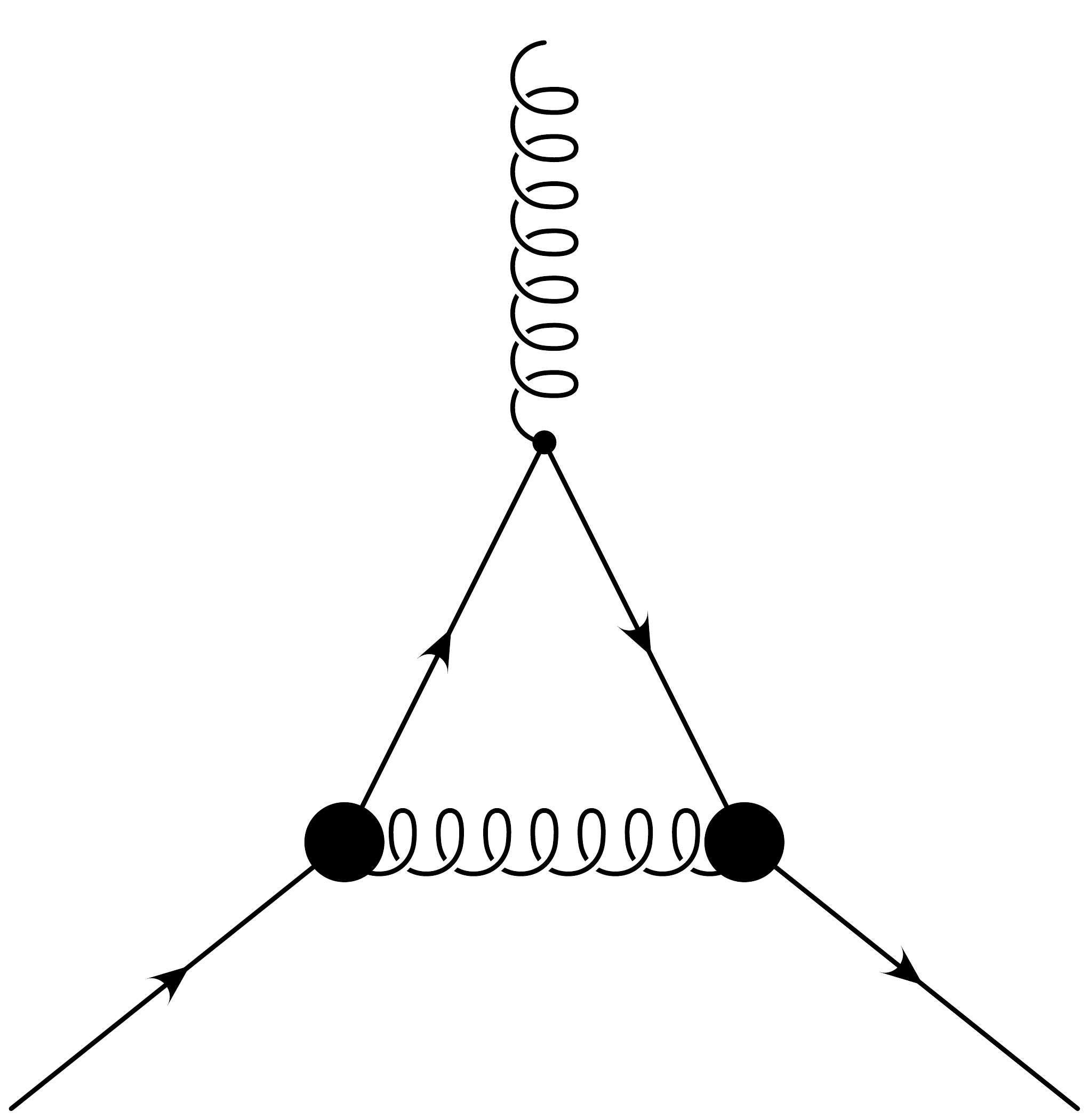}
\end{minipage}
\begin{minipage}{0.72\linewidth}
     \begin{center}
  \begin{tabular}{| l | l | l | l | l |}
    \hline
	      & group invariant & SU(2) & SU(3) & G$_2$ \\ \hline
    adjoint  & $\frac{1}{2}C_{A}f^{abc}$ & $f^{abc}$ & $\frac{3}{2}f^{abc}$ & $f^{abc}$ \\ 
    fund.  & $(C_{F}-\frac{1}{2}C_{A})t^a_{ij}$ & $-\frac{1}{4}t^a_{ij}$ & $-\frac{1}{6}t^a_{ij}$ & 0 \\[1mm]
\hline
  \end{tabular}
\end{center}
\end{minipage}

\vspace{0.2cm}

\no there is a structural difference in the fundamental case between G$_2$ and the SU($N$) case, as would be desired.

\subsection{4-point function DSEs in the quenched approximation}

The only tree-level 4-point function in the pure gauge sector is the 4-gluon vertex. Here, the coefficients in order of the base tensors are given, and for the different Lorentz tensors appearing in the third diagram separately. In some of the cases the effect of the symmetry and anti-symmetry of the base tensors is nicely reflected by some of the coefficients, which are zero at this level of truncation for all gauge groups.

\vspace{0.4cm}

\no\begin{minipage}{0.2\linewidth}
  \includegraphics[width=1.00\linewidth]{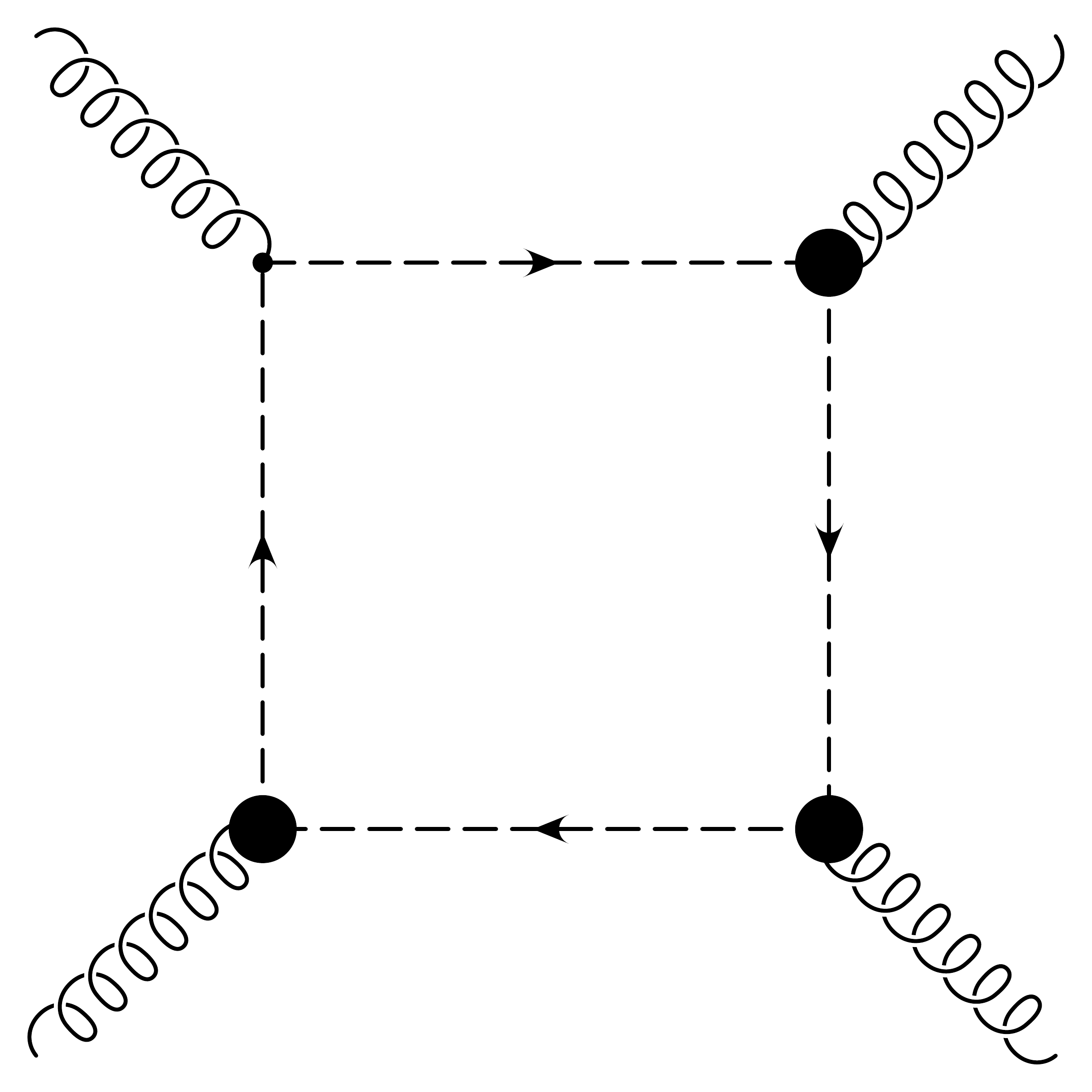}
\end{minipage}
\begin{minipage}{0.7\linewidth}
  \begin{tabular}{| l | l | l | l | l |}
    \hline
	adj.      & group invariant & SU(2) & SU(3) & G$_2$ \\ \hline
    $1^{st}$  & $\frac{1}{4}N_{A}C_{A}^{3}$ & $6$ & $54$ & $28$ \\[1mm]
    $2^{nd}$  & $-\frac{1}{4}N_{A}C_{A}^{3}$  & $-6$ & $-54$ & $-28$ \\[1mm]  
    $3^{rd}$  & $\frac{5}{2}N_{A}C_{A}^{2}$  & $30$ & $180$ & $140$ \\[1mm]  
    $4^{th}$  & $\frac{1}{4}N_{A}C_{A}^{3}(2N_{A}-3)$   & $18$ & $702$ & $700$ \\[1mm] 
    $5^{th}$  & $-\frac{1}{12}N_{A}C_{A}^{3}(-2N_{A}+1)$ & $-10$ & $-270$ & $-252$ \\[1mm]
    $6^{th}$  & $\frac{1}{2T_{R}(C_{A}-6C_{F})}\left(5N_{A}T_{R}\times \right. $ & $\frac{3825}{128}$ 
	      & $\frac{1435}{8}$ & $\frac{1113}{8}$ \\
	      & $\left. \times C_{A}^{2}(C_{A}-6C_{F})+\right.$ &  & & \\
	      & $\left.+12d_{44}(2+N_{A}) \right)$ & & & \\
    \hline
  \end{tabular}
\end{minipage}

\vspace{0.2cm}

\no\begin{minipage}{0.18\linewidth}
     \includegraphics[width=1.0\linewidth]{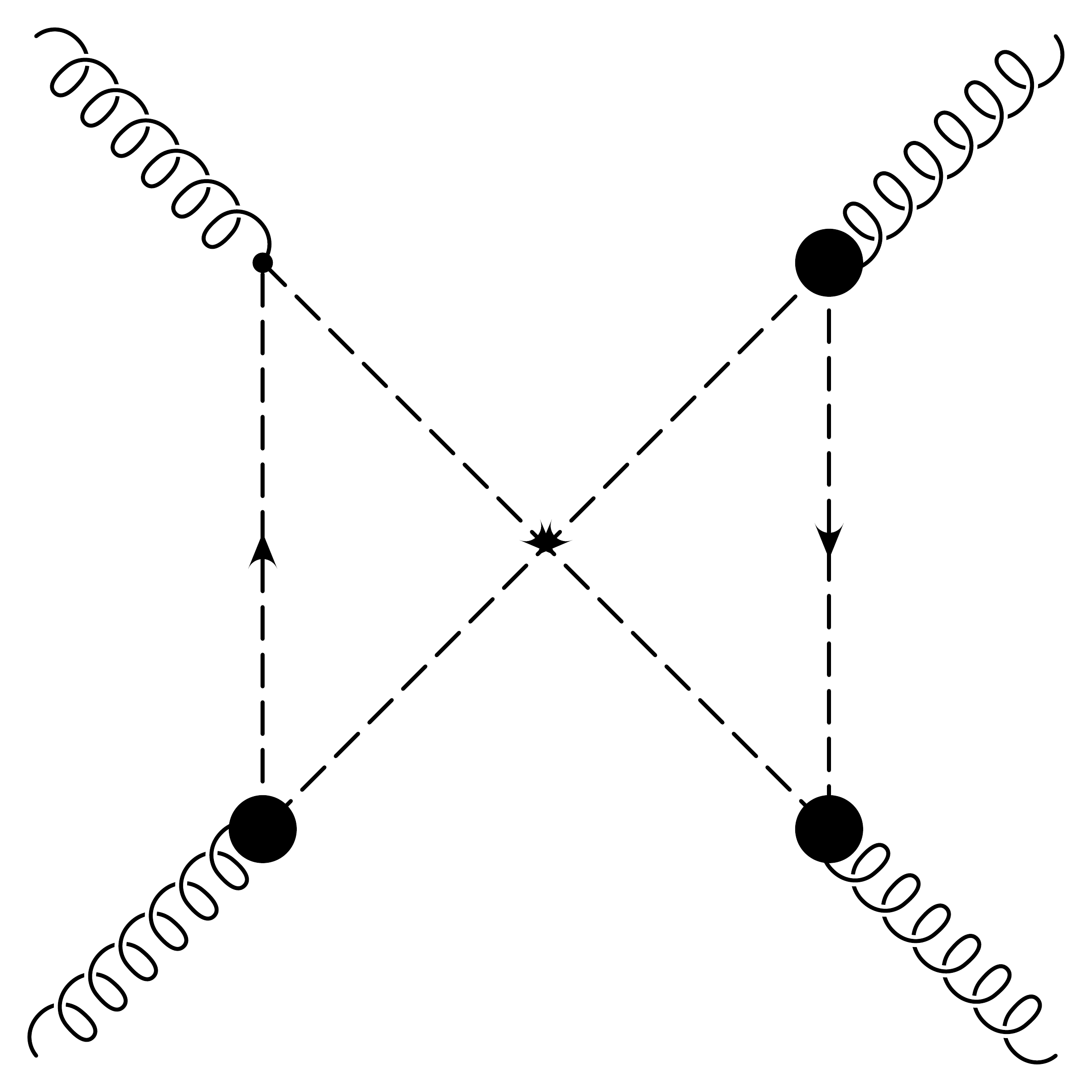}
\end{minipage}
\begin{minipage}{0.72\linewidth}
     \begin{center}
  \begin{tabular}{| l | l | l | l | l |}
    \hline
	adj.      & group invariant & SU(2) & SU(3) & G$_2$ \\ \hline
    $1^{st}$  & $-\frac{1}{2}N_{A}C_{A}^{3}$ & $-12$ & $-108$ & $-56$ \\ 
    $2^{nd}$  & 0  & 0 & 0 & 0 \\  
    $3^{rd}$  & $\frac{5}{2}N_{A}C_{A}^{2}$  & $30$ & $180$ & $140$ \\  
    $4^{th}$  & 0   & 0 & 0 & 0 \\ 
    $5^{th}$  & $\frac{1}{6}N_{A}C_{A}^{3}(2N_{A}-1)$ & $20$ & $540$ & $504$ \\[1mm]
    $6^{th}$  & $\frac{1}{2T_{R}(C_{A}-6C_{F})}\left(5N_{A}T_{R}C_{A}^{2}\times \right. $ & $\frac{3825}{128}$ 
	      & $\frac{1435}{8}$ & $\frac{1113}{8}$ \\
	      & $\left. \times(C_{A}-6C_{F})+\right.$ &  & & \\
	      & $\left.+12d_{44}(2+N_{A}) \right)$ & & & \\
    \hline
  \end{tabular}
\end{center}
\end{minipage}

\vspace{0.2cm}

\no\begin{minipage}{0.2\linewidth}
    \includegraphics[width=1.00\linewidth]{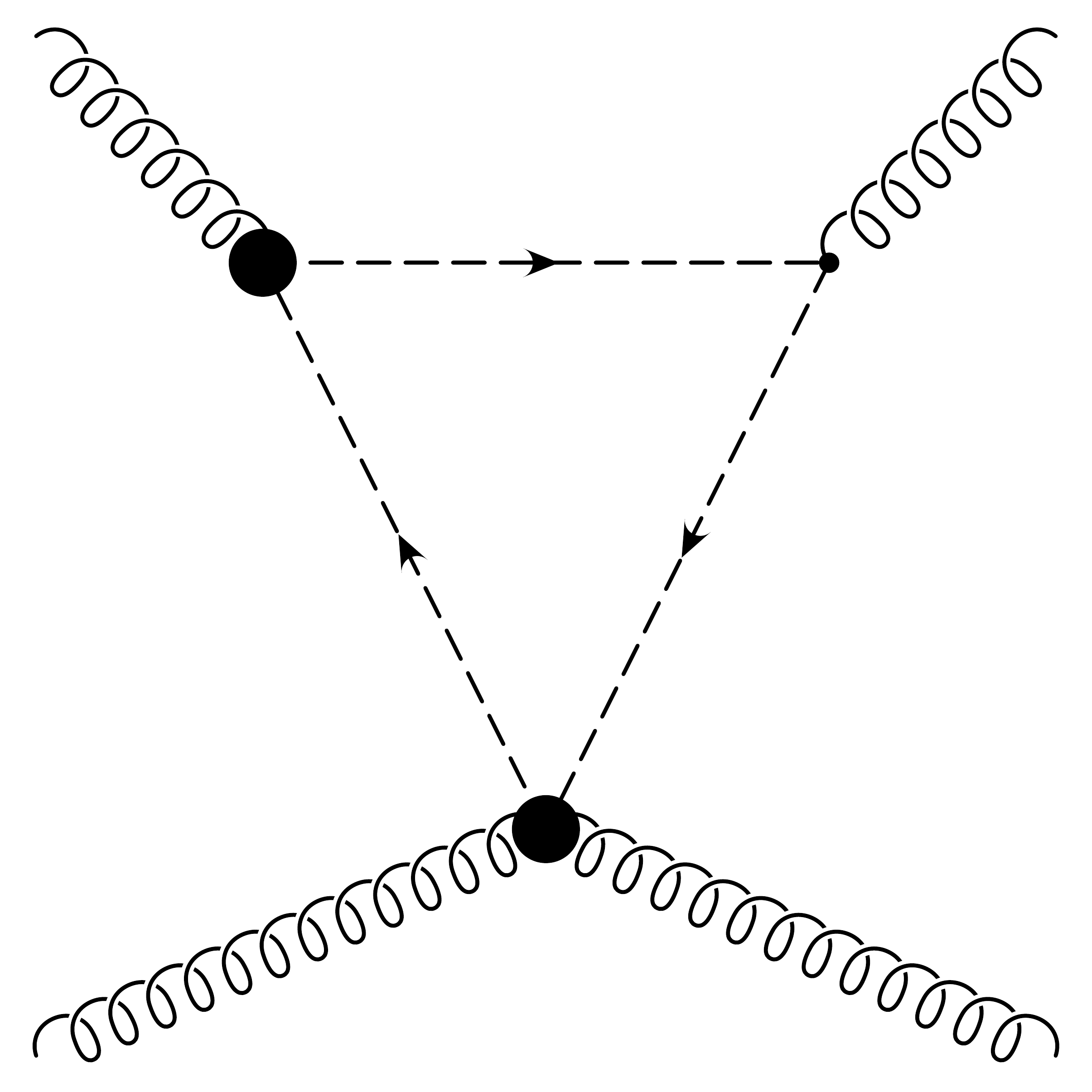}
\end{minipage}
\begin{minipage}{0.7\linewidth}
  \begin{tabular}{| l | l | l | l | l |}
    \hline
	$T_{1}$      & group invariant & SU(2) & SU(3) & G$_2$ \\ \hline
    $1^{st}$  & $\frac{1}{4}N_{A}C_{A}^{3}$ & 6 & 54 & 28 \\[1mm] 
    $2^{nd}$  & $-\frac{1}{4}N_{A}C_{A}^{3}$  & -6 & -54 & -28 \\[1mm]  
    $3^{rd}$  & $\frac{5}{2}N_{A}C_{A}^{2}$  & 30 & 180 & 140 \\[1mm]  
    $4^{th}$  & $\frac{1}{4}N_{A}C_{A}^{3}(2N_{A}-3)$   & 18 & 702 & 700 \\[1mm] 
    $5^{th}$  & $\frac{1}{12}N_{A}C_{A}^{3}(-2N_{A}+1)$ & -10 & -270 & -252 \\[1mm]
    $6^{th}$  & $\frac{1}{2T_{R}(C_{A}-6C_{F})}\left(5N_{A}T_{R}C_{A}^{2}\times \right. $ & $\frac{3825}{128}$ 
	      & $\frac{1435}{8}$ & $\frac{1113}{8}$ \\
	      & $\left. \times(C_{A}-6C_{F})+\right.$ &  & & \\
	      & $\left.+12d_{44}(2+N_{A}) \right)$ & & & \\
    \hline
  \end{tabular}

\vspace{0.2cm}

\begin{tabular}{| l | l | l | l | l |}
    \hline
	$T_{2}$      & group invariant & SU(2) & SU(3) & G$_2$ \\ \hline
    $1^{st}$  & $-\frac{3}{4}N_{A}C_{A}^{3}$ & -18 & -162 & -84 \\[1mm] 
    $2^{nd}$  & $\frac{1}{4}N_{A}C_{A}^{3}$  & 6 & 54 & 28 \\  
    $3^{rd}$  & 0  & 0 & 0 & 0 \\  
    $4^{th}$  & $\frac{1}{4}N_{A}C_{A}^{3}(-2N_{A}+3)$   & -18 & -702 & -700 \\[1mm] 
    $5^{th}$  & $\frac{1}{4}N_{A}C_{A}^{3}(2N_{A}-1)$ & 30 & 810 & 756 \\
    $6^{th}$  & 0 & 0 & 0 & 0 \\
    \hline
  \end{tabular}
\end{minipage}

\vspace{0.2cm}

\no For these three diagrams no qualitative difference is obtained for the different gauge groups.

The remaining two four-point vertices involve directly the scalar fields. At the current level, according to power-counting, the diagrams concerning the ghost diagrams play the relevant role for the two-scalar-two-gluon vertex. This is only one,

\vspace{0.2cm}

\no\begin{minipage}{0.2\linewidth}
     \includegraphics[width=1.0\linewidth]{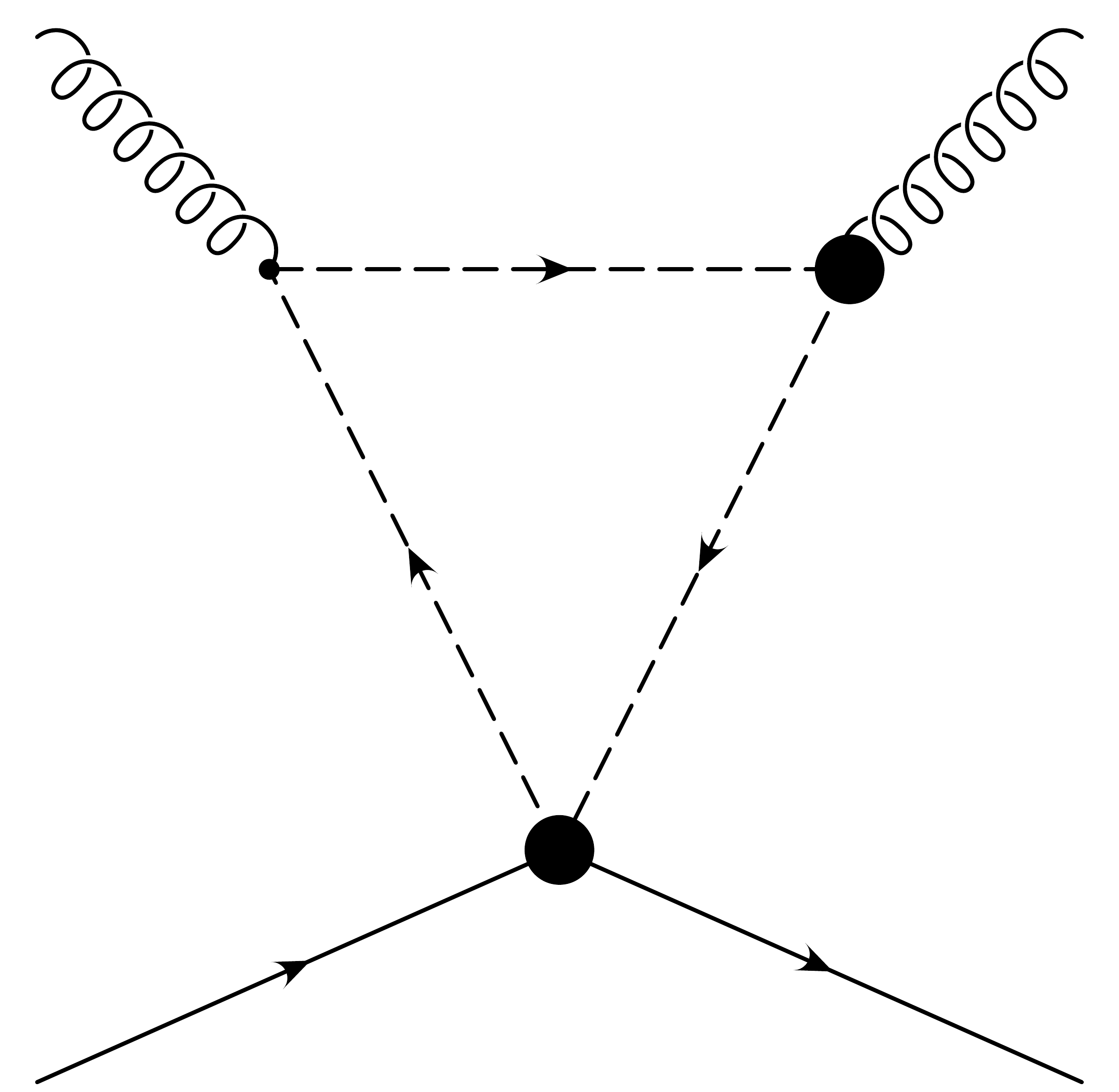}
\end{minipage}
\begin{minipage}{0.7\linewidth}
     \begin{center}
  \begin{tabular}{| l | l | l | l | l |}
    \hline
	adj.      & group invariant  & SU(2) & SU(3) & G$_2$ \\ \hline
    $1^{st}$  & 0 & 0 & 0 & 0\\ 
    $2^{nd}$  & $-\frac{1}{2}N_{A}C_{A}^{3}$  & -12 & -108 & -56 \\ 
    $3^{rd}$  & 0   & 0 & 0 & 0 \\ 
    $4^{th}$  & $\frac{1}{2}N_{A}C_{A}^{3}(2N_{A}-3)$   & 36 & 1404 & 1400 \\ 
    $5^{th}$  & 0   & 0 & 0 & 0 \\
    $6^{th}$  & 0 & 0 & 0 & 0 \\
    \hline
  \end{tabular}
\end{center}
\no\begin{center}
  \begin{tabular}{| l | l | l | l | l |}
    \hline
	fund.		& group invariant  & SU(2) & SU(3) & G$_2$ \\ \hline
    $1^{st}$  & $-\frac{1}{8}N_{A}T_{R}\left(3C_{A}^{2}-10C_{A}C_{F}+\right.$ & $-\frac{9}{32}$ & $-\frac{11}{18}$ & 0\\  
	      & $\left.+8C_{F}^{2}\right)$ & & & \\
    $2^{nd}$  & 0  & 0 & 0 & 0 \\ 
    $3^{rd}$  & $\frac{1}{8}N_{A}T_{R}\left(-3C_{A}^{2}+10C_{A}C_{F}+\right.$ & $\frac{27}{32}$ & $\frac{205}{18}$ 
	      & $\frac{105}{4}$ \\ 
	      & $\left.+4C_{F}(-2C_{F}+T_{R}+T_{R}N_{A})\right)$ & & & \\
    $4^{th}$  & 0   & 0 & 0 & 0 \\ 
    $5^{th}$  & $\frac{1}{8}N_{A}T_{R}\left(-3C_{A}^{2}+10C_{A}C_{F}+\right.$ & $\frac{27}{32}$ & $\frac{205}{18}$ 
	      & $\frac{105}{4}$ \\
	      & $\left.+4C_{F}(-2C_{F}+T_{R}+T_{R}N_{A})\right)$ & & & \\
    \hline
  \end{tabular}
\end{center}
\end{minipage}

\vspace{0.2cm}

\no Here, some more differences emerge. Due to the symmetries, in the adjoint case the coefficient for the tree-level tensor is zero for all groups. This symmetry imprints also to other tensors. On the other hand, the fundamental color structure is proportional to a finite value, though it is still zero for the group G$_2$. This is exactly the type of pattern desired for the quenched case. However, this pattern does not replicate beyond the tree-level tensor. Furthermore, an important ingredient here is the assumed form for the ghost-scalar scattering kernel, which has not been included self-consistently.

After encountering this difference, it is worthwhile to investigate something beyond the current truncation. In the power-counting scheme, the second leading contribution in the 2-scalar-2-gluon DSE is a higher-order diagram containing a 5-point function. Expanding this 5-point function in a skeleton expansion \cite{Alkofer:2004it,Fister:2010yw} is giving the same result according to power counting. Investigating these two diagrams yields

\vspace{0.2cm}

\no\begin{minipage}{0.2\linewidth}
     \includegraphics[width=1.0\linewidth]{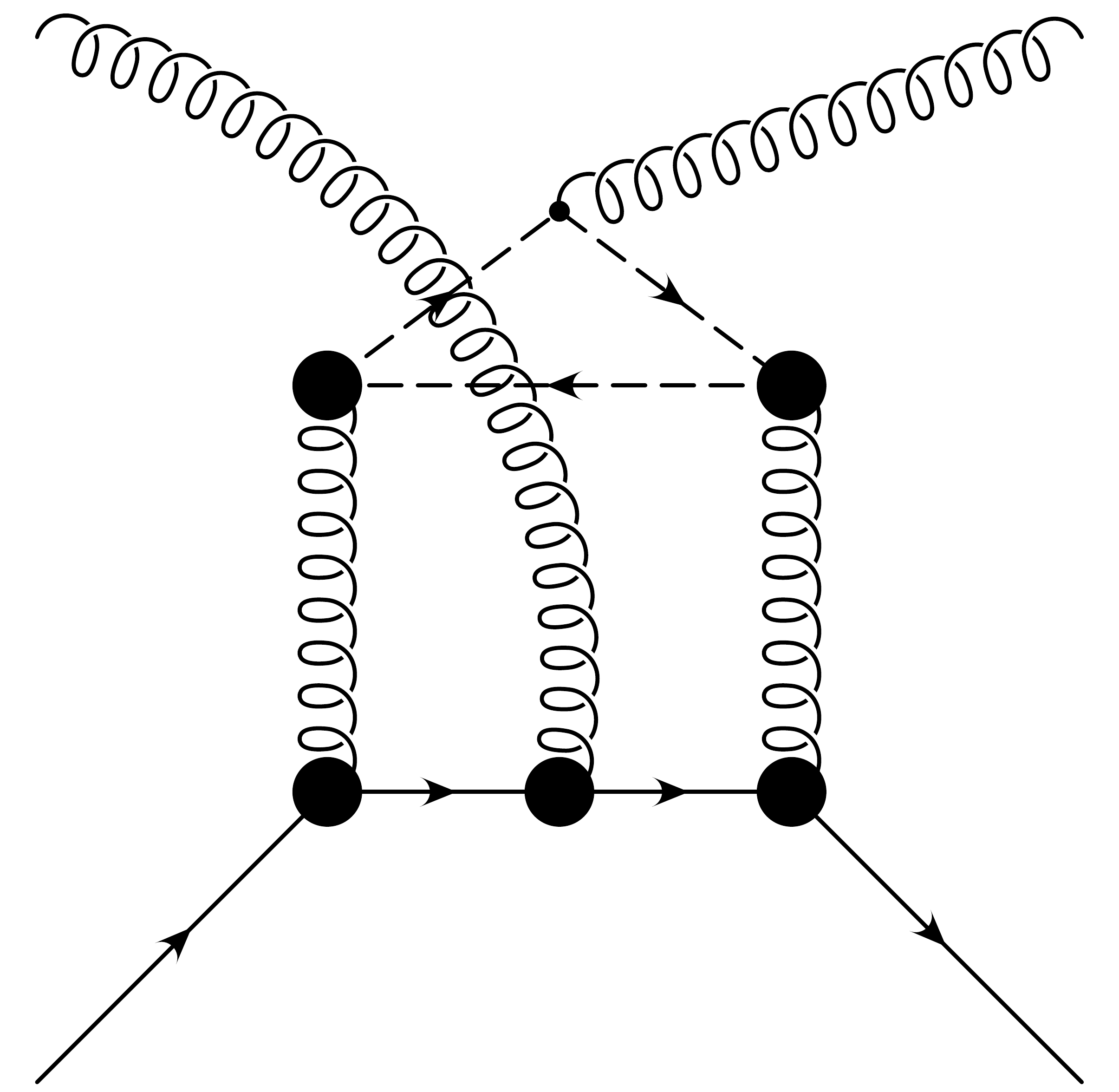}
\end{minipage}
\begin{minipage}{0.7\linewidth}
     \begin{center}
  \begin{tabular}{| l | l | l | l | l |}
    \hline
	adj.		& group invariant  & SU(2) & SU(3) & G$_2$ \\ \hline
    $1^{st}$  & $-\frac{1}{8}N_{A}C_{A}^{4}$ & -6 & -81 & -28 \\[1mm]  
    $2^{nd}$  & $-\frac{1}{8}N_{A}C_{A}^{4}$  & -6 & -81 & -28 \\[1mm]  
    $3^{rd}$  & $-\frac{5}{4} N_{A}C_{A}^{3}$   & -30 & -270 & -140 \\[1mm] 
    $4^{th}$  & $\frac{1}{8}N_{A}C_{A}^{4}(2N_{A}-3)$   & 18 & 1053 & 700 \\[1mm] 
    $5^{th}$  & $\frac{1}{24}N_{A}C_{A}^{4}(-1+2N_{A})$   & 10 & 405 & 252 \\[1mm]
    $6^{th}$  & $-\frac{C_{A}}{4T_{R}(C_{A}-6C_{F})}\left(5N_{A}T_{R}C_{A}^{2}\times \right.$ & $-\frac{3825}{128}$ 
	      & $-\frac{4305}{16}$ & $-\frac{1113}{8}$ \\
	      & $\left.\times (C_{A}-6C_{F})+ \right.$ & & & \\
	      & $\left. +12d_{44}(2+N_{A}) \right)$ & & & \\
    \hline
  \end{tabular}
\end{center}
\begin{center}
  \begin{tabular}{| l | l | l | l | l |}
    \hline
	fund.		& group invariant  & SU(2) & SU(3) & G$_2$ \\ \hline
    $1^{st}$  & $-\frac{1}{4}iN_{A}T_{R}C_{A}^{2}(C_{A}-2C_{F})$ & $-\frac{3i}{4}$ & $-3i$ & 0 \\  
    $2^{nd}$  & 0  & 0 & 0 & 0 \\ 
    $3^{rd}$  & $\frac{1}{4}iN_{A}T_{R}C_{A}^{2}\left(-C_{A}+2C_{F}+\right.$ & $\frac{9i}{4}$ 
	      & $-\frac{75i}{2}$ & $-\frac{105i}{2}$ \\ 
	      & $\left. +T_{R}+T_{R}N_{A}\right)$ & & & \\
    $4^{th}$  & 0  & 0 & 0 & 0 \\ 
    $5^{th}$  & $\frac{iN_{A}T_{R}C_{A}^{2}}{4N_{F}(1+N_{A})}\left(-C_{A}\left(-1+N_{F}+ \right.\right.$  & $\frac{9i}{16}$ 
	      & $\frac{20i}{9}$  & $\frac{28i}{15}$ \\
	      & $\left.\left.+N_{A}N_{F}\right)+2(2T_{R}(1+N_{A}))+\right.$ &  & & \\
	      & $\left.+C_{F}(-2+N_{F}+N_{A}N_{F})\right)$ & & & \\
    \hline
  \end{tabular}
\end{center}
\end{minipage}

\vspace{0.2cm}

\no\begin{minipage}{0.2\linewidth}
    \includegraphics[width=1.0\linewidth]{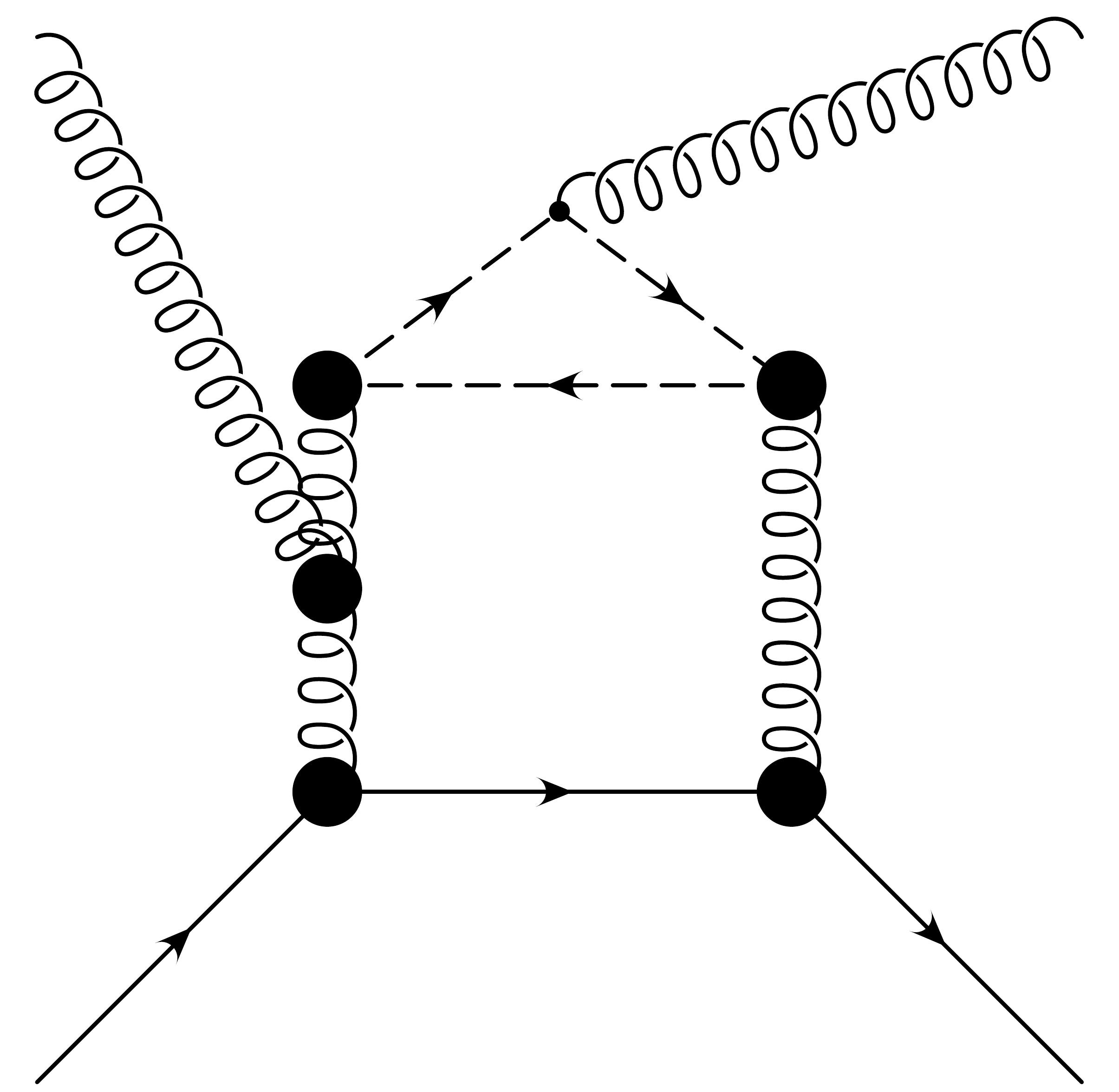}
\end{minipage}
\begin{minipage}{0.7\linewidth}
     \begin{center}
  \begin{tabular}{| l | l | l | l | l |}
    \hline
	adj.		& group invariant  & SU(2) & SU(3) & G$_2$ \\ \hline
    $1^{st}$  & $-\frac{1}{8}N_{A}C_{A}^{4}$ & -6 & -81 & -28 \\[1mm]  
    $2^{nd}$  & $\frac{1}{8}N_{A}C_{A}^{4}$  & 6 & 81 & 28 \\[1mm]  
    $3^{rd}$  & $-\frac{5}{4} N_{A}C_{A}^{3}$   & -30 & -270 & -140 \\[1mm] 
    $4^{th}$  & $\frac{1}{8}N_{A}C_{A}^{4}(-2N_{A}+3)$   & -18 & -1053 & -700 \\[1mm] 
    $5^{th}$  & $\frac{1}{24}N_{A}C_{A}^{4}(-1+2N_{A})$   & 10 & 405 & 252 \\[1mm]
    $6^{th}$  & $-\frac{C_{A}}{4T_{R}(C_{A}-6C_{F})}\left(5N_{A}T_{R}C_{A}^{2}\times \right.$ & $-\frac{3825}{128}$ 
	      & $-\frac{4305}{16}$ & $-\frac{1113}{8}$ \\
	      & $\left.\times (C_{A}-6C_{F})+ \right.$ & & & \\
	      & $\left. +12d_{44}(2+N_{A}) \right)$ & & & \\
    \hline
  \end{tabular}
\end{center}
\begin{center}
  \begin{tabular}{| l | l | l | l | l |}
    \hline
	fund.		& group invariant  & SU(2) & SU(3) & G$_2$ \\ \hline
    $1^{st}$  & $\frac{1}{8}N_{A}T_{R}C_{A}^{3}$ & $\frac{3}{2}$ & $\frac{27}{2}$ & 7 \\[1mm]  
    $2^{nd}$  & $\frac{1}{8}N_{A}T_{R}C_{A}^{3}$  & $\frac{3}{2}$ & $\frac{27}{2}$ & 7 \\[1mm]  
    $3^{rd}$  & $\frac{1}{8}N_{A}T_{R}C_{A}^{2}(C_{A}-4T_{R}(1+N_{A}))$ & $-\frac{9}{2}$ 
	      & $-\frac{135}{2}$ & -98 \\[1mm] 
    $4^{th}$  & $-\frac{1}{8}N_{A}T_{R}C_{A}^{3}\left(C_{A}-2T_{R}\times \right.$  & 0 & 54 & 77 \\ 
	      & $\left.\times (-1+N_{A})\right)$ & & & \\[1mm]
    $5^{th}$  & $\frac{N_{A}T_{R}C_{A}^{2}}{8N_{F}(1+N_{A})}\left(8C_{F}+C_{A}\times \right.$  & $-\frac{45}{16}$ 
	      & $-\frac{290}{9}$ & $-\frac{1021}{30}$ \\
	      & $\left.\times (-2+N_{F}+N_{F}N_{A})+\right.$ & & & \\
	      & $\left.-2T_{R}\left(N_{A}^{2}N_{F}+4+N_{F}+\right.\right.$ &  & & \\
	      & $\left.\left.+2N_{A}(1+N_{F})\right)\right)$ & & & \\
    \hline
  \end{tabular}
\end{center}
\end{minipage}

\vspace{0.2cm}

\no In the first of these diagrams also the same structure is observed as in the previous case, matching the pattern of Wilson loops. This is not the case for the second diagram. Thus indeed such a pattern may be influencing the contributions, in particular given that the tree-level tensors are usually held to be the most important ones. Still, the level of approximations made here should be kept in mind.

The final vertex to be investigated is the 4-scalar vertex. According to power-counting, here a difference between a massive and massless scalar is expected, though lattice results indicate that at least for the quenched case scalars massless at tree-level will acquire a spontaneously generated screening mass \cite{Maas:2011yx}.

In the massless case, the leading diagram is the bare 4-scalar vertex, having no qualitative difference for different gauge groups, see appendix \ref{app:tl}.

However, if the scalar is massive, the leading contribution contains a 5-point function. Using again a skeleton expansion, this yields

\vspace{0.2cm}

\no\begin{minipage}{0.2\linewidth}
     \includegraphics[width=1.0\linewidth]{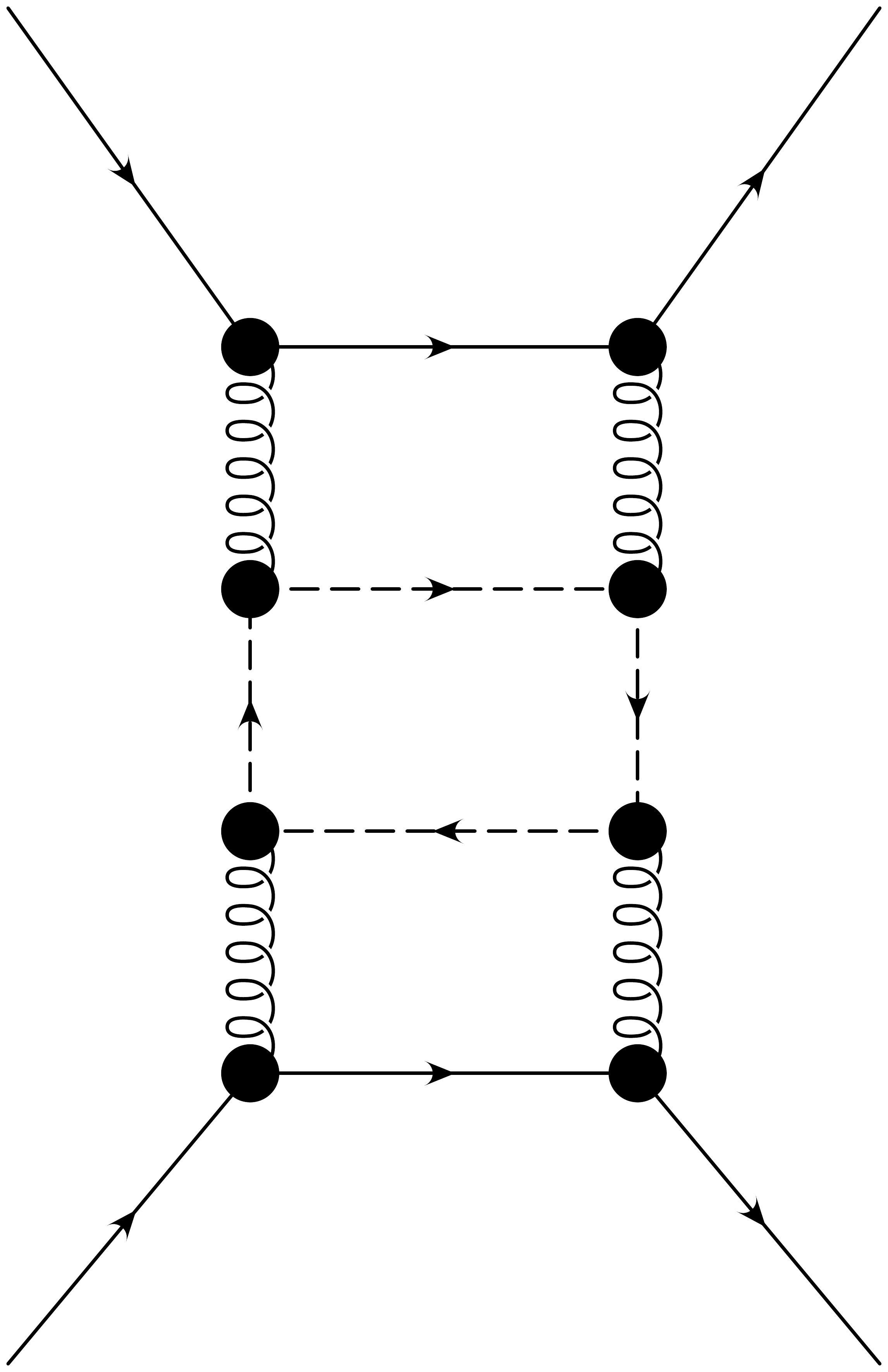}
\end{minipage}
\begin{minipage}{0.7\linewidth}
     \begin{center}
  \begin{tabular}{| l | l | l | l | l |}
    \hline
	adj.		& group invariant  & SU(2) & SU(3) & G$_2$ \\ \hline
    $1^{st}$  & $\frac{1}{24}\left(48d_{44}+25N_{A}C_{A}^{4}\right)$ & $\frac{25605}{512}$ & $\frac{64805}{96}$ 
	      & $\frac{44807}{192}$ \\[1mm] 
    $2^{nd}$  & $\frac{1}{144}C_{A}\left(-96d_{44}+13N_{A}C_{A}^{4}\right)$  & $\frac{2217}{256}$ & $\frac{16893}{96}$ 
	      & $\frac{11641}{288}$ \\[1mm] 
    $3^{rd}$  & $-\frac{1}{144}C_{A}\left(96d_{44}+5N_{A}C_{A}^{4}\right)$  & $-\frac{855}{256}$ & $-\frac{6485}{96}$ 
	      & $-\frac{4487}{288}$ \\[1mm]  
    $4^{th}$  & $\frac{1}{144}C_{A}\left(-48d_{44}(3N_{A}-1)+\right.$  & $\frac{12585}{64}$ & $\frac{2708525}{192}$ 
	      & $\frac{3485153}{576}$ \\
	      & $\left.5N_{A}C_{A}^{4}(30N_{A}-31)\right)$ & & & \\[1mm]
    $5^{th}$  & $\frac{1}{144}C_{A}\left(-48d_{44}(3N_{A}+1)+\right.$ & $\frac{28155}{256}$ & $\frac{920105}{192}$ 
	      & $\frac{1044617}{576}$ \\
	      & $\left.N_{A}C_{A}^{4}(38N_{A}-51)\right)$ & & & \\[1mm]
    $6^{th}$  & $\frac{1}{24T_{R}(C_{A}-6C_{F})}\left(48d_{44}C_{A}T_{R}+ \right.$ & $\frac{204555}{4096}$ & $\frac{48545}{72}$
	      & $\frac{178829}{768}$ \\
	      & $\left.+N_{A}T_{R}C_{A}^{4}(25C_{A}-150C_{F})+\right.$ & & & \\
	      & $\left.+20C_{A}^{2}d_{44}(2+N_{A}) \right.$ & & & \\
	      & $\left.144\left(-2C_{F}T_{R}d_{44}+\right.\right.$ & & & \\
	      & $\left.\left.+\mbox{ d444}(2+N_{A})\right)\right)$ & & & \\
    \hline
  \end{tabular}
\end{center}
     \begin{center}
  \begin{tabular}{| l | l | l | l | l |}
    \hline
	fund.		& group invariant  & SU(2) & SU(3) & G$_2$ \\ \hline
    $1^{st}$  & $\frac{1}{12}\left(12d_{44}+N_{A}T_{R}C_{A}^{2}(C_{A}+12T_{R})\right)$ & $\frac{4101}{1024}$ & 27 
	      & $\frac{56}{3}$\\[1mm] 
    $2^{nd}$  & $\frac{1}{12}\left(12d_{44}+N_{A}T_{R}C_{A}^{2}(C_{A}-12T_{R})\right)$  & $-\frac{2043}{1024}$ & -9 
	      & $-\frac{28}{3}$ \\[1mm] 
    $3^{rd}$  & $\frac{1}{48}\left(-48d_{44}\left(T_{R}N_{A}+(C_{A}-C_{F})\times \right.\right.$ & $-\frac{11565}{1024}$ 
	      & $-\frac{1265}{8}$ & $-\frac{37779}{128}$ \\ 
	      & $\left.\left.\times N_{F}(1+N_{F})\right)+ \right.$ & & & \\
	      & $\left.+C_{A}^{2}\left(-4C_{A}^{2}T_{R}N_{A}N_{F}(1+N_{F})\right)+\right.$ & & & \\
	      & $\left.+12\left(T_{R}^{3}N_{A}^{2}+\mbox{ d33}N_{F}(1+N_{F})+\right)\right.$ & & &  \\ 
	      & $\left.C_{A}T_{R}N_{A}\left(4C_{F}N_{F}(1+N_{F})+\right.\right.$ & & & \\
	      & $\left.\left.+T_{R}(-4N_{A}+3N_{F}(1+N_{F}))\right)\right)$ & & & \\[1mm]
    $4^{th}$  & $\frac{1}{48}\left(-48d_{44}\left(T_{R}N_{A}-(C_{A}-C_{F})\times \right.\right.$ & $\frac{6405}{1024}$ 
	      & $\frac{6125}{32}$ & $\frac{128821}{384}$ \\ 
	      & $\left.\left.\times N_{F}(-1+N_{F})\right)\right)+$ & & & \\
	      & $\left.+C_{A}^{2}\left(-4C_{A}^{2}T_{R}N_{A}N_{F}(-1+N_{F})\right)+\right.$ & & & \\
	      & $\left.+12\left(T_{R}^{3}N_{A}^{2}+\mbox{ d33}N_{F}(-1+N_{F})+\right)\right.$ & & &  \\ 
	      & $\left.C_{A}T_{R}N_{A}\left(-4C_{F}N_{F}(-1+N_{F})+\right.\right.$ & & & \\
	      & $\left.\left.+T_{R}(-4N_{A}+3N_{F}(-1+N_{F}))\right)\right)$ & & & \\
    \hline
  \end{tabular}
\end{center}
\end{minipage}

\vspace{0.4cm}

\no which also does not manifest a difference between different gauge groups. To study whether this may be an artifact of the approximations employed also the DSEs for the non-primitive divergent vertices, which have been assumed so far, have been studied for the quenched case. The results, presented in appendix \ref{app:res2}, do not show the patterns searched for. Therefore, at the current time, it does not appear likely that this may be the resolution of the question where the differences between the gauge groups are located.

To investigate whether unquenching may resolve the problem, the diagrams neglected so far have been investigated.

\newpage

\subsection{Unquenching the three-point vertices}

In the ghost-gluon vertex equation the diagram

\vspace{0.2cm}

\no\begin{minipage}{0.18\linewidth}
     \includegraphics[width=1.0\linewidth]{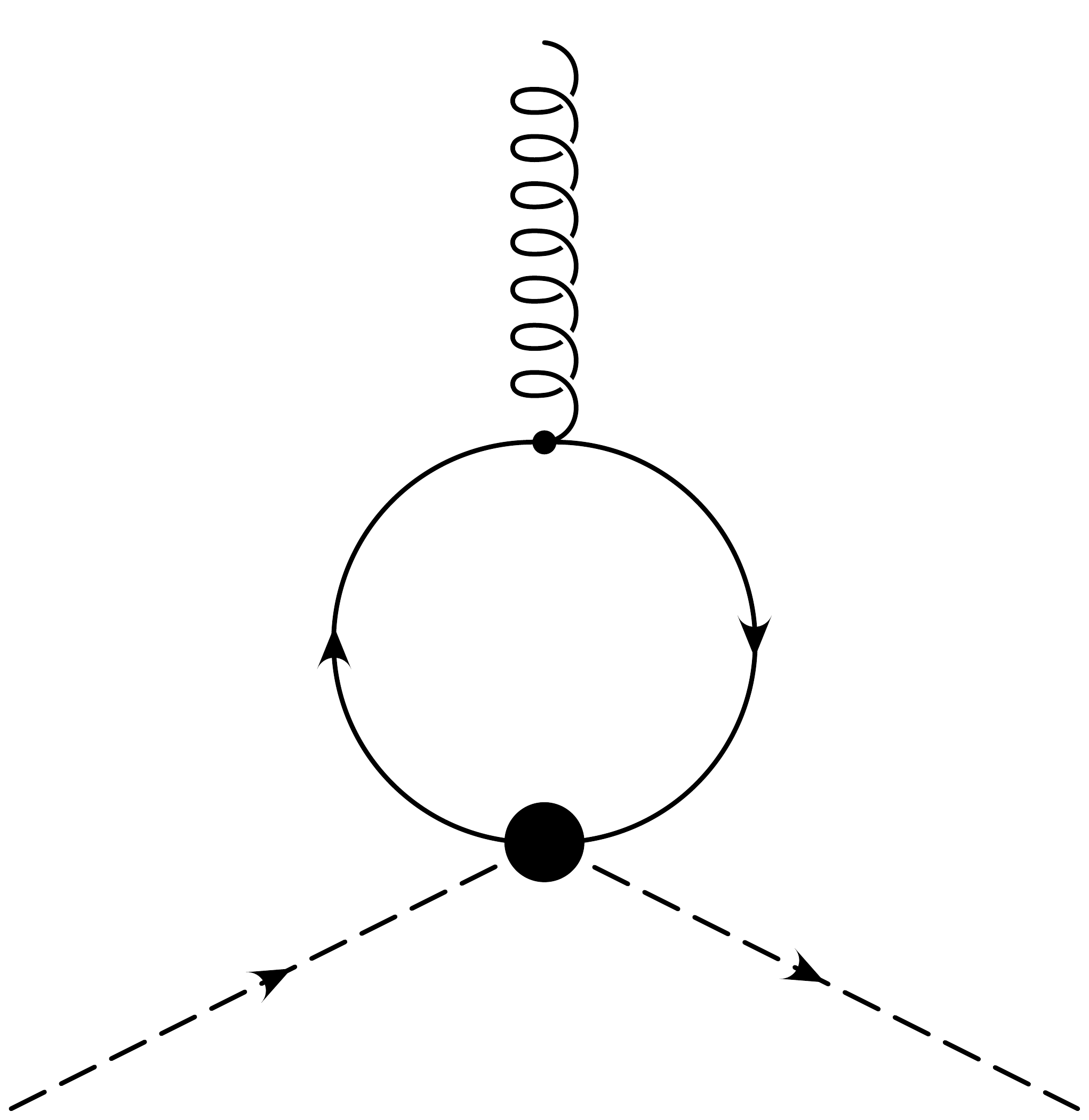}
\end{minipage}
\begin{minipage}{0.72\linewidth}
     \begin{center}
  \begin{tabular}{| l | l | l | l | l |}
    \hline
	      & group invariant & SU(2) & SU(3) & G$_2$ \\ \hline
    adjoint  & $-C_{A}f^{abc}$ & $-2f^{abc}$ & $-3f^{abc}$ & $-2f^{abc}$ \\ 
    fundamental  & $T_{R}f^{abc}$ & $\frac{1}{2}f^{abc}$ & $\frac{1}{2}f^{abc}$ & $\frac{1}{2}f^{abc}$ \\[1mm]
\hline
  \end{tabular}
\end{center}
\end{minipage}

\vspace{0.2cm}

\no appears, exhibiting no profound differences.

For the three-gluon vertex the appearing graph is given by

\vspace{0.2cm}

\no\begin{minipage}{0.18\linewidth}
     \includegraphics[width=1.0\linewidth]{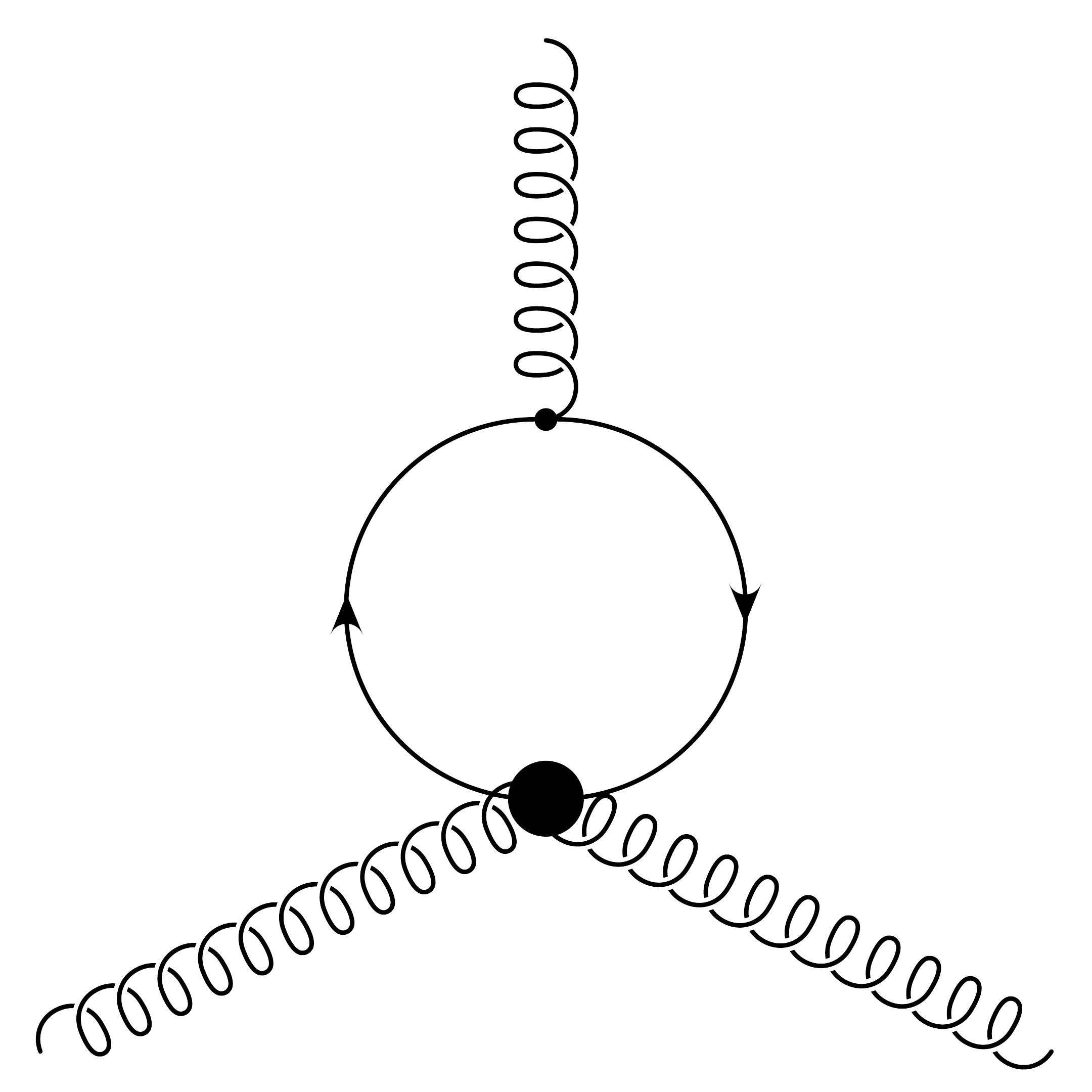}
\end{minipage}
\begin{minipage}{0.72\linewidth}
     \begin{center}
  \begin{tabular}{| l | l | l | l | l |}
    \hline
	      & group invariant & SU(2) & SU(3) & G$_2$ \\ \hline
	 adjoint     & 0 & 0 & 0 & 0 \\ 
	 fundamental     & $\frac{1}{2}d^{abc}$ & 0 & $\frac{1}{2}d^{abc}$ & 0 \\[1mm]
\hline
  \end{tabular}
\end{center}
\end{minipage}

\vspace{0.2cm}

\no where a change is visible due to the appearance of the $d^{abc}$ structure. However, this is no explanation for the Wilson line behavior in SU(2). If at all, this may be relevant only for quantitative differences between SU(2) and SU(3).

These results support the assumption that the gluonic sector is basically not affected by unquenching, for sufficiently few flavors.

There is a sudden change for the scalar-gluon vertex,

\vspace{0.2cm}

\no\begin{minipage}{0.18\linewidth}
     \includegraphics[width=1.0\linewidth]{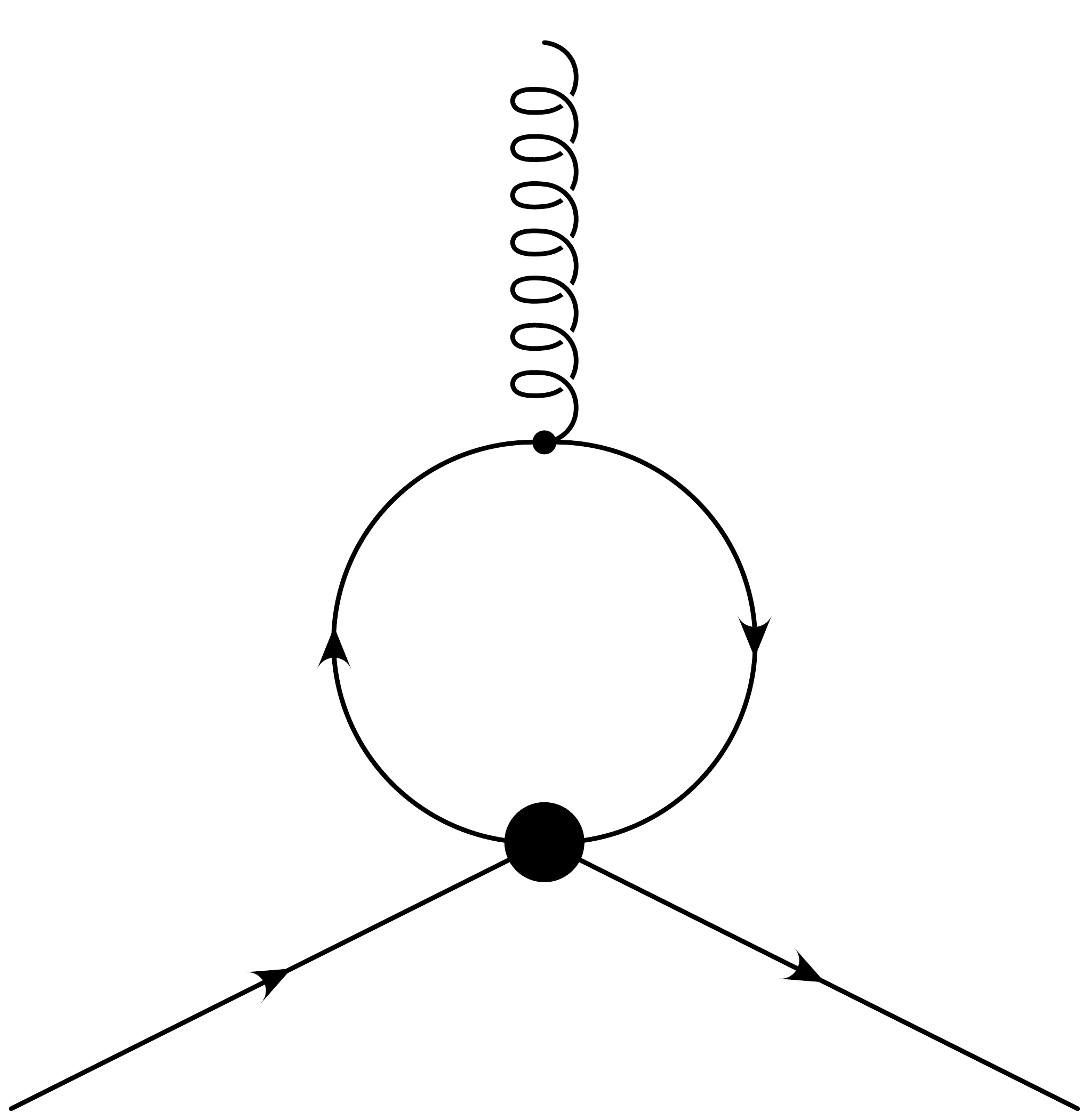}
\end{minipage}
\begin{minipage}{0.72\linewidth}
     \begin{center}
  \begin{tabular}{| l | l | l | l | l |}
    \hline
	      & group invariant & SU(2) & SU(3) & G$_2$ \\ \hline
    adjoint  & 0 & 0 & 0 & 0 \\ 
    fundamental  & $\frac{1}{6}t^{a}_{ij}$ & $\frac{1}{6}t^{a}_{ij}$ & $\frac{1}{6}t^{a}_{ij}$ 
		  & $\frac{1}{6}t^{a}_{ij}$ \\[1mm]
\hline
  \end{tabular}
\end{center}
\end{minipage}

\vspace{0.2cm}

\no showing a profound difference between the adjoint and the fundamental case, in particular, no change for the adjoint case due to symmetries present. This may be helpful for understanding why the fundamental Wilson loop changes by unquenching, if this vertex would play an important role in the screening process. However, this diagram is sub-leading in the infrared by power-counting \cite{Fister:2010yw}, and thus would not be helpful if the power-counting analysis would be accurate.  To make it relevant would require cancellations.

At the level of primitive four-point functions, no relevant changes occur, as can be seen in appendix \ref{app:res}. Thus, if at all, only the two-scalar-gluon vertex could play a role at this level of truncation.

Concerning the difference between SU($N$) and G$_2$, there are still some remarkable differences in the listing in the appendix \ref{app:res} for the tree-level coefficients. It can be seen that there are several 
diagrams whose color structures have finite values for SU($N$) but are zero for G$_2$, but this kind of difference is not observed in the gluonic sector. In the matter sector, this behavior occurs only in the fundamental representation. This supports the idea that the difference between G$_2$ and SU($N$) may be traceable by algebraic results.

Some further cases emerge for various diagrams listed in the appendices \ref{app:res} and \ref{app:res2}, though none of them provide a striking pattern. However, selective differences between groups seems to occur primarily in the tree-level tensor, while differences between adjoint and fundamental also occurs for other tensor structures, but then irrespective of the gauge group. This is the most striking pattern seen in this analysis.

\section{Summary}

Summarizing, there are very little obvious patterns for the differences between gauge groups at this level of truncation. Though some diagrams have been identified which show a pattern corresponding to the expected one, there are also some diagrams which behave in contradiction to expectation. The most striking pattern is that differences between the gauge groups is manifest primarily in the tree-level color tensor, while differences between different representations for the matter fields result in different tensor structure being (non-) zero. However, this may be an artifact of the chosen tensor basis. Still, since these effects are mainly due to the different symmetry properties of the fundamental case and the adjoint case, this may be an interesting approach to this problem.

On the other hand, the differences between the gauge groups usually emerge as a direct consequence of the different Casimirs and dimensionality. How to track those back to the generic possibilities of screening is not obvious. However, it is in general observed that the gluonic sector is less sensitive to these effects than the matter sector.

Finally, to establish whether these patterns may be relevant, the current results should be extended to other gauge groups, like e.\ g.\ F$_4$ and large $N$ SU($N$) groups. Of course, improvements in the truncation may also play a role. However, these results, together with lattice results, strongly suggest that the different behavior due to the gauge group structure may be rather intricate to identify, at least in Landau gauge.\\

\no{\bf Acknowledgments}\\
We are grateful to L.\ Fister and M.\ Q.\ Huber for helpful discussions. V.\ M.\ acknowledges support by the Paul-Urban foundation. A.\ M.\ was supported by the FWF under grant number M1099-N16 and the DFG under grant number MA 3935/5-1.

\newpage

\appendix

\section{The tree-level vertices}\label{app:tl}

In the truncated versions of the DSEs used here, the following tree-level vertices appear:

\no Ghost-gluon vertex:
\vspace{0.2cm}

\begin{minipage}{0.4\linewidth}
 \includegraphics[width=0.4\linewidth]{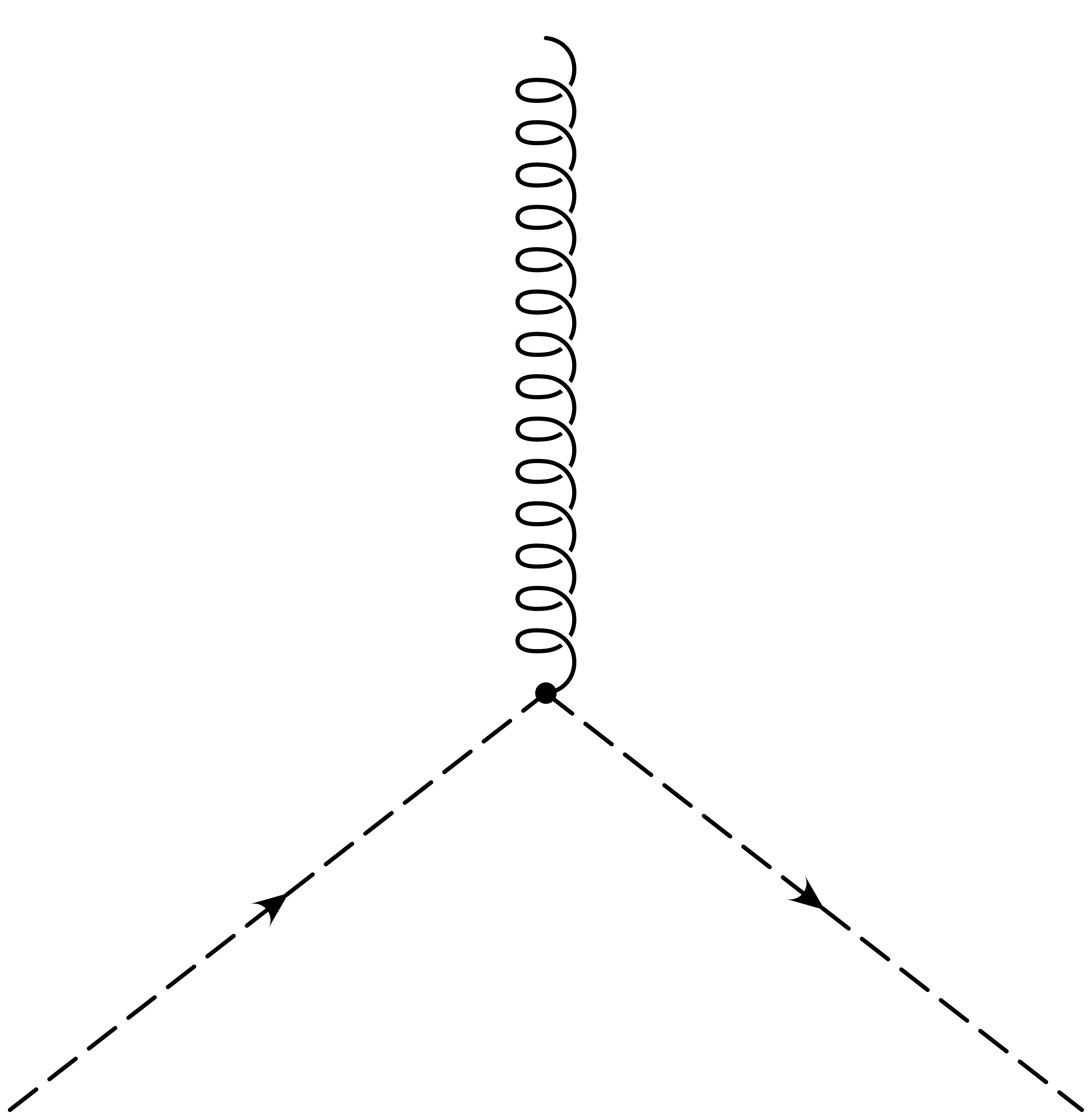}
\end{minipage}
\begin{minipage}{0.5\linewidth}
     $\sim igf^{abc}p^{\mu}$

\end{minipage}

\vspace{0.3cm}

\noindent
3-gluon vertex:
\vspace{0.2cm}

\begin{minipage}{0.4\linewidth}
     \includegraphics[width=0.4\linewidth]{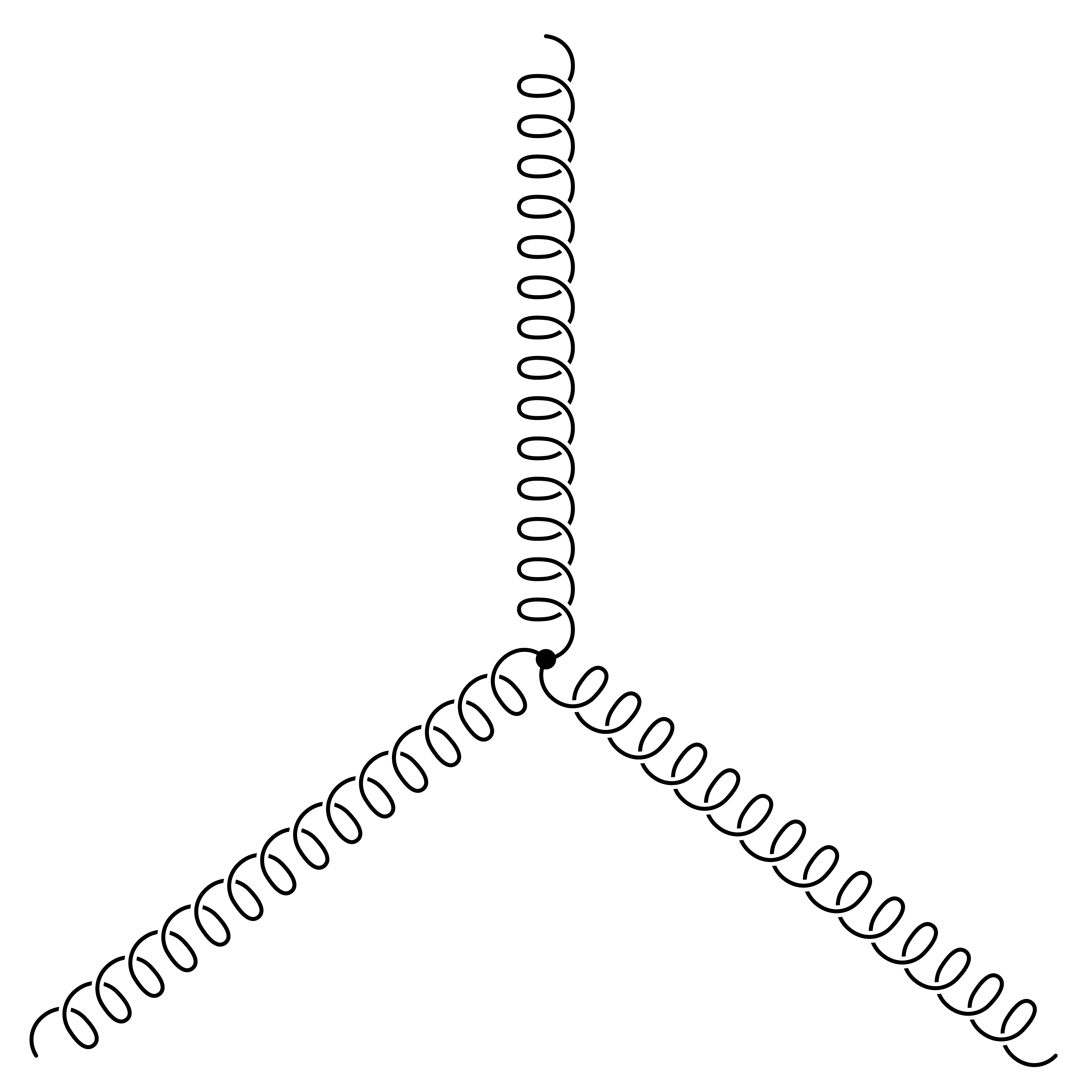}%
\end{minipage}
\begin{minipage}{0.5\linewidth}
     $\sim -igf^{abc}\left((q-k)_{\mu}\delta_{\nu\rho} +(k-p)_{\nu}\delta_{\mu\rho} +(p-q)_{\rho}\delta_{\mu\nu}\right)$
\end{minipage}

\vspace{0.3cm}

\noindent
4-gluon vertex:
\vspace{0.2cm}

\begin{minipage}{0.4\linewidth}
     \includegraphics[width=0.4\linewidth]{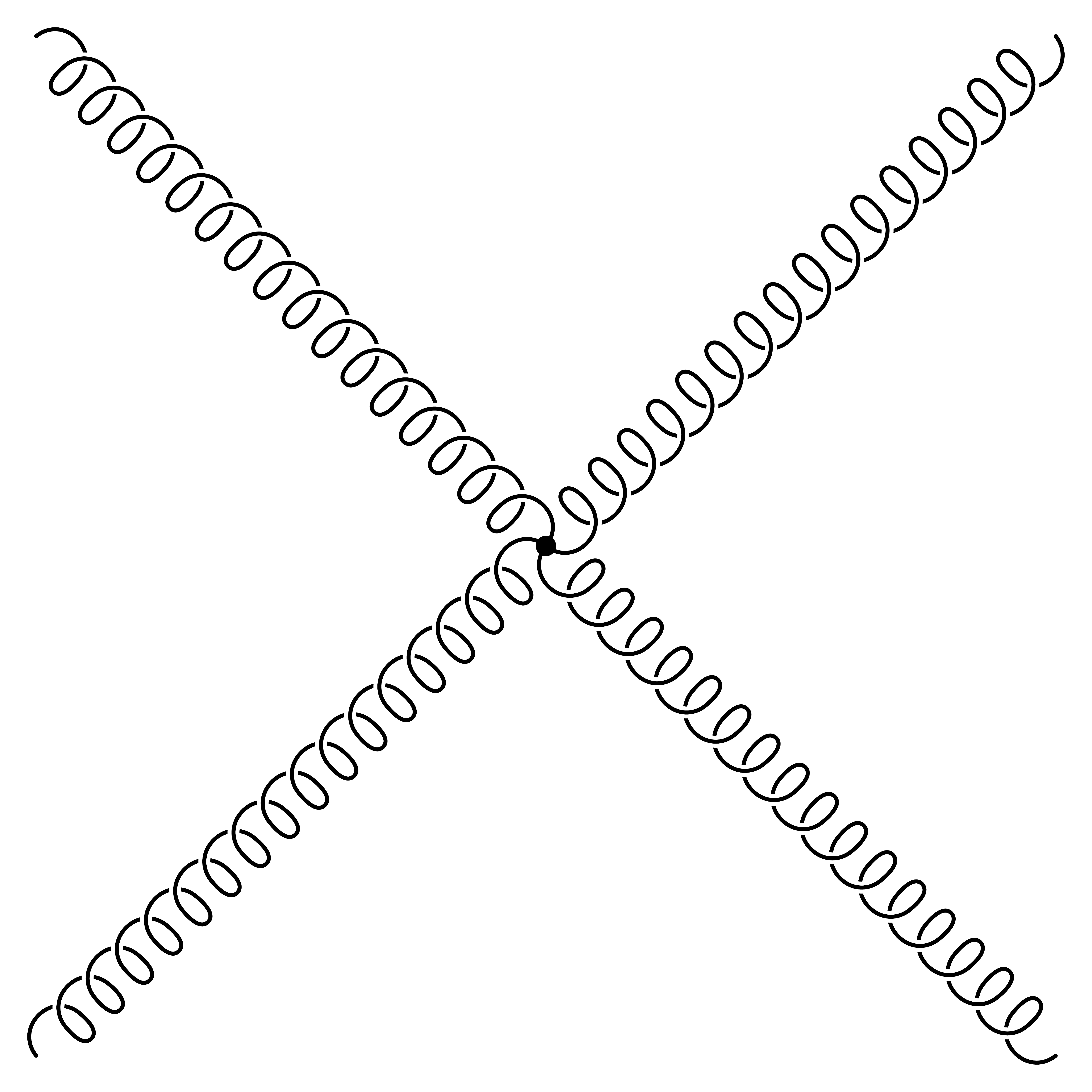}%
\end{minipage}
\begin{minipage}{0.5\linewidth}
     $$\sim g^{2} (f^{eab}f^{ecd}(\delta_{\mu\sigma}\delta_{\nu\rho}-\delta_{\mu\rho}\delta_{\nu\sigma})
			+f^{eac}f^{ebd}(\delta_{\mu\nu}\delta_{\sigma\rho}-\delta_{\mu\rho}\delta_{\nu\sigma})$$
			$$+f^{ead}f^{ebc}(\delta_{\mu\nu}\delta_{\sigma\rho}-\delta_{\mu\sigma}\delta_{\nu\rho}) ) $$
    $$=g^{2} (f^{eab}f^{ecd}A+f^{eac}f^{ebd}B+f^{ead}f^{ebc}C ) $$
\end{minipage}

\vspace{0.3cm}

\no 2-adjoint-scalar-gluon vertex:
\vspace{0.2cm}

\begin{minipage}{0.4\linewidth}
     \includegraphics[width=0.4\linewidth]{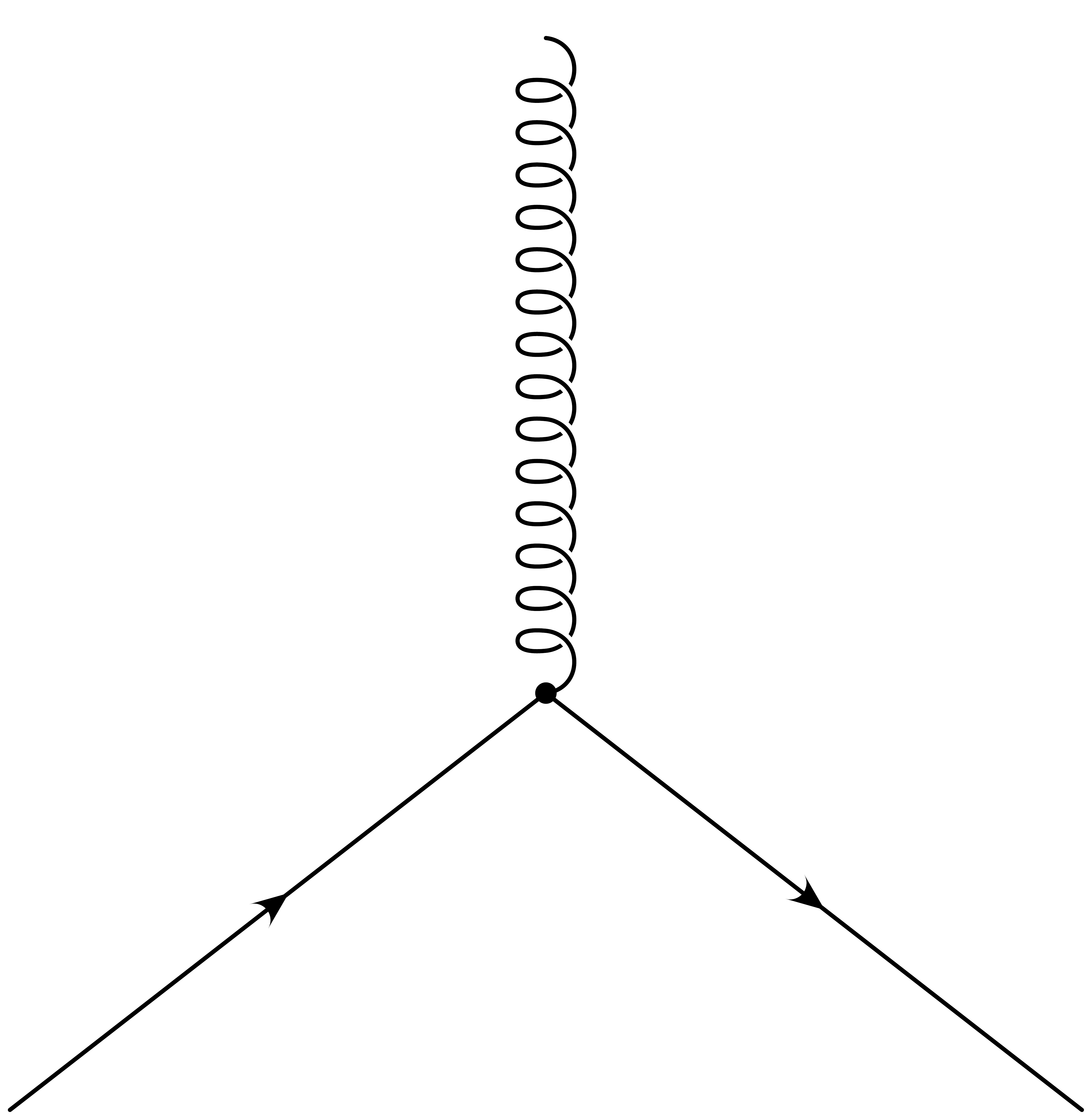}%
\end{minipage}
\begin{minipage}{0.5\linewidth}
     $\sim igf^{abc}(q-k)_{\mu} $
\end{minipage}

\vspace{0.3cm}

\noindent
2-adjoint-scalar-2-gluon vertex:
\vspace{0.2cm}

\begin{minipage}{0.4\linewidth}
     \includegraphics[width=0.4\linewidth]{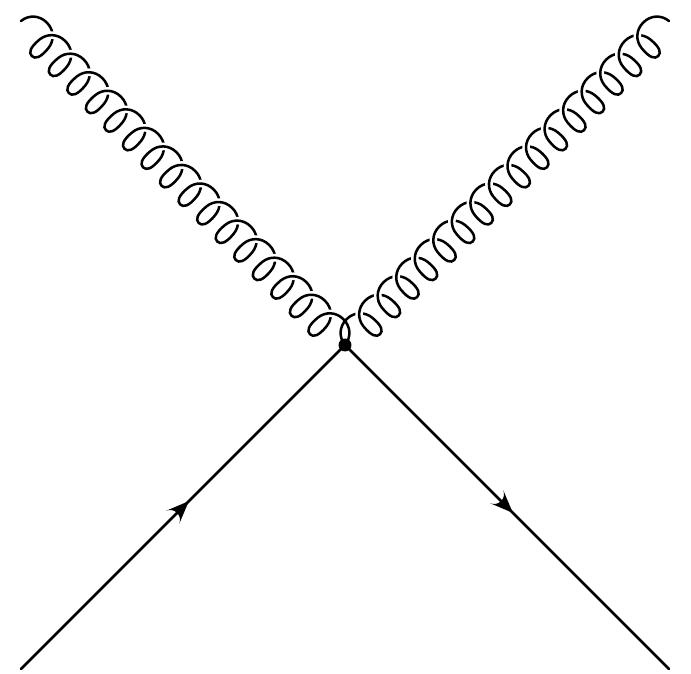}%
\end{minipage}
\begin{minipage}{0.5\linewidth}
     $\sim g^{2}\left(f^{eac}f^{ebd}+f^{ead}f^{ebc}\right)\delta_{\mu\nu}$
\end{minipage}

\vspace{0.3cm}

\noindent
4-adjoint-scalar vertex:
\vspace{0.2cm}

\begin{minipage}{0.4\linewidth}
     \includegraphics[width=0.4\linewidth]{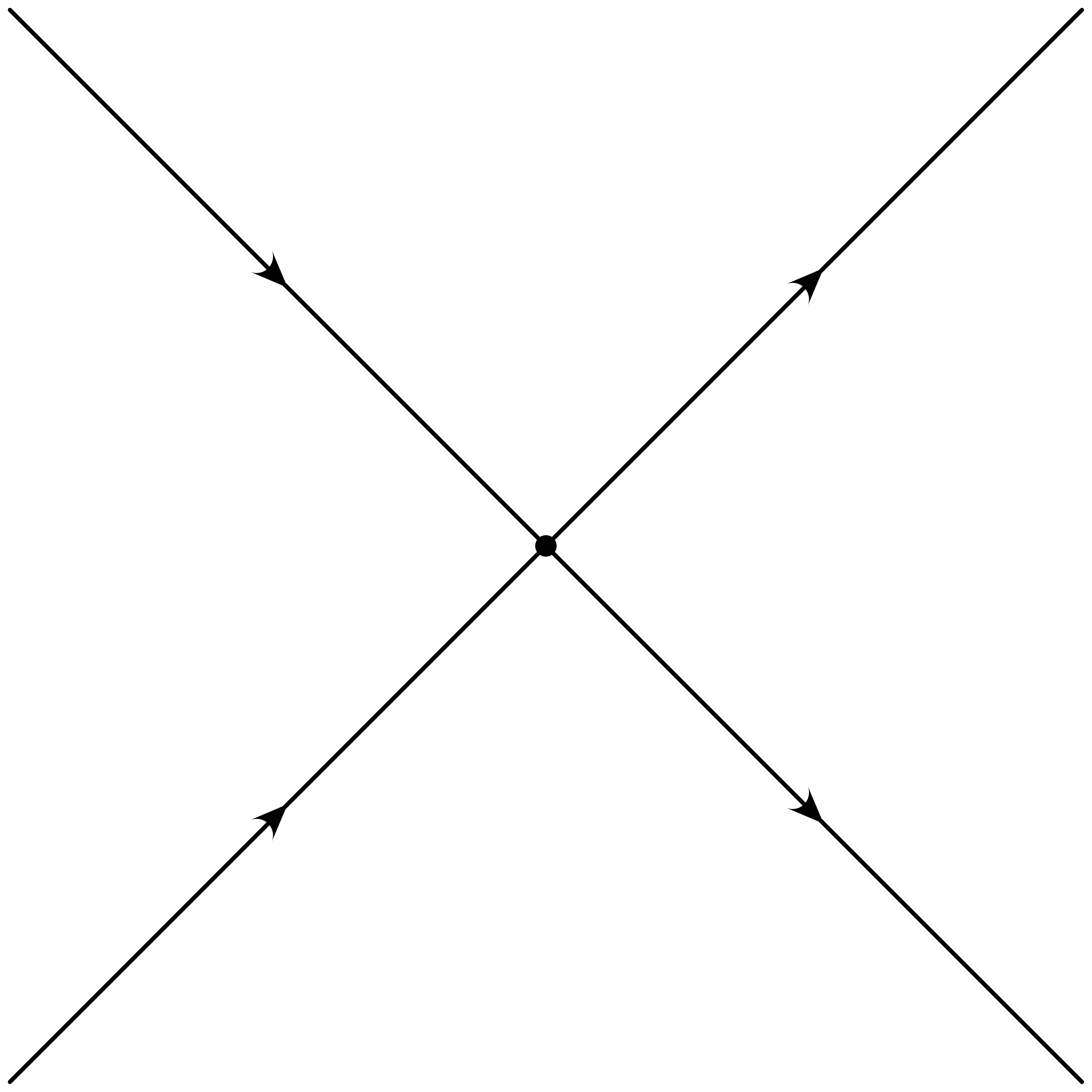}%
\end{minipage}
\begin{minipage}{0.5\linewidth}
     $\sim 2h \left(\delta_{ab}\delta_{cd}+\delta_{ac}\delta_{bd}+\delta_{ad}\delta_{bc}\right) $
\end{minipage}

\vspace{0.3cm}

\newpage

\no 2-fundamental-scalar-gluon vertex:

\begin{minipage}{0.4\linewidth}
     \includegraphics[width=0.4\linewidth]{sg.pdf}%
\end{minipage}
\begin{minipage}{0.5\linewidth}
     $\sim igt^{a}_{ij}q_{\mu} $
\end{minipage}

\vspace{0.3cm}

\noindent
2-fundamental-scalar-2-gluon vertex:
\vspace{0.2cm}

\begin{minipage}{0.4\linewidth}
     \includegraphics[width=0.4\linewidth]{ssgg.jpg}%
\end{minipage}
\begin{minipage}{0.5\linewidth}
     $\sim g^{2}\left[ \frac{t^{a}_{ik}}{2}\frac{t^{b}_{kj}}{2}+
			\frac{t^{b}_{ik}}{2}\frac{t^{a}_{kj}}{2}  \right]\delta_{\mu\nu}$
\end{minipage}

\vspace{0.3cm}

\noindent
4-fundamental-scalar vertex:
\vspace{0.2cm}

\begin{minipage}{0.4\linewidth}
     \includegraphics[width=0.4\linewidth]{4s.pdf}
\end{minipage}
\begin{minipage}{0.5\linewidth}
     $\sim -\frac{h}{3!} \left(\delta_{ab}\delta_{cd}+\delta_{ad}\delta_{bc}\right) $
\end{minipage}

\section{Further results for primitive divergent vertices}\label{app:res}\

\no Capital Latin letters denote Lorentz tensor structures.

\subsection{Propagator equations}\label{app:res:prop}

\subsubsection{Ghost propagator DSE}

\begin{minipage}{0.18\linewidth}
     \includegraphics[width=1.0\linewidth]{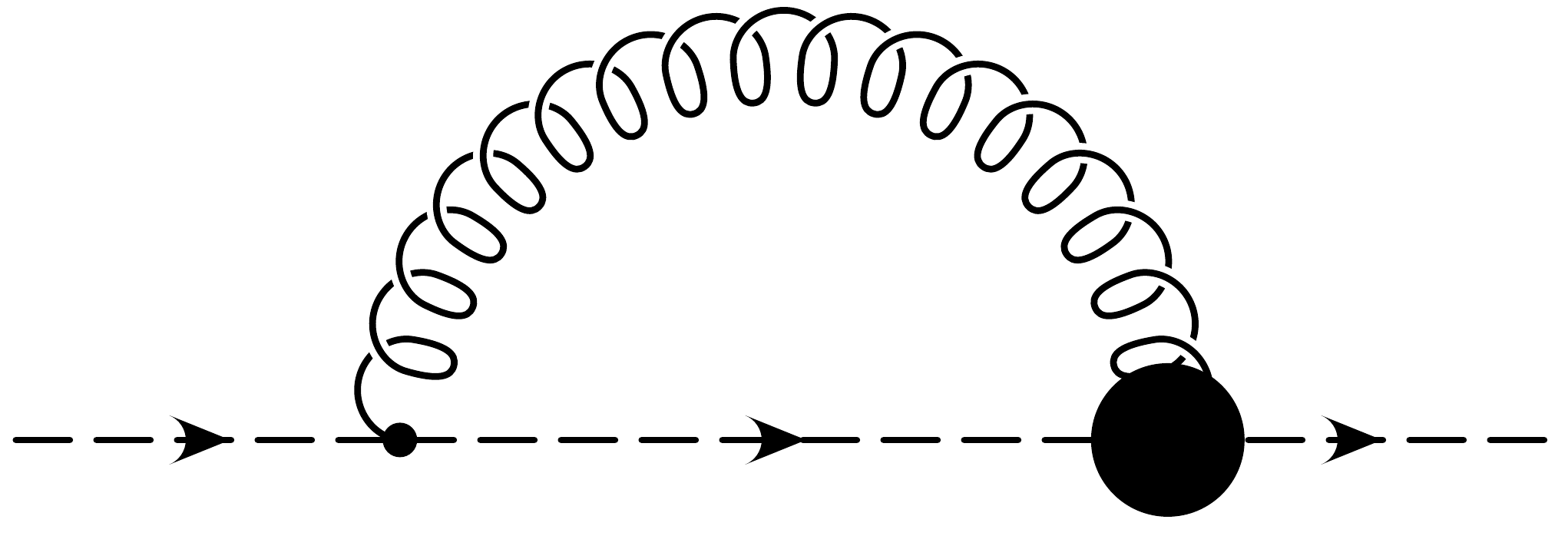}

\end{minipage}
\begin{minipage}{0.72\linewidth}
     \begin{center}
\begin{tabular}{| l | l | l | l | l |}
    \hline
	      & group invariant & SU(2) & SU(3) & G$_2$ \\ \hline
    adjoint  & $C_{A}\delta^{ab}$ & $2\delta^{ab}$ & $3\delta^{ab}$ & $2\delta^{ab}$ \\ 
\hline
  \end{tabular}
\end{center}
\end{minipage}

\subsubsection{Gluon propagator DSE}

\begin{minipage}{0.18\linewidth}
     \includegraphics[width=1.0\linewidth]{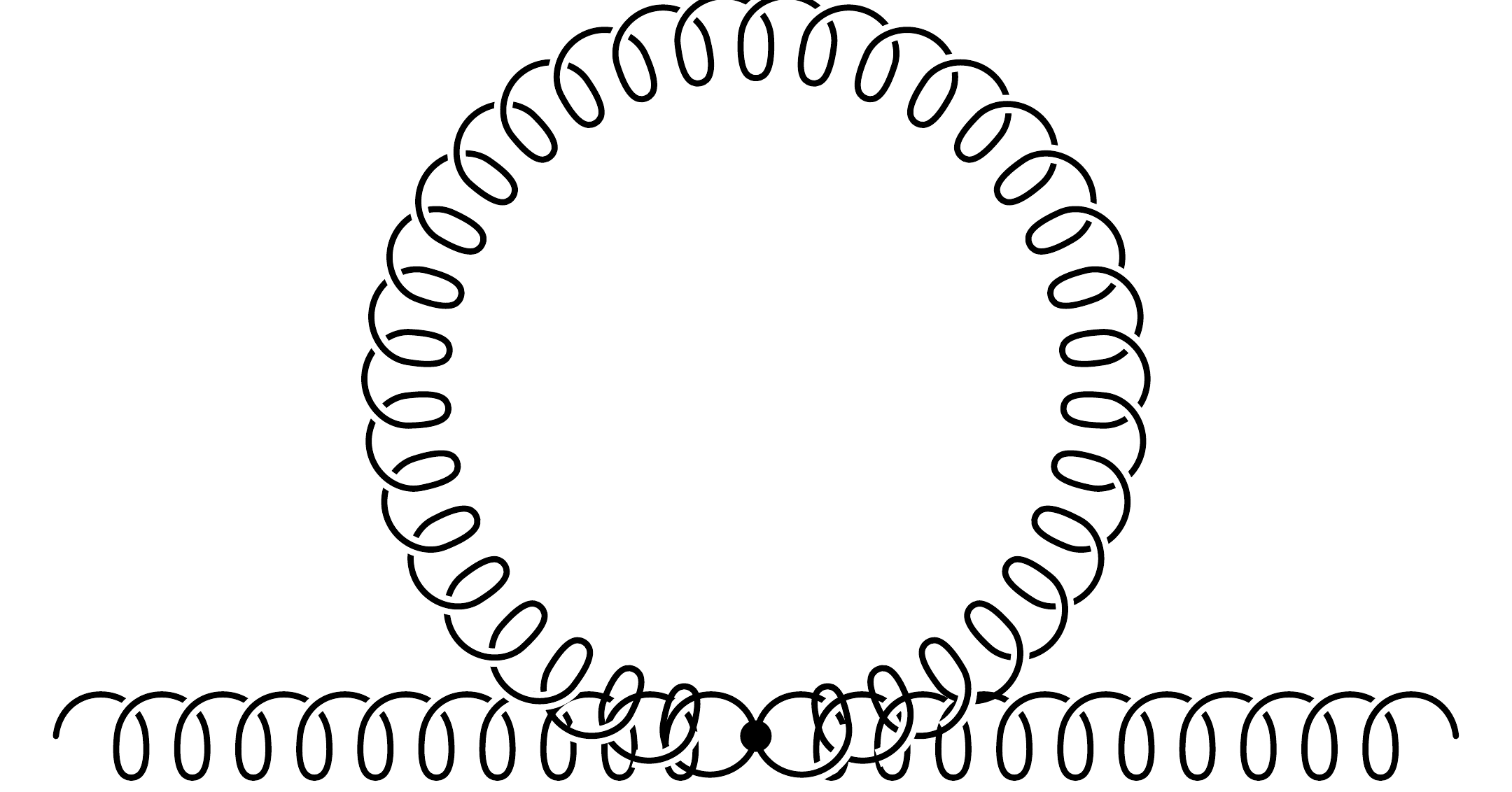}
\end{minipage}
\begin{minipage}{0.72\linewidth}
     \begin{center}
  \begin{tabular}{| l | l | l | l | l |}
    \hline
	      & group invariant & SU(2) & SU(3) & G$_2$ \\ \hline
    adjoint  & $2C_{A}\delta^{ab}$ & $4\delta^{ab}$ & $6\delta^{ab}$ & $4\delta^{ab}$ \\ 
\hline
  \end{tabular}
\end{center}
\end{minipage}

\no\begin{minipage}{0.18\linewidth}
     \includegraphics[width=1.0\linewidth]{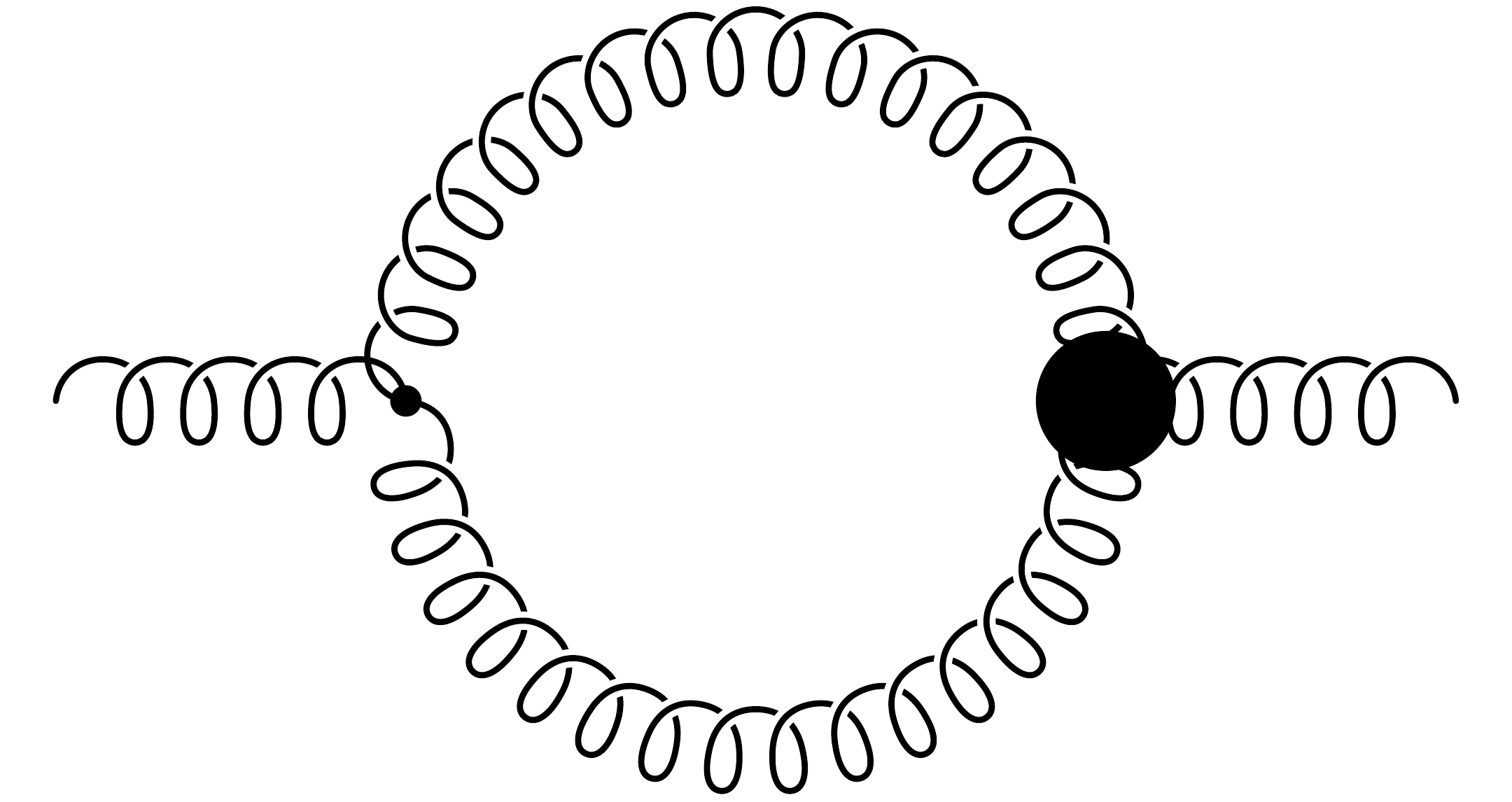}
\end{minipage}
\begin{minipage}{0.72\linewidth}
     \begin{center}
  \begin{tabular}{| l | l | l | l | l |}
    \hline
	      & group invariant & SU(2) & SU(3) & G$_2$ \\ \hline
    adjoint  & $C_{A}\delta^{ab}$ & $2\delta^{ab}$ & $3\delta^{ab}$ & $2\delta^{ab}$ \\ 
\hline
  \end{tabular}
\end{center}
\end{minipage}

\no\begin{minipage}{0.18\linewidth}
     \includegraphics[width=1.0\linewidth]{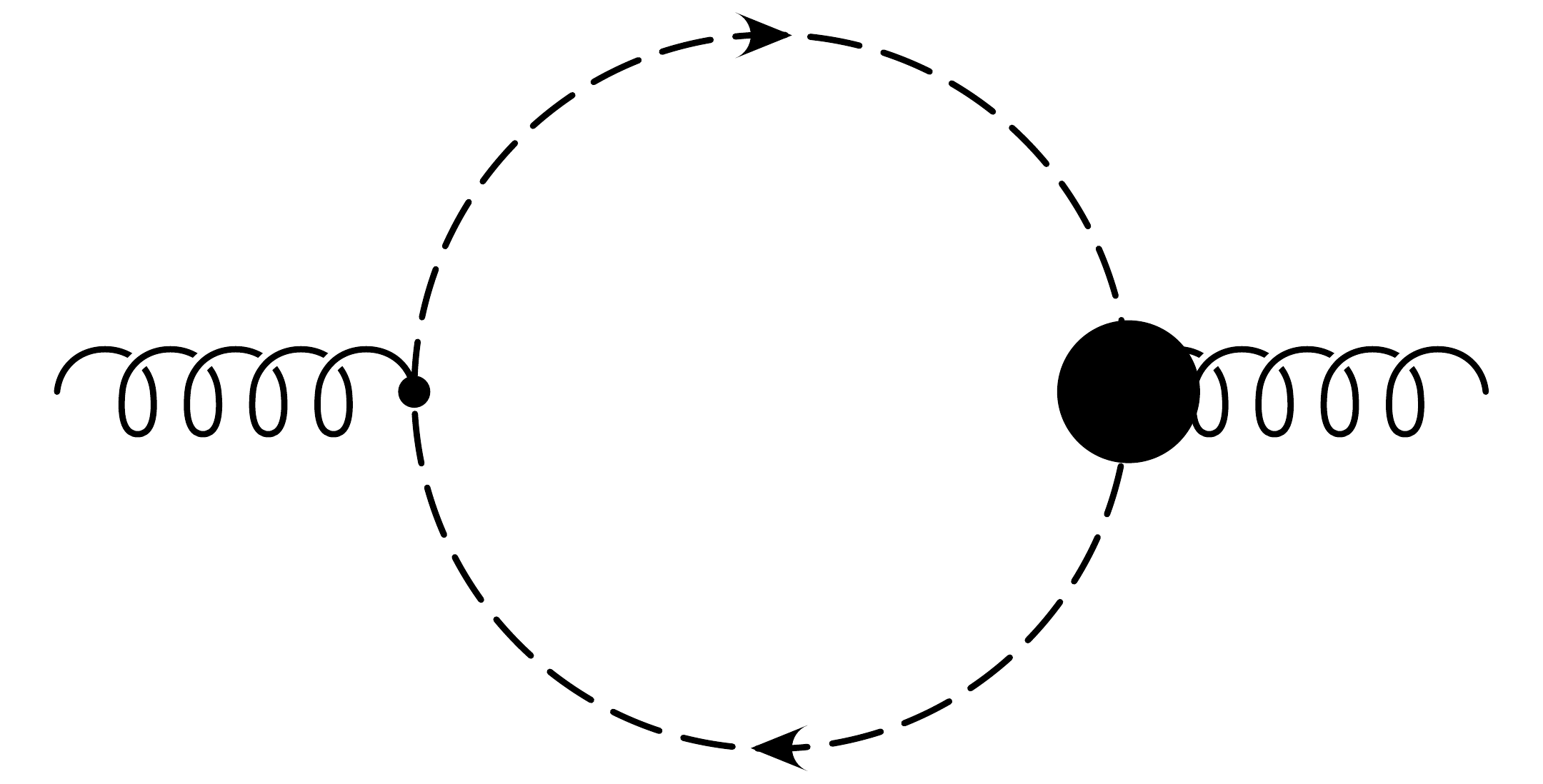}
\end{minipage}
\begin{minipage}{0.72\linewidth}
     \begin{center}
  \begin{tabular}{| l | l | l | l | l |}
    \hline
	      & group invariant & SU(2) & SU(3) & G$_2$ \\ \hline
    adjoint  & $C_{A}\delta^{ab}$ & $2\delta^{ab}$ & $3\delta^{ab}$ & $2\delta^{ab}$ \\ 
\hline
  \end{tabular}
\end{center}
\end{minipage}

\no\begin{minipage}{0.18\linewidth}
     \includegraphics[width=1.0\linewidth]{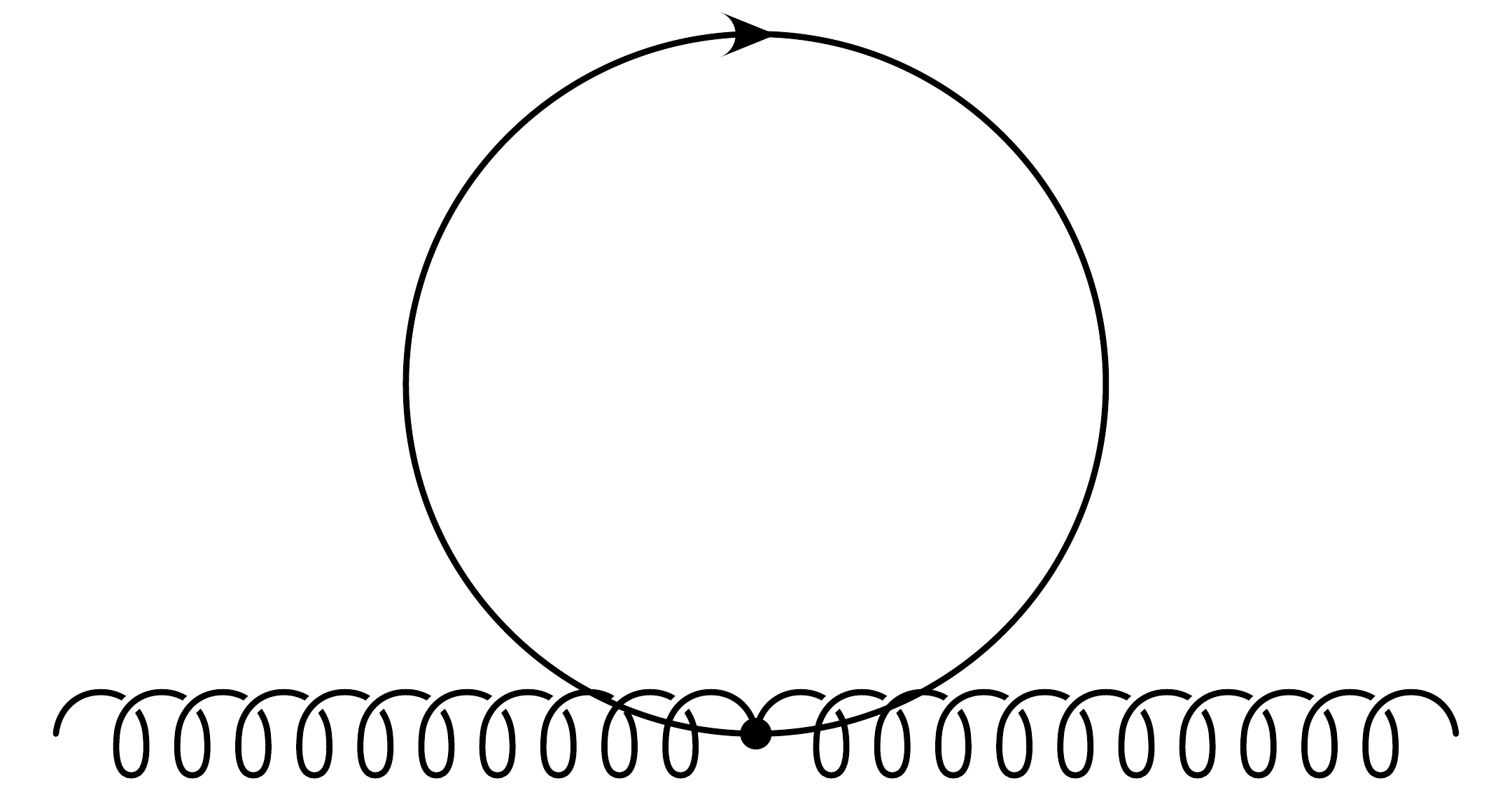}
\end{minipage}
\begin{minipage}{0.72\linewidth}
     \begin{center}
  \begin{tabular}{| l | l | l | l | l |}
    \hline
	      & group invariant & SU(2) & SU(3) & G$_2$ \\ \hline
    adjoint  & $2C_{A}\delta^{ab}$ & $4\delta^{ab}$ & $6\delta^{ab}$ & $4\delta^{ab}$ \\ 
    fundamental & $T_{R}\delta^{ab}$ & $\frac{1}{2}\delta^{ab}$ & $\frac{1}{2}\delta^{ab}$ & $\frac{1}{2}\delta^{ab}$ \\[1mm]
\hline
  \end{tabular}
\end{center}
\end{minipage}

\no\begin{minipage}{0.18\linewidth}
     \includegraphics[width=1.0\linewidth]{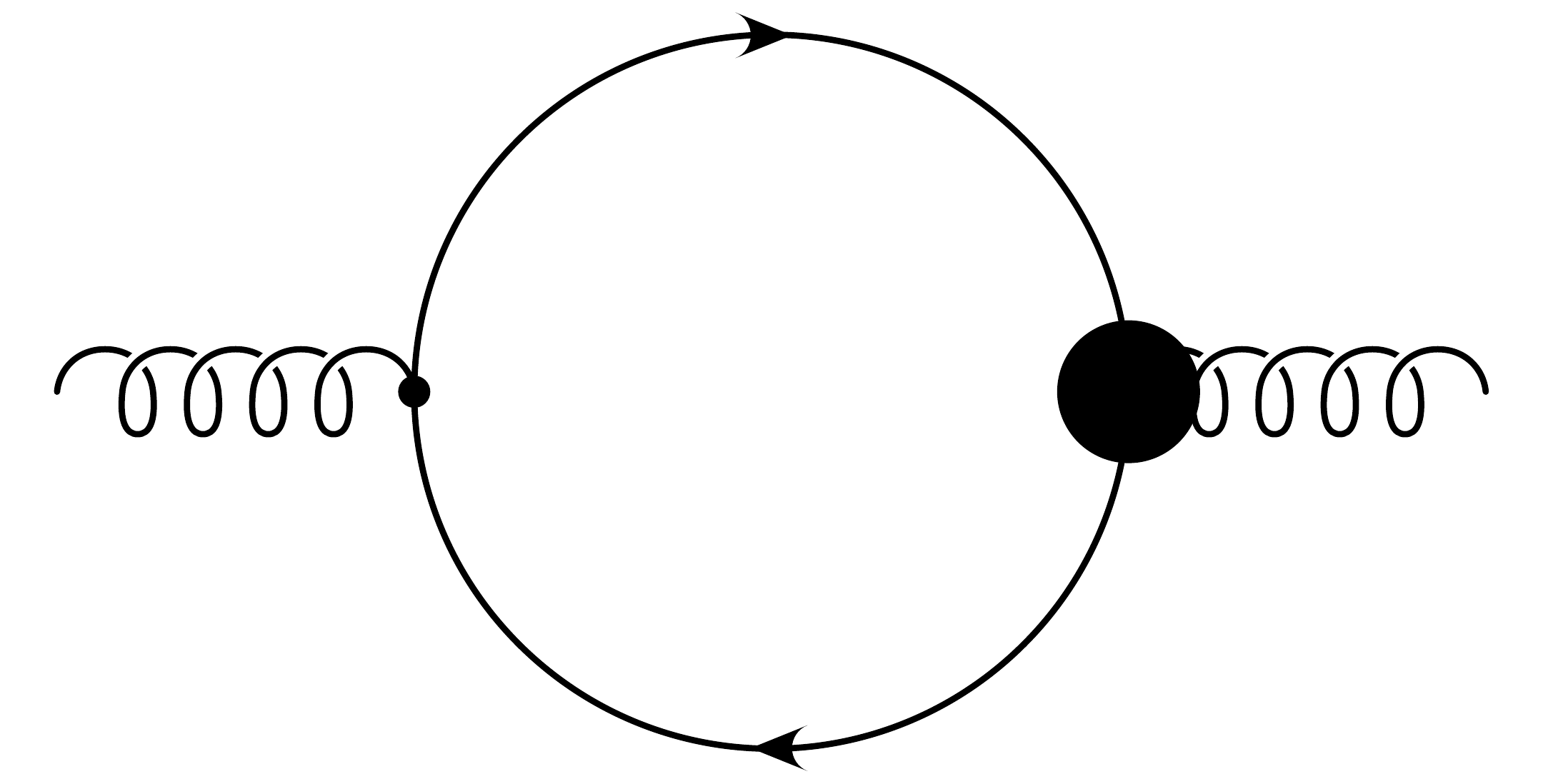}
\end{minipage}
\begin{minipage}{0.72\linewidth}
     \begin{center}
  \begin{tabular}{| l | l | l | l | l |}
    \hline
	      & group invariant & SU(2) & SU(3) & G$_2$ \\ \hline
    adjoint  & $C_{A}\delta^{ab}$ & $2\delta^{ab}$ & $3\delta^{ab}$ & $2\delta^{ab}$ \\ 
    fundamental & $-T_{R}\delta^{ab}$ & $-\frac{1}{2}\delta^{ab}$ & $-\frac{1}{2}\delta^{ab}$ & $-\frac{1}{2}\delta^{ab}$ \\[1mm]
\hline
  \end{tabular}
\end{center}
\end{minipage}

\subsubsection{Scalar propagator DSE}

\begin{minipage}{0.18\linewidth}
     \includegraphics[width=1.0\linewidth]{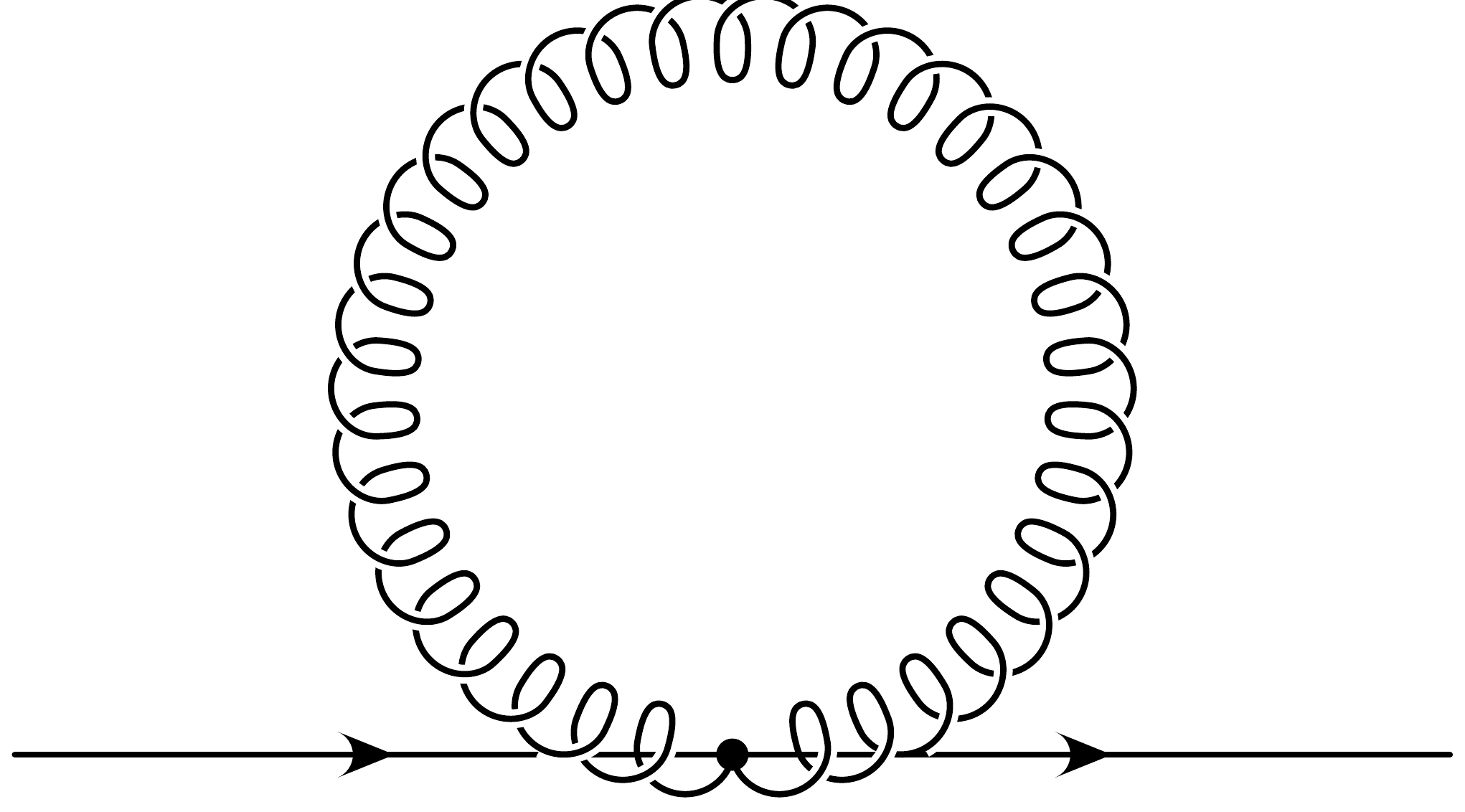}
\end{minipage}
\begin{minipage}{0.72\linewidth}
     \begin{center}
  \begin{tabular}{| l | l | l | l | l |}
    \hline
	      & group invariant & SU(2) & SU(3) & G$_2$ \\ \hline
    adjoint  & $2C_{A}\delta^{ab}$ & $4\delta^{ab}$ & $6\delta^{ab}$ & $4\delta^{ab}$ \\ 
    fundamental & $\frac{1}{2}C_{F}\delta^{ij}$ & $\frac{3}{8}\delta^{ij}$ & $\frac{4}{6}\delta^{ij}$ & $\frac{1}{2}\delta^{ij}$ \\[1mm]
\hline
  \end{tabular}
\end{center}
\end{minipage}

\no\begin{minipage}{0.18\linewidth}
     \includegraphics[width=1.0\linewidth]{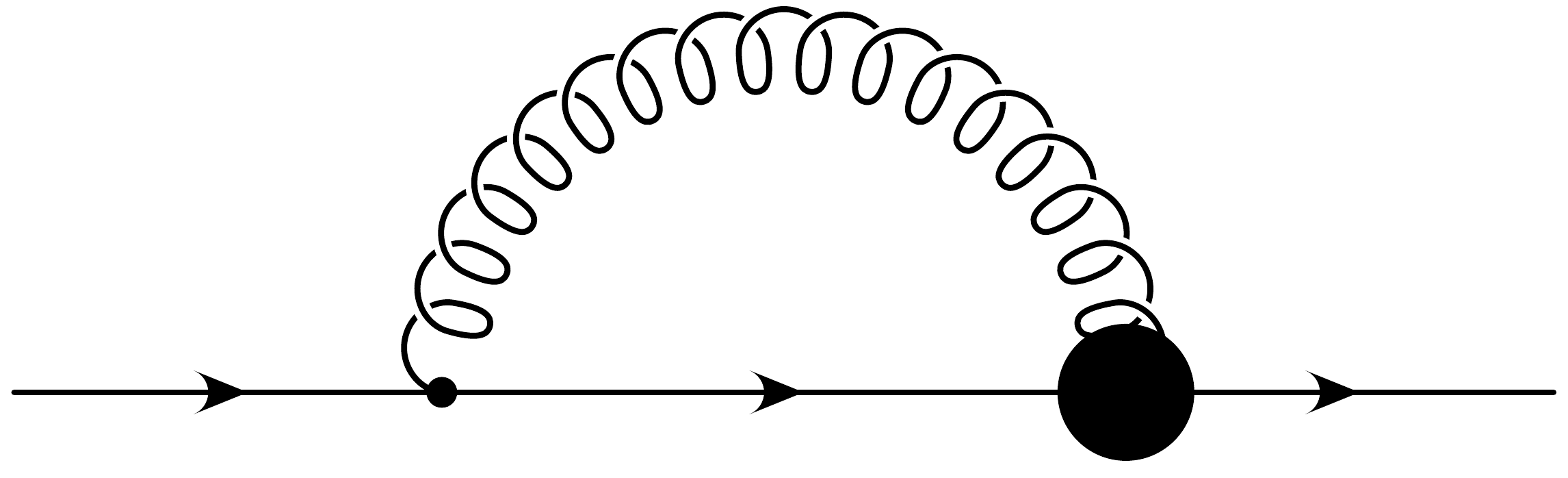}
\end{minipage}
\begin{minipage}{0.72\linewidth}
     \begin{center}
  \begin{tabular}{| l | l | l | l | l |}
    \hline
	      & group invariant & SU(2) & SU(3) & G$_2$ \\ \hline
    adjoint  & $C_{A}\delta^{ab}$ & $2\delta^{ab}$ & $3\delta^{ab}$ & $2\delta^{ab}$ \\ 
    fundamental & $-C_{F}\delta^{ij}$ & $-\frac{3}{4}\delta^{ij}$ & $-\frac{4}{3}\delta^{ij}$ & $-\delta^{ij}$ \\[1mm]
\hline
  \end{tabular}
\end{center}
\end{minipage}

\no\begin{minipage}{0.18\linewidth}
     \includegraphics[width=1.0\linewidth]{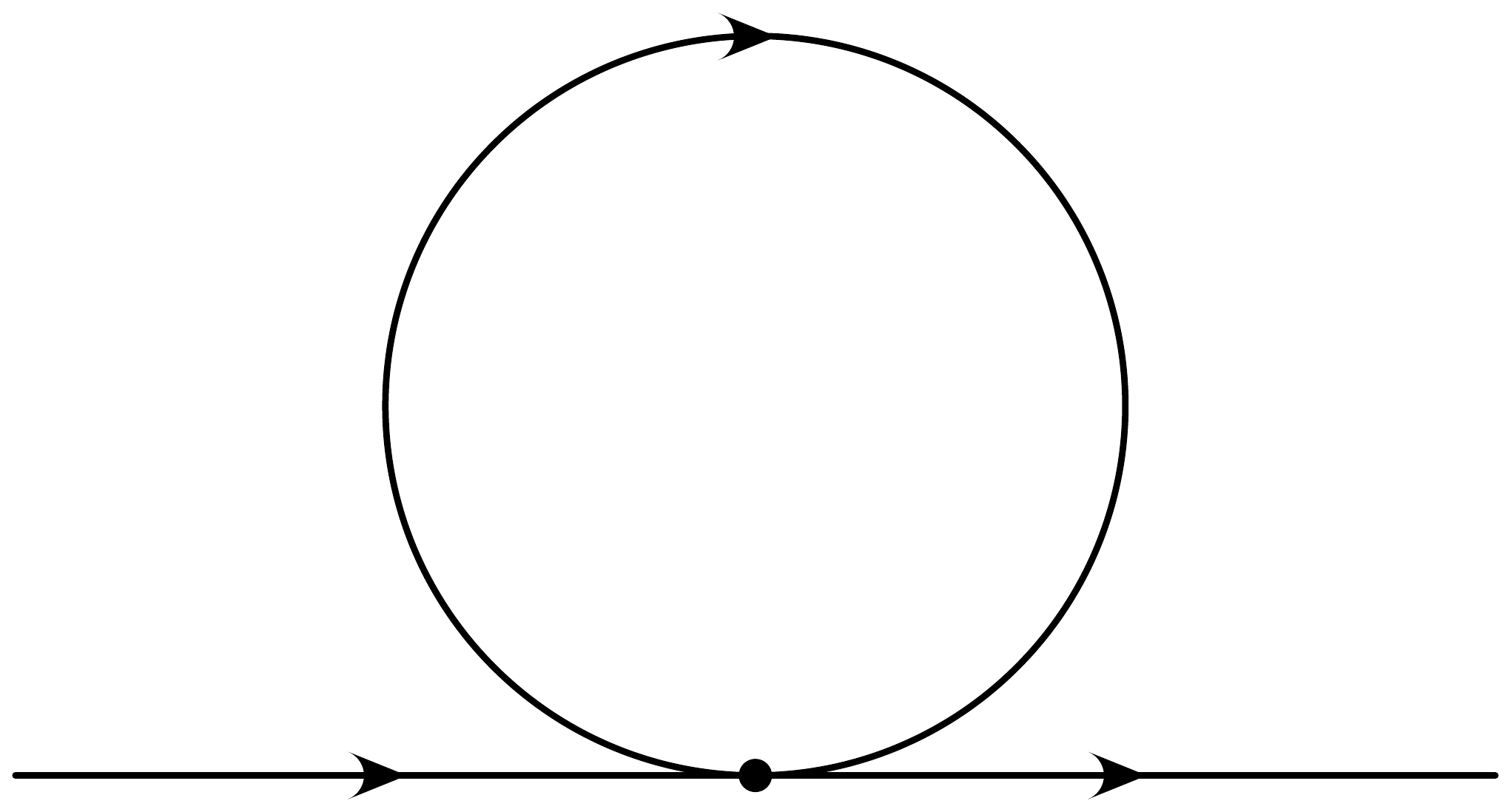}
\end{minipage}
\begin{minipage}{0.72\linewidth}
     \begin{center}
  \begin{tabular}{| l | l | l | l | l |}
    \hline
	      & group invariant & SU(2) & SU(3) & G$_2$ \\ \hline
    adjoint  & $2\left(N_{A}+2\right)\delta^{ab}$ & $10\delta^{ab}$ & $20\delta^{ab}$ & $32\delta^{ab}$ \\ 
    fundamental & $(N_{F}+1)\delta^{ij}$ & $3\delta^{ij}$ & $4\delta^{ij}$ & $8\delta^{ij}$ \\
\hline
  \end{tabular}
\end{center}
\end{minipage}

\subsection{Three-point equations}

\subsubsection{Ghost-gluon vertex DSE}

\begin{minipage}{0.18\linewidth}
     \includegraphics[width=1.0\linewidth]{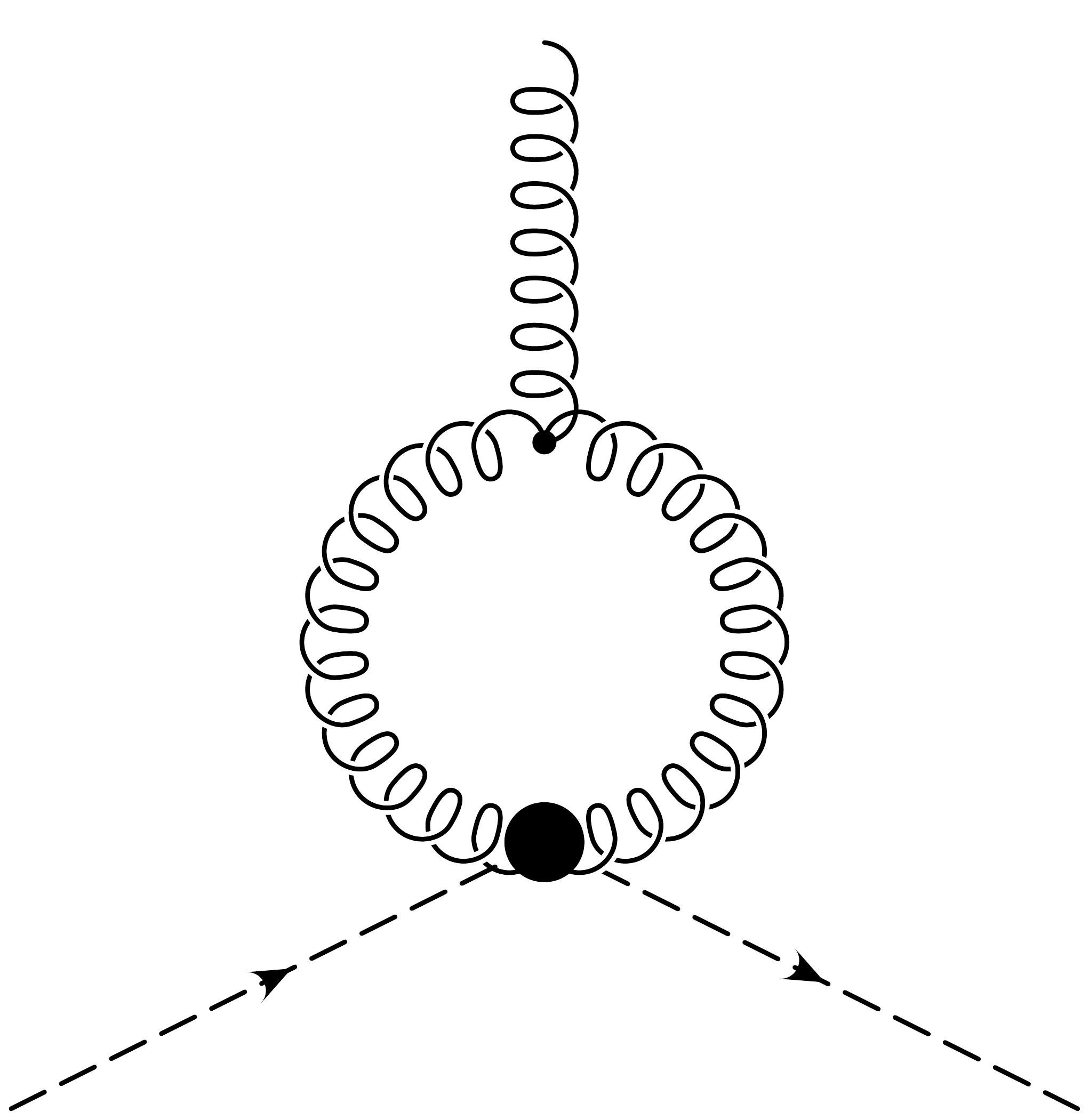}
\end{minipage}
\begin{minipage}{0.72\linewidth}
     \begin{center}
  \begin{tabular}{| l | l | l | l | l |}
    \hline
	      & group invariant & SU(2) & SU(3) & G$_2$ \\ \hline
    L  & $-\frac{1}{2}C_{A}f^{abc}$ & $-f^{abc}$ & $\frac{3}{2}f^{abc}$ & $-f^{abc}$ \\ 
    K  & $C_{A}f^{abc}$ & $2f^{abc}$ & $3f^{abc}$ & $2f^{abc}$ \\
\hline
  \end{tabular}
\end{center}
\end{minipage}

\no\begin{minipage}{0.18\linewidth}
     \includegraphics[width=1.0\linewidth]{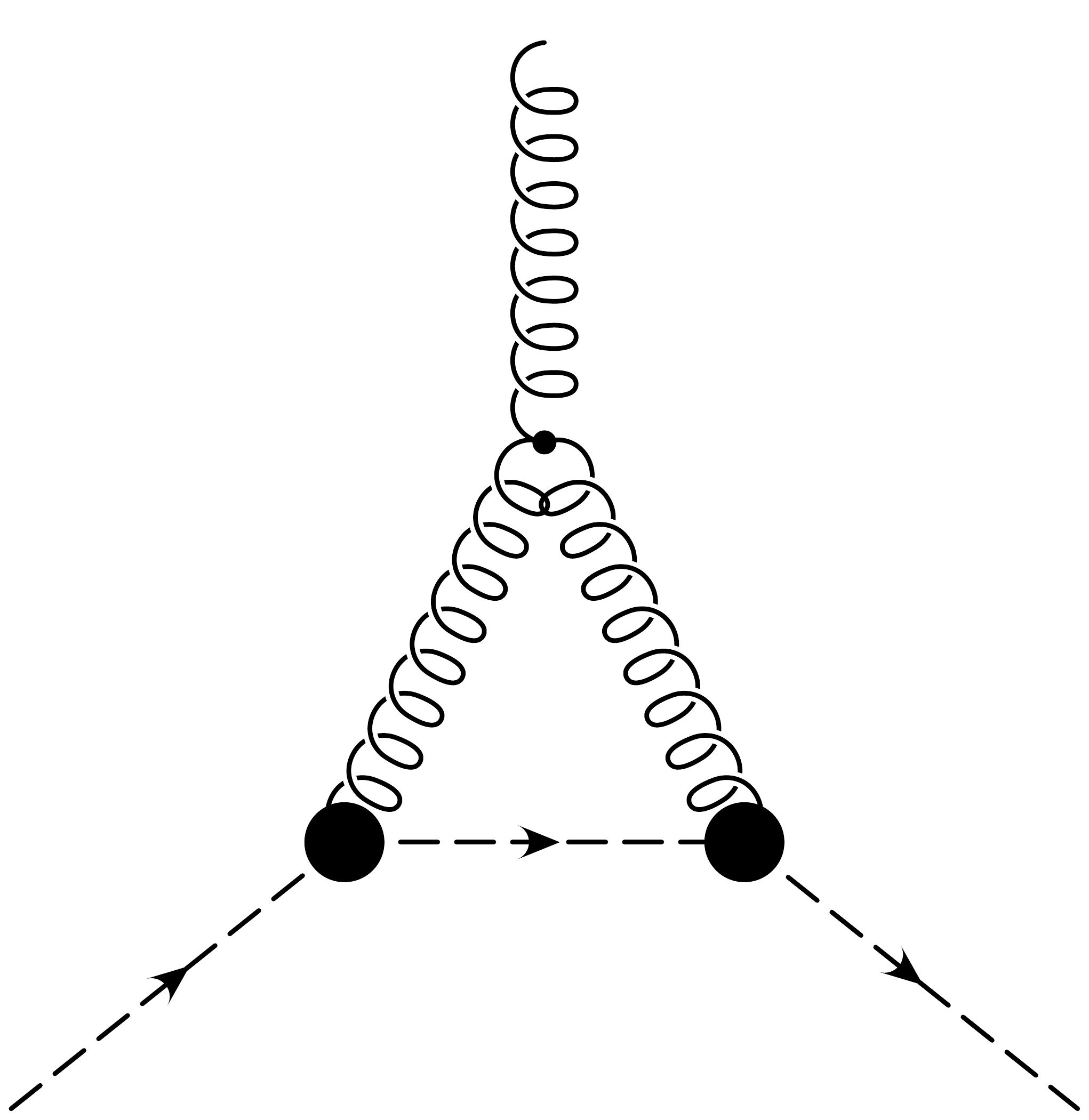}
\end{minipage}
\begin{minipage}{0.72\linewidth}
     \begin{center}
  \begin{tabular}{| l | l | l | l | l |}
    \hline
	      & group invariant & SU(2) & SU(3) & G$_2$ \\ \hline
    adjoint  & $-\frac{1}{2}C_{A}f^{abc}$ & $-f^{abc}$ & $\frac{3}{2}f^{abc}$ & $-f^{abc}$ \\[1mm] 
\hline
  \end{tabular}
\end{center}
\end{minipage}

\subsubsection{3-gluon vertex DSE}

\begin{minipage}{0.18\linewidth}
     \includegraphics[width=1.0\linewidth]{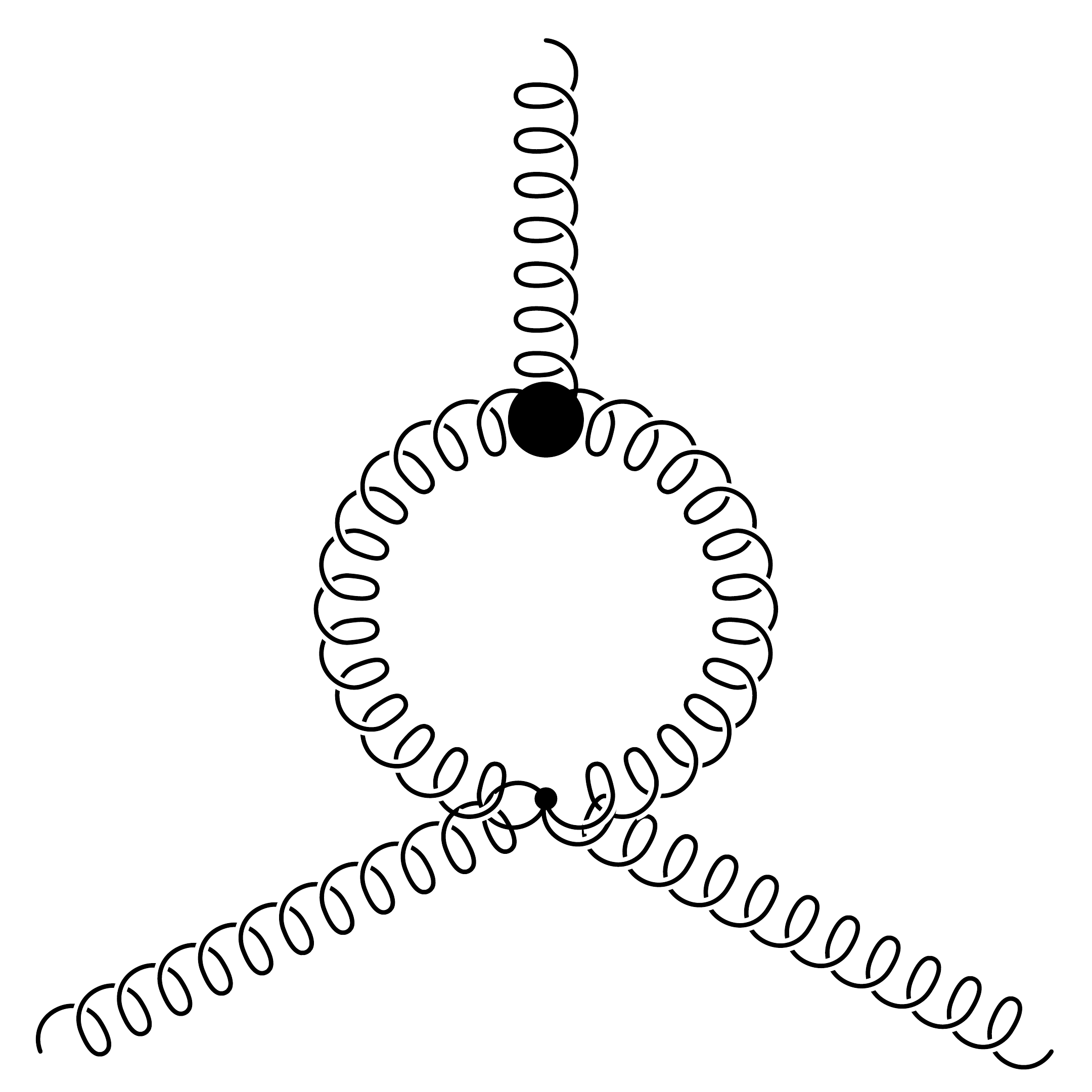}
\end{minipage}
\begin{minipage}{0.72\linewidth}
     \begin{center}
  \begin{tabular}{| l | l | l | l | l |}
    \hline
	      & group invariant & SU(2) & SU(3) & G$_2$ \\ \hline
	      & $\frac{D}{2}C_{A}f^{abc}\times$ & $f^{abc}D\times$ & $\frac{3}{2}Df^{abc}\times$ & $f^{abc}D\times$ \\ 
	      & $\times \left(A-B\right.$ & $\times \left(A-B\right.$ & $\times \left(A-B\right.$ & $\times \left(A-B\right.$ \\
	      & $\left.-2C\right)$ & $\left.-2C\right)$ & $\left.-2C\right)$ & $\left.-2C\right)$ \\
\hline
  \end{tabular}
\end{center}
\end{minipage}

\no\begin{minipage}{0.18\linewidth}
     \includegraphics[width=1.0\linewidth]{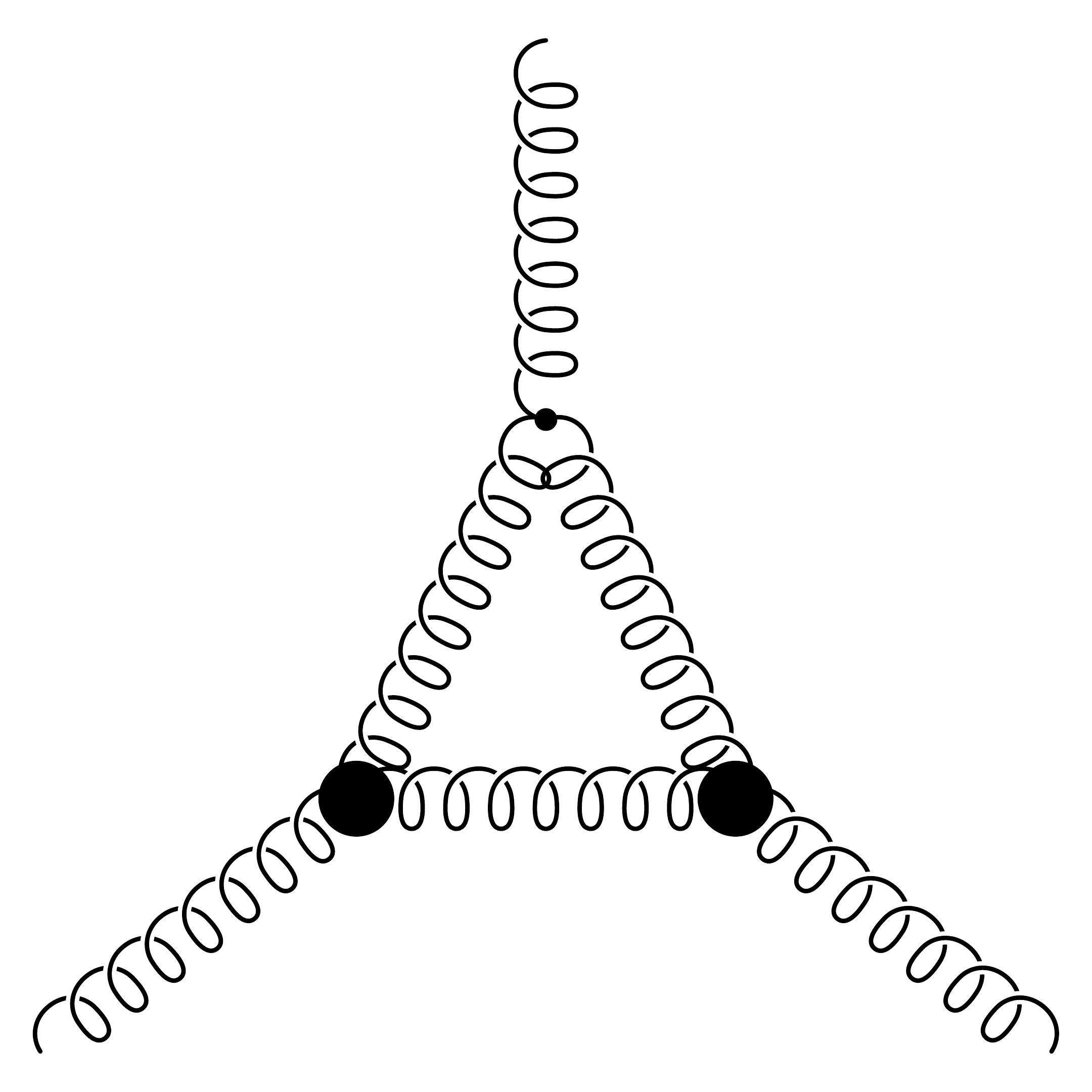}
\end{minipage}
\begin{minipage}{0.72\linewidth}
     \begin{center}
  \begin{tabular}{| l | l | l | l | l |}
    \hline
	      & group invariant & SU(2) & SU(3) & G$_2$ \\ \hline
	      & $-\frac{1}{2}C_{A}f^{abc}$ & $-f^{abc}$ & $-\frac{3}{2}f^{abc}$ & $-f^{abc}$ \\[1mm] 
\hline
  \end{tabular}
\end{center}
\end{minipage}

\no\begin{minipage}{0.18\linewidth}
     \includegraphics[width=1.0\linewidth]{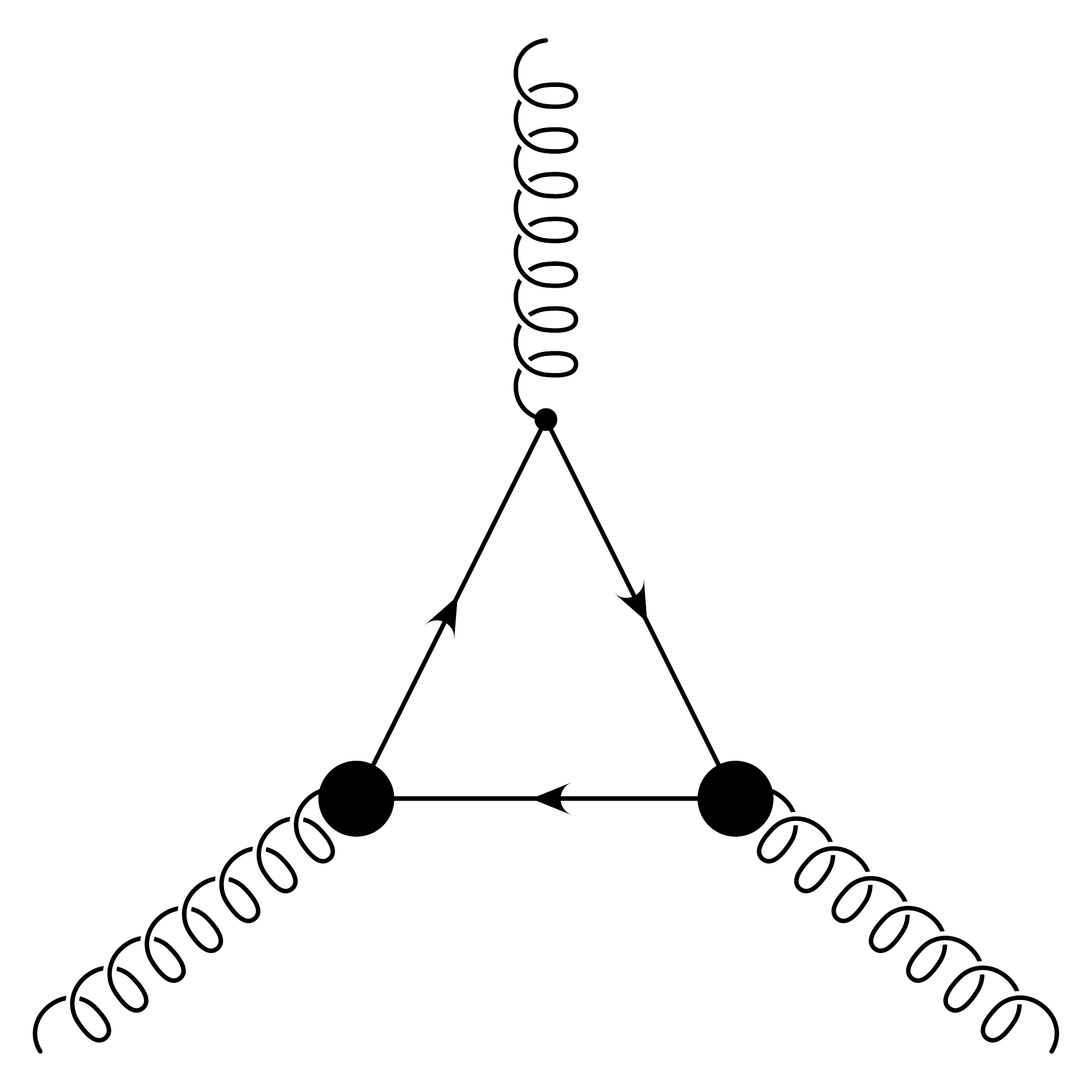}
\end{minipage}
\begin{minipage}{0.72\linewidth}
     \begin{center}
  \begin{tabular}{| l | l | l | l | l |}
    \hline
	      & group invariant & SU(2) & SU(3) & G$_2$ \\ \hline
	  adjoint    & $\frac{1}{2}C_{A}f^{abc}$ & $f^{abc}$ & $\frac{3}{2}f^{abc}$ & $f^{abc}$ \\ 
	  fundamental    & $-d^{abc}+\frac{1}{2}f^{abc}T_{R}$ & $\frac{1}{4}f^{abc}$ & $-d^{abc}+\frac{1}{4}f^{abc}$ & $\frac{1}{4}f^{abc}$ \\[1mm]
\hline
  \end{tabular}
\end{center}
\end{minipage}

\subsubsection{Two-scalar-gluon DSE}

\begin{minipage}{0.18\linewidth}
     \includegraphics[width=1.0\linewidth]{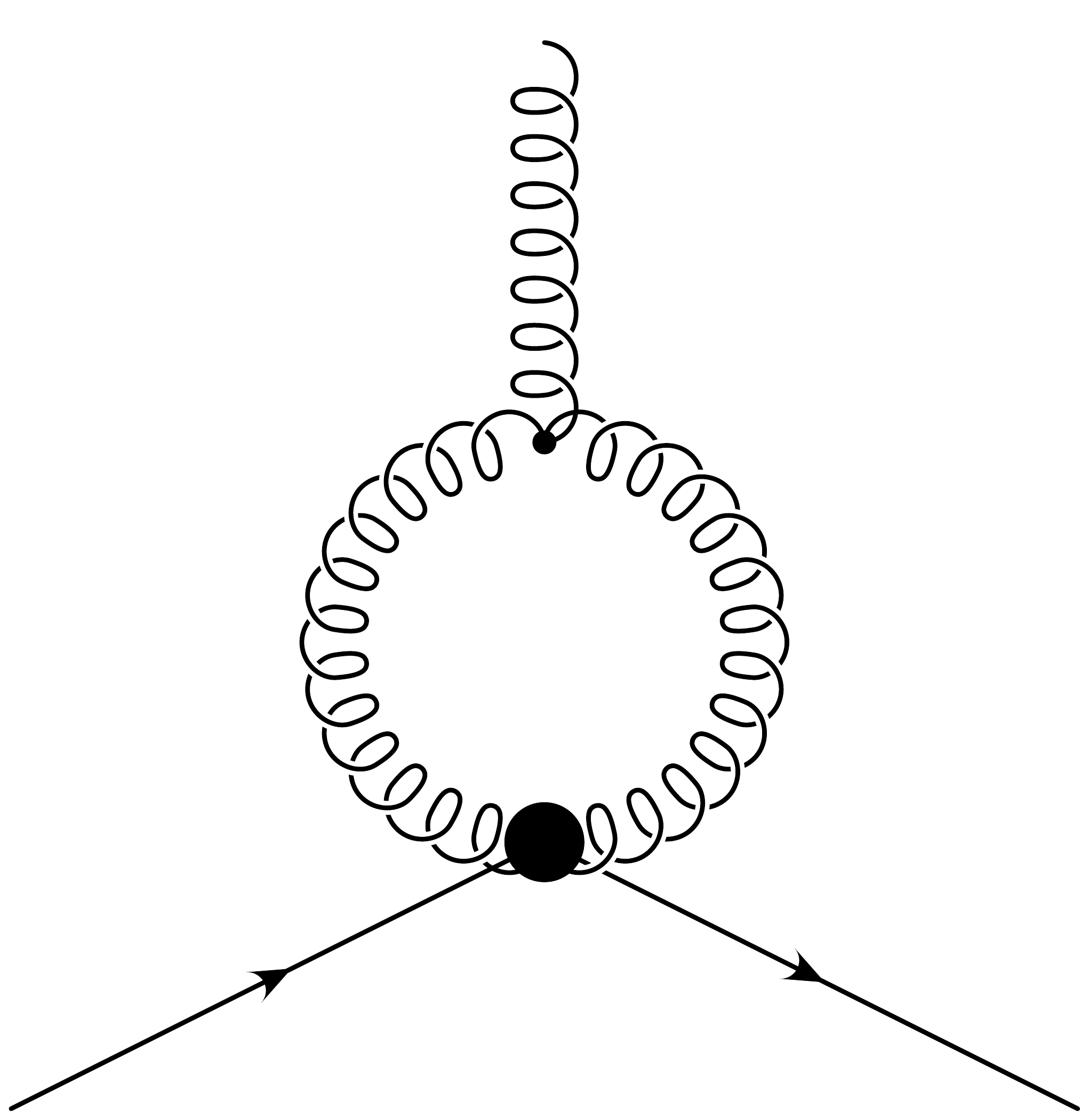}
\end{minipage}
\begin{minipage}{0.72\linewidth}
     \begin{center}
  \begin{tabular}{| l | l | l | l | l |}
    \hline
	      & group invariant & SU(2) & SU(3) & G$_2$ \\ \hline
    adjoint  & 0 & 0 & 0 & 0 \\ 
    fundamental  & 0 & 0 & 0 & 0 \\
\hline
  \end{tabular}
\end{center}
\end{minipage}

\no\begin{minipage}{0.18\linewidth}
     \includegraphics[width=1.0\linewidth]{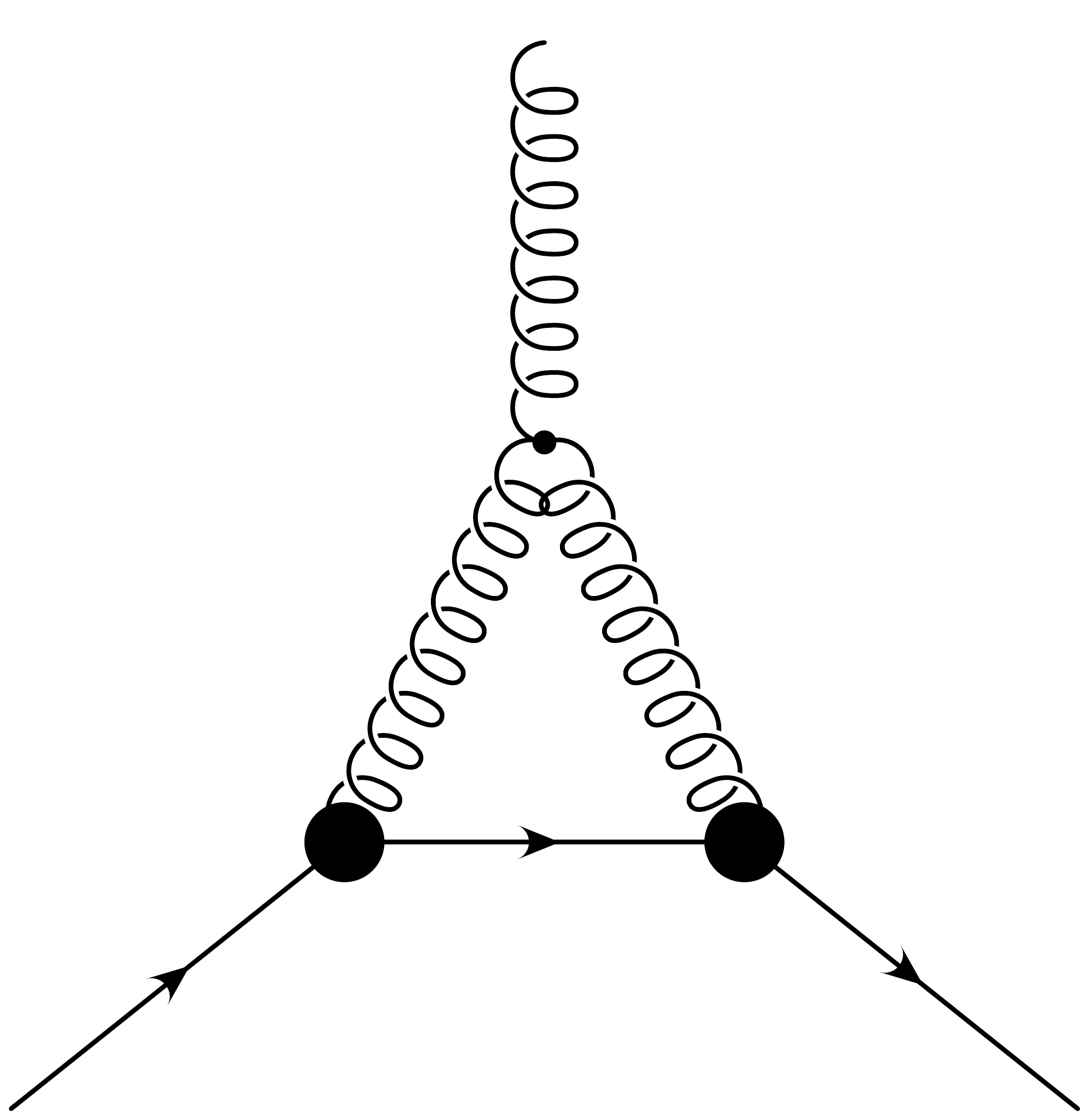}
\end{minipage}
\begin{minipage}{0.72\linewidth}
     \begin{center}
  \begin{tabular}{| l | l | l | l | l |}
    \hline
	      & group invariant & SU(2) & SU(3) & G$_2$ \\ \hline
    adjoint  & $-\frac{1}{2}C_{A}f^{abc}$ & $-f^{abc}$ & $-\frac{3}{2}f^{abc}$ & $-f^{abc}$ \\[1mm] 
    fund.  & $-\frac{1}{8}C_{A}t^a_{ij}$ & $-\frac{1}{4}t^a_{ij}$ & $-\frac{3}{8}t^a_{ij}$ & $-\frac{1}{4}t^a_{ij}$ \\[1mm]
\hline
  \end{tabular}
\end{center}
\end{minipage}

\no\begin{minipage}{0.18\linewidth}
     \includegraphics[width=1.0\linewidth]{sgv2.pdf}
\end{minipage}
\begin{minipage}{0.72\linewidth}
     \begin{center}
  \begin{tabular}{| l | l | l | l | l |}
    \hline
	      & group invariant & SU(2) & SU(3) & G$_2$ \\ \hline
    adjoint  & 0 & 0 & 0 & 0 \\ 
    fundamental  & $\frac{1}{6}t^a_{ij}$ & $\frac{1}{6}t^a_{ij}$ & $\frac{1}{6}t^a_{ij}$ 
		  & $\frac{1}{6}t^a_{ij}$ \\[1mm]
\hline
  \end{tabular}
\end{center}
\end{minipage}

\subsection{Four-point equations}

\subsubsection{4-gluon vertex DSE}

\no\begin{minipage}{0.18\linewidth}
     \includegraphics[width=1.0\linewidth]{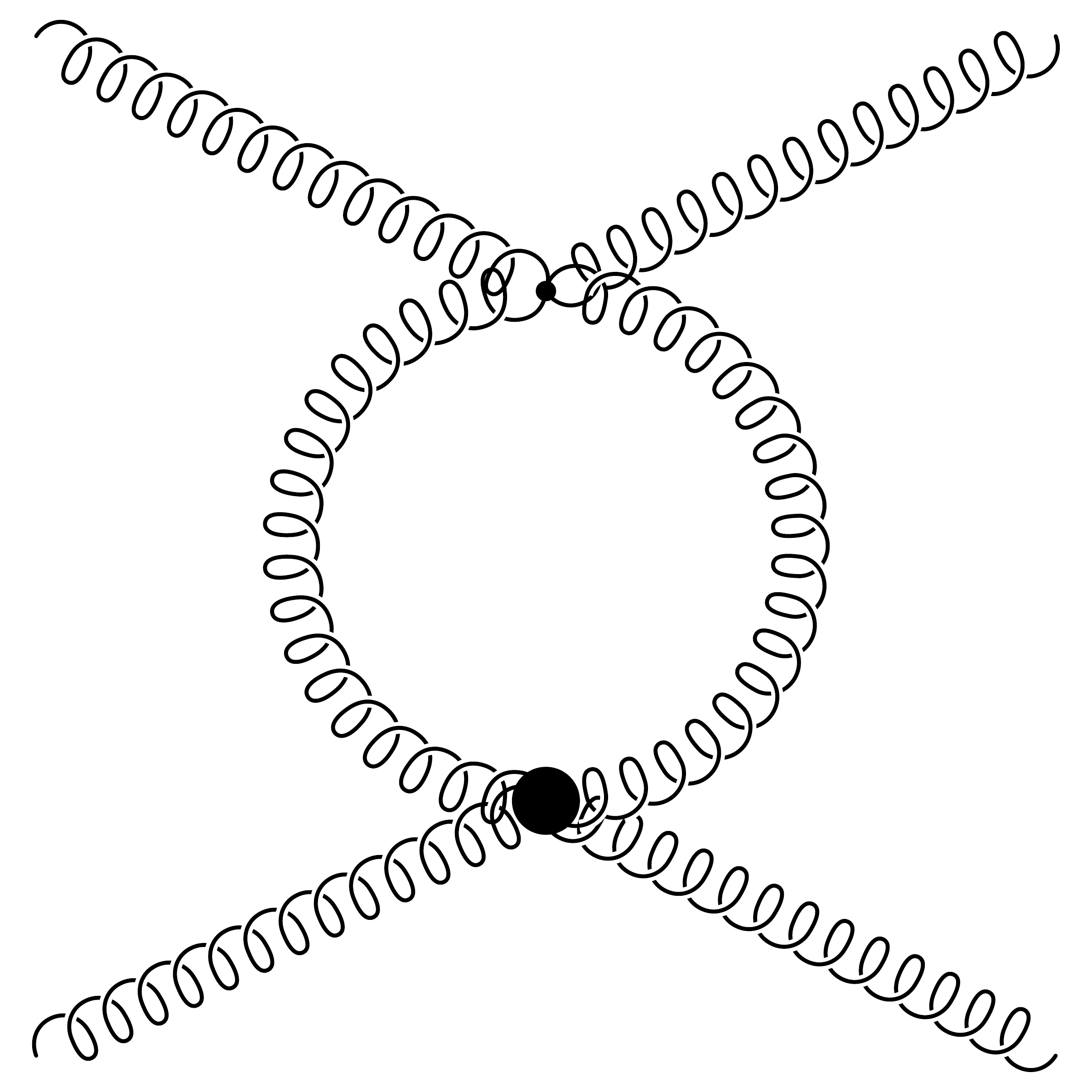}
\end{minipage}
\begin{minipage}{0.72\linewidth}
     \begin{center}
  \begin{tabular}{| l | l | l | l | l |}
    \hline
	adj.      & group invariant & SU(2)  \\ \hline
    $1^{st}$  & $\frac{N_{A}C_{A}^{3}}{4}\left(A(D-3F-3G)+\right.$ & $6\left(A(D-3F-3G)+\right.$ \\
	      & $\left.+3C(-D+2F+G)+\right.$ & $\left.+3C(-D+2F+G)+\right.$ \\
	      & $\left.+B(-2D+3F+G)\right)$ & $\left.+B(-2D+3F+G)\right)$  \\[1mm]

    $2^{nd}$  & $-\frac{1}{2}N_{A}C_{A}^{3}\left(A(D-F)+\right.$  & $-6\left(A(D-F)+\right.$  \\ 
	      & $\left.+B(G+F)+\right.$ & $\left.+B(G+F)+\right.$  \\
	      &	$\left.+C(-D+G+2F)\right)$ & $\left.+C(-D+G+2F)\right)$  \\
    $3^{rd}$  & $\frac{5}{2}N_{A}C_{A}^{2}(A+B)(D+G)$  & $30(A+B)(D+G)$  \\[1mm]

    $4^{th}$  & $\frac{1}{4}N_{A}C_{A}^{3}(2N_{A}-3)\left(A(D-F)+\right.$  & $18\left(A(D-F)+B(G+F)+\right.$  \\ 
	      & $\left.+B(G+F)+C(-D+G+2F)\right)$ & $\left.+C(-D+G+2F)\right)$ \\[1mm]

    $5^{th}$  & $-\frac{1}{12}N_{A}C_{A}^{3}(2N_{A}-1)\times$ 
	      & $-10\left(A(D-2G-3F)+\right.$ \\
	      & $\left. \times(A(D-2G-3F)+\right.$ & $\left.+3C(-D+G+2F)+\right.$  \\
	      & $\left.+3C(-D+G+2F)+\right.$ & $\left.+B(-2D+G+3F)\right)$  \\
	      & $\left.+B(-2D+G+3F)\right)$ &  \\[1mm]

    $6^{th}$  & $\frac{(A+B)(D+G)}{2T_{R}(C_{A}-6C_{F})}\left(5N_{A}T_{R}C_{A}^{2}\times \right. $ & $\frac{3825}{128}(A+B)(D+G)$  \\
	      & $\left. \times(C_{A}-6C_{F})+12d_{44}(2+N_{A}) \right)$ &   \\
    \hline
  \end{tabular}
\end{center}

\begin{center}
 \begin{tabular} {| l | l | l | l | l |}
     \hline
	      & SU(3) & G$_2$ \\ \hline
    $1^{st}$  & $54\left(A(D-3F-3G)+\right.$ & $28\left(A(D-3F-3G)+\right.$ \\ 
	      & $\left.+3C(-D+2F+G)+\right.$ & $\left.+3C(-D+2F+G)+\right.$ \\
	      & $\left.+B(-2D+3F+G)\right)$ & $\left.+B(-2D+3F+G)\right)$ \\[1mm]

    $2^{nd}$  & $-54\left(A(D-F)+\right.$ & $-28\left(A(D-F)+\right.$ \\ 
	      & $\left.+B(G+F)+\right.$ & $\left.+B(G+F)+\right.$ \\
	      & $\left.+C(-D+G+2F)\right)$ & $\left.+C(-D+G+2F)\right)$ \\
    $3^{rd}$  & $180(A+B)(D+G)$ & $140(A+B)(D+G)$ \\[1mm]

    $4^{th}$  & $702\left(A(D-F)+B(G+F)+\right.$ & $700\left(A(D-F)+B(G+F)+\right.$ \\ 
	      & $\left.+C(-D+G+2F)\right)$ & $\left.+C(-D+G+2F)\right)$ \\[1mm]

    $5^{th}$  & $-270\left(A(D-2G-3F)+\right.$ & $-252\left(A(D-2G-3F)+\right.$ \\
	      & $\left.+3C(-D+G+2F)+\right.$ & $\left.+3C(-D+G+2F)+\right.$ \\
	      & $\left.+B(-2D+G+3F)\right)$ & $\left.+B(-2D+G+3F)\right)$ \\[1mm]

    $6^{th}$  & $\frac{1435}{8}(A+B)(D+G)$ & $\frac{1113}{8}(A+B)(D+G)$ \\[1mm]
    \hline
  \end{tabular}
\end{center}
\end{minipage}

\no\begin{minipage}{0.18\linewidth}
     \includegraphics[width=1.0\linewidth]{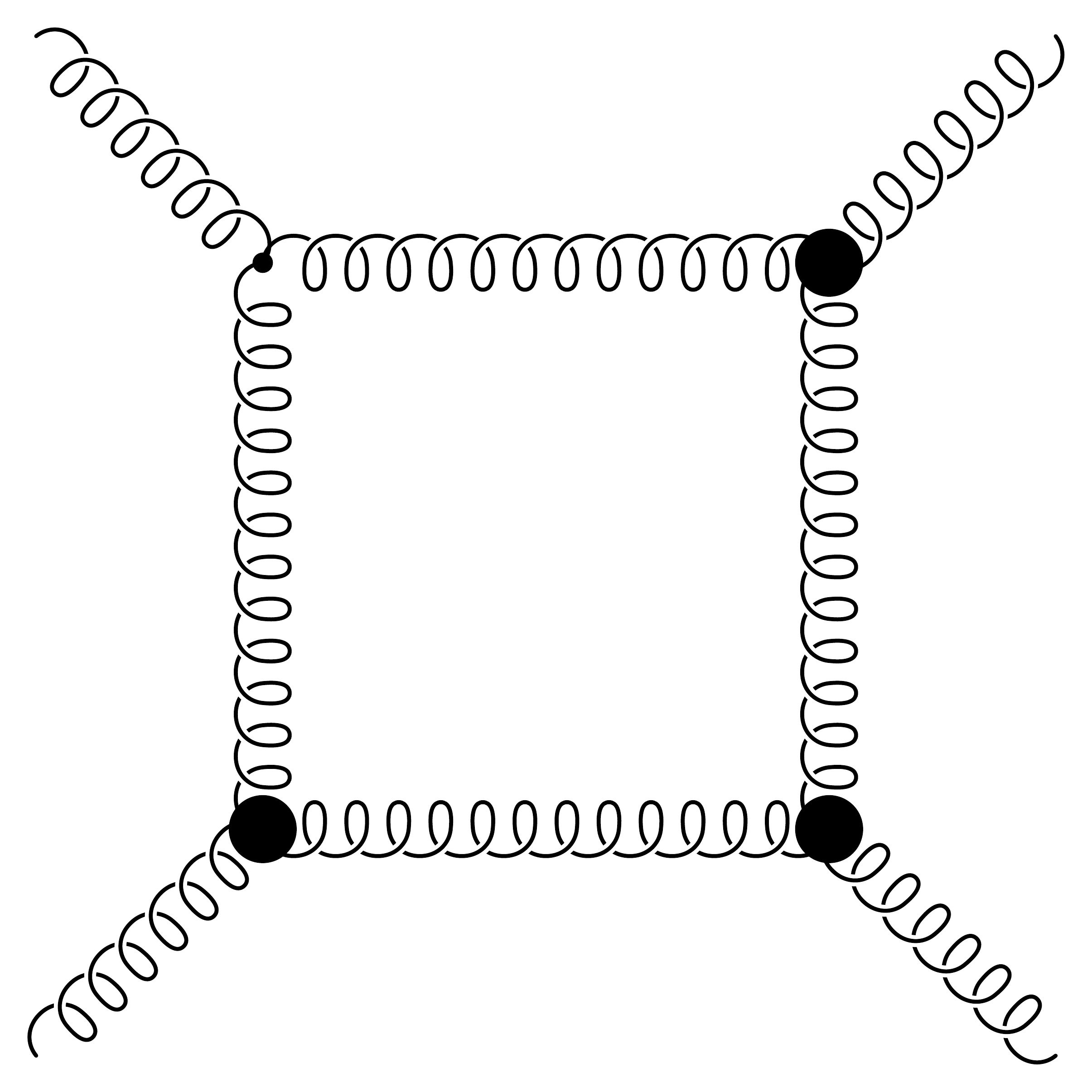}
\end{minipage}
\begin{minipage}{0.72\linewidth}
     \begin{center}
  \begin{tabular}{| l | l | l | l | l |}
    \hline
	adj.      & group invariant & SU(2) & SU(3) & G$_2$ \\ \hline
    $1^{st}$  & $\frac{1}{4}N_{A}C_{A}^{3}$ & $6$ & $54$ & $28$ \\ 
    $2^{nd}$  & $-\frac{1}{4}N_{A}C_{A}^{3}$  & $-6$ & $-54$ & $-28$ \\  
    $3^{rd}$  & $\frac{5}{2}N_{A}C_{A}^{2}$  & $30$ & $180$ & $140$ \\[1mm]  
    $4^{th}$  & $\frac{1}{4}N_{A}C_{A}^{3}(2N_{A}-3)$   & $18$ & $702$ & $700$ \\ 
    $5^{th}$  & $-\frac{(1)}{12}N_{A}C_{A}^{3}(-2N_{A}+1)$ & $-10$ & $-270$ & $-252$ \\
    $6^{th}$  & $\frac{1}{2T_{R}(C_{A}-6C_{F})}\left(5N_{A}T_{R}C_{A}^{2}\times \right. $ & $\frac{3825}{128}$ 
	      & $\frac{1435}{8}$ & $\frac{1113}{8}$ \\
	      & $\left. \times(C_{A}-6C_{F})+12d_{44}(2+N_{A}) \right)$ &  & & \\
    \hline
  \end{tabular}
\end{center}
\end{minipage}

\no\begin{minipage}{0.18\linewidth}
     \includegraphics[width=1.0\linewidth]{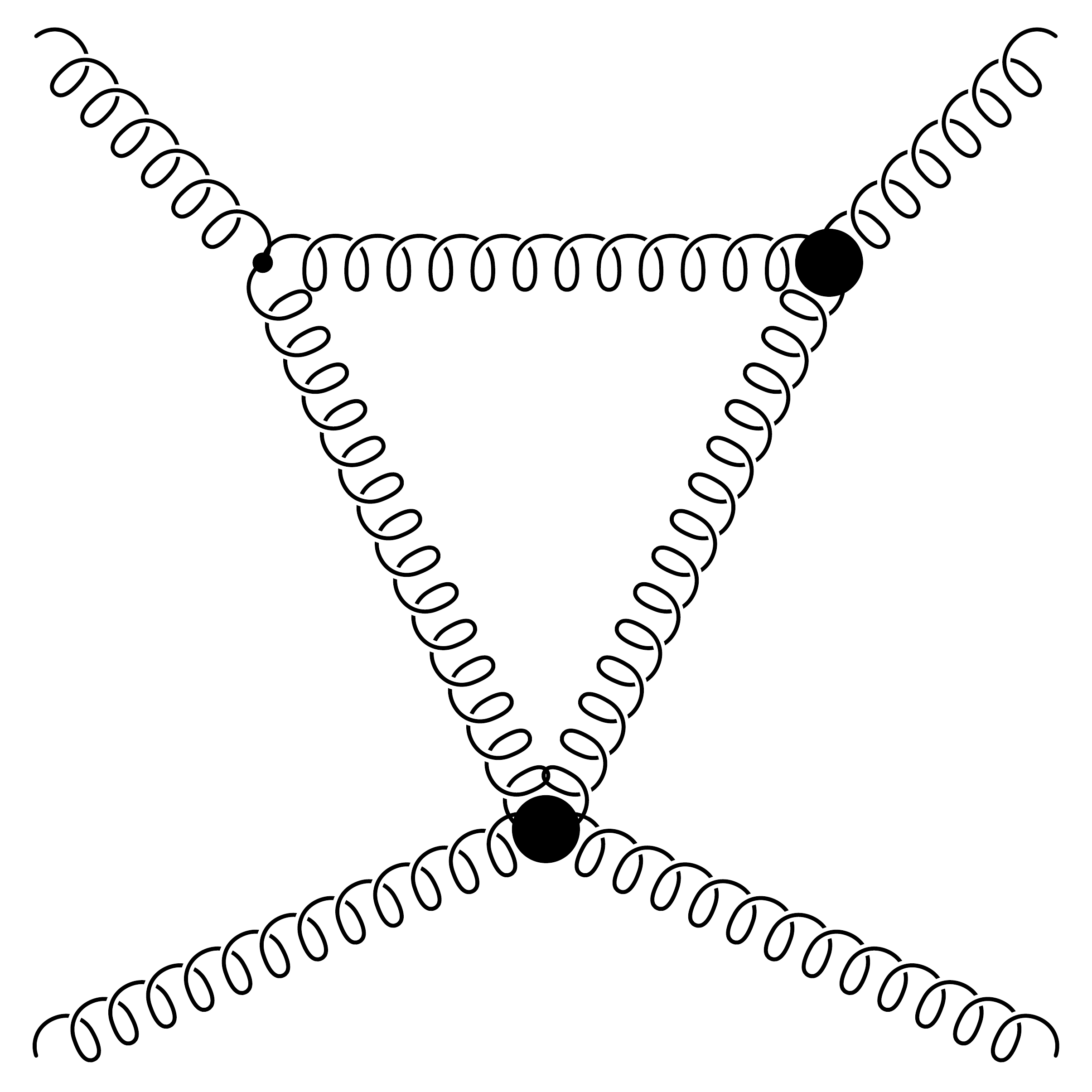}
\end{minipage}
\begin{minipage}{0.72\linewidth}
     \begin{center}
  \begin{tabular}{| l | l | l | l | l |}
    \hline
	adj.      & group invariant & SU(2) & SU(3) & G$_2$ \\ \hline
    $1^{st}$  & $\frac{(A-2B-3C)}{4}N_{A}C_{A}^{3}$ & $6\left(A-2B+\right.$ & $54\left(A-2B+\right.$ & $28\left(A-2B+\right.$ \\ 
	      &					& $\left.-3C\right) $ & $\left.-3C\right) $ & $\left. -3C\right)$ \\
    $2^{nd}$  & $\frac{(C-A)}{4}N_{A}C_{A}^{3}$  & $6(C-A)$ & $54(C-A)$ & $28(C-A)$ \\[1mm]  
    $3^{rd}$  & $\frac{5}{2}N_{A}C_{A}^{2}(A+B)$  & $30(A+B)$ & $180(A+B)$ & $140(A+B)$ \\[1mm]  
    $4^{th}$  & $\frac{(A-C)}{4}N_{A}C_{A}^{3}(2N_{A}-3)$   & $18(A-C)$ & $702(A-C)$ & $700(A-C)$ \\[1mm] 
    $5^{th}$  & $-\frac{(A-2B-3C)}{12}N_{A}C_{A}^{3}\times$ & $-10\left(A+\right.$ & $-270\left(A+\right.$ & $-252\left(A+\right.$ \\
	      &	$\times (2N_{A}-1)$		& $\left.-2B-3C\right)$ & $\left.-2B-3C\right)$ & $\left.-2B-3C\right)$ \\[1mm]
    $6^{th}$  & $\frac{(A+B)}{2T_{R}(C_{A}-6C_{F})}\times $ & $\frac{3825}{128}(A+B)$ 
	      & $\frac{1435}{8}(A+B)$ & $\frac{1113}{8}(A+B)$ \\
	      & $\times \left(5N_{A}T_{R}C_{A}^{2}\right.$ & & & \\
	      & $\left. \times(C_{A}-6C_{F})+\right.$ & & & \\
	      & $\left.+12d_{44}(2+N_{A}) \right)$ &  & & \\
    \hline
  \end{tabular}
\end{center}
\end{minipage}

\no\begin{minipage}{0.18\linewidth}
     \includegraphics[width=1.0\linewidth]{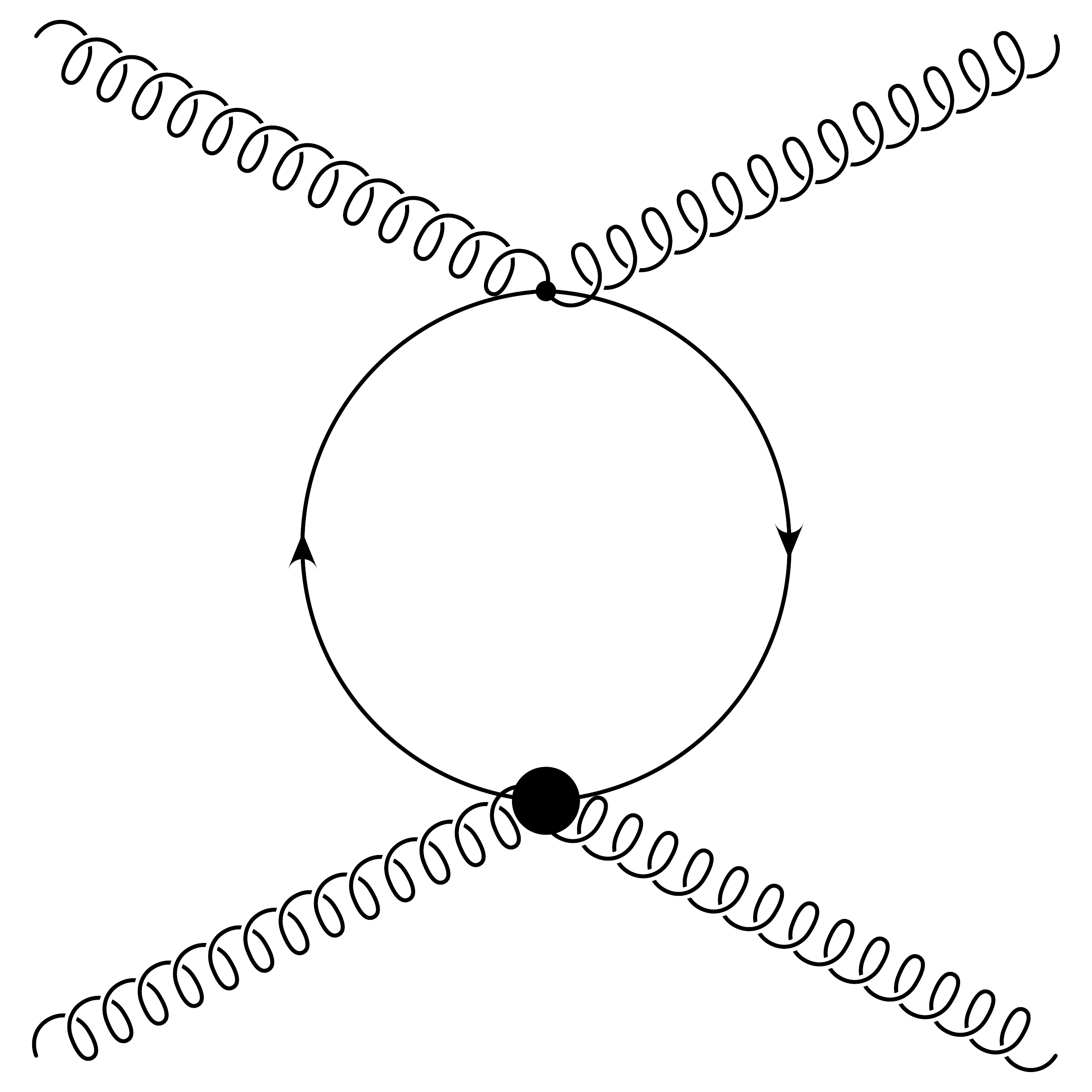}
\end{minipage}
\begin{minipage}{0.72\linewidth}
     \begin{center}
  \begin{tabular}{| l | l | l | l | l |}
    \hline
	adj.      & group invariant & SU(2) & SU(3) & G$_2$ \\ \hline
    $1^{st}$  & $-\frac{1}{2}N_{A}C_{A}^{3}$ & -12 & -108 & -56 \\[1mm] 
    $2^{nd}$  & $-\frac{1}{2}N_{A}C_{A}^{3}$  & -12 & -108 & -56 \\  
    $3^{rd}$  & $10N_{A}C_{A}^{2}$  & 120 & 720 & 560 \\  
    $4^{th}$  & $\frac{1}{2}N_{A}C_{A}^{3}(2N_{A}-3)$   & 36 & 1404 & 1400 \\[1mm] 
    $5^{th}$  & $\frac{1}{6}N_{A}C_{A}^{3}(2N_{A}-1)$ & 20 & 540 & 504 \\[1mm]
    $6^{th}$  & $\frac{2}{T_{R}(C_{A}-6C_{F})}\left(5N_{A}T_{R}C_{A}^{2}\times \right. $ & $\frac{3825}{32}$ 
	      & $\frac{1435}{2}$ & $\frac{1113}{2}$ \\
	      & $\left. \times(C_{A}-6C_{F})+12d_{44}(2+N_{A}) \right)$ &  & & \\
    \hline
  \end{tabular}
\end{center}

\begin{center}
  \begin{tabular}{| l | l | l | l | l |}
    \hline
	fund.      & group invariant & SU(2) & SU(3) & G$_2$ \\ \hline
    $1^{st}$  & $-\frac{1}{16}N_{A}T_{R}C_{A}^{2}$ & $-\frac{3}{8}$ & $-\frac{9}{4}$ & $-\frac{7}{4}$ \\[1mm] 
    $2^{nd}$  & $-\frac{1}{16}N_{A}T_{R}C_{A}^{2}$  & $-\frac{3}{8}$ & $-\frac{9}{4}$ & $-\frac{7}{4}$ \\[1mm]  
    $3^{rd}$  & $-\frac{1}{4}N_{A}T_{R}(C_{A}-6C_{F})$  & $\frac{15}{16}$ & 5 & 7 \\[1mm]  
    $4^{th}$  & $\frac{1}{16}N_{A}T_{R}C_{A}^{2}(2N_{A}-3)$   & $\frac{9}{8}$ & $\frac{117}{4}$ & $\frac{175}{4}$ \\[1mm] 
    $5^{th}$  & $\frac{1}{48}N_{A}T_{R}C_{A}^{2}(2N_{A}-1)$ & $\frac{5}{8}$ & $\frac{45}{4}$ & $\frac{63}{4}$ \\[1mm]
    $6^{th}$  & $\frac{1}{192T_{R}(C_{A}-6C_{F})}\left(48d_{44}(2+N_{A})+  \right.$ & $\frac{2235}{1024}$ 
	      & $\frac{3425}{576}$ & $-\frac{455}{192}$ \\
	      & $\left.+N_{A}T_{R}\left(-59C_{A}^{3}(2+N_{A})+\right.\right.$ & & & \\
	      & $\left.\left.+576C_{F}^{2}\left(-3T_{R}+C_{F}(2+N_{A})\right)\right.\right.$ & & & \\
	      & $\left.\left.-72C_{A}C_{F}\left(-8T_{R}+9C_{F}(2+N_{A})\right)+\right.\right.$ &  & & \\
	      & $\left.\left.+6C_{A}^{2}(-8T_{R}+51C_{F}(2+N_{A}))\right)\right)$ & & & \\
    \hline
  \end{tabular}
\end{center}
\end{minipage}

\no\begin{minipage}{0.18\linewidth}
     \includegraphics[width=1.0\linewidth]{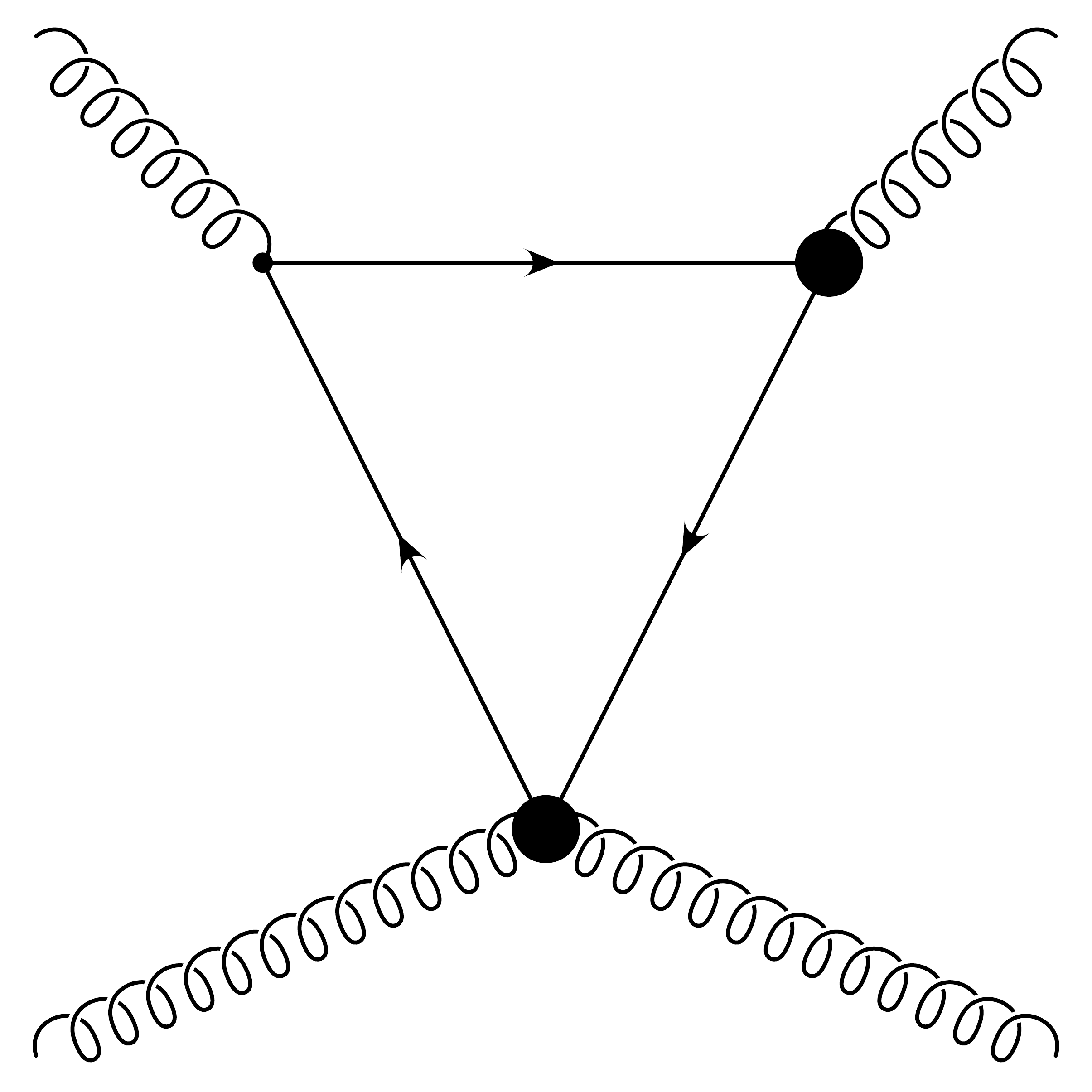}
\end{minipage}
\begin{minipage}{0.72\linewidth}
     \begin{center}
  \begin{tabular}{| l | l | l | l | l |}
    \hline
	adj.      & group invariant & SU(2) & SU(3) & G$_2$ \\ \hline
    $1^{st}$  & $-\frac{1}{4}N_{A}C_{A}^{3}$ & -6 & -54 & -28 \\[1mm] 
    $2^{nd}$  & $-\frac{1}{4}N_{A}C_{A}^{3}$  & -6 & -54 & -28 \\[1mm]  
    $3^{rd}$  & $5N_{A}C_{A}^{2}$  & 60 & 360 & 280 \\  
    $4^{th}$  & $\frac{1}{4}N_{A}C_{A}^{3}(2N_{A}-3)$   & 18 & 702 & 700 \\[1mm] 
    $5^{th}$  & $\frac{1}{12}N_{A}C_{A}^{3}(2N_{A}-1)$ & 10 & 270 & 252 \\[1mm]
    $6^{th}$  & $\frac{1}{T_{R}(C_{A}-6C_{F})}\left(5N_{A}T_{R}C_{A}^{2}\times \right. $ & $\frac{3825}{64}$ 
	      & $\frac{1435}{4}$ & $\frac{1113}{4}$ \\
	      & $\left. \times(C_{A}-6C_{F})+12d_{44}(2+N_{A}) \right)$ &  & & \\
    \hline
  \end{tabular}
\end{center}

\begin{center}
  \begin{tabular}{| l | l | l | l | l |}
    \hline
	fund.      & group invariant & SU(2) & SU(3) & G$_2$ \\ \hline
    $1^{st}$  & $\frac{1}{16}N_{A}T_{R}C_{A}^{2}$ & $\frac{3}{8}$ & $\frac{9}{4}$ & $\frac{7}{4}$ \\[1mm] 
    $2^{nd}$  & $\frac{1}{16}N_{A}T_{R}C_{A}^{2}$  & $\frac{3}{8}$ & $\frac{9}{4}$ & $\frac{7}{4}$ \\[1mm]  
    $3^{rd}$  & $\frac{1}{4}N_{A}T_{R}(C_{A}-6C_{F})$  & $-\frac{15}{16}$ & -5 & -7 \\[1mm]  
    $4^{th}$  & $\frac{1}{16}N_{A}T_{R}C_{A}^{2}(-2N_{A}+3)$   & $-\frac{9}{8}$ & $-\frac{117}{4}$ & $-\frac{175}{4}$ \\[1mm] 
    $5^{th}$  & $\frac{1}{48}N_{A}T_{R}C_{A}^{2}(-2N_{A}+1)$ & $-\frac{5}{8}$ & $-\frac{45}{4}$ & $-\frac{63}{4}$ \\[1mm]
    $6^{th}$  & $\frac{1}{96T_{R}(C_{A}-6C_{F})}\left(-24d_{44}(2+N_{A})+ N_{A}T_{R}\times \right.$ & $-\frac{2139}{1024}$ 
	      & $-\frac{3089}{576}$ & $\frac{455}{192}$ \\
	      & $\left.\left(31C_{A}^{3}(2+N_{A})-288C_{F}^{2}\left(-3T_{R}+C_{F}(2+N_{A})\right)+\right.\right.$ & & & \\
	      & $\left.\left.+48C_{A}C_{F}\left(-6T_{R}+7C_{F}(2+N_{A})\right)+\right.\right.$ &  & & \\
	      & $\left.\left.-6C_{A}^{2}(-4T_{R}+27C_{F}(2+N_{A}))\right)\right)$ & & & \\
    \hline
  \end{tabular}
\end{center}
\end{minipage}

\no\begin{minipage}{0.18\linewidth}
     \includegraphics[width=1.0\linewidth]{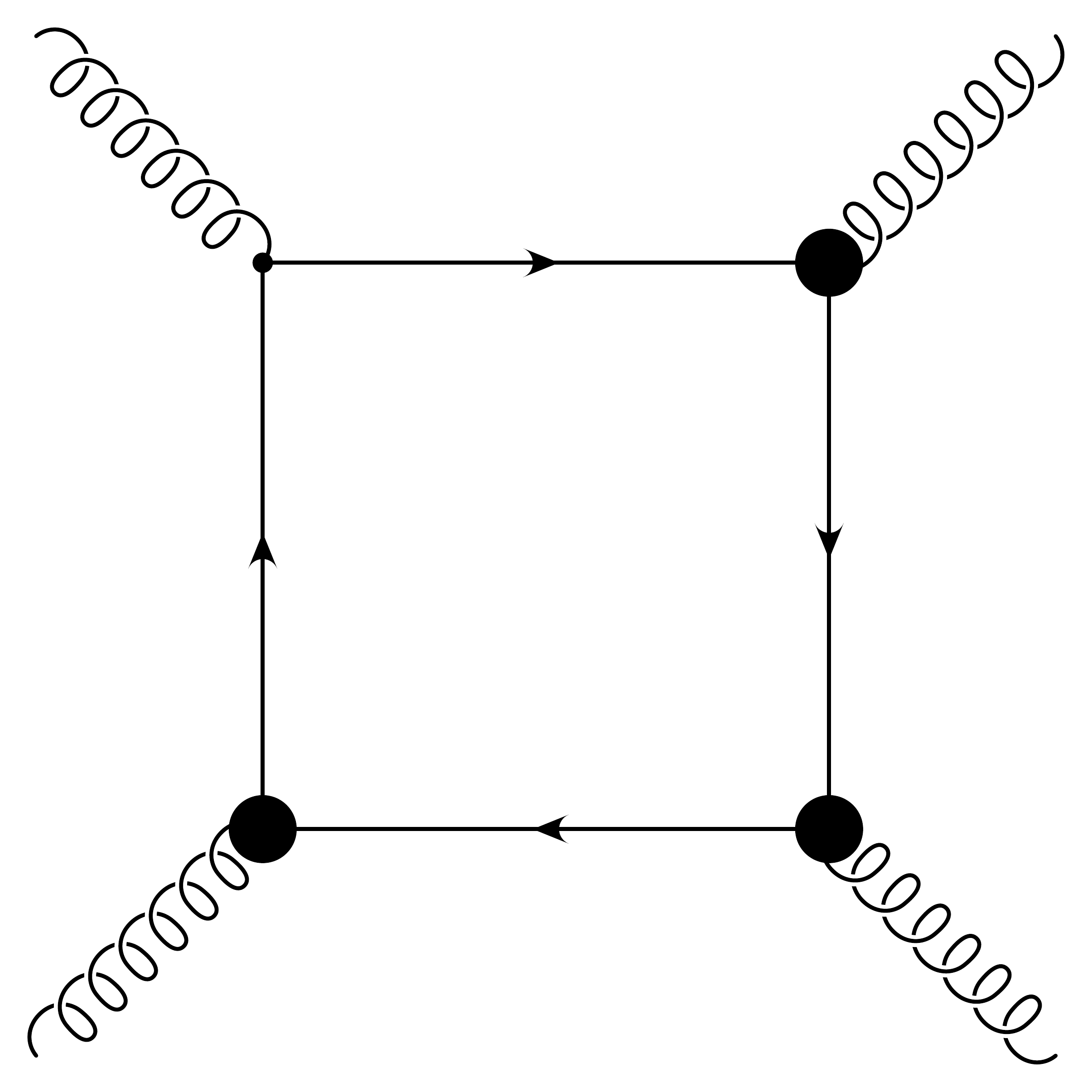}
\end{minipage}
\begin{minipage}{0.72\linewidth}
     \begin{center}
  \begin{tabular}{| l | l | l | l | l |}
    \hline
	adj.      & group invariant & SU(2) & SU(3) & G$_2$ \\ \hline
    $1^{st}$  & $\frac{1}{4}N_{A}C_{A}^{3}$ & $6$ & $54$ & $28$ \\[1mm] 
    $2^{nd}$  & $-\frac{1}{4}N_{A}C_{A}^{3}$  & $-6$ & $-54$ & $-28$ \\[1mm]  
    $3^{rd}$  & $\frac{5}{2}N_{A}C_{A}^{2}$  & $30$ & $180$ & $140$ \\[1mm]  
    $4^{th}$  & $\frac{1}{4}N_{A}C_{A}^{3}(2N_{A}-3)$   & $18$ & $702$ & $700$ \\[1mm] 
    $5^{th}$  & $-\frac{1}{12}N_{A}C_{A}^{3}(-2N_{A}+1)$ & $-10$ & $-270$ & $-252$ \\[1mm]
    $6^{th}$  & $\frac{1}{2T_{R}(C_{A}-6C_{F})}\left(5N_{A}T_{R}C_{A}^{2}\times \right. $ & $\frac{3825}{128}$ 
	      & $\frac{1435}{8}$ & $\frac{1113}{8}$ \\
	      & $\left. \times(C_{A}-6C_{F})+12d_{44}(2+N_{A}) \right)$ &  & & \\
    \hline
  \end{tabular}
\end{center}

 \begin{center}
  \begin{tabular}{| l | l | l | l | l |}
    \hline
	fund.      & group invariant & SU(2) & SU(3) & G$_2$ \\ \hline
    $1^{st}$  & $\frac{1}{4}N_{A}T_{R}C_{A}^{2}$ & $\frac{3}{2}$ & 9 & 7 \\[1mm] 
    $2^{nd}$  & $-\frac{1}{4}N_{A}T_{R}C_{A}^{2}$  & $-\frac{3}{2}$ & -9 & -7 \\[1mm]  
    $3^{rd}$  & $-\frac{1}{2}N_{A}T_{R}(C_{A}-6C_{F})$  & $\frac{15}{8}$ & 10 & 14 \\[1mm]  
    $4^{th}$  & $\frac{1}{4}N_{A}T_{R}C_{A}^{2}(2N_{A}-3)$   & $\frac{9}{2}$ & 117 & 175 \\[1mm] 
    $5^{th}$  & $\frac{1}{12}N_{A}T_{R}C_{A}^{2}(-2N_{A}+1)$ & $-\frac{5}{2}$ & -45 & -63 \\[1mm]
    $6^{th}$  & $\frac{1}{12T_{R}(C_{A}-6C_{F})}\left(12d_{44}(2+N_{A})+ N_{A}T_{R}\times \right.$ & $\frac{9675}{256}$ 
	      & $-\frac{39505}{144}$ & $\frac{12313}{48}$ \\
	      & $\left.\left(-35C_{A}^{3}(2+N_{A})+72C_{F}^{2}\left(-3T_{R}+C_{F}(2+N_{A})\right)+\right.\right.$ & & & \\
	      & $\left.\left.+18C_{A}C_{F}\left(-4T_{R}+9C_{F}(2+N_{A})\right)+\right.\right.$ &  & & \\
	      & $\left.\left.+6C_{A}^{2}(-T_{R}+20C_{F}(2+N_{A}))\right)\right)$ & & & \\
    \hline
  \end{tabular}
\end{center}
\end{minipage}

\no\begin{minipage}{0.18\linewidth}
     \includegraphics[width=1.0\linewidth]{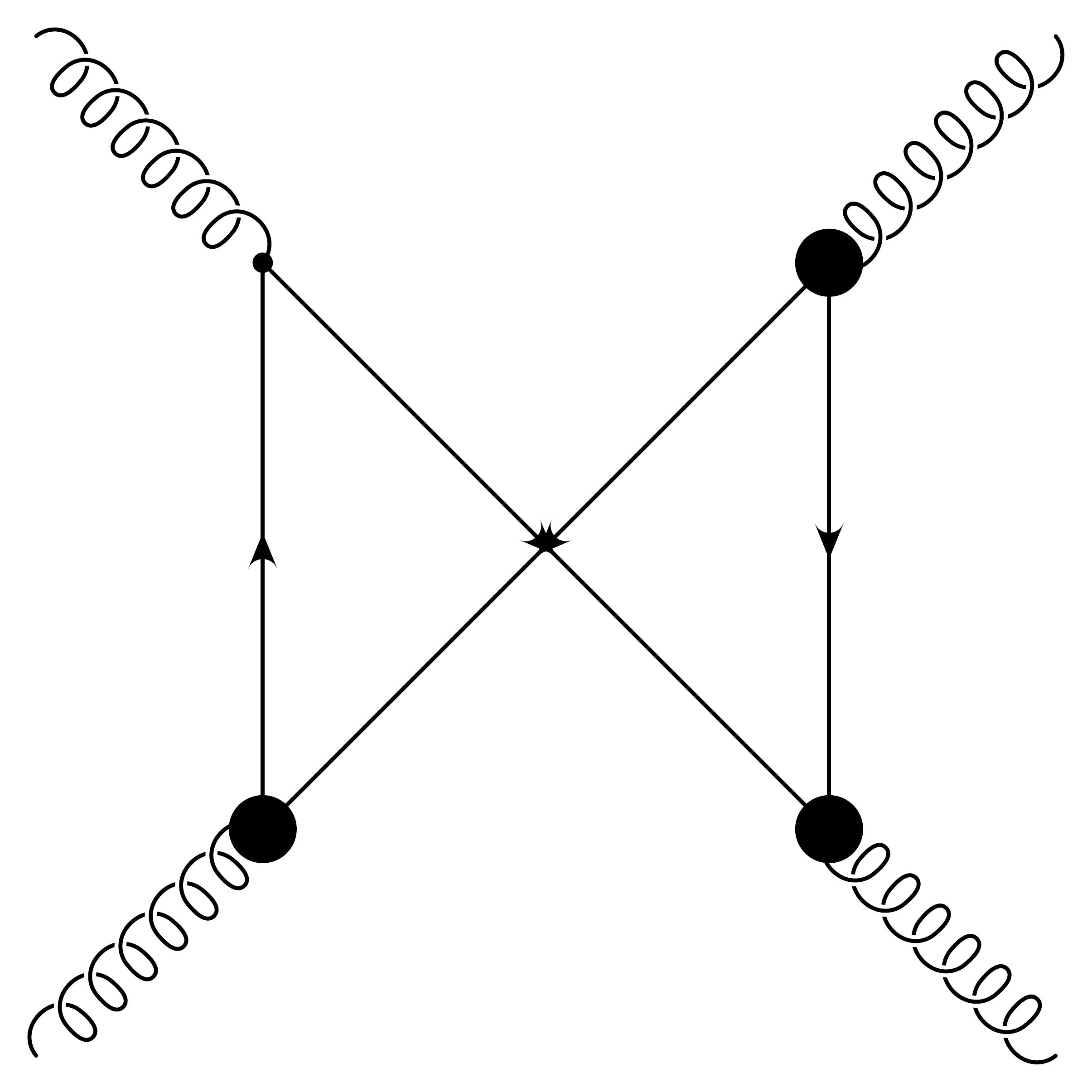}
\end{minipage}
\begin{minipage}{0.72\linewidth}
     \begin{center}
  \begin{tabular}{| l | l | l | l | l |}
    \hline
	adj.      & group invariant & SU(2) & SU(3) & G$_2$ \\ \hline
    $1^{st}$  & $-\frac{1}{2}N_{A}C_{A}^{3}$ & $-12$ & $-108$ & $-56$ \\ 
    $2^{nd}$  & 0  & 0 & 0 & 0 \\  
    $3^{rd}$  & $\frac{5}{2}N_{A}C_{A}^{2}$  & $30$ & $180$ & $140$ \\  
    $4^{th}$  & 0   & 0 & 0 & 0 \\ 
    $5^{th}$  & $\frac{1}{6}N_{A}C_{A}^{3}(2N_{A}-1)$ & $20$ & $540$ & $504$ \\[1mm]
    $6^{th}$  & $\frac{1}{2T_{R}(C_{A}-6C_{F})}\left(5N_{A}T_{R}C_{A}^{2}\times \right. $ & $\frac{3825}{128}$ 
	      & $\frac{1435}{8}$ & $\frac{1113}{8}$ \\
	      & $\left. \times(C_{A}-6C_{F})+12d_{44}(2+N_{A}) \right)$ &  & & \\
    \hline
  \end{tabular}
\end{center}

 \begin{center}
  \begin{tabular}{| l | l | l | l | l |}
    \hline
	fund.      & group invariant & SU(2) & SU(3) & G$_2$ \\ \hline
    $1^{st}$  & $-\frac{1}{2}N_{A}T_{R}C_{A}^{2}$ & -3 & -18 & -14 \\ 
    $2^{nd}$  & 0  & 0 & 0 & 0 \\  
    $3^{rd}$  & $-\frac{1}{2}N_{A}T_{R}(C_{A}-6C_{F})$  & $\frac{15}{8}$ & 10 & 14 \\  
    $4^{th}$  & 0   & 0 & 0 & 0 \\ 
    $5^{th}$  & $\frac{1}{6}N_{A}T_{R}C_{A}^{2}(2N_{A}-1)$ & 5 & 90 & 126 \\[1mm]
    $6^{th}$  & $\frac{1}{2T_{R}(C_{A}-6C_{F})}\left(12d_{44}(2+N_{A})+ N_{A}T_{R}\times \right.$ & $\frac{225}{128}$ 
	      & $-\frac{75}{8}$ & $\frac{105}{8}$ \\
	      & $\left.\left(-35C_{A}^{3}(2+N_{A})+72C_{F}^{2}\left(-3T_{R}+C_{F}(2+N_{A})\right)+\right.\right.$ & & & \\
	      & $\left.\left.+18C_{A}C_{F}\left(-4T_{R}+9C_{F}(2+N_{A})\right)+\right.\right.$ &  & & \\
	      & $\left.\left.+6C_{A}^{2}(-T_{R}+20C_{F}(2+N_{A}))\right)\right)$ & & & \\
    \hline
  \end{tabular}
\end{center}
\end{minipage}

\subsubsection{2-scalar-2-gluon vertex DSE}


\begin{minipage}{0.2\linewidth}
     \includegraphics[width=1.0\linewidth]{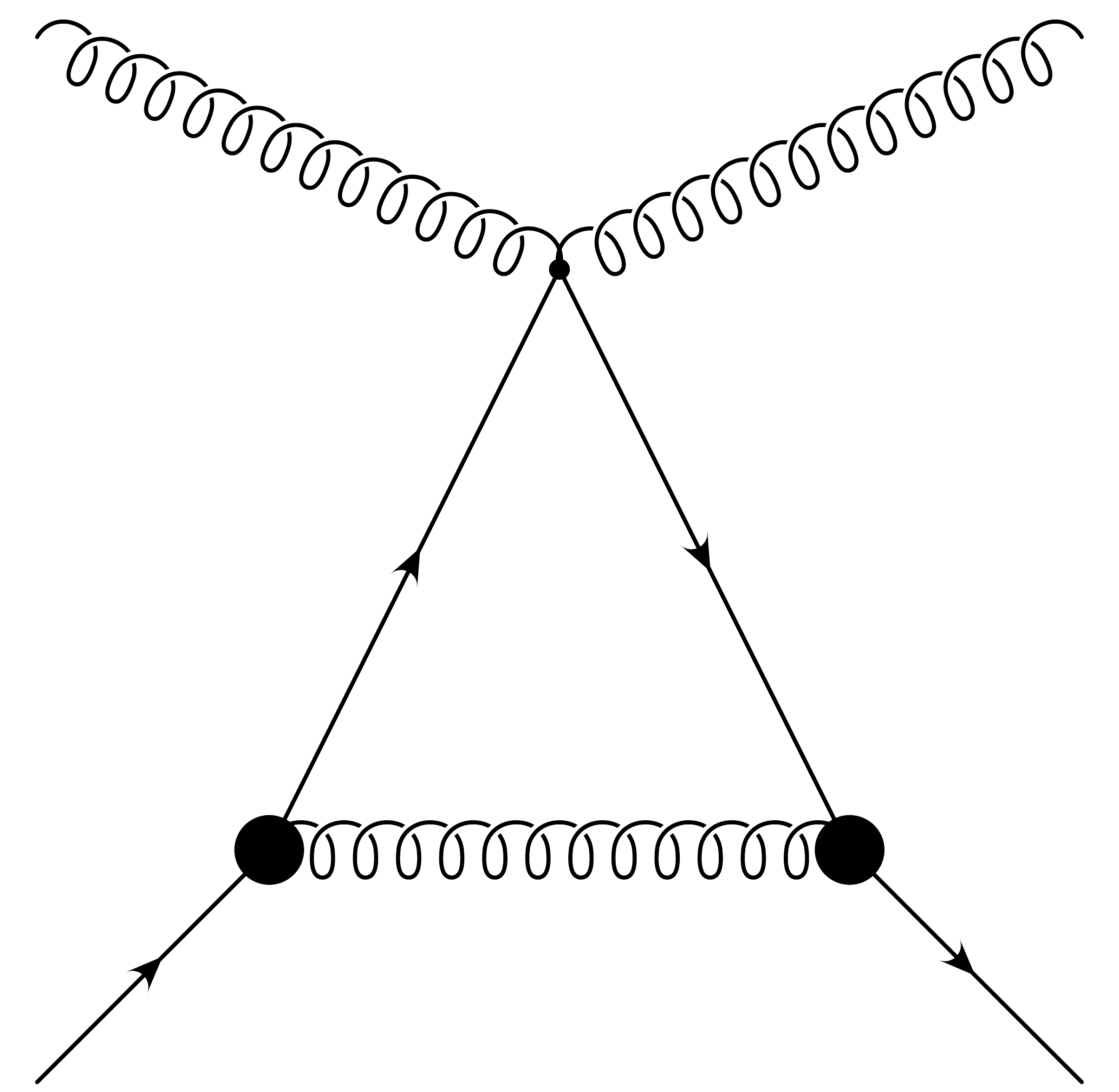}
\end{minipage}
\begin{minipage}{0.7\linewidth}
     \begin{center}
  \begin{tabular}{| l | l | l | l | l |}
    \hline
	adj.	& group invariant  & SU(2) & SU(3) & G$_2$ \\ \hline
    $1^{st}$  & $\frac{1}{2}N_{A}C_{A}^{3}$ & 12 & 108 & 56 \\  
    $2^{nd}$  & 0  & 0 & 0 & 0 \\  
    $3^{rd}$  & $5N_{A}C_{A}^{2}$   & 60 & 360 & 280 \\ 
    $4^{th}$  & $\frac{1}{6}N_{A}C_{A}^{3}(1-2N_{A})$   & -20 & -540 & -504 \\ 
    $5^{th}$  & 0   & 0 & 0 & 0 \\
    $6^{th}$  & $\frac{1}{T_{R}(C_{A}-6C_{F})}\left(5N_{A}T_{R}C_{A}^{2}\times \right. $ & $\frac{3825}{64}$ 
	      & $\frac{1435}{4}$ & $\frac{1113}{4}$ \\
	      & $\left. \times(C_{A}-6C_{F})+\right.$ &  & & \\
	      & $\left.+12d_{44}(2+N_{A}) \right)$ & & & \\
    \hline
  \end{tabular}
\end{center}
\begin{center}
  \begin{tabular}{| l | l | l | l | l |}
    \hline
	fund.	& group invariant  & SU(2) & SU(3) & G$_2$ \\ \hline
    $1^{st}$  & $-\frac{1}{4}N_{A}T_{R}C_{A}^{2}$ & $-\frac{3}{2}$ & $-9$ & $-7$\\[1mm]  
    $2^{nd}$  & $-\frac{1}{4}N_{A}T_{R}C_{A}^{2}$  & $-\frac{3}{2}$ & $-9$ & $-7$ \\[1mm] 
    $3^{rd}$  & $-\frac{1}{4}N_{A}T_{R}C_{A}\left(C_{A}+\right.$  & $\frac{9}{2}$ & 45 & 98 \\ 
	      & $\left.-4T_{R}(1+N_{A})\right)$ & & & \\[1mm]
    $4^{th}$  & $\frac{1}{4}N_{A}T_{R}C_{A}^{3}$   & 3 & 27 & 14 \\[1mm]  
    $5^{th}$  & $-\frac{N_{A}T_{R}C_{A}}{4N_{F}(1+N_{A})}\left(16\left(C_{F}+\right.\right.$ & $\frac{9}{8}$ & $-\frac{55}{27}$ 
	      & $-\frac{49}{15}$ \\
	      & $\left.\left.-T_{R}(1+N_{A})\right)+\right.$ & & & \\
	      & $\left.+C_{A}(N_{F}+N_{A}N_{F}-4)\right)$ & & & \\
    \hline
  \end{tabular}
\end{center}
\end{minipage}

\no\begin{minipage}{0.2\linewidth}
     \includegraphics[width=1.0\linewidth]{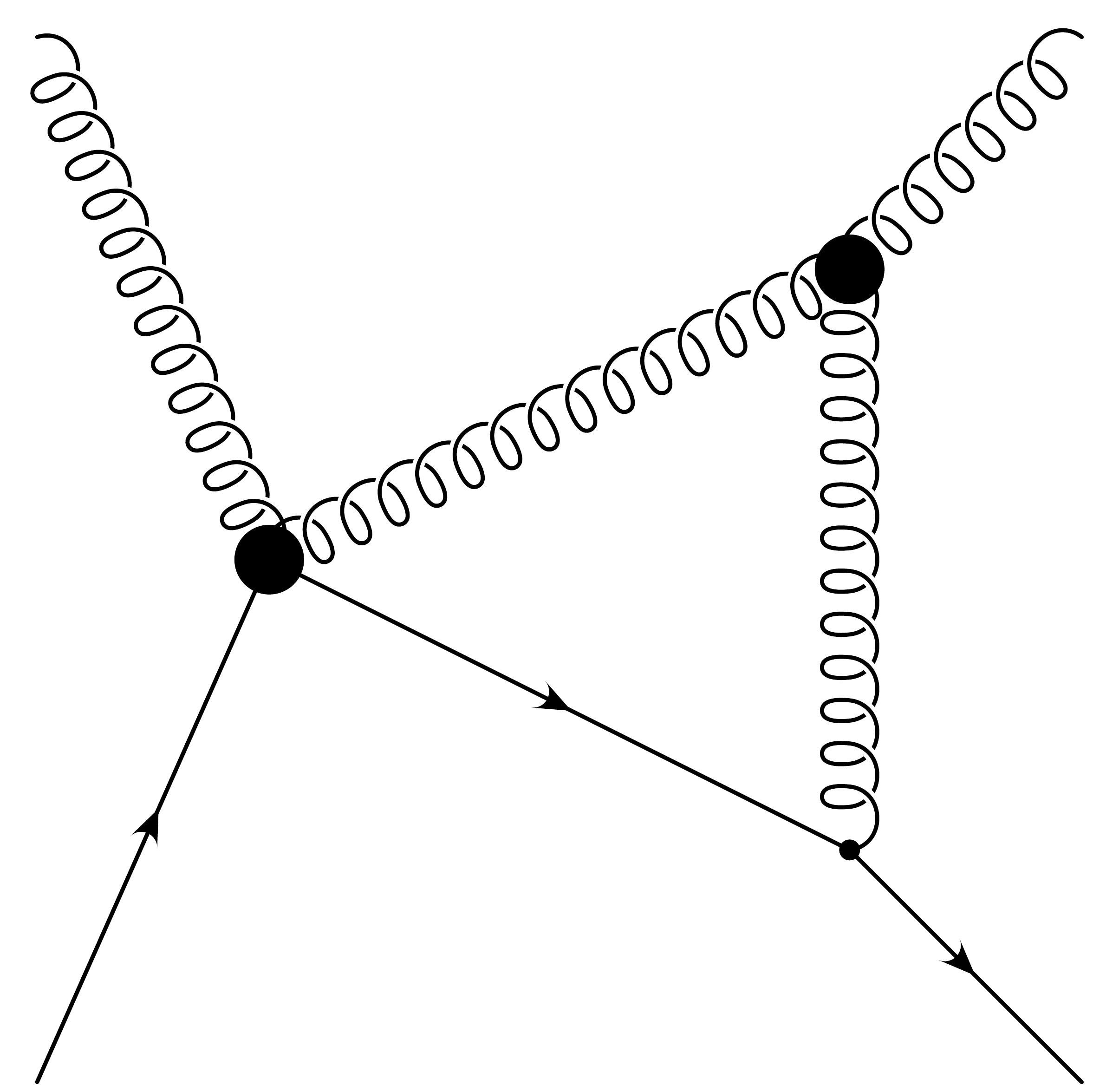}
\end{minipage}
\begin{minipage}{0.7\linewidth}
     \begin{center}
  \begin{tabular}{| l | l | l | l | l |}
    \hline
	adj.	& group invariant  & SU(2) & SU(3) & G$_2$ \\ \hline
    $1^{st}$  & $-\frac{5}{4} N_{A}C_{A}^{3}$ & -30 & -270 & -140 \\[1mm] 
    $2^{nd}$  & $\frac{1}{4} N_{A}C_{A}^{3}$  & 6 & 54 & 28 \\[1mm] 
    $3^{rd}$  & $\frac{5}{2} N_{A}C_{A}^{2}$  & 30 & 180 & 140 \\[1mm] 
    $4^{th}$  & $\frac{1}{4} N_{A}C_{A}^{3}(3-2N_{A})$   & -18 & -702 & -700 \\[1mm] 
    $5^{th}$  & $\frac{5}{12}N_{A}C_{A}^{3}(2N_{A}-1)$   & 50 & 1350 & 1260 \\[1mm]
    $6^{th}$  & $\frac{1}{2T_{R}(C_{A}-6C_{F})}\left(5N_{A}T_{R}\times \right.$ & $\frac{3825}{128}$ 
	      & $\frac{1435}{8}$ & $\frac{1113}{8}$ \\
	      & $\left.\times C_{A}^{2}(C_{A}-6C_{F})+\right.$ & & & \\
	      & $\left. +12d_{44}(2+N_{A}) \right)$ & & & \\
    \hline
  \end{tabular}
\end{center}
\begin{center}
  \begin{tabular}{| l | l | l | l | l |}
    \hline
	fund.	& group invariant  & SU(2) & SU(3) & G$_2$ \\ \hline
    $1^{st}$  & $\frac{1}{16}iN_{A}T_{R}C_{A}(3C_{A}-8C_{F})$ & 0 & $-\frac{5i}{4}$ & $-\frac{7i}{4}$\\[1mm] 
    $2^{nd}$  & $-\frac{1}{16}iN_{A}T_{R}C_{A}^{2}$  & $-\frac{3i}{8}$ & $-\frac{9i}{4}$ & $-\frac{7i}{4}$ \\[1mm] 
    $3^{rd}$  & $\frac{1}{16}iN_{A}T_{R}C_{A}(3C_{A}-8C_{F})$   & 0 & $-\frac{5i}{4}$ & $-\frac{7i}{4}$ \\[1mm] 
    $4^{th}$  & $\frac{1}{16}iN_{A}T_{R}C_{A}\left(C_{A}^{2}+\right.$ & $\frac{3i}{2}$ & $\frac{125i}{4}$ 
	      & $49i$ \\  
	      & $\left.-4C_{A}T_{R}(N_{A}-1)+\right.$ & & & \\
	      & $\left. +16C_{F}T_{R}(N_{A}-1)\right)$ & & & \\
    $5^{th}$  & $\frac{1}{16}iN_{A}T_{R}C_{A}(3C_{A}-8C_{F})$ & 0 & $-\frac{5i}{4}$ & $-\frac{7i}{4}$ \\
    \hline
  \end{tabular}
\end{center}
\end{minipage}

\no\begin{minipage}{0.2\linewidth}
     \includegraphics[width=1.0\linewidth]{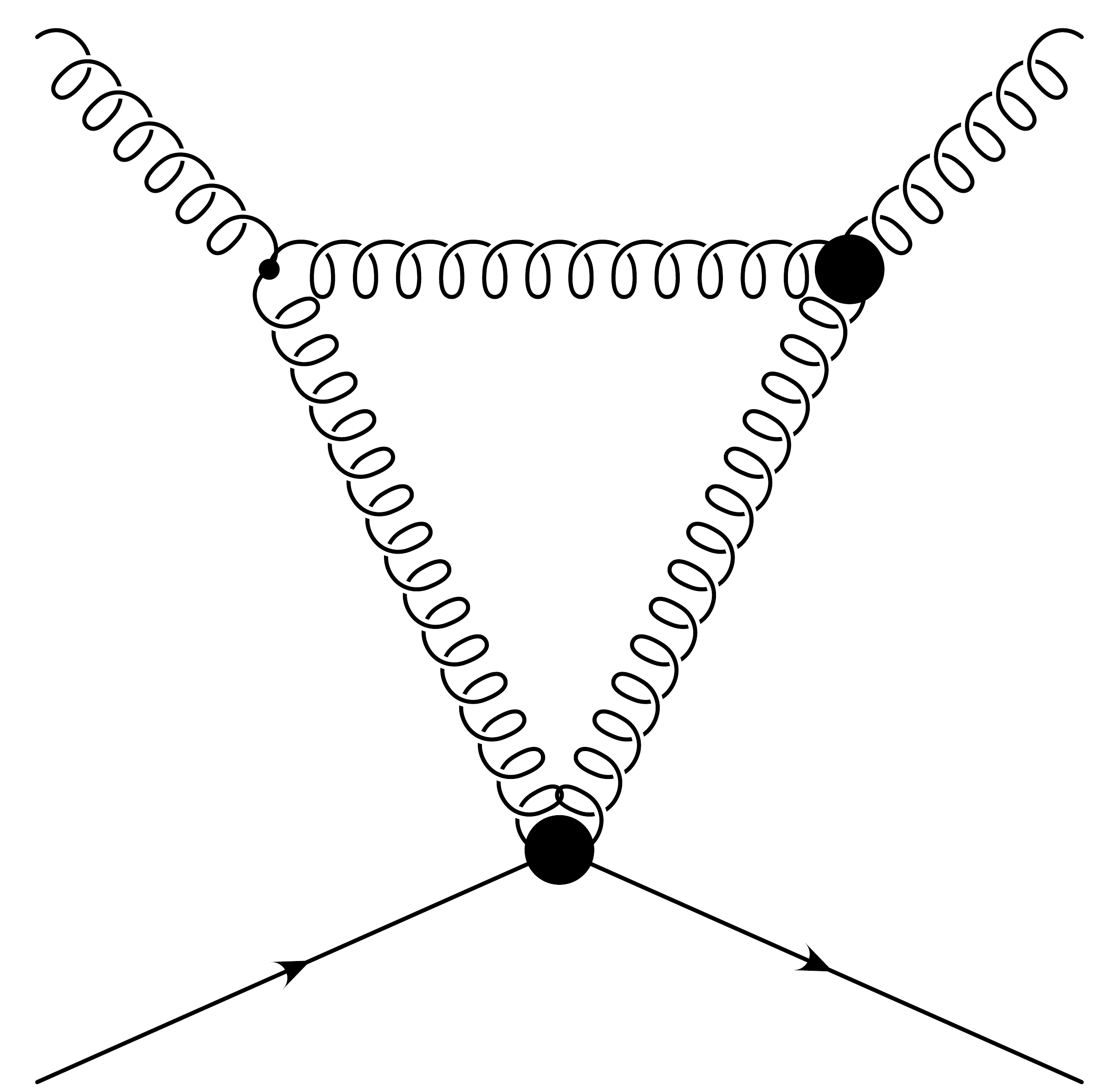}
\end{minipage}
\begin{minipage}{0.7\linewidth}
     \begin{center}
  \begin{tabular}{| l | l | l | l | l |}
    \hline
	adj.	& group invariant  & SU(2) & SU(3) & G$_2$ \\ \hline
    $1^{st}$  & $-\frac{1}{2}N_{A}C_{A}^{3}$ & -12 & -108 & -56 \\ 
    $2^{nd}$  & 0  & 0 & 0 & 0 \\  
    $3^{rd}$  & $-5N_{A}C_{A}^{2}$   & -60 & -360 & -280 \\ 
    $4^{th}$  & 0  & 0 & 0 & 0 \\ 
    $5^{th}$  & $\frac{1}{6}N_{A}C_{A}^{3}(2N_{A}-1)$   & 20 & 540 & 504 \\
    $6^{th}$  & $-\frac{1}{T_{R}(C_{A}-6C_{F})}\left(5N_{A}T_{R}C_{A}^{2}\times \right. $ & $-\frac{3825}{64}$ 
	      & $-\frac{1435}{4}$ & $-\frac{1113}{4}$ \\
	      & $\left. \times(C_{A}-6C_{F})+\right.$ &  & & \\
	      & $\left.+12d_{44}(2+N_{A}) \right)$ & & & \\
    \hline
  \end{tabular}
\end{center}
\begin{center}
  \begin{tabular}{| l | l | l | l | l |}
    \hline
	fund.	& group invariant  & SU(2) & SU(3) & G$_2$ \\ \hline
    $1^{st}$  & $-\frac{1}{2}iN_{A}T_{R}C_{A}(C_{A}-2C_{F})$ & $-\frac{3i}{4}$ & $-2i$ & 0\\ 
    $2^{nd}$  & 0  & 0 & 0 & 0 \\ 
    $3^{rd}$  & $-\frac{1}{2}iN_{A}T_{R}C_{A}\left(C_{A}-2C_{F}-T_{R}+\right.$  & $\frac{9i}{4}$ & $25i$ & $\frac{105i}{2}$ \\ 
	      &$\left.-T_{R}N_{A}\right)$ & & & \\
    $4^{th}$  & $-2iN_{A}T_{R}^{2}C_{A}C_{F}(N_{A}-1)$   & $-\frac{9i}{2}$ & $-112i$ & $-182i$ \\[1mm]
    $5^{th}$  & $-\frac{N_{A}T_{R}C_{A}}{2N_{F}(1+N_{A})}\left(C_{A}(N_{F}+N_{A}N_{F}-1)+\right.$ & $\frac{9i}{16}$ 
	      & $\frac{40i}{27}$ & $\frac{28i}{15}$ \\
	      & $\left.-4T_{R}(1+N_{A})+\right.$ & & & \\
	      & $\left.+2C_{F}(N_{F}+N_{A}N_{F}-2)\right)$ & & & \\
    \hline
  \end{tabular}
\end{center}
\end{minipage}

\no\begin{minipage}{0.2\linewidth}
     \includegraphics[width=1.0\linewidth]{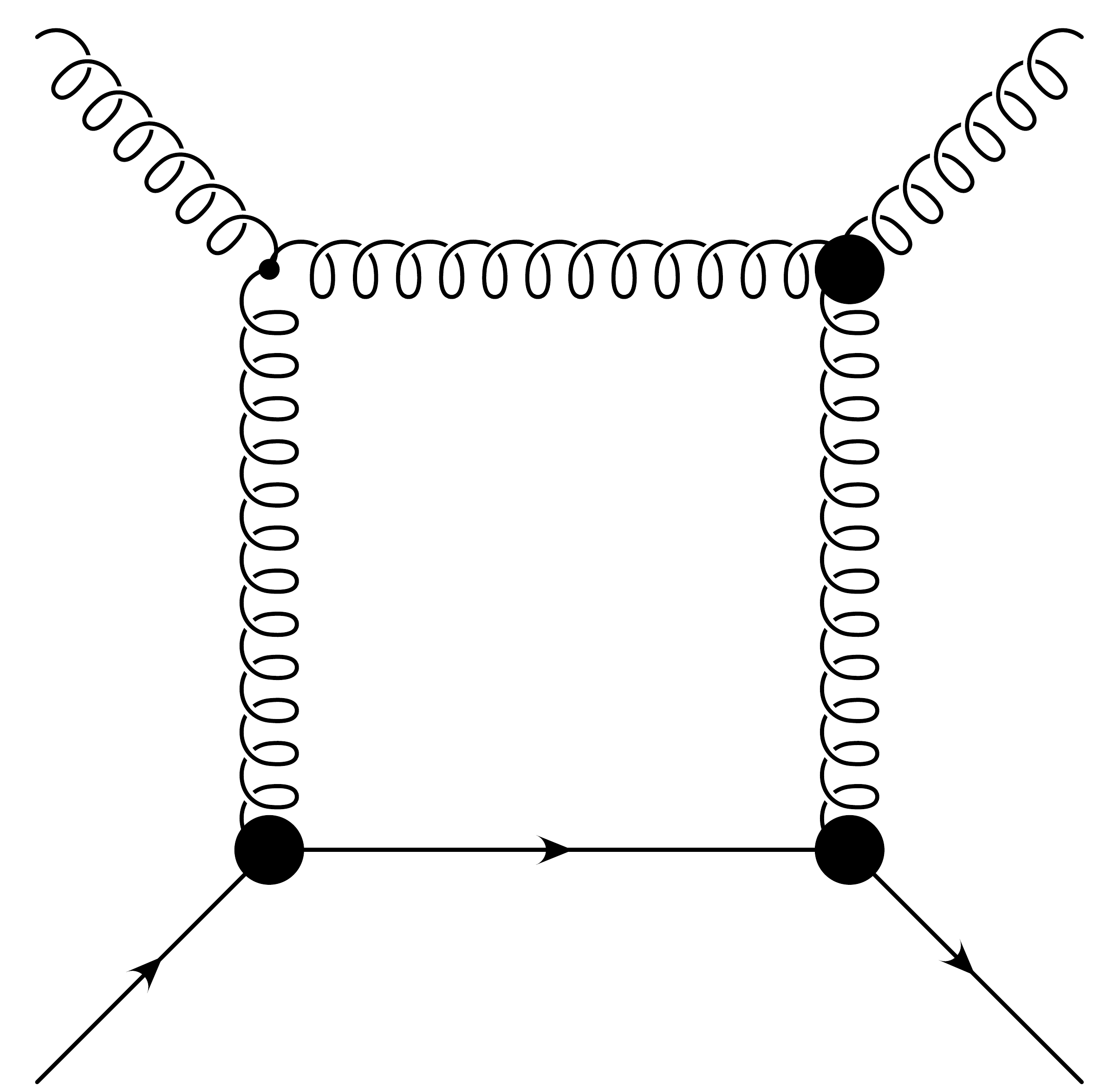}
\end{minipage}
\begin{minipage}{0.7\linewidth}
     \begin{center}
  \begin{tabular}{| l | l | l | l | l |}
    \hline
	adj.	& group invariant  & SU(2) & SU(3) & G$_2$ \\ \hline
    $1^{st}$  & $-\frac{1}{4}N_{A}C_{A}^{3}$ & -6 & -54 & -28 \\  
    $2^{nd}$  & $\frac{1}{4}N_{A}C_{A}^{3}$  & 6 & 54 & 28 \\ 
    $3^{rd}$  & $-\frac{5}{2} N_{A}C_{A}^{2}$   & -30 & -180 & -140 \\ 
    $4^{th}$  & $\frac{1}{4}N_{A}C_{A}^{3}(3-2N_{A})$   & -18 & -702 & -700 \\[1mm] 
    $5^{th}$  & $\frac{1}{12}N_{A}C_{A}^{3}(2N_{A}-1)$    & 10 & 270 & 252 \\
    $6^{th}$  & $-\frac{1}{2T_{R}(C_{A}-6C_{F})}\left(5N_{A}T_{R}\times \right.$ & $-\frac{3825}{128}$ 
	      & $-\frac{1435}{8}$ & $-\frac{1113}{8}$ \\
	      & $\left.\times C_{A}^{2}(C_{A}-6C_{F})+\right.$ & & & \\
	      & $\left. +12d_{44}(2+N_{A}) \right)$ & & & \\
    \hline
  \end{tabular}
\end{center}
\begin{center}
  \begin{tabular}{| l | l | l | l | l |}
    \hline
	fund.	& group invariant  & SU(2) & SU(3) & G$_2$ \\ \hline
    $1^{st}$  & $\frac{1}{4}N_{A}T_{R}(3C_{A}^{2}-10C_{A}C_{F}+8C_{F}^{2})$ & $\frac{9}{16}$ & $\frac{11}{9}$ & 0\\[1mm] 
    $2^{nd}$  & $-\frac{1}{4}N_{A}T_{R}C_{A}(C_{A}-2C_{F})$  & $-\frac{3}{8}$ & $-1$ & 0 \\[1mm] 
    $3^{rd}$  & $-\frac{1}{4}N_{A}T_{R}\left(-3C_{A}^{2}+10C_{A}C_{F}+\right.$ & $-\frac{27}{16}$ & $-\frac{205}{9}$ 
	      & $-\frac{105}{2}$ \\ 
	      & $\left.+4C_{F}(-2C_{F}+T_{R}+T_{R}N_{A})\right)$ & & & \\
    $4^{th}$  & $\frac{1}{4}N_{A}T_{R}C_{A}\left(C_{A}^{2}-2C_{A}C_{F}+\right.$ & 3 & 59 & 91 \\ 
	      & $\left.+4C_{F}T_{R}(N_{A}-1)\right)$ & & & \\
    $5^{th}$  & $-\frac{N_{A}T_{R}}{4}\left(-3C_{A}^{2}+10C_{A}C_{F}+\right.$ & $-\frac{27}{16}$ & $-\frac{205}{9}$ 
	      & $-\frac{105}{2}$ \\
	      & $\left.+4C_{F}(-2C_{F}+T_{R}+T_{R}N_{A})\right)$ & & & \\
    \hline
  \end{tabular}
\end{center}
\end{minipage}

\no\begin{minipage}{0.2\linewidth}
     \includegraphics[width=1.0\linewidth]{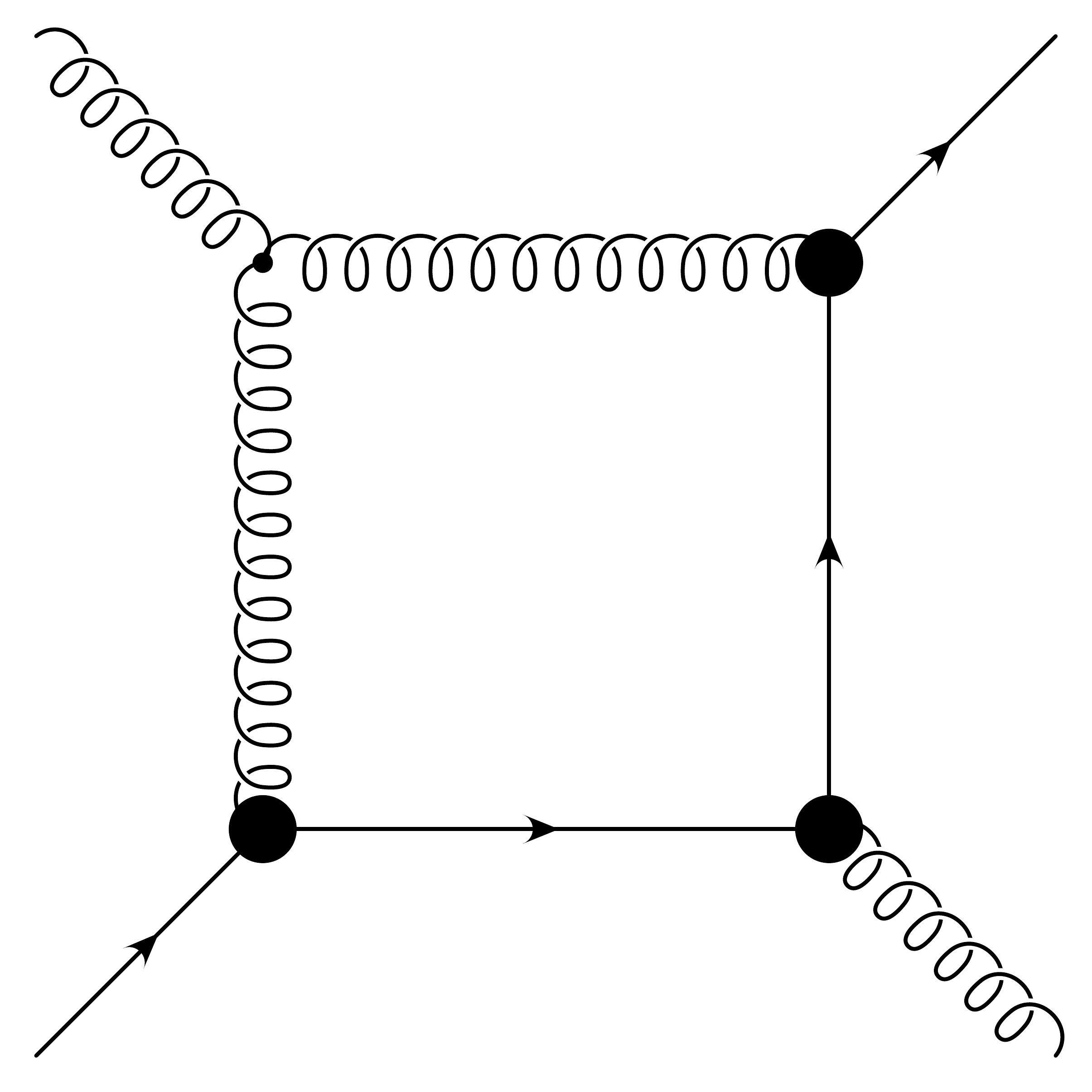}
\end{minipage}
\begin{minipage}{0.7\linewidth}
     \begin{center}
  \begin{tabular}{| l | l | l | l | l |}
    \hline
	adj.	& group invariant  & SU(2) & SU(3) & G$_2$ \\ \hline
    $1^{st}$  & $\frac{1}{2}N_{A}C_{A}^{3}$ & 12 & 108 & 56 \\ 
    $2^{nd}$  & 0  & 0 & 0 & 0 \\ 
    $3^{rd}$  & $-\frac{5}{2} N_{A}C_{A}^{2}$   & -30 & -180 & -140 \\  
    $4^{th}$  & 0   & 0 & 0 & 0 \\ 
    $5^{th}$  & $\frac{1}{6}N_{A}C_{A}^{3}(1-2N_{A})$   & -20 & -540 & -504 \\
    $6^{th}$  & $-\frac{1}{2T_{R}(C_{A}-6C_{F})}\left(5N_{A}T_{R}\times \right.$ & $-\frac{3825}{128}$ 
	      & $-\frac{1435}{8}$ & $-\frac{1113}{8}$ \\
	      & $\left.\times C_{A}^{2}(C_{A}-6C_{F})+\right.$ & & & \\
	      & $\left. +12d_{44}(2+N_{A}) \right)$ & & & \\
    \hline
  \end{tabular}
\end{center}
\begin{center}
  \begin{tabular}{| l | l | l | l | l |}
    \hline
	fund.	& group invariant  & SU(2) & SU(3) & G$_2$ \\ \hline
    $1^{st}$  & 0 & 0 & 0 & 0\\ 
    $2^{nd}$  & $\frac{1}{2}iN_{A}T_{R}C_{A}^{2}$  & $3i$ & $18i$ & $14i$ \\ 
    $3^{rd}$  & 0 & 0 & 0 & 0\\ 
    $4^{th}$  & $-\frac{1}{1}iN_{A}T_{R}C_{A}^{2}\left(C_{A}+\right.$   & $12i$ & $954i$ & $2520i$ \\ 
	      & $\left.-2T_{R}N_{A}(N_{A}-1)\right)$ & & & \\
    $5^{th}$  & 0 & 0 & 0 & 0\\
    \hline
  \end{tabular}
\end{center}
\end{minipage}

\no\begin{minipage}{0.2\linewidth}
     \includegraphics[width=1.0\linewidth]{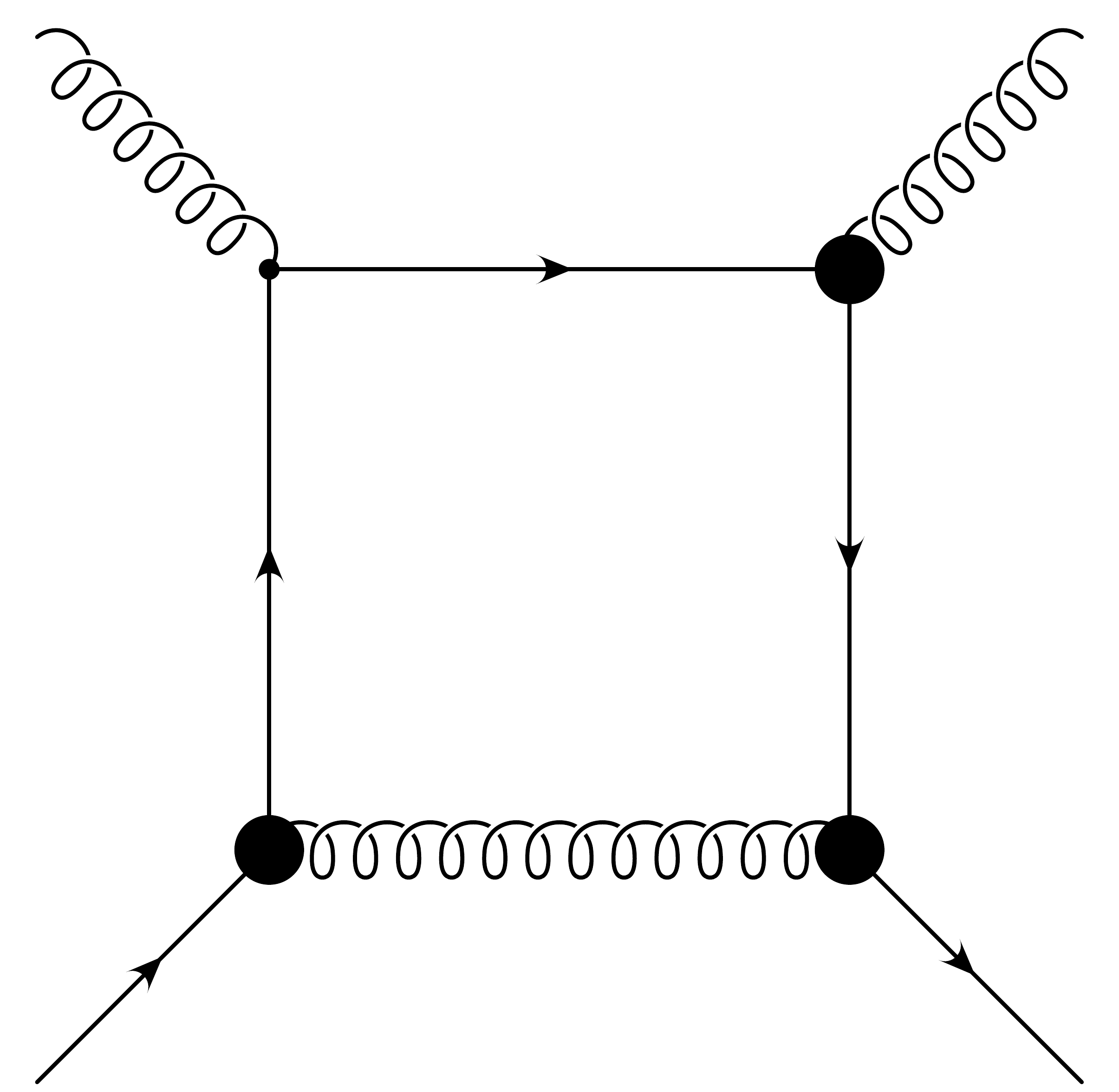}
\end{minipage}
\begin{minipage}{0.7\linewidth}
     \begin{center}
  \begin{tabular}{| l | l | l | l | l |}
    \hline
	adj.	& group invariant  & SU(2) & SU(3) & G$_2$ \\ \hline
    $1^{st}$  & $\frac{1}{4}N_{A}C_{A}^{3}$ & 6 & 54 & 28 \\  
    $2^{nd}$  & $-\frac{1}{4}N_{A}C_{A}^{3}$  & -6 & -54 & -28 \\  
    $3^{rd}$  & $\frac{5}{2} N_{A}C_{A}^{2}$   & 30 & 180 & 140 \\[1mm] 
    $4^{th}$  & $\frac{1}{4}N_{A}C_{A}^{3}(2N_{A}-3)$   & 18 & 702 & 700 \\[1mm] 
    $5^{th}$  & $\frac{1}{12}N_{A}C_{A}^{3}(1-2N_{A})$   & -10 & -270 & -252 \\
    $6^{th}$  & $\frac{1}{2T_{R}(C_{A}-6C_{F})}\left(5N_{A}T_{R}\times \right.$ & $\frac{3825}{128}$ 
	      & $\frac{1435}{8}$ & $\frac{1113}{8}$ \\
	      & $\left.\times C_{A}^{2}(C_{A}-6C_{F})+\right.$ & & & \\
	      & $\left. +12d_{44}(2+N_{A}) \right)$ & & & \\
    \hline
  \end{tabular}
\end{center}
\begin{center}
  \begin{tabular}{| l | l | l | l | l |}
    \hline
	fund.	& group invariant  & SU(2) & SU(3) & G$_2$ \\ \hline
    $1^{st}$  & $\frac{1}{8}N_{A}T_{R}C_{A}^{2}$ & $\frac{3}{4}$ & $\frac{9}{2}$ & $\frac{7}{2}$\\  
    $2^{nd}$  & 0  & 0 & 0 & 0 \\ 
    $3^{rd}$  & $\frac{1}{8}N_{A}T_{R}C_{A}(C_{A}-4T_{R}(1+N_{A}))$ & $-\frac{9}{4}$ & $-\frac{45}{2}$ & $-49$ \\ 
    $4^{th}$  & 0  & 0 & 0 & 0 \\ 
    $5^{th}$  & $\frac{N_{A}T_{R}C_{A}}{8N_{F}(1+N_{A})}\left(16\left(C_{F}+\right.\right.$  & $-\frac{9}{16}$ & $\frac{55}{54}$ 
	      & $\frac{49}{30}$ \\
	      & $\left.\left.-T_{R}(1+N_{A})\right)+\right.$ & & & \\
	      & $\left.+C_{A}(N_{F}+N_{A}N_{F}-4)\right)$ & & & \\
    \hline
  \end{tabular}
\end{center}
\end{minipage}

\no\begin{minipage}{0.2\linewidth}
     \includegraphics[width=1.0\linewidth]{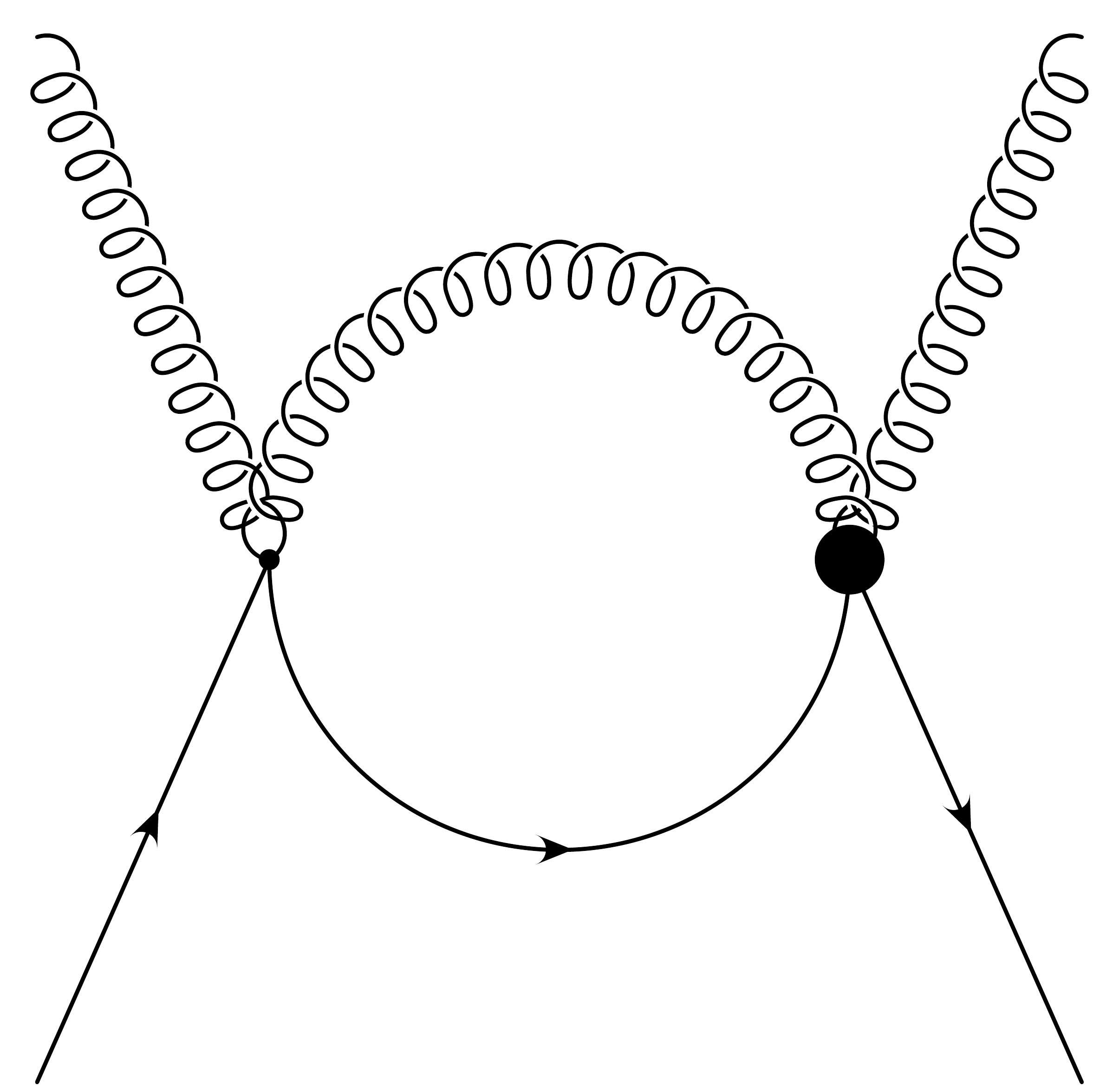}
\end{minipage}
\begin{minipage}{0.7\linewidth}
     \begin{center}
  \begin{tabular}{| l | l | l | l | l |}
    \hline
	adj.	& group invariant  & SU(2) & SU(3) & G$_2$ \\ \hline
    $1^{st}$  & $\frac{13}{4} N_{A}C_{A}^{3}$ & 78 & 702 & 364\\ 
    $2^{nd}$  & $-\frac{5}{4} N_{A}C_{A}^{3}$  & -30 & 270 & -140 \\ 
    $3^{rd}$  & $\frac{5}{2} N_{A}C_{A}^{2}$   & 30 & 180 & 140 \\[1mm]  
    $4^{th}$  & $\frac{5}{4} N_{A}C_{A}^{3}(2N_{A}-3)$   & 90 & 3510 & 3500 \\[1mm] 
    $5^{th}$  & $\frac{13}{12}N_{A}C_{A}^{3}(1-2N_{A})$   & -130 & -3510 & -3276 \\
    $6^{th}$  & $\frac{1}{2T_{R}(C_{A}-6C_{F})}\left(5N_{A}T_{R}\times \right.$ & $\frac{3825}{128}$ 
	      & $\frac{1435}{8}$ & $\frac{1113}{8}$ \\
	      & $\left.\times C_{A}^{2}(C_{A}-6C_{F})+\right.$ & & & \\
	      & $\left. +12d_{44}(2+N_{A}) \right)$ & & & \\
    \hline
  \end{tabular}
\end{center}
 \begin{center}
  \begin{tabular}{| l | l | l | l | l |}
    \hline
	fund.	& group invariant  & SU(2) & SU(3) & G$_2$ \\ \hline
    $1^{st}$  & $\frac{1}{32}N_{A}T_{R}(5C_{A}^{2}-24C_{A}C_{F}+32C_{F}^{2})$ & $\frac{3}{32}$ & $\frac{53}{72}$ & $\frac{7}{8}$\\[1mm] 
    $2^{nd}$  & $\frac{1}{32}N_{A}T_{R}C_{A}(8C_{F}-3C_{A})$  & 0 & $\frac{5}{8}$ & $\frac{7}{8}$ \\[1mm] 
    $3^{rd}$  & $\frac{1}{32}N_{A}T_{R}\left(5C_{A}^{2}+\right.$  & $-\frac{9}{32}$ 
	      & $-\frac{325}{72}$ & $-\frac{49}{4}$ \\
	      & $\left.+4C_{A}(T_{R}+T_{R}N_{A}-6C_{F})+ \right.$ & & & \\
	      & $\left. +16C_{F}(2C_{F}-T_{R}(1+N_{A})\right) )$ & & & \\
    $4^{th}$  & $\frac{1}{32}N_{A}T_{R}C_{A}^{2}(-8C_{F}+3C_{A})$   & 0 & $-\frac{15}{8}$ & $-\frac{7}{4}$ \\[1mm] 
    $5^{th}$  & $\frac{1}{32(1+N_{A})N_{F}}N_{A}T_{R} \times$  & $-\frac{45}{64}$ 
	      & $-\frac{2245}{216}$ & $-\frac{2989}{120}$ \\
	      & $\times \left(-16C_{F}(1+N_{A})\times \right.$ & & & \\
	      & $\left.\times \left(-2C_{F}+T_{R}+T_{R}N_{A}\right)+\right. $ & & & \\
	      & $\left.+C_{A}^{2}(4+5N_{F}(1+N_{A}))+ \right.$ & & & \\
	      &	$\left. -8C_{A}\left(-2T_{R}(1+N_{A})+\right. \right. $ & & & \\
	      & $\left.\left.+C_{F}(2+3N_{F}(1+N_{A}))\right)\right)$ & & & \\
    \hline
  \end{tabular}
\end{center}
\end{minipage}

\no\begin{minipage}{0.2\linewidth}
     \includegraphics[width=1.0\linewidth]{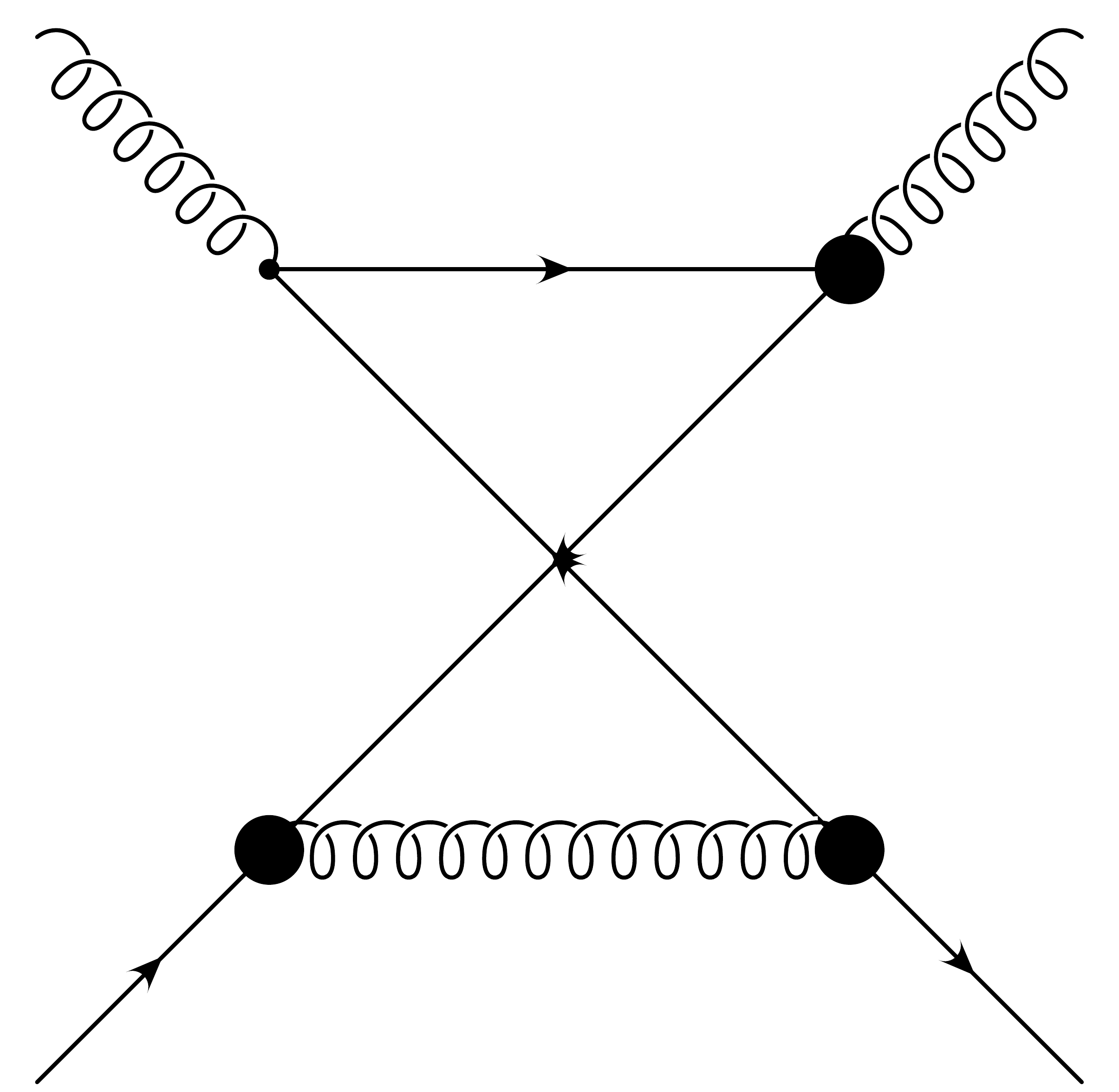}
\end{minipage}
\begin{minipage}{0.7\linewidth}
     \begin{center}
  \begin{tabular}{| l | l | l | l | l |}
    \hline
	adj.	& group invariant  & SU(2) & SU(3) & G$_2$ \\ \hline
    $1^{st}$  & $\frac{1}{4}N_{A}C_{A}^{3}$ & 6 & 54 & 28 \\[1mm]  
    $2^{nd}$  & $\frac{1}{4}N_{A}C_{A}^{3}$  & 6 & 54 & 28 \\[1mm]  
    $3^{rd}$  & $\frac{5}{2} N_{A}C_{A}^{2}$   & 30 & 180 & 140 \\[1mm] 
    $4^{th}$  & $\frac{1}{4}N_{A}C_{A}^{3}(-2N_{A}+3)$   & -18 & -702 & -700 \\[1mm] 
    $5^{th}$  & $\frac{1}{12}N_{A}C_{A}^{3}(1-2N_{A})$   & -10 & -270 & -252 \\[1mm]
    $6^{th}$  & $\frac{1}{2T_{R}(C_{A}-6C_{F})}\left(5N_{A}T_{R}\times \right.$ & $\frac{3825}{128}$ 
	      & $\frac{1435}{8}$ & $\frac{1113}{8}$ \\
	      & $\left.\times C_{A}^{2}(C_{A}-6C_{F})+\right.$ & & & \\
	      & $\left. +12d_{44}(2+N_{A}) \right)$ & & & \\
    \hline
  \end{tabular}
\end{center}
\begin{center}
  \begin{tabular}{| l | l | l | l | l |}
    \hline
	fund.	& group invariant  & SU(2) & SU(3) & G$_2$ \\ \hline
    $1^{st}$  & $\frac{1}{4}N_{A}T_{R}(3C_{A}^{2}-10C_{A}C_{F}+8C_{F}^{2})$ & $\frac{9}{16}$ & $\frac{11}{9}$ & 0\\[1mm]  
    $2^{nd}$  & $\frac{1}{4}N_{A}T_{R}C_{A}(C_{A}-2C_{F})$  & $\frac{3}{8}$ & 1 & 0 \\[1mm] 
    $3^{rd}$  & $-\frac{1}{4}N_{A}T_{R}\left(-3C_{A}^{2}+10C_{A}C_{F}+\right.$ & $-\frac{27}{16}$ 
	      & $-\frac{205}{9}$ & $-\frac{105}{2}$ \\ 
	      & $\left.+4C_{F}^{2}(-2C_{F}+T_{R}+T_{R}N_{A})\right)$ & & & \\
    $4^{th}$  & $-\frac{1}{4}N_{A}T_{R}C_{A}\left(C_{A}^{2}-2C_{A}C_{F}+\right.$  & -3 & -59 & -91 \\ 
	      & $\left.+4C_{F}T_{R}(N_{A}-1)\right)$ & & & \\
    $5^{th}$  & $\frac{N_{A}T_{R}C_{A}}{8N_{F}(1+N_{A})}\left(16(C_{F}-T_{R}(1+N_{A}))+\right.$  & $-\frac{27}{16}$ & $-\frac{205}{9}$ 
	      & $-\frac{105}{2}$ \\
	      & $-\frac{1}{4}N_{A}T_{R}\left(-3C_{A}^{2}+10C_{A}C_{F}+\right.$ &  & & \\
	      & $\left.+4C_{F}^{2}(-2C_{F}+T_{R}+T_{R}N_{A})\right)$ & & & \\
    \hline
  \end{tabular}
\end{center}
\end{minipage}

\no\begin{minipage}{0.2\linewidth}
     \includegraphics[width=1.0\linewidth]{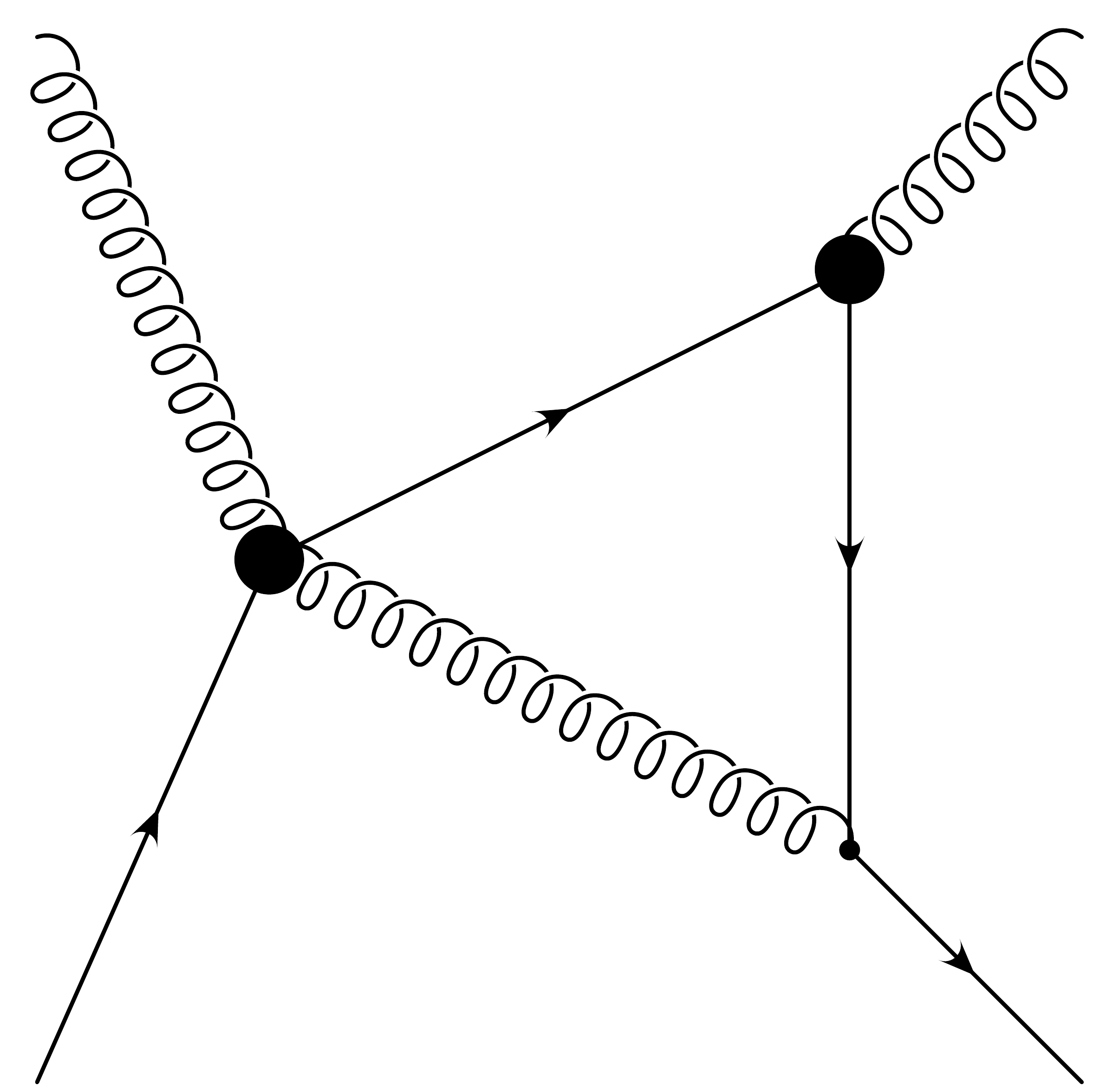}
\end{minipage}
\begin{minipage}{0.7\linewidth}
     \begin{center}
  \begin{tabular}{| l | l | l | l | l |}
    \hline
	adj.	& group invariant  & SU(2) & SU(3) & G$_2$ \\ \hline
    $1^{st}$  & $-N_{A}C_{A}^{3}$ & -24 & -216 & -112 \\ 
    $2^{nd}$  & $\frac{1}{2}N_{A}C_{A}^{3}$  & 12 & 108 & 56 \\ 
    $3^{rd}$  & $-\frac{5}{2} N_{A}C_{A}^{2}$   & -30 & -180 & -140 \\ 
    $4^{th}$  & $\frac{1}{2}N_{A}C_{A}^{3}(3-2N_{A})$   & -36 & -1404 & -1400 \\[1mm] 
    $5^{th}$  & $\frac{1}{3}N_{A}C_{A}^{3}(2N_{A}-1)$   & 40 & 1080 & 1008 \\
    $6^{th}$  & $-\frac{1}{2T_{R}(C_{A}-6C_{F})}\left(5N_{A}T_{R}\times \right.$ & $-\frac{3825}{128}$ 
	      & $-\frac{1435}{8}$ & $-\frac{1113}{8}$ \\
	      & $\left.\times C_{A}^{2}(C_{A}-6C_{F})+ \right.$ & & & \\
	      & $\left. +12d_{44}(2+N_{A}) \right)$ & & & \\
    \hline
  \end{tabular}
\end{center}
 \begin{center}
  \begin{tabular}{| l | l | l | l | l |}
    \hline
	fund.	& group invariant  & SU(2) & SU(3) & G$_2$ \\ \hline
    $1^{st}$  & $-\frac{1}{4}N_{A}T_{R}(C_{A}-2C_{F})^{2}$ & $-\frac{3}{32}$ & $-\frac{1}{9}$ & 0\\[1mm] 
    $2^{nd}$  & $\frac{1}{8}N_{A}T_{R}C_{A}(C_{A}-2C_{F})$  & $\frac{3}{16}$ & $\frac{1}{2}$ & 0 \\[1mm]  
    $3^{rd}$  & $-\frac{1}{8}N_{A}T_{R}\left(2C_{A}^{2}+\right.$ & $\frac{9}{32}$ & $\frac{185}{36}$ 
	      & $\frac{105}{8}$ \\  
	      & $\left.+C_{A}(-8C_{F}+T_{R}+T_{R}N_{A})+\right.$ & & & \\
	      & $\left.+4C_{F}(2C_{F}-T_{R}(1+N_{A}))\right)$ & & & \\
    $4^{th}$  & $-\frac{1}{8}N_{A}T_{R}C_{A}^{2}(C_{A}-2C_{F})$  & $-\frac{3}{8}$ & $-\frac{3}{2}$ & 0 \\[1mm]  
    $5^{th}$  & $-\frac{T_{R}N_{A}}{8N_{F}(1+N_{A})}\left(4C_{F}N_{F}(1+N_{A})\times \right.$  & $\frac{45}{64}$ 
	      & $\frac{595}{54}$ & $\frac{1547}{60}$ \\
	      & $\left.\times \left(-2C_{F}+T_{R}+T_{R}N_{A}\right)+\right.$ & & & \\
	      & $\left.-C_{A}^{2}\left(1+2(1+N_{A})N_{F}\right)+\right.$ & & & \\
              & $\left. +4C_{A}(-T_{R}(1+N_{A})+\right.$ & & & \\
	      & $\left. +C_{F}(1+2N_{F}(1+N_{A})))\right)$ & & & \\
    \hline
  \end{tabular}
\end{center}
\end{minipage}

\no\begin{minipage}{0.18\linewidth}
     \includegraphics[width=1.0\linewidth]{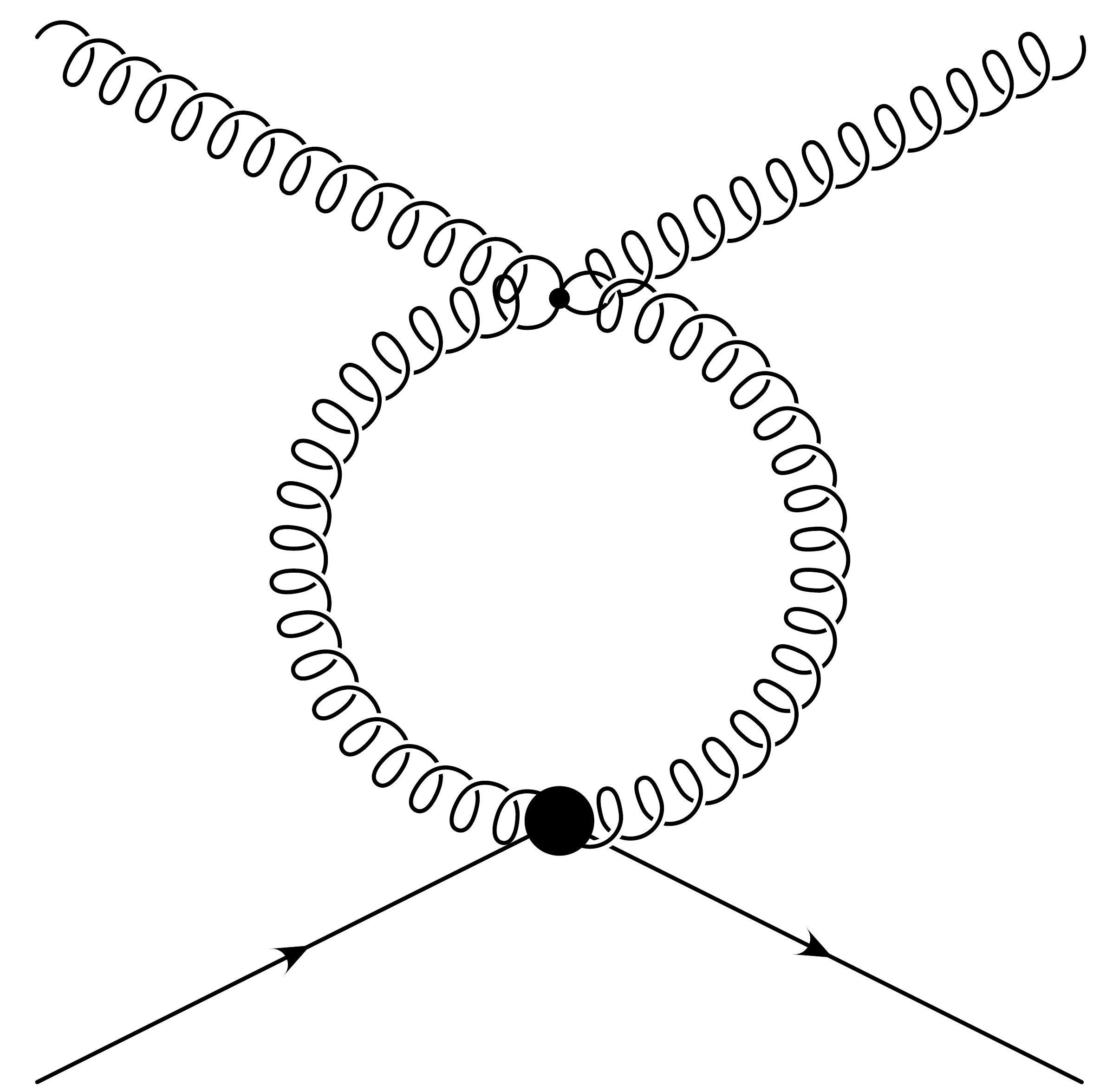}
\end{minipage}
\begin{minipage}{0.72\linewidth}
     \begin{center}
  \begin{tabular}{| l | l | p{2.1cm} | p{2.3cm} | p{2.3cm} |}
    \hline
	adj.      & group invariant & SU(2) & SU(3) & G$_2$ \\ \hline
    $1^{st}$  & $-\frac{1}{2}N_{A}C_{A}^{3}(C+B)$ & $12(C+B)$ & $108(C+B)$ & $56(C+B)$\\[1mm] 
    $2^{nd}$  & 0  & 0 & 0 & 0 \\  
    $3^{rd}$  & $5N_{A}C_{A}^{2}(C+B)$   & $60(C+B)$ & $360(C+B)$ & $280(C+B)$ \\  
    $4^{th}$  & 0   & 0 & 0 & 0 \\ 
    $5^{th}$  & $-\frac{1}{6}N_{A}C_{A}^{3}(2N_{A}-1)(B+C)$ & $-20(C+B)$ & $-540(C+B)$ & $-504(C+B)$ \\[1mm]
    $6^{th}$  & $\frac{(C+B)}{T_{R}(C_{A}-6C_{F})}\left(5N_{A}T_{R}C_{A}^{2}\times \right. $ & $\frac{3825}{64}(C+B)$ 
	      & $\frac{1435}{4}(C+B)$ & $\frac{1113}{4}(C+B)$ \\
	      & $\left. \times(C_{A}-6C_{F})+12d_{44}(2+N_{A}) \right)$ &  & & \\
    \hline
  \end{tabular}
\end{center}
\begin{center}
  \begin{tabular}{| l | l | l | l | l |}
    \hline
	fund.	& group invariant  & SU(2) & SU(3) & G$_2$ \\ \hline
    $1^{st}$  & $-\frac{1}{8}N_{A}T_{R}C_{A}^{2}(A+B)$ & $-\frac{3}{4}(A+B)$ & $-\frac{9}{2}(A+B)$ & $-\frac{7}{2}(A+B)$\\  
    $2^{nd}$  & 0  & 0 & 0 & 0 \\ 
    $3^{rd}$  & $-\frac{1}{8}N_{A}T_{R}C_{A}(C_{A}-4T_{R}(1+N_{A}))$ & $\frac{9}{4}(A+B)$ & $\frac{45}{2}(A+B)$ & $49(A+B)$ \\ 
	      & $(A+B)$ & & & \\[1mm]
    $4^{th}$  & $\frac{1}{2}N_{A}T_{R}^{2}C_{A}^{2}(N_{A}-1)C$   & $3C$ & $63C$ & $91C$ \\[1mm]  
    $5^{th}$  & $-\frac{A+B}{8N_{F}(1+N_{A})}N_{A}T_{R}C_{A} \times$   & $\frac{9}{16}(A+B)$ & $-\frac{55}{54}(A+B)$ 
	      & $-\frac{49}{30}(A+B)$ \\[1mm]
	      & $\times\left(16(C_{F}-T_{R}(1+N_{A}))+ \right.$ & & & \\
	      & $ \left. +C_{A}(N_{F}+N_{F}N_{A}-4)\right)$ & & & \\
    \hline
  \end{tabular}
\end{center}
\end{minipage}

\no\begin{minipage}{0.1\linewidth}
     \includegraphics[width=1.0\linewidth]{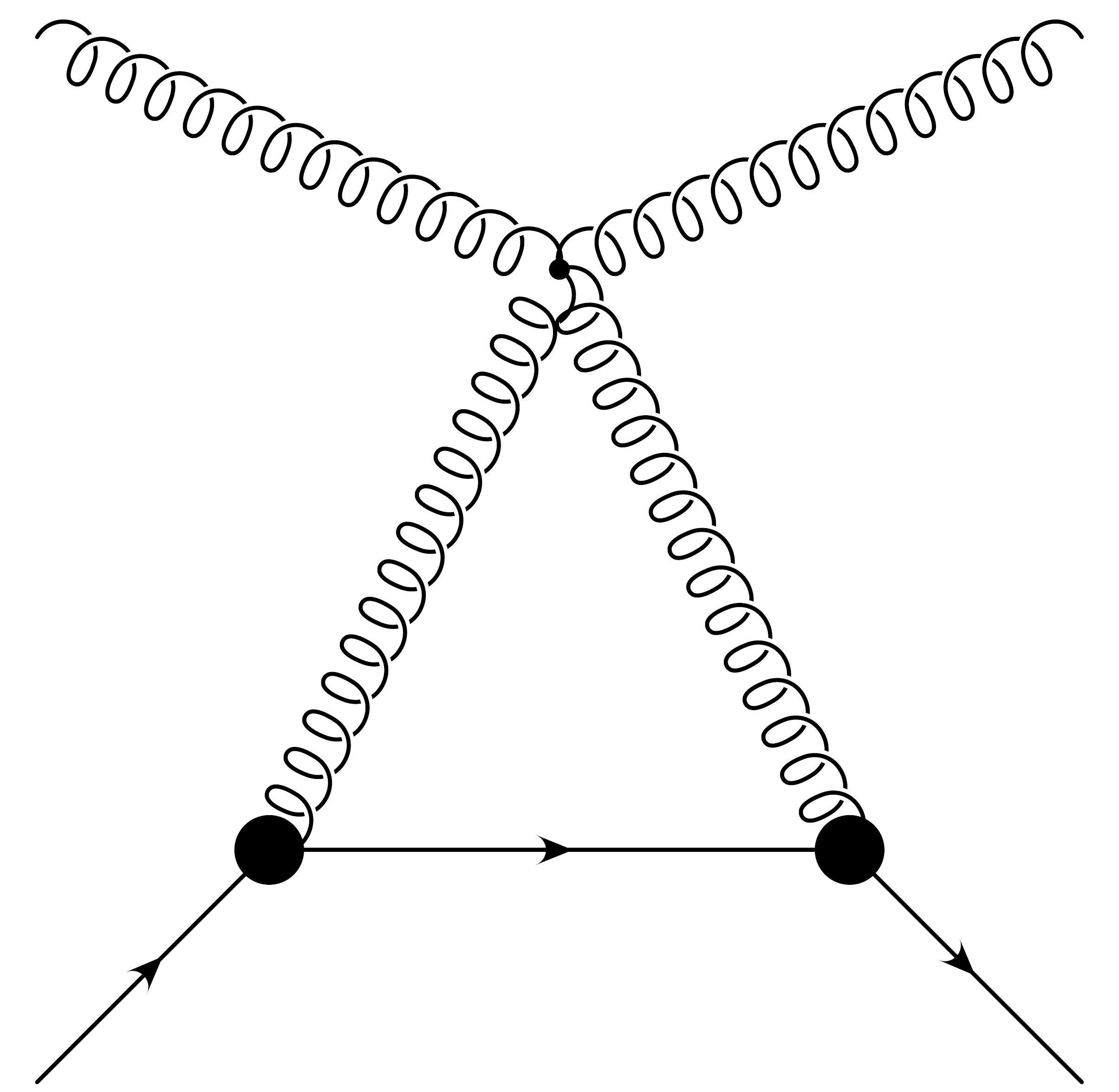}
\end{minipage}
\begin{minipage}{0.74\linewidth}
     \begin{center}
  \begin{tabular}{| l | p{5.7cm} | p{2.3cm} | p{2.5cm} | p{2.3cm} |}
    \hline
	adj.	& group invariant  & SU(2) & SU(3) & G$_2$ \\ \hline
    $1^{st}$  & $-\frac{1}{4} N_{A}C_{A}^{3}(A+B)$ & -6(A+B) & -54(A+B) & -28(A+B) \\[1mm] 
    $2^{nd}$  & $\frac{1}{4} N_{A}C_{A}^{3}(A-B-2C)$  & 6(A-B-2C) & 54(A-B-2C) & 28(A-B-2C) \\[1mm] 
    $3^{rd}$  & $-\frac{5}{2} N_{A}C_{A}^{2}(A+B)$   & -30(A+B) & -180(A+B) & -140(A+B) \\[1mm] 
    $4^{th}$  & $-\frac{1}{4}N_{A}C_{A}^{3}(2N_{A}-3)(A-B-2C)$   & 18(A-B-2C) & -702(A-B-2C) & -700(A-B-2C) \\[1mm] 
    $5^{th}$  & $\frac{1}{12}N_{A}C_{A}^{3}(2N_{A}-1)(A+B)$   & 10(A+B) & 270(A+B) & 252(A+B) \\[1mm]
    $6^{th}$  & $-\frac{(A+B)}{2T_{R}(C_{A}-6C_{F})}\left(5N_{A}T_{R}C_{A}^{2}\times \right.$ & $-\frac{3825}{128}(A+B)$ 
	      & $-\frac{1435}{8}(A+B)$ & $-\frac{1113}{8}(A+B)$ \\
	      & $\left. \times(C_{A}-6C_{F})+12d_{44}(2+N_{A}) \right)$ & & & \\
    \hline
  \end{tabular}
\end{center}

\begin{center}
  \begin{tabular}{| l | l | l | l | l |}
    \hline
	fund.	& group invariant  & SU(2) & SU(3) & G$_2$ \\ \hline
    $1^{st}$  & $\frac{1}{4}N_{A}T_{R}C_{A}^{2}(A+B)$ & $\frac{3}{2}(A+B)$ & $9(A+B)$ & $7(A+B)$\\[1mm] 
    $2^{nd}$  & $\frac{1}{4}N_{A}T_{R}C_{A}^{2}(A-B-2C)$  & $\frac{3}{2}(A-B-2C)$ & $9(A-B-2C)$ & $7(A-B-2C)$ \\[1mm] 
    $3^{rd}$  & $\frac{1}{4}N_{A}T_{R}C_{A}\left(C_{A}+\right.$  & $-\frac{9}{2}(A+B)$ & $-45(A+B)$ & $-98(A+B)$ \\ 
	      & $\left.-4T_{R}(1+N_{A})\right)(A+B)$ & & & \\[1mm]
    $4^{th}$  & $\frac{1}{4}N_{A}T_{R}C_{A}^{3}(B-A+2C)$   & $3(B-A+2C)$ & $27(-A+B+2C)$ & $14(-A+B+2C)$ \\[1mm] 
    $5^{th}$  & $\frac{T_{R}N_{A}C_{A}}{4N_{F}(1+N_{A})}\times$  & $-\frac{9}{8}(A+B)$ & $\frac{55}{27}(A+B)$ 
	      & $\frac{49}{15}(A+B)$ \\
	      & $\times(16\left(C_{F}-T_{R}(1+N_{A}))+\right.$ & & & \\
	      & $\left.+C_{A}(-4+N_{F}+N_{F}N_{A})\right)$ & & & \\
    \hline
  \end{tabular}
\end{center}
\end{minipage}


\no\begin{minipage}{0.2\linewidth}
     \includegraphics[width=1.0\linewidth]{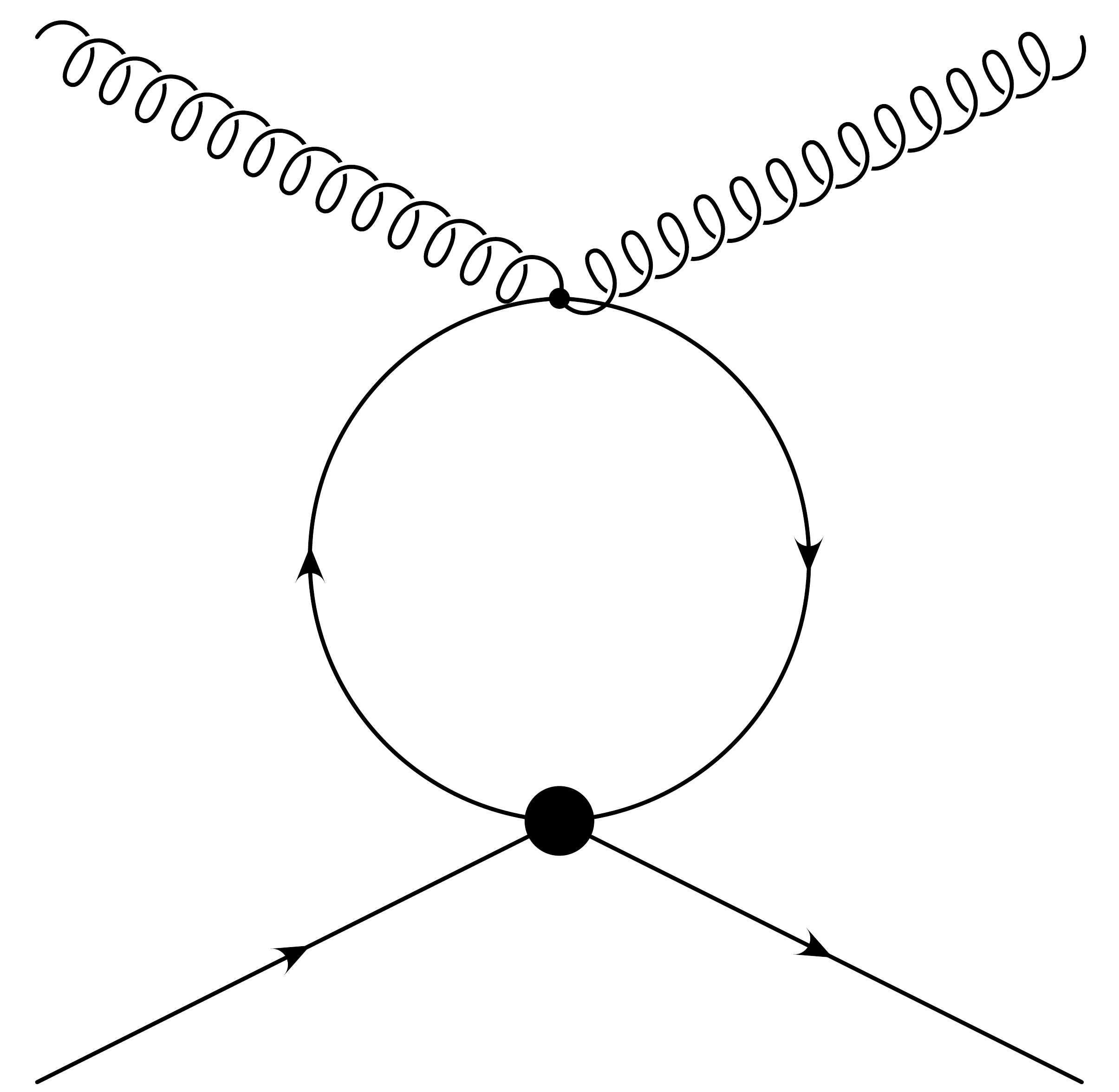}
\end{minipage}
\begin{minipage}{0.7\linewidth}
     \begin{center}
  \begin{tabular}{| l | l | l | l | l |}
    \hline
	adj.	& group invariant  & SU(2) & SU(3) & G$_2$ \\ \hline
    $1^{st}$  & $20N_{A}C_{A}^{2}$ & 240 & 1440 & 1120 \\ 
    $2^{nd}$  & 0  & 0 & 0 & 0 \\ 
    $3^{rd}$  & $4N_{A}C_{A}(N_{A}+2)$   & 120 & 960 & 1792 \\ 
    $4^{th}$  & 0   & 0 & 0 & 0 \\  
    $5^{th}$  & $-\frac{4}{3}N_{A}C_{A}^{2}(2N_{A}^{2}+6N_{A}-11)$   & -400 & -15840 & -34720 \\
    $6^{th}$  & 0 & 0 & 0 & 0 \\
    \hline
  \end{tabular}
\end{center}
\begin{center}
  \begin{tabular}{| l | l | l | l | l |}
    \hline
	fund.	& group invariant  & SU(2) & SU(3) & G$_2$ \\ \hline
    $1^{st}$  & $\frac{1}{24}N_{A}T_{R}(C_{A}-4(C_{F}+T_{R}))$ & $-\frac{3}{16}$ & $-\frac{13}{18}$ & $-\frac{7}{6}$\\ 
    $2^{nd}$  & 0  & 0 & 0 & 0 \\ 
    $3^{rd}$  & $\frac{1}{24}N_{A}T_{R}\left(C_{A}+2\left(-2C_{F}+ \right. \right.$   & $\frac{9}{16}$ & $\frac{95}{18}$ 
	      & $\frac{203}{6}$ \\  
	      & $\left. \left. +T_{R}(N_{A}+N_{F}+N_{A}N_{F}-1)\right)\right)$ & & & \\
    $4^{th}$  & 0   & 0 & 0 & 0 \\ 
    $5^{th}$  & $\frac{1}{24(1+N_{A})}N_{A}T_{R}\left((3+N_{A})\times \right.$ & $\frac{9}{32}$ & $\frac{110}{81}$ 
	      & $\frac{1547}{360}$\\
	      & $\left. \times(C_{A}+2(-2C_{F}+T_{R}+T_{R}N_{A}))\right)$ & & & \\
    \hline
  \end{tabular}
\end{center}
\end{minipage}

\no\begin{minipage}{0.2\linewidth}
     \includegraphics[width=1.0\linewidth]{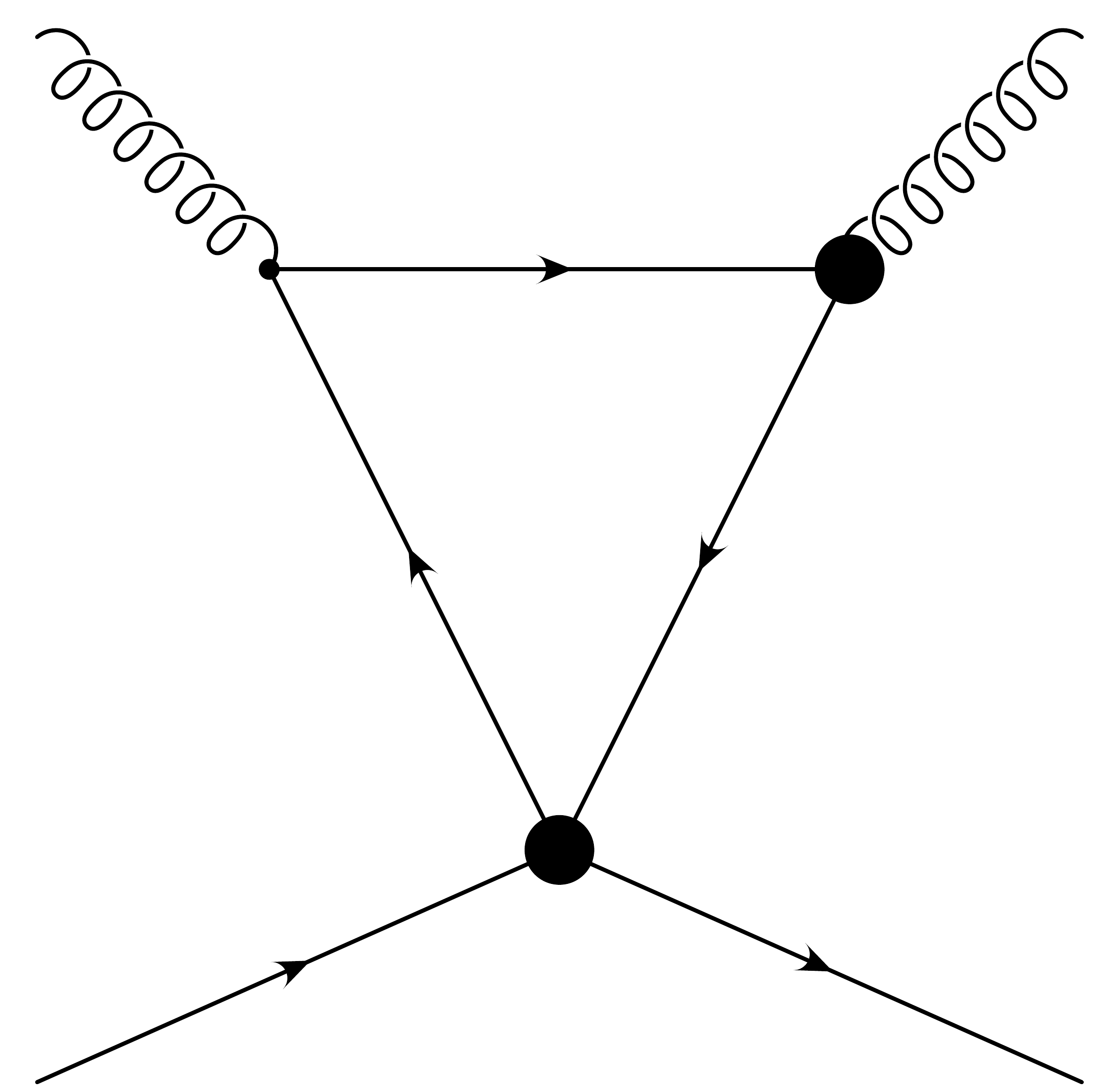}
\end{minipage}
\begin{minipage}{0.7\linewidth}
     \begin{center}
  \begin{tabular}{| l | l | l | l | l |}
    \hline
	adj.	& group invariant  & SU(2) & SU(3) & G$_2$ \\ \hline
    $1^{st}$  & $10 N_{A}C_{A}^{2}$ & 120 & 720 & 560\\ 
    $2^{nd}$  & 0  & 0 & 0 & 0 \\ 
    $3^{rd}$  & $2 N_{A}C_{A}(N_{A}+2)$   & 60 & 480 & 896 \\ 
    $4^{th}$  & 0   & 0 & 0 & 0 \\ 
    $5^{th}$  & $-\frac{2}{3}N_{A}C_{A}^{2}(2N_{A}^{2}+6N_{A}-11)$   & -200 & -7920 & -17360 \\
    $6^{th}$  & 0 & 0 & 0 & 0 \\
    \hline
  \end{tabular}
\end{center}
\begin{center}
  \begin{tabular}{| l | l | l | l | l |}
    \hline
	fund.	& group invariant  & SU(2) & SU(3) & G$_2$ \\ \hline
    $1^{st}$  & $\frac{1}{12}N_{A}T_{R}(-C_{A}+4(C_{F}+T_{R}))$ & $\frac{3}{8}$ & $\frac{13}{9}$ & $\frac{7}{3}$\\[1mm] 
    $2^{nd}$  & $\frac{1}{12}N_{A}T_{R}C_{A}$  & $\frac{1}{4}$ & 1 & $\frac{7}{6}$ \\[1mm]  
    $3^{rd}$  & $-\frac{1}{12}N_{A}T_{R}\left(C_{A}+2(-2C_{F}+ \right.$    & $-\frac{9}{8}$ & $-\frac{95}{9}$ & $-\frac{203}{3}$ \\ 
	      & $\left.+T_{R}(N_{A}+N_{F}+N_{A}N_{F}-1))\right)$ & & & \\[1mm]
    $4^{th}$  & $-\frac{1}{12}N_{A}T_{R}C_{A}(C_{A}-2T_{R}(N_{A}-1))$   & 0 & 4 & $\frac{77}{6}$ \\[1mm] 
    $5^{th}$  & $-\frac{3+N_{A}}{12(1+N_{A})}N_{A}T_{R}\left(C_{A}+2(-2C_{F}+\right.$ & $-\frac{9}{16}$ & $-\frac{220}{81}$ 
	      & $-\frac{1547}{180}$ \\
	      & $\left.+T_{R}+T_{R}N_{A})\right)$ & & & \\
    \hline
  \end{tabular}
\end{center}
\end{minipage}

\no\begin{minipage}{0.18\linewidth}
     \includegraphics[width=1.0\linewidth]{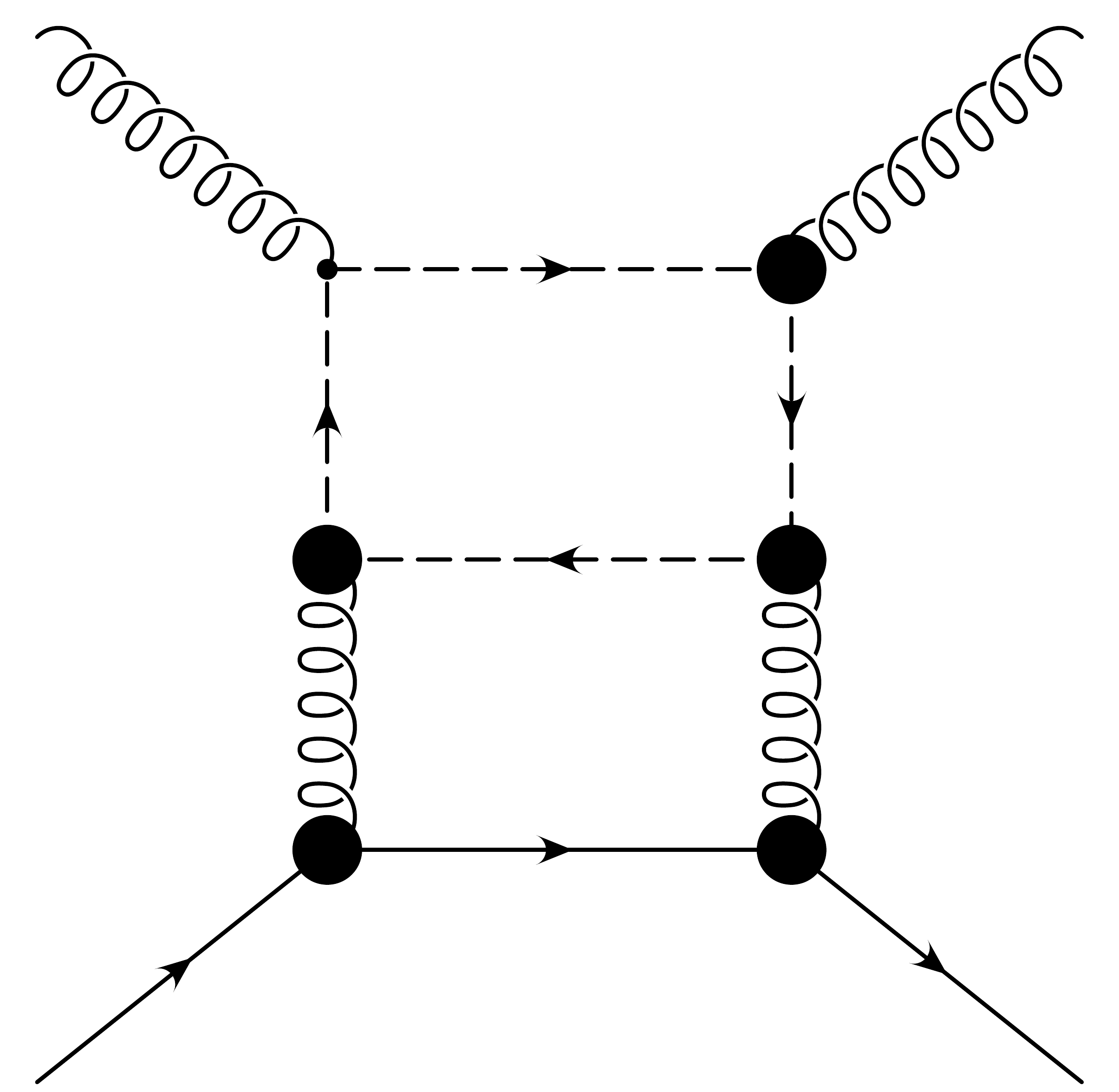}
\end{minipage}
\begin{minipage}{0.72\linewidth}
     \begin{center}
  \begin{tabular}{| l | l | l | l | l |}
    \hline
	adj.	& group invariant  & SU(2) & SU(3) & G$_2$ \\ \hline
    $1^{st}$  & $\frac{1}{24}(N_{A}C_{A}^{4}+48d_{44})$ & $\frac{1029}{512}$ & $\frac{2597}{96}$ & $\frac{1799}{192}$ \\[1mm] 
    $2^{nd}$  & $-\frac{1}{8}N_{A}C_{A}^{4}$ & -6 & -81 & -28 \\[1mm] 
    $3^{rd}$  & $\frac{5}{4}N_{A}C_{A}^{3}$  & 30 & 270 & 140 \\[1mm] 
    $4^{th}$  & $\frac{1}{8}N_{A}C_{A}^{4}(2N_{A}-3)$ & 18 & 1053 & 700 \\[1mm] 
    $5^{th}$  & $\frac{1}{24}(N_{A}C_{A}^{4}(13+14N_{A})+48d_{44})$ & $-\frac{56315}{512}$ & $-\frac{323995}{96}$ 
	      & $-\frac{374521}{192}$  \\[1mm]
    $6^{th}$  & $\frac{C_{A}}{4T_{R}(C_{A}-6C_{F})}\left(5N_{A}T_{R}C_{A}^{2}(C_{A}-6C_{F})+ \right.$ & $\frac{3825}{128}$ 
	      & $\frac{4305}{16}$ & $\frac{1113}{8}$ \\
	      & $\left. +12d_{44}(2+N_{A}) \right)$ & & & \\
    \hline
  \end{tabular}
\end{center}

\begin{center}
  \begin{tabular}{| l | l | l | l | l |}
    \hline
	fund.	& group invariant  & SU(2) & SU(3) & G$_2$ \\ \hline
    $1^{st}$  & $-\frac{1}{24}(N_{A}T_{R}C_{A}^{3}+48 d_{44})$ & $-\frac{261}{512}$ & $-\frac{437}{96}$ & $-\frac{455}{192}$\\[1mm] 
    $2^{nd}$  & $-\frac{1}{8}N_{A}T_{R}C_{A}^{3}$  & $-\frac{3}{2}$ & $\frac{27}{2}$ & -7 \\[1mm] 
    $3^{rd}$  & $-\frac{1}{24}(N_{A}T_{R}C_{A}^{2}\left(C_{A}-24T_{R}(1+N_{A}))+\right.$  & $\frac{5883}{512}$ & $\frac{15115}{96}$ 
	      & $\frac{39865}{192}$ \\ 
	      & $\left.+48d_{44}\right)$  & & & \\[1mm]
    $4^{th}$  & $\frac{1}{8}N_{A}T_{R}C_{A}^{3}(C_{A}-4T_{R}(N_{A}-1))$ & -3 & $-\frac{297}{2}$ & -168 \\[1mm]  
    $5^{th}$  & $\frac{1}{24N_{F}(1+N_{A})}\left(-48d_{44} N_{F}(1+N_{A})-N_{A}T_{R}C_{A}^{3}\times\right.$ 
	      & $\frac{3291}{512}$ & $\frac{4955}{96}$ & $\frac{53501}{960}$ \\
	      & $\left.\times(N_{F}+N_{A}N_{F}-18)+6C_{A}^{2}N_{A}T_{R}(-12C_{F}+\right.$ &  & & \\
	      & $\left.+T_{R}(1+N_{A})(12+N_{F}+N_{A}N_{F}))\right)$ & & & \\
    \hline
  \end{tabular}
\end{center}
\end{minipage}

\subsubsection{4-scalar vertex}

\no\begin{minipage}{0.2\linewidth}
     \includegraphics[width=1.0\linewidth]{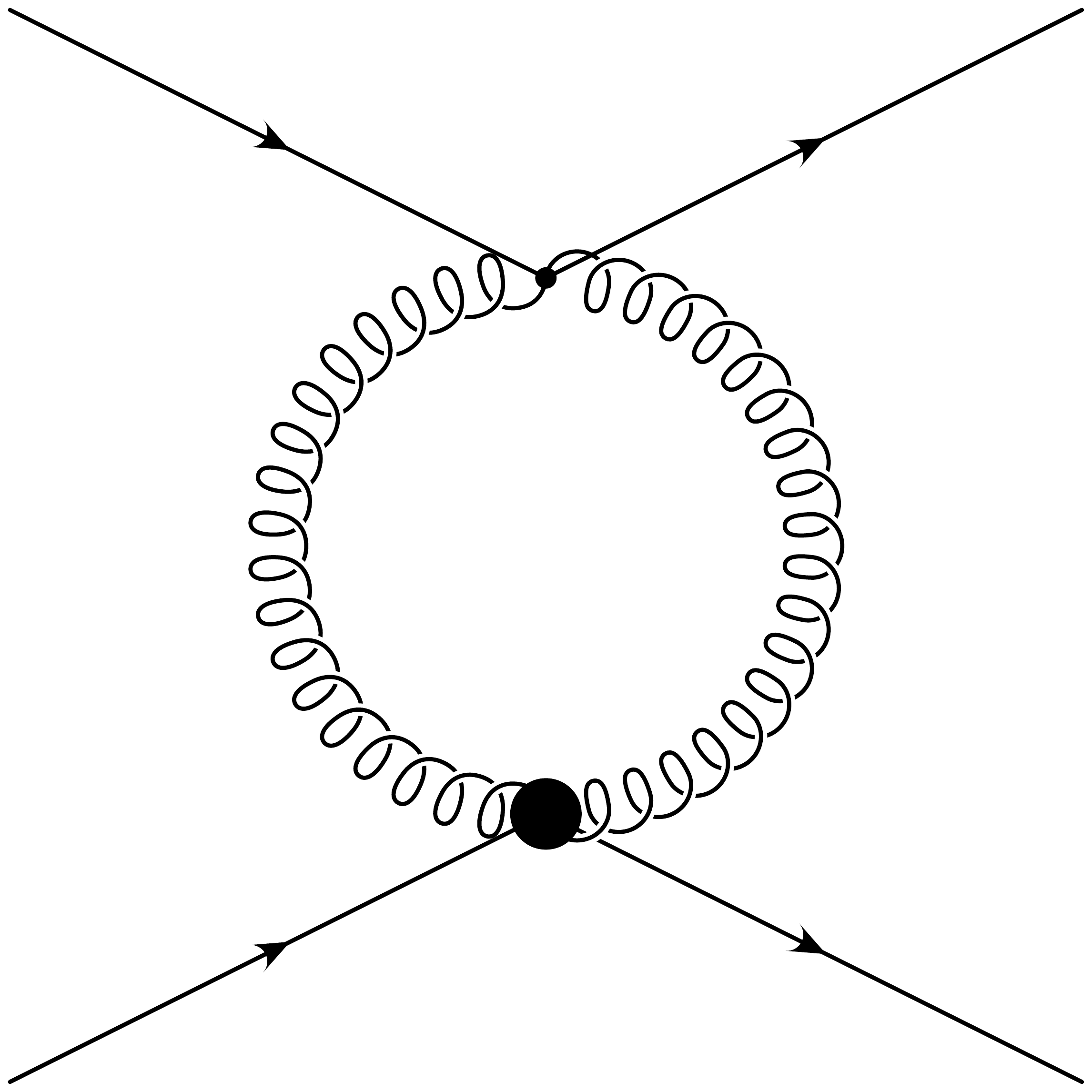}
\end{minipage}
\begin{minipage}{0.7\linewidth}
     \begin{center}
  \begin{tabular}{| l | l | l | l | l |}
    \hline
	adj.	& group invariant  & SU(2) & SU(3) & G$_2$ \\ \hline
    $1^{st}$  & $10N_{A}C_{A}^{2}$ & 120 & 720 & 560\\ 
    $2^{nd}$  & $-\frac{1}{2}N_{A}C_{A}^{3}$  & -12 & -108 & -56 \\ 
    $3^{rd}$  & $-\frac{1}{2}N_{A}C_{A}^{3}$   & -12 & -108 & -56 \\ 
    $4^{th}$  & $-\frac{1}{2}N_{A}C_{A}^{3}(2N_{A}-3)$   & 36 & 1404 & 1400 \\ 
    $5^{th}$  & $\frac{1}{6}N_{A}C_{A}^{3}(2N_{A}-1)$ & 20 & 540 & 504 \\
    $6^{th}$  & $\frac{2}{T_{R}(C_{A}-6C_{F})}\left(5N_{A}T_{R}\times \right.$ & $\frac{3825}{32}$ & $\frac{1435}{2}$ 
	      & $\frac{1113}{2}$ \\
	      & $\left.\times C_{A}^{2}(C_{A}-6C_{F})+\right.$ & & & \\
	      & $\left.+12d_{44}(2+N_{A}) \right)$ & & & \\
    \hline
  \end{tabular}
\end{center}
     \begin{center}
  \begin{tabular}{| l | l | l | l | l |}
    \hline
	fund.	& group invariant  & SU(2) & SU(3) & G$_2$ \\ \hline
    $1^{st}$  & $\frac{1}{8}N_{A}T_{R}(-C_{A}+4(C_{F}+T_{R}))$ & $\frac{9}{16}$ & $\frac{13}{6}$ & $\frac{7}{2}$\\[1mm]  
    $2^{nd}$  & $-\frac{1}{8}N_{A}T_{R}(-C_{A}-4(C_{F}-T_{R}))$  & $-\frac{3}{16}$ & $\frac{1}{6}$ & 0 \\[1mm]  
    $3^{rd}$  & $-\frac{1}{8}N_{A}T_{R}\left(4C_{F}T_{R}N_{A}+4T_{R}^{2}N_{A}+\right.$ & 0 & 10 & $\frac{147}{2}$ \\ 
	      & $\left.-4C_{F}^{2}N_{F}(1+N_{F})+\right.$ & & & \\ 
	      & $\left.+C_{A}(-T_{R}N_{A}+C_{F}N_{F}(1+N_{F}))\right)$  & & & \\ 
    $4^{th}$  & $\frac{1}{8}N_{A}T_{R}\left(-4C_{F}T_{R}N_{A}+4T_{R}^{2}N_{A}+\right.$ & 0 & -10 & $-\frac{147}{2}$ \\
	      & $\left.-4C_{F}^{2}N_{F}(-1+N_{F})+\right.$ & & & \\
	      & $\left.+C_{A}(T_{R}N_{A}+C_{F}N_{F}(-1+N_{F}))\right)$  & & & \\ 
    \hline
  \end{tabular}
\end{center}
\end{minipage}
 
\no\begin{minipage}{0.2\linewidth}
     \includegraphics[width=1.0\linewidth]{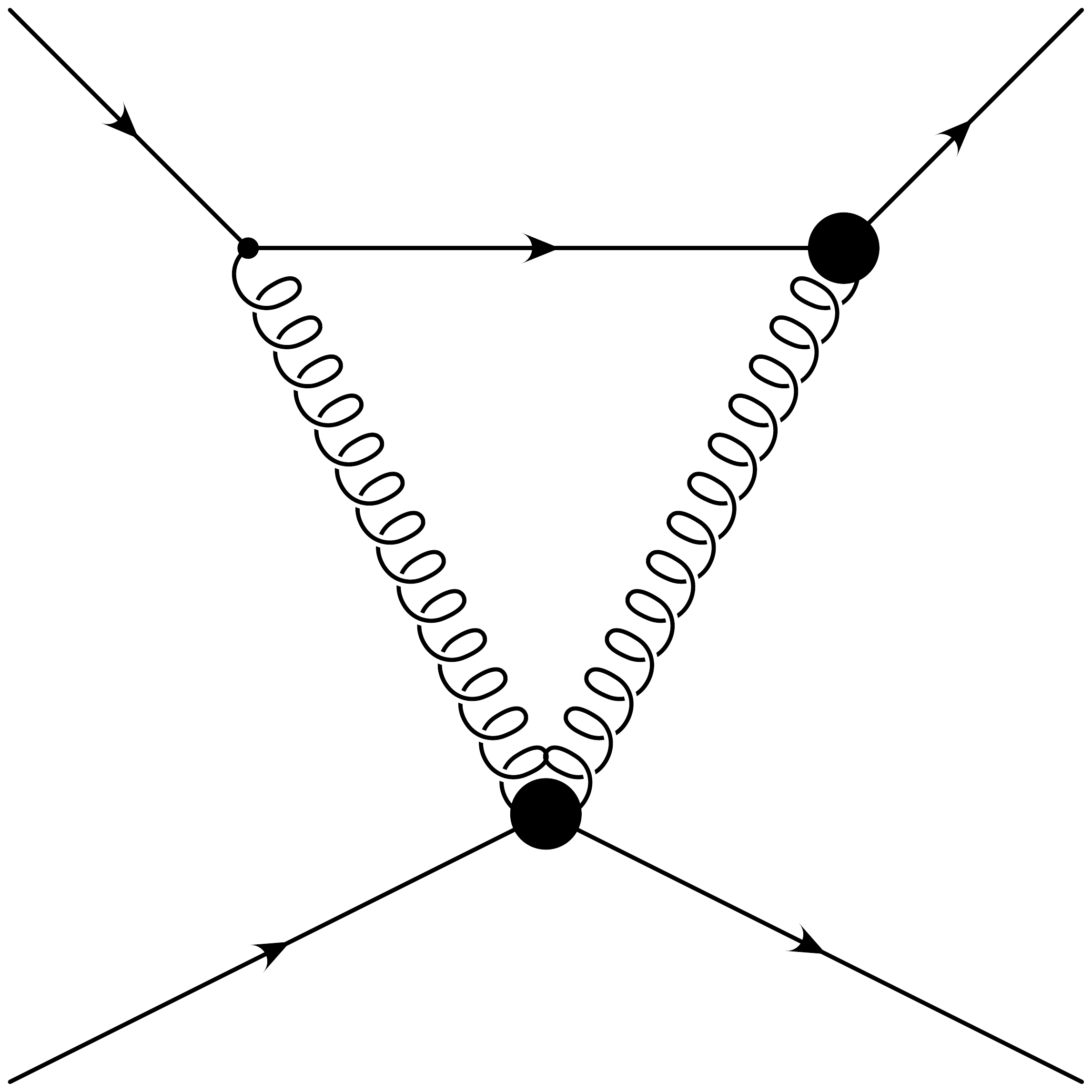}
\end{minipage}
\begin{minipage}{0.7\linewidth}
     \begin{center}
  \begin{tabular}{| l | l | l | l | l |}
    \hline
	adj.	& group invariant  & SU(2) & SU(3) & G$_2$ \\ \hline
    $1^{st}$  & $5N_{A}C_{A}^{2}$ & 60 & 360 & 280\\ 
    $2^{nd}$  & $-\frac{1}{4} N_{A}C_{A}^{3}$  & -6 & -54 & -28 \\[1mm] 
    $3^{rd}$  & $-\frac{1}{4} N_{A}C_{A}^{3}$    & -6 & -54 & -28 \\ 
    $4^{th}$  & $\frac{1}{4} N_{A}C_{A}^{3}(2N_{A}-3)$   & 18 & 702 & 700 \\[1mm] 
    $5^{th}$  & $\frac{1}{12}N_{A}C_{A}^{3}(2N_{A}-1)$ & 10 & 270 & 252 \\
    $6^{th}$  & $\frac{1}{T_{R}(C_{A}-6C_{F})}\left(5N_{A}T_{R}\times \right.$ & $\frac{3825}{64} $ & $\frac{1435}{4}$ 
	      & $\frac{1113}{4}$ \\
	      & $\left.\times C_{A}^{2}(C_{A}-6C_{F})+ \right.$ & & & \\
	      & $\left.+12d_{44}(2+N_{A}) \right)$ & & & \\
    \hline
  \end{tabular}
\end{center}
     \begin{center}
  \begin{tabular}{| l | l | l | l | l |}
    \hline
	fund.	& group invariant  & SU(2) & SU(3) & G$_2$ \\ \hline
    $1^{st}$  & $\frac{1}{8}N_{A}T_{R}(C_{A}-4(C_{F}+T_{R}))$ & $-\frac{9}{16}$ & $-\frac{13}{6}$ & $-\frac{7}{2}$\\[1mm]  
    $2^{nd}$  & $\frac{1}{8}N_{A}T_{R}(C_{A}-4(C_{F}-T_{R}))$  & $\frac{3}{16}$ & $-\frac{1}{6}$ & 0 \\  
    $3^{rd}$  & $-\frac{1}{16}\left(3N_{A}T_{R}N_{F}C_{A}^{2}(1+N_{F})+\right.$ & 0 & -15 & 0 \\ 
	      & $\left.+2N_{A}T_{R}C_{A}\left(N_{A}T_{R}+\right.\right.$ & & & \\
	      & $\left.\left.-5N_{F}C_{F}(1+N_{F})\right)+\right.$ & & & \\
	      & $\left.+8\left(-N_{A}^{2}T_{R}^{2}C_{F}-N_{A}^{2}T_{R}^{3}+\right.\right.$ & & & \\
	      & $\left.\left.+\mbox{ d33}N_{F}(1+N_{F})+\right.\right.$ & & & \\
	      & $\left.\left.+N_{A}T_{R}N_{F}C_{F}^{2}(1+N_{F}) \right) \right)$  & & & \\ 
    $4^{th}$  & $\frac{1}{16}\left(3N_{A}T_{R}N_{F}C_{A}^{2}(-1+N_{F})+\right.$ & 0 & $-\frac{15}{2}$ & 0 \\ 
	      & $\left.-2N_{A}T_{R}C_{A}\left(N_{A}T_{R}+\right.\right.$ & & & \\
	      & $\left.\left.+5N_{F}C_{F}(-1+N_{F})\right)+\right.$ & & & \\
	      & $\left.\left.+8\left(-N_{A}^{2}T_{R}^{2}C_{F}-N_{A}^{2}T_{R}^{3}+\right.\right.\right.$ & & & \\
	      & $\left.\left.-\mbox{ d33}N_{F}(-1+N_{F})+\right.\right.$ & & & \\
	      & $\left.\left.+N_{A}T_{R}N_{F}C_{F}^{2}(-1+N_{F}) \right) \right)$ & & & \\ 
    \hline
  \end{tabular}
\end{center}
\end{minipage}

\no\begin{minipage}{0.2\linewidth}
     \includegraphics[width=1.0\linewidth]{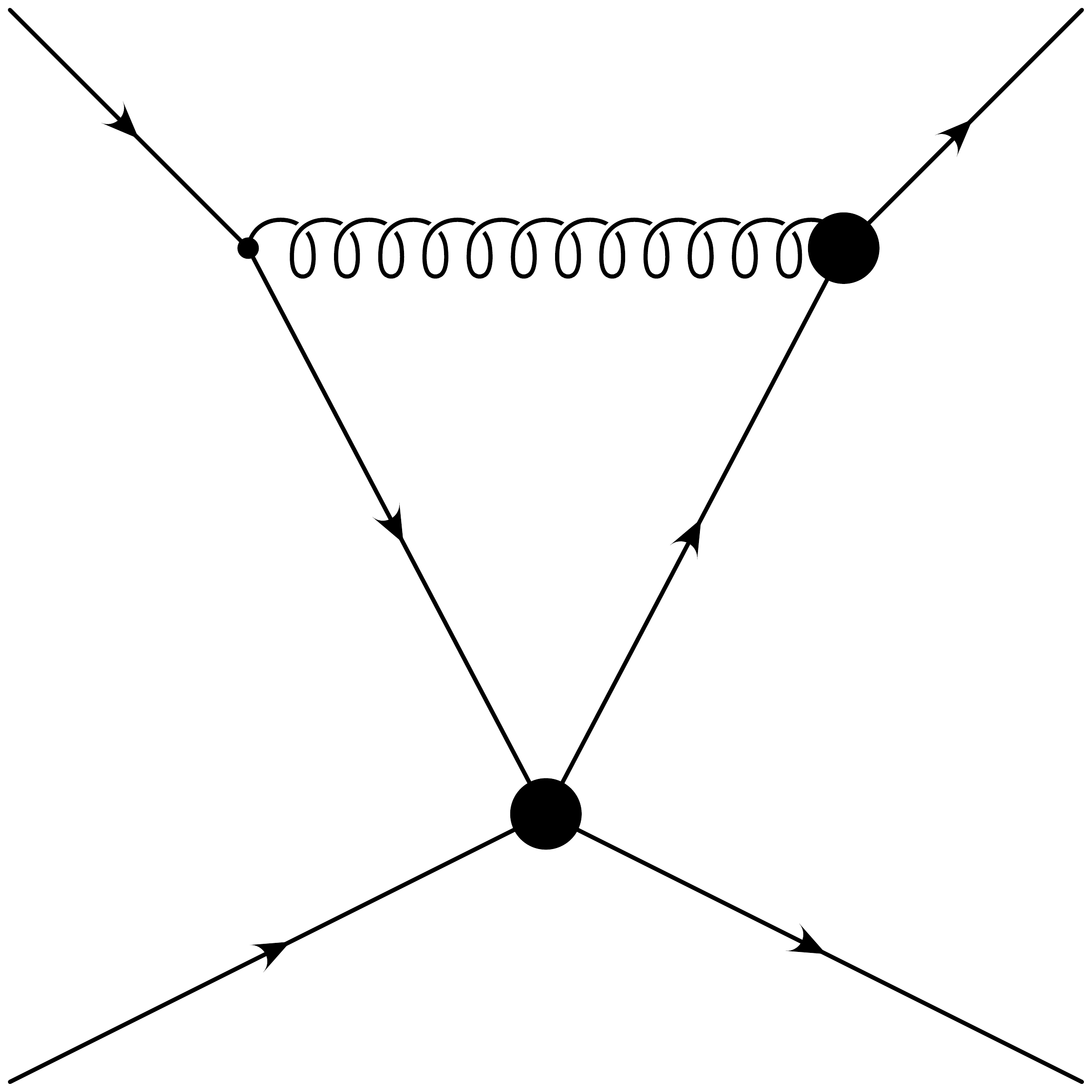}
\end{minipage}
\begin{minipage}{0.7\linewidth}
     \begin{center}
  \begin{tabular}{| l | l | l | l | l |}
    \hline
	adj.	& group invariant  & SU(2) & SU(3) & G$_2$ \\ \hline
    $1^{st}$  & $2N_{A}C_{A}(2+N_{A})$ & 60 & 480 & 896\\  
    $2^{nd}$  & $-5N_{A}C_{A}^{2}$   & -60 & -360 & -280 \\ 
    $3^{rd}$  & $-5N_{A}C_{A}^{2}$  & -6 & -360 & -280 \\ 
    $4^{th}$  & $N_{A}C_{A}^{2}(2N_{A}^{2}+2N_{A}-9)$   & 180 & 9720 & 23016 \\  
    $5^{th}$  & $\frac{1}{3}N_{A}C_{A}^{2}(2N_{A}^{2}+6N_{A}-11)$ & 100 & 3960 & 8680 \\
    $6^{th}$  & 0  & 0 & 0 & 0 \\
    \hline
  \end{tabular}
\end{center}
     \begin{center}
  \begin{tabular}{| l | l | l | l | l |}
    \hline
	fund.	& group invariant  & SU(2) & SU(3) & G$_2$ \\ \hline
    $1^{st}$  & $-\frac{1}{6}N_{A}T_{R}(2+N_{F})$ & -1 & $-\frac{10}{3}$ & $-\frac{21}{2}$\\[1mm]  
    $2^{nd}$  & $\frac{1}{6}N_{A}T_{R}N_{F}$  & $\frac{1}{2}$ & 2 & $\frac{49}{6}$ \\ [1mm]
    $3^{rd}$  & $\frac{1}{12}N_{A}T_{R}(N_{F}\left(C_{A}-4C_{F})\times \right.$ & $-\frac{3}{2}$ & -28 & $-\frac{2695}{6}$ \\
	      & $\left.\times (1+N_{F})+\right.$ & & & \\
	      & $\left.-2N_{A}T_{R}(-2+N_{F}^{2})\right)$ & & &  \\ 
    $4^{th}$  & $\frac{1}{12}N_{A}T_{R}\left(2N_{A}T_{R}(-2+N_{F})+\right.$  & $\frac{1}{2}$ & 14 & $\frac{2009}{6}$ \\ 
	      & $\left.+C_{A}(-1+N_{F})\right)$ & & & \\
    \hline
  \end{tabular}
\end{center}
\end{minipage}

\no\begin{minipage}{0.2\linewidth}
     \includegraphics[width=1.0\linewidth]{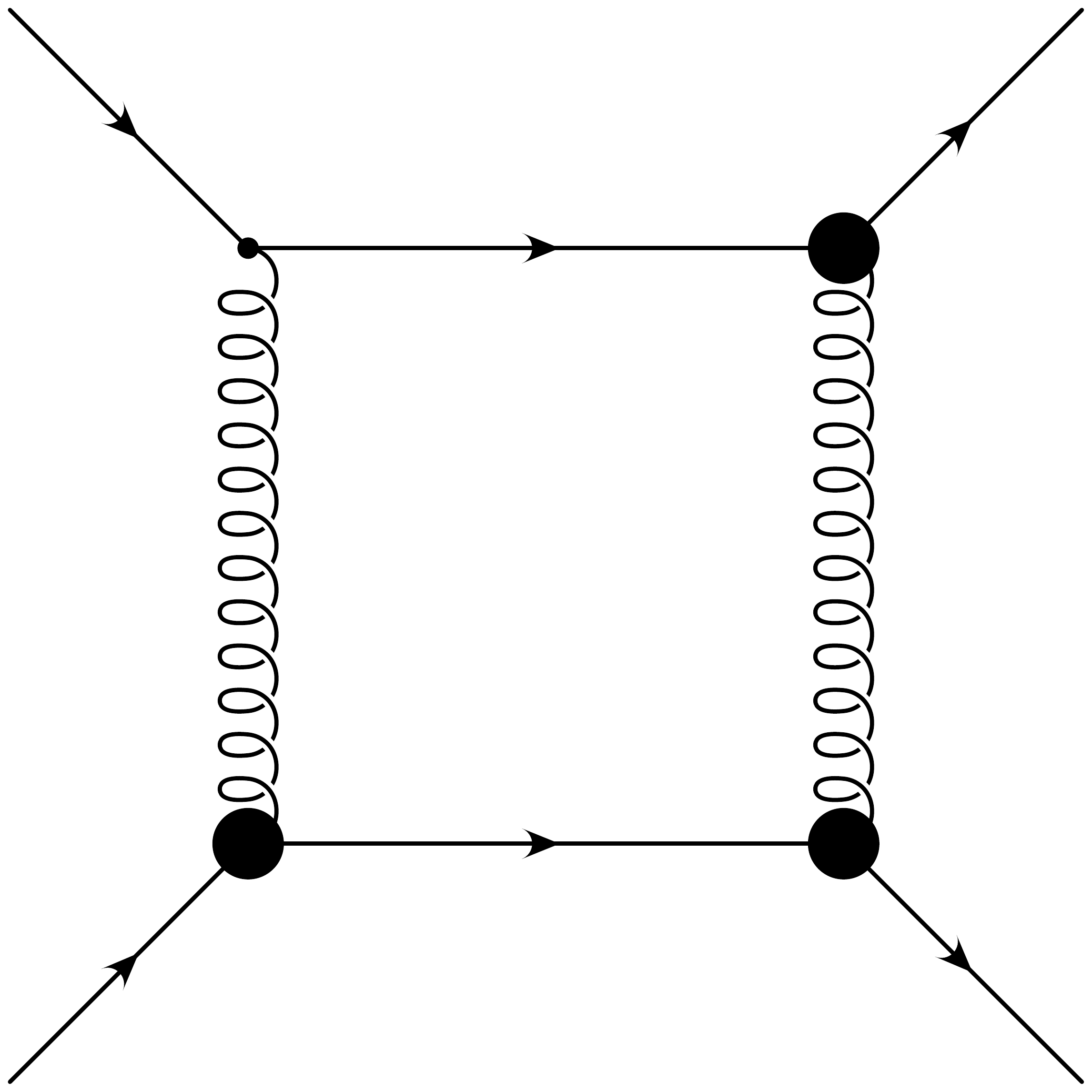}
\end{minipage}
\begin{minipage}{0.7\linewidth}
     \begin{center}
  \begin{tabular}{| l | l | l | l | l |}
    \hline
	adj.	& group invariant  & SU(2) & SU(3) & G$_2$ \\ \hline
    $1^{st}$  & $\frac{5}{2} N_{A}C_{A}^{2}$ & 30 & 180 & 140\\[1mm] 
    $2^{nd}$  & $\frac{1}{4} N_{A}C_{A}^{3}$  & 6 & 54 & 28 \\ 
    $3^{rd}$  & $-\frac{1}{4} N_{A}C_{A}^{3}$  & -6 & -54 & -28 \\[1mm]  
    $4^{th}$  & $-\frac{1}{4} N_{A}C_{A}^{3}(2N_{A}-3)$  & 18 & 702 & 700 \\[1mm] 
    $5^{th}$  & $-\frac{1}{12} N_{A}C_{A}^{3}(1-N_{A})$ & -10 & -270 & -252 \\
    $6^{th}$  & $\frac{1}{2T_{R}(C_{A}-6C_{F})}\left(5N_{A}T_{R}\times \right.$ & $\frac{3825}{128}$ & $\frac{1435}{8}$ 
	      & $\frac{1113}{8}$ \\
	      & $\left.\times C_{A}^{2}(C_{A}-6C_{F})+\right.$ & & & \\
	      & $\left.+12d_{44}(2+N_{A}) \right)$ & & & \\
    \hline
  \end{tabular}
\end{center}
     \begin{center}
  \begin{tabular}{| l | l | l | l | l |}
    \hline
	fund.	& group invariant  & SU(2) & SU(3) & G$_2$ \\ \hline
    $1^{st}$  & $N_{A}T_{R}(C_{F}+T_{R})$ & $\frac{15}{8}$ & $\frac{22}{3}$ & $\frac{21}{2}$\\[1mm] 
    $2^{nd}$  & $N_{A}T_{R}(-C_{F}+T_{R})$  & $-\frac{3}{8}$ & $-\frac{10}{3}$ & $-\frac{7}{2}$ \\[1mm] 
    $3^{rd}$  & $-\frac{1}{2}N_{A}T_{R}\left(2T_{R}^{2}N_{A}+\right.$ & $-\frac{9}{2}$ & -40 
	      & $-\frac{147}{2}$ \\ 
	      & $\left.-2C_{F}^{2}N_{F}(1+N_{F})+\right.$ & & & \\
	      & $\left.+C_{F}(2T_{R}N_{A}+C_{A}N_{F}(1+N_{F}))\right)$ & & &  \\[1mm] 
    $4^{th}$  & $-\frac{1}{2}N_{A}T_{R}\left(2T_{R}^{2}N_{A}+\right.$ & 0 & 8 & $\frac{49}{2}$ \\  
	      & $\left.-2C_{F}^{2}N_{F}(-1+N_{F})+ \right.$ & & & \\
	      & $\left.+C_{F}(-2T_{R}N_{A}+\right.$ & & & \\ 
	      & $\left.+C_{A}N_{F}(-1+N_{F}))\right)$ & & & \\
    \hline
  \end{tabular}
\end{center}
\end{minipage}


\no\begin{minipage}{0.2\linewidth}
      \includegraphics[width=1.0\linewidth]{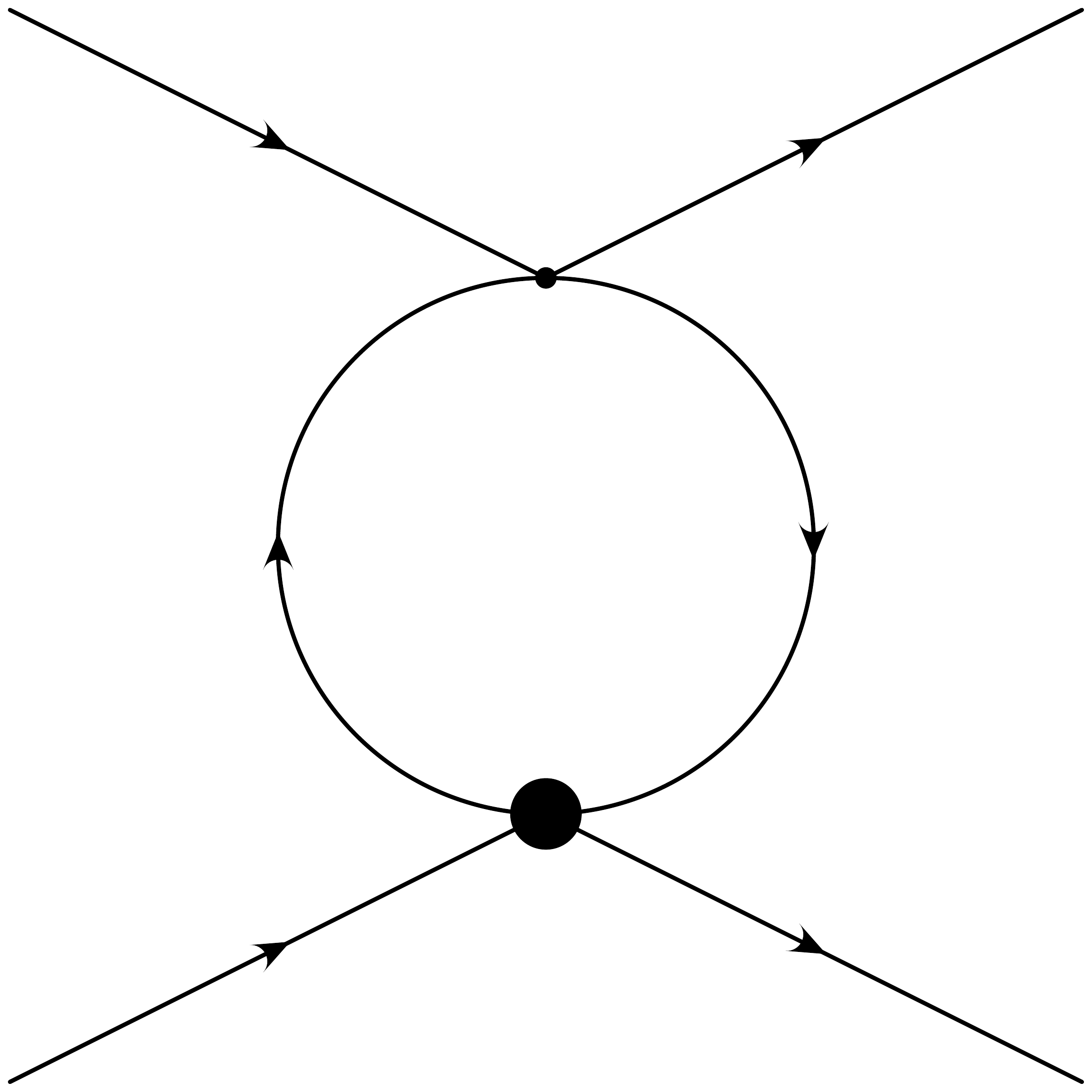}
 \end{minipage}
 \begin{minipage}{0.7\linewidth}
      \begin{center}
   \begin{tabular}{| l | l | l | l | l |}
    \hline
	adj.	& group invariant  & SU(2) & SU(3) & G$_2$ \\ \hline
    $1^{st}$  & $4N_{A}(16+10N_{A}+N_{A}^{2})$ & 660 & 5120 & 19712\\ 
    $2^{nd}$  & $-4N_{A}C_{A}(2+N_{A})$  & -120 & -960 & -1792 \\ 
    $3^{rd}$  & $-4N_{A}C_{A}(2+N_{A})$   & -120 & -960 & -1792 \\ 
    $4^{th}$  & $-4N_{A}^{2}C_{A}(-4+N_{A}^{2})$  & 360 & 46080 & 301056 \\ 
    $5^{th}$  & $\frac{4}{3}N_{A}C_{A}(-2+N_{A})(2+N_{A})^{2}$ & 200 & 19200 & 114688 \\
    $6^{th}$  & 0 & 0 & 0 & 0 \\
    \hline
   \end{tabular}
 \end{center}
     \begin{center}
  \begin{tabular}{| l | l | l | l | l |}
    \hline
	fund.	& group invariant  & SU(2) & SU(3) & G$_2$ \\ \hline
    $1^{st}$  & $\frac{1}{36}N_{F}(3+4N_{F}+N_{F}^{2})$ & $\frac{5}{6}$ & 2 & $\frac{140}{9}$\\[1mm]  
    $2^{nd}$  & $-\frac{1}{36}N_{F}(-1+N_{F}^{2})$  & $-\frac{1}{6}$ & $-\frac{2}{3}$ & $-\frac{28}{3}$ \\  
    $3^{rd}$  & 0  & 0 & 0 & 0 \\  
    $4^{th}$  & 0  & 0 & 0 & 0 \\ 
    \hline
  \end{tabular}
\end{center}
 \end{minipage}

\section{Further results for non-primitive divergent vertices}\label{app:res2}

\subsection{Gluonic sector}

\subsubsection{4-ghost vertex DSE}

\begin{minipage}{0.2\linewidth}
     \includegraphics[width=1.0\linewidth]{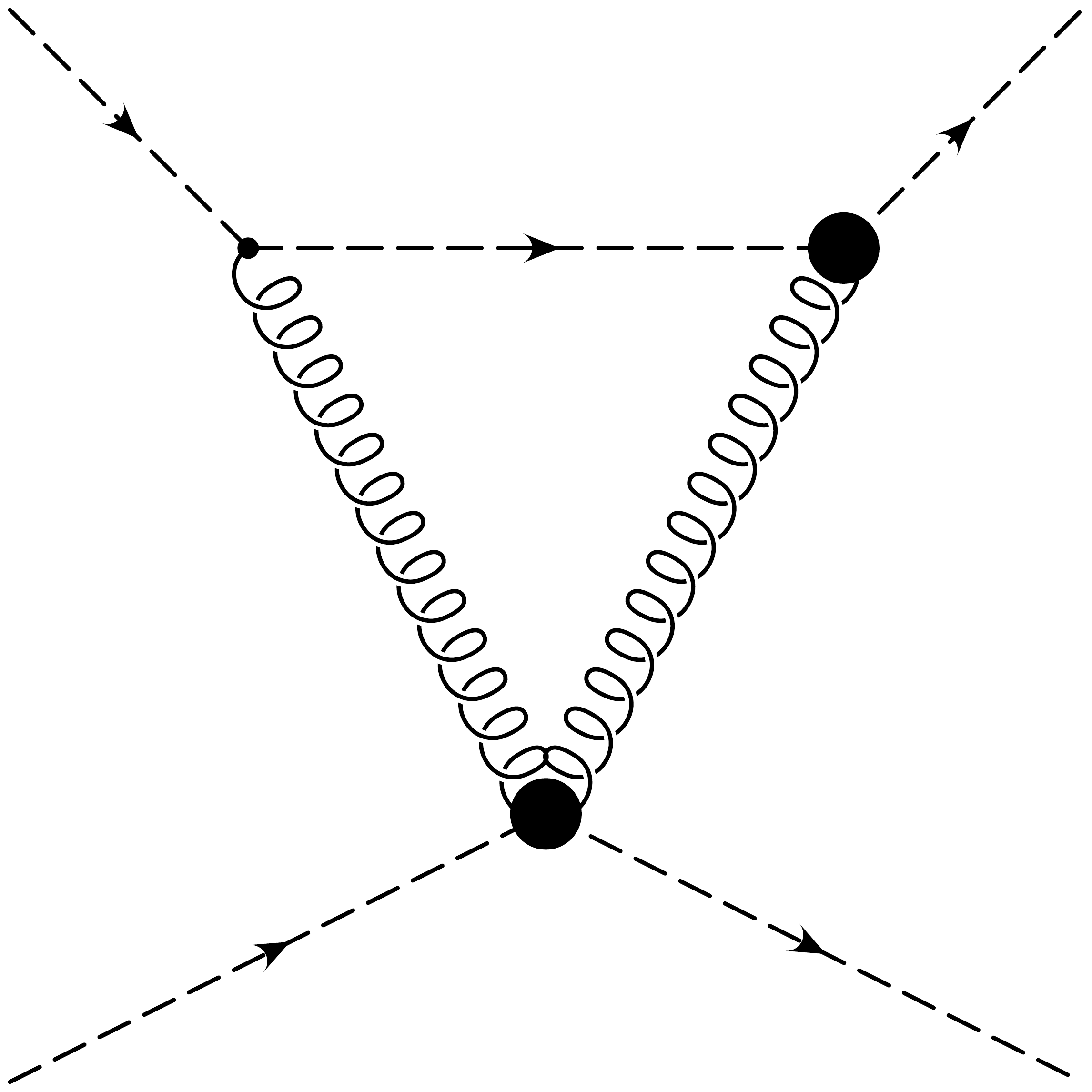}
\end{minipage}
\begin{minipage}{0.7\linewidth}
     \begin{center}
  \begin{tabular}{| l | l | l | l | l |}
    \hline
	$L_{1}$      & group invariant  & SU(2) & SU(3) & G$_2$ \\ \hline
    $1^{st}$  & $\frac{1}{4}N_{A}C_{A}^{3}$ & 6 & 54 & 28\\ 
    $2^{nd}$  & $-\frac{1}{4}N_{A}C_{A}^{3}$  & -6 & -54 & -28 \\ 
    $3^{rd}$  & $\frac{5}{2}N_{A}C_{A}^{2}$  & 30 & 180 & 140 \\[1mm]  
    $4^{th}$  & $\frac{1}{4} N_{A}C_{A}^{3}(2N_{A}-3)$  & 18 & 702 & 700 \\[1mm] 
    $5^{th}$  & $\frac{1}{12} N_{A}C_{A}^{3}(1-2N_{A})$ & -10 & -270 & -252 \\
    $6^{th}$  & $\frac{1}{2T_{R}(C_{A}-6C_{F})}\left(5N_{A}T_{R}\times \right.$ & $\frac{3825}{128}$ & $\frac{1435}{8}$ 
	      & $\frac{1113}{8}$ \\
	      & $\left.\times C_{A}^{2}(C_{A}-6C_{F})+\right.$ & & & \\
	      & $\left.+12d_{44}(2+N_{A}) \right)$ & & & \\
    \hline
  \end{tabular}
\end{center} \begin{center}
  \begin{tabular}{| l | l | l | l | l |}
    \hline
	$L_{2}$		& group invariant  & SU(2) & SU(3) & G$_2$ \\ \hline
    $1^{st}$  & $-\frac{3}{4}N_{A}C_{A}^{3}$ & -18 & -162 & -84 \\ 
    $2^{nd}$  & $\frac{1}{4}N_{A}C_{A}^{3}$  & 6 & 54 & 28 \\ 
    $3^{rd}$  & 0 & 0 & 0 & 0 \\ 
    $4^{th}$  & $\frac{1}{4} N_{A}C_{A}^{3}(-2N_{A}+3)$ & -18 & -702 & -700 \\[1mm]  
    $5^{th}$  & $\frac{1}{4} N_{A}C_{A}^{3}(-1+2N_{A})$ & 30 & 810 & 756 \\
    $6^{th}$  & 0 & 0 & 0  & 0 \\
    \hline
  \end{tabular}
\end{center}
\end{minipage}

\no\begin{minipage}{0.2\linewidth}
     \includegraphics[width=1.0\linewidth]{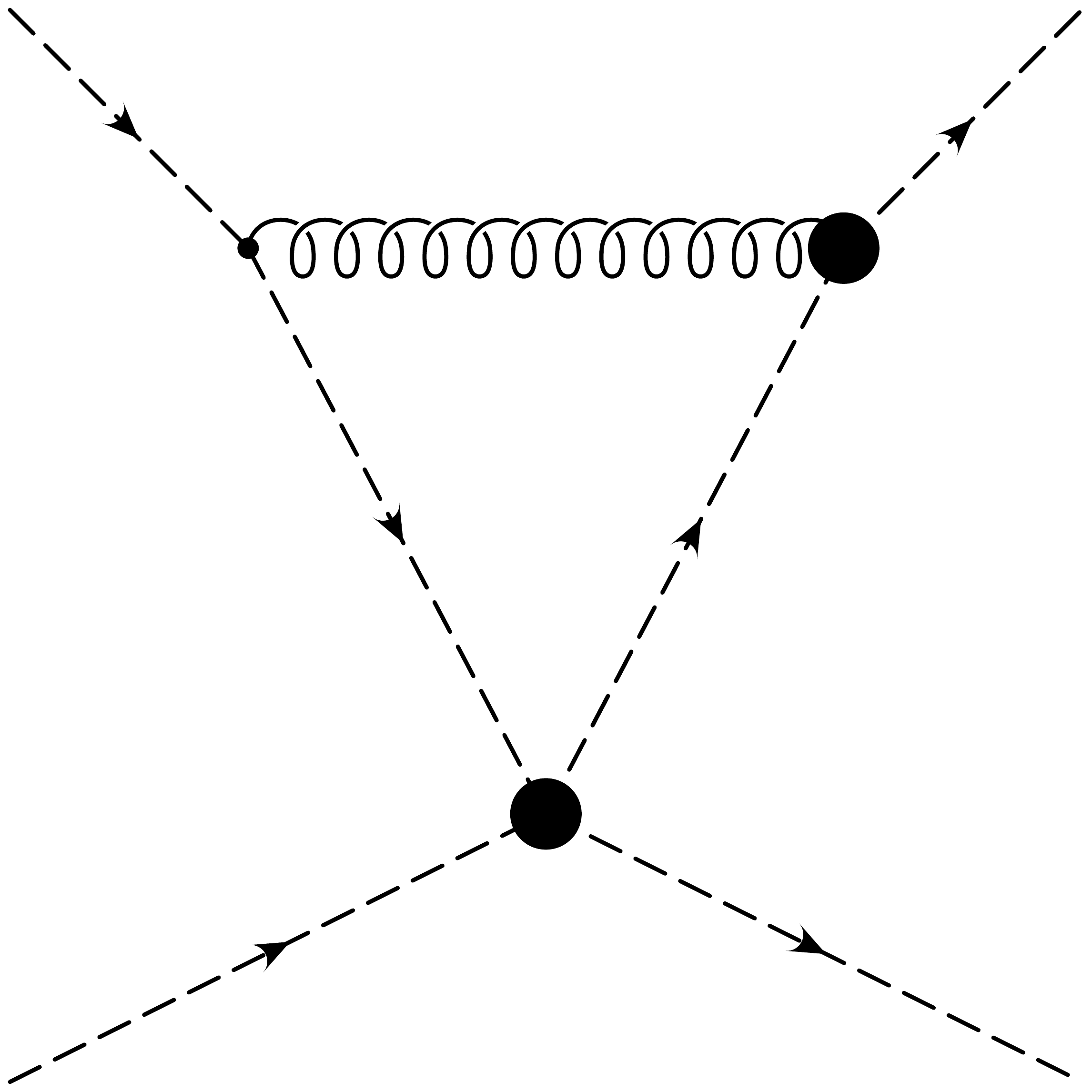}
\end{minipage}
\begin{minipage}{0.7\linewidth}
     \begin{center}
  \begin{tabular}{| l | l | l | l | l |}
    \hline
	adj.     & group invariant  & SU(2) & SU(3) & G$_2$ \\ \hline
    $1^{st}$  & $\frac{3}{4}N_{A}C_{A}^{3}$ & 18 & 162 & 84\\ 
    $2^{nd}$  & $-\frac{1}{4}N_{A}C_{A}^{3}$  & -6 & -54 & -28 \\ 
    $3^{rd}$  & 0  & 0 & 0 & 0 \\  
    $4^{th}$  & $\frac{1}{4} N_{A}C_{A}^{3}(2N_{A}-3)$  & 18 & 702 & 700 \\[1mm] 
    $5^{th}$  & $\frac{1}{4} N_{A}C_{A}^{3}(1-2N_{A})$ & -30 & -810 & -756 \\
    $6^{th}$  & 0 & 0 & 0  & 0 \\
    \hline
  \end{tabular}
\end{center}
\end{minipage}

\no\begin{minipage}{0.2\linewidth}
     \includegraphics[width=1.0\linewidth]{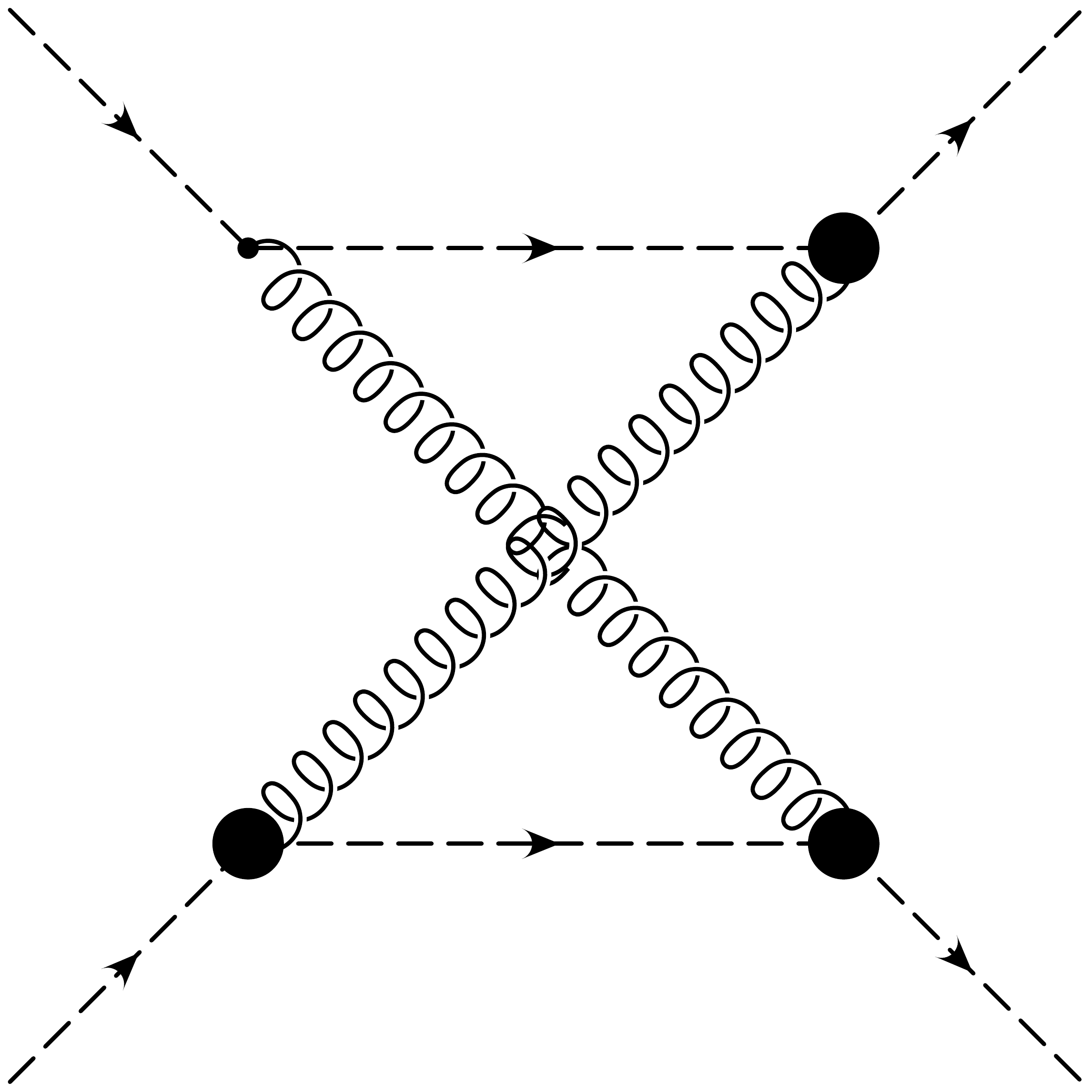}
\end{minipage}
\begin{minipage}{0.7\linewidth}
     \begin{center}
  %
\end{center}
\end{minipage}

\no\begin{minipage}{0.2\linewidth}
     \includegraphics[width=1.0\linewidth]{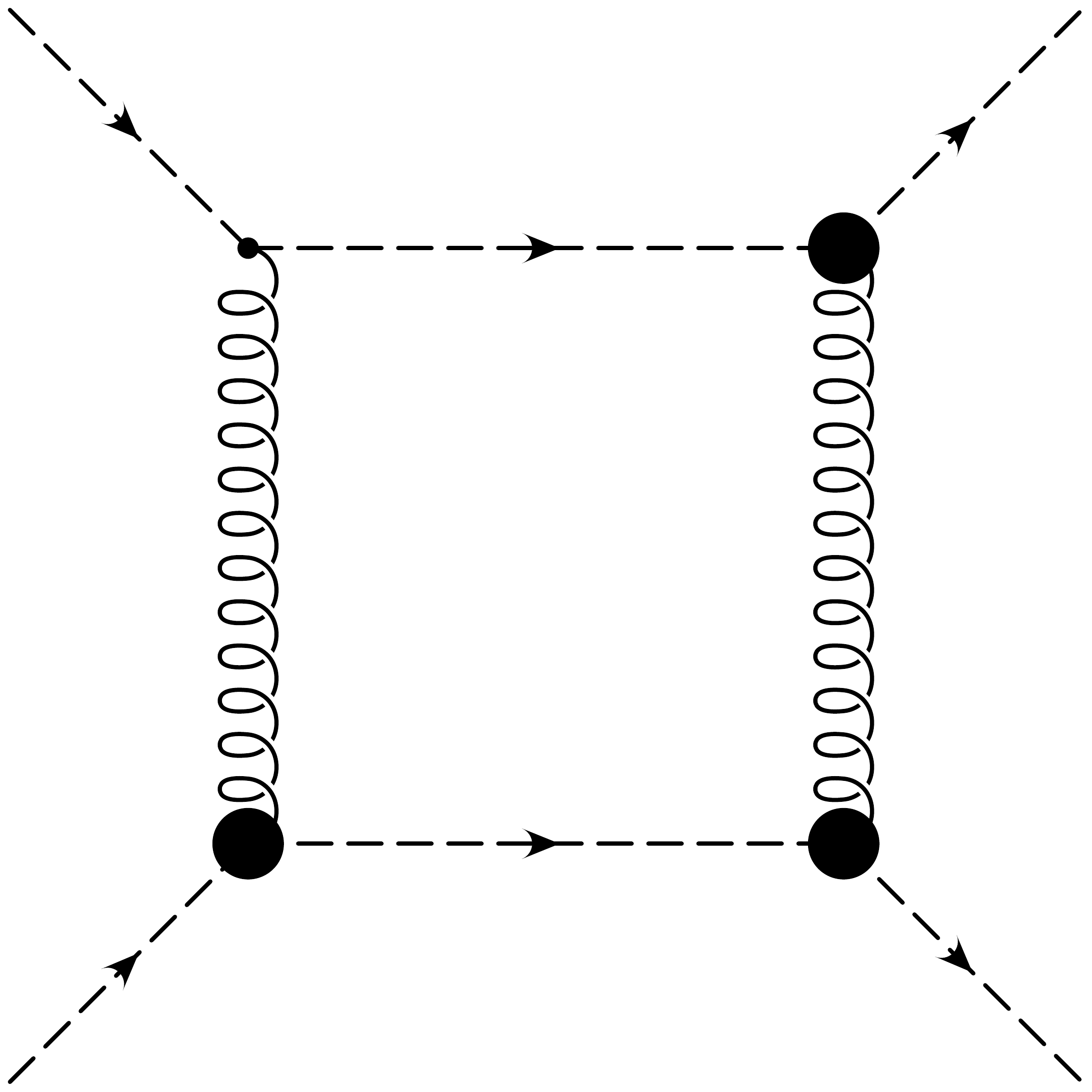}
\end{minipage}
\begin{minipage}{0.7\linewidth}
     \begin{center}
  %
\end{center}
\end{minipage}

\subsubsection{2-gluon-2-ghost vertex DSE}

\begin{minipage}{0.2\linewidth}
     \includegraphics[width=1.0\linewidth]{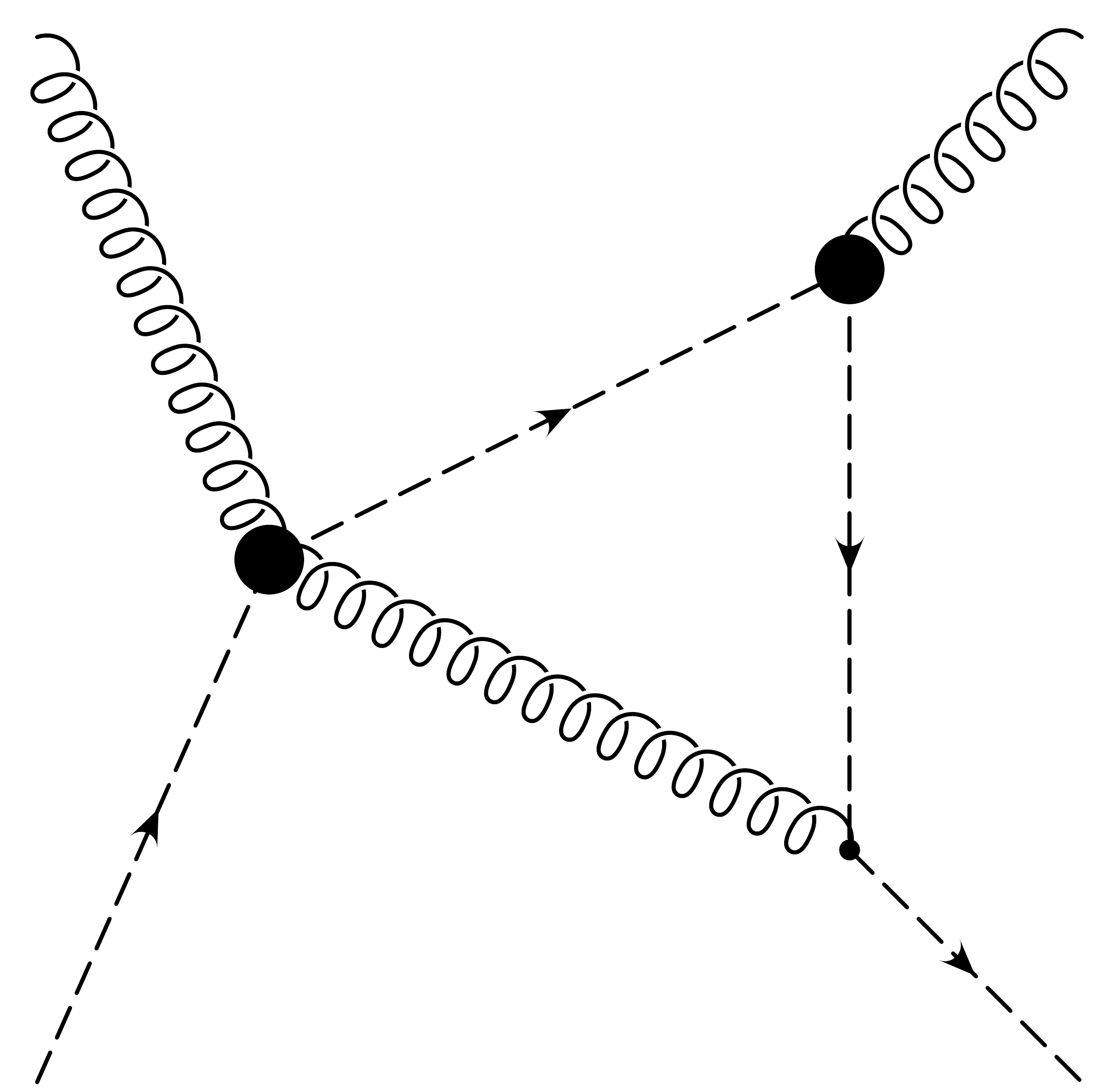}
\end{minipage}
\begin{minipage}{0.7\linewidth}
     \begin{center}
  %
\end{center}
\end{minipage}

\no\begin{minipage}{0.2\linewidth}
     \includegraphics[width=1.0\linewidth]{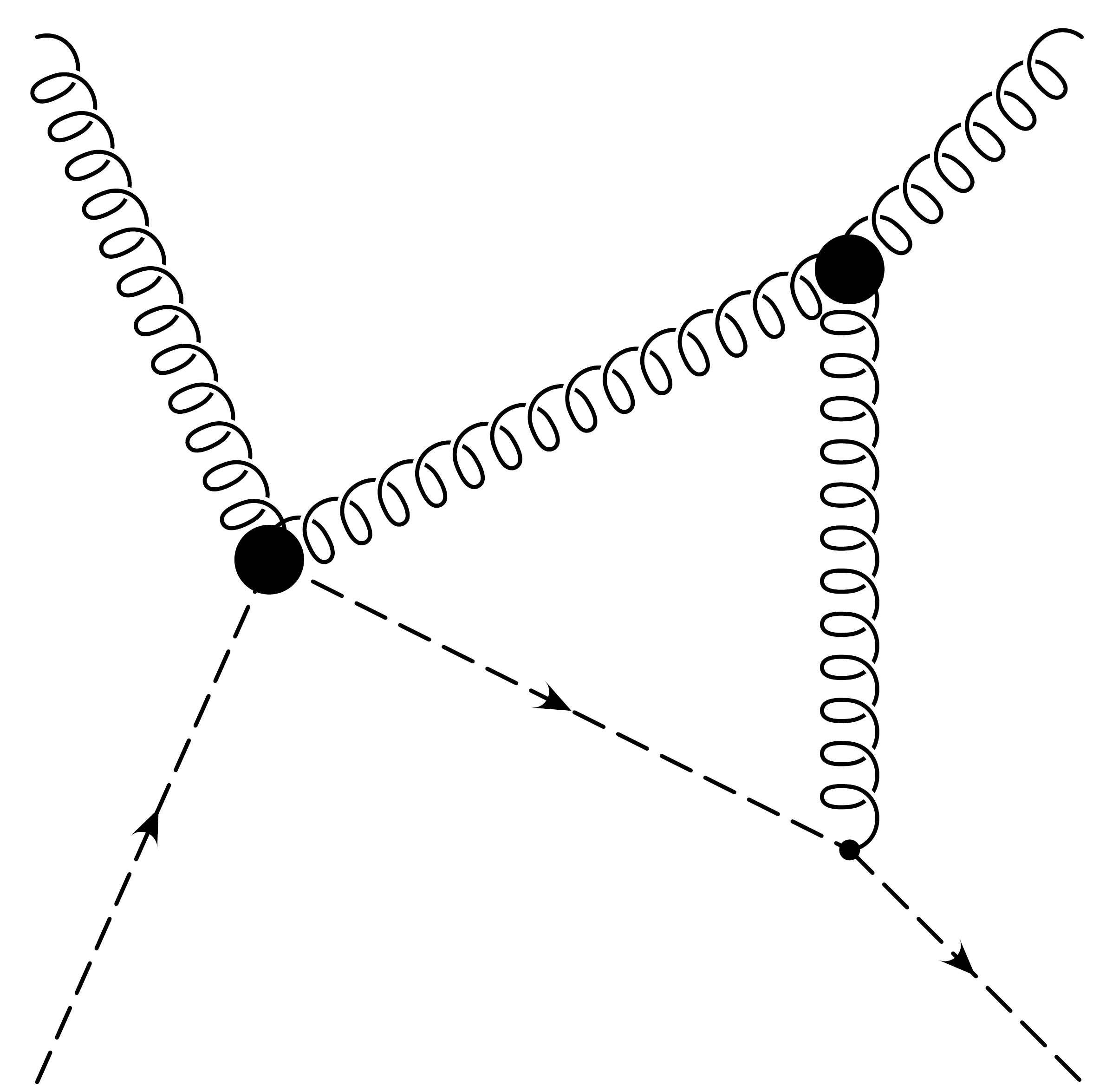}
\end{minipage}
\begin{minipage}{0.7\linewidth}
     \begin{center}
  %
\end{center}
\end{minipage}

\no\begin{minipage}{0.2\linewidth}
     \includegraphics[width=1.0\linewidth]{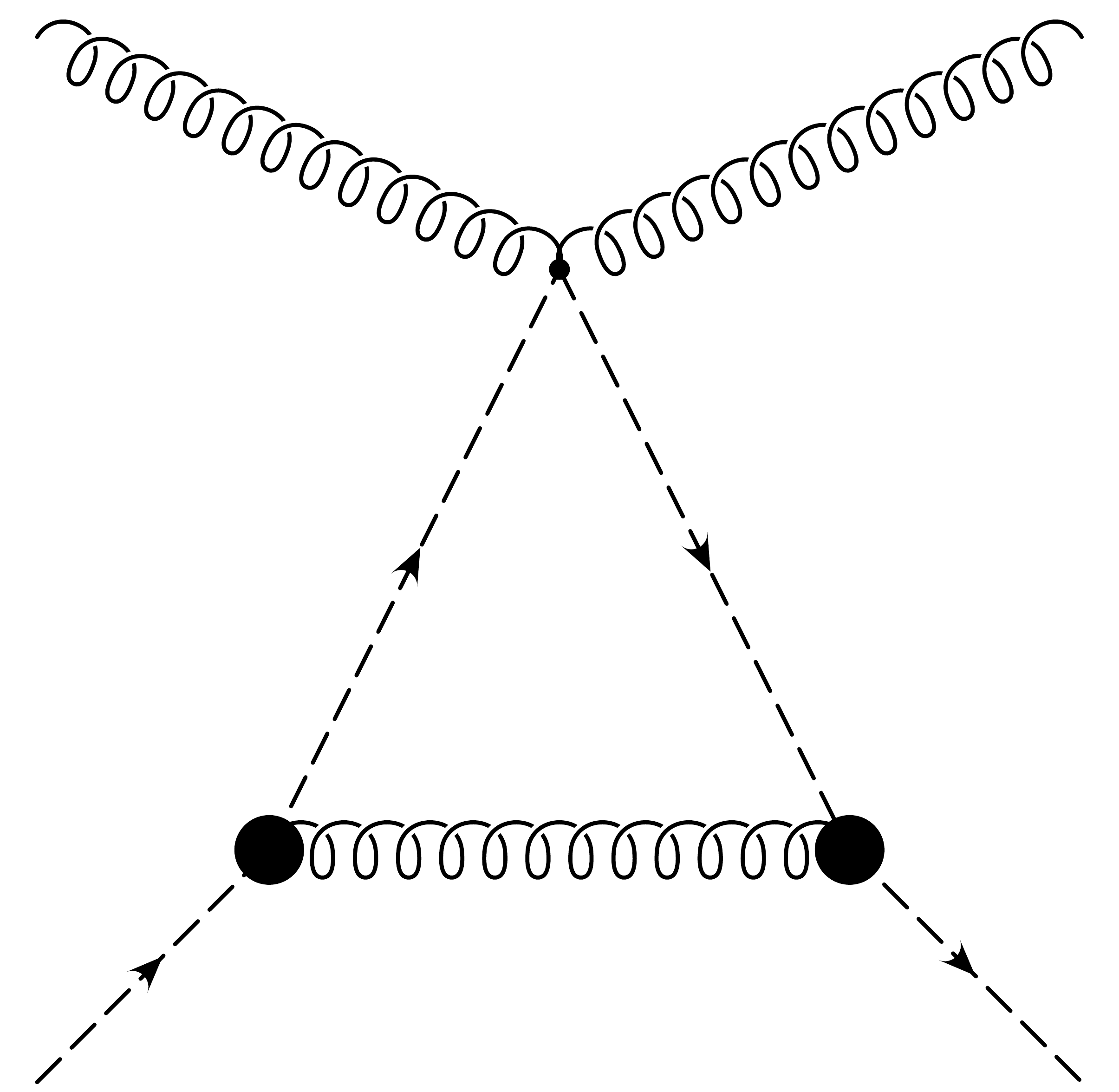}
\end{minipage}
\begin{minipage}{0.7\linewidth}
     \begin{center}
  %
\end{center}
\end{minipage}

\no\begin{minipage}{0.2\linewidth}
     \includegraphics[width=1.0\linewidth]{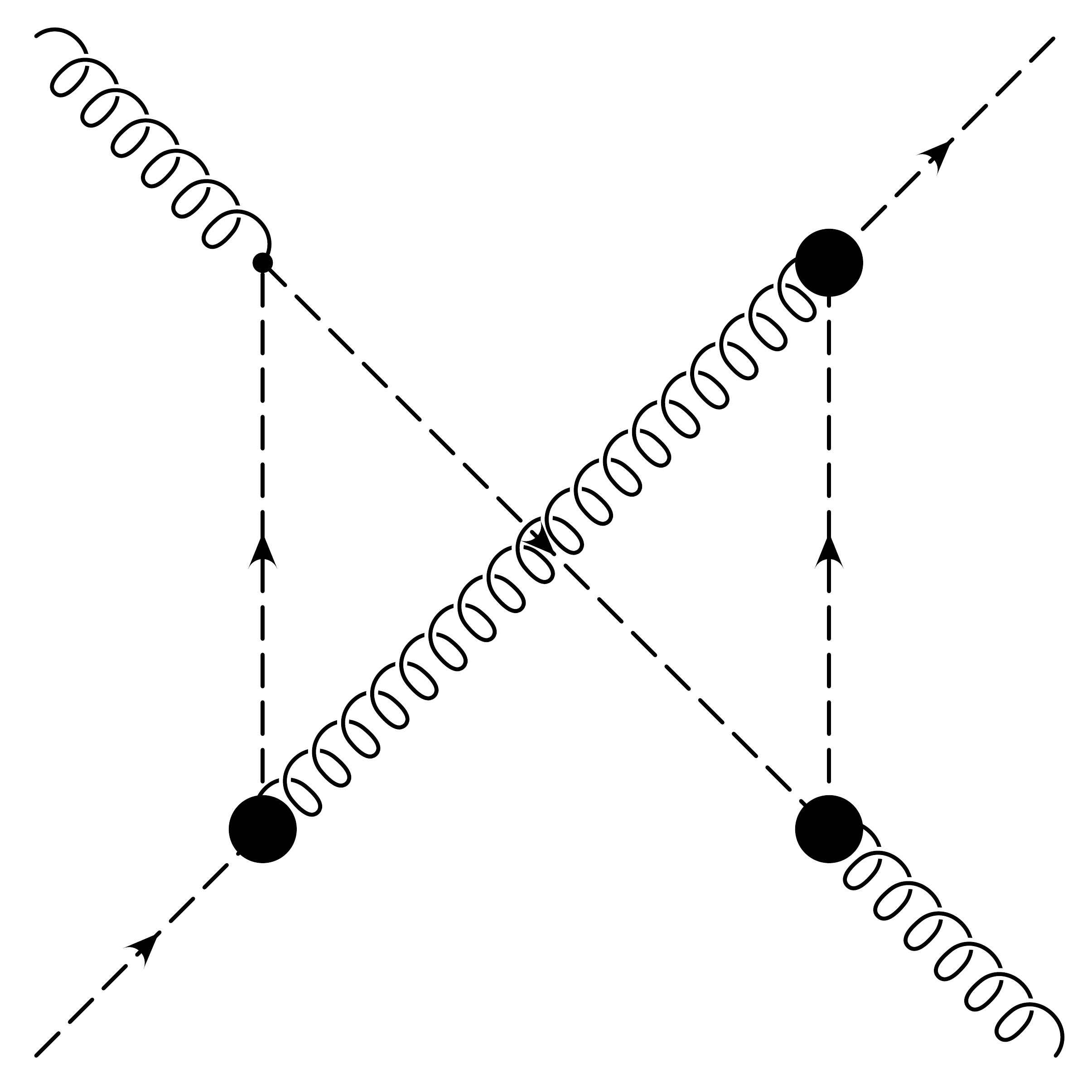}
\end{minipage}
\begin{minipage}{0.7\linewidth}
     \begin{center}
  %
\end{center}
\end{minipage}

\no\begin{minipage}{0.2\linewidth}
     \includegraphics[width=1.0\linewidth]{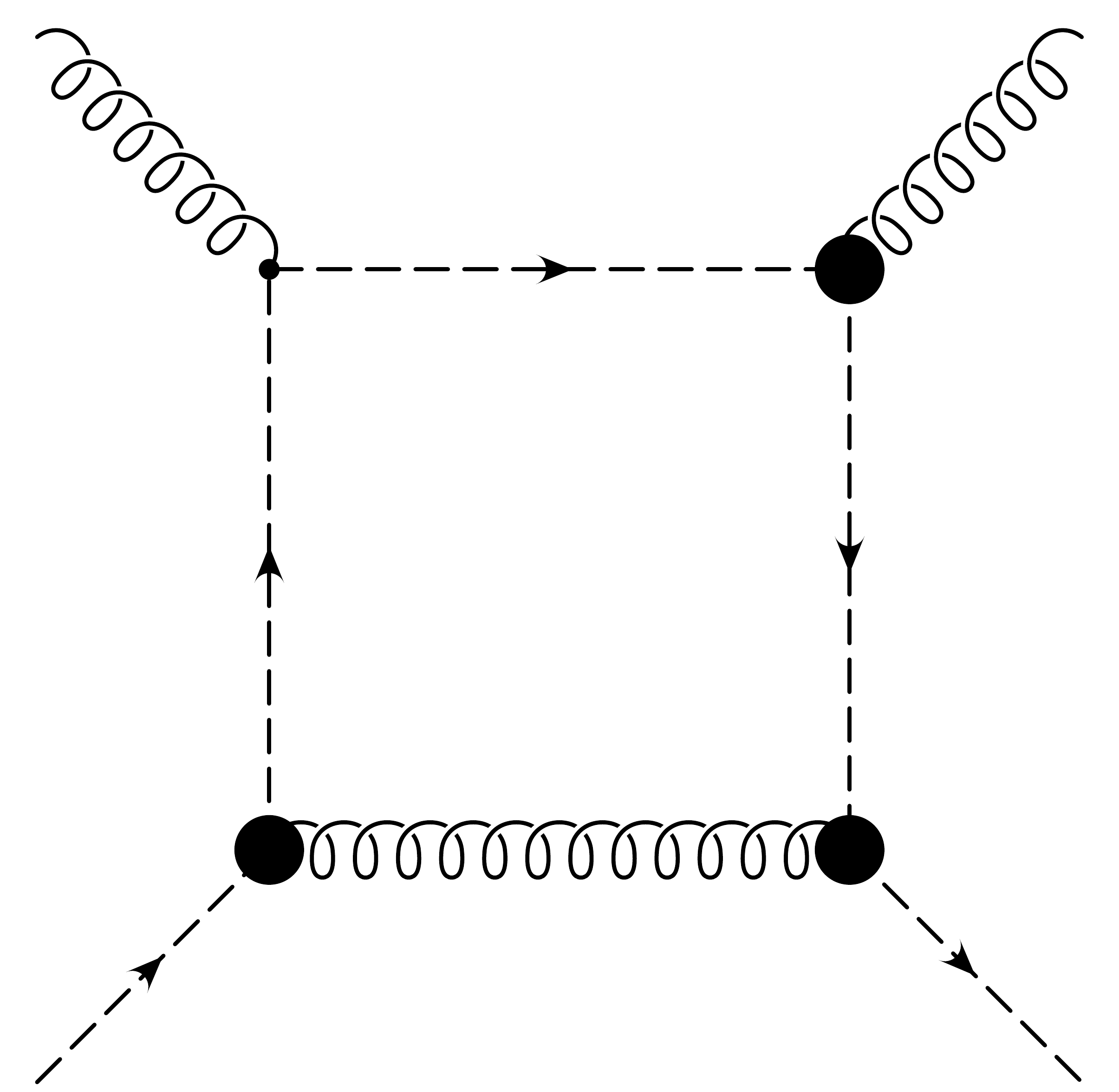}
\end{minipage}
\begin{minipage}{0.7\linewidth}
     \begin{center}
  %
\end{center}
\end{minipage}

\no\begin{minipage}{0.2\linewidth}
     \includegraphics[width=1.0\linewidth]{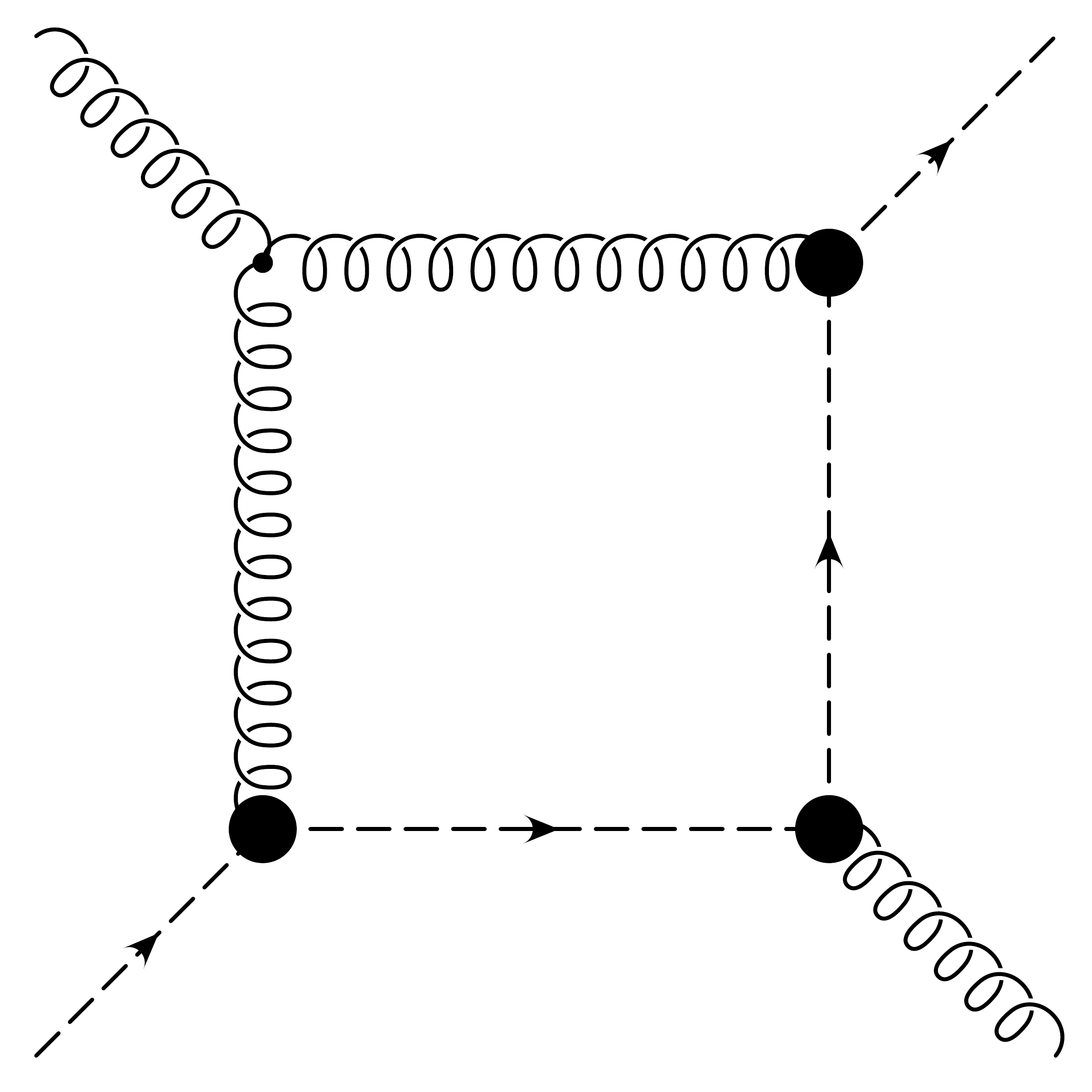}
\end{minipage}
\begin{minipage}{0.7\linewidth}
     \begin{center}
  %
\end{center}
\end{minipage}

\no\begin{minipage}{0.2\linewidth}
     \includegraphics[width=1.0\linewidth]{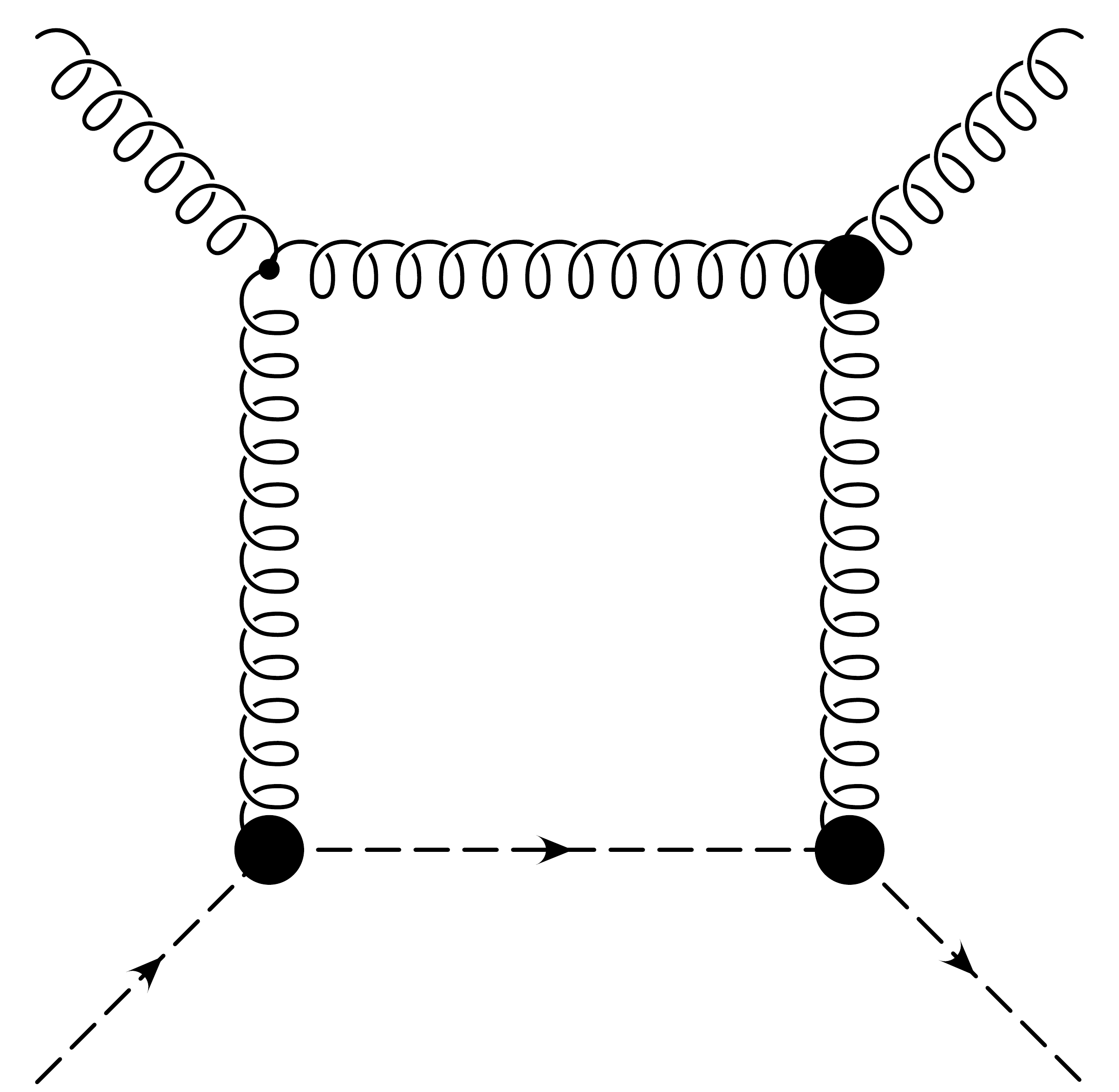}
\end{minipage}
\begin{minipage}{0.7\linewidth}
     \begin{center}
  %
\end{center}
\end{minipage}

\no\begin{minipage}{0.2\linewidth}
     \includegraphics[width=1.0\linewidth]{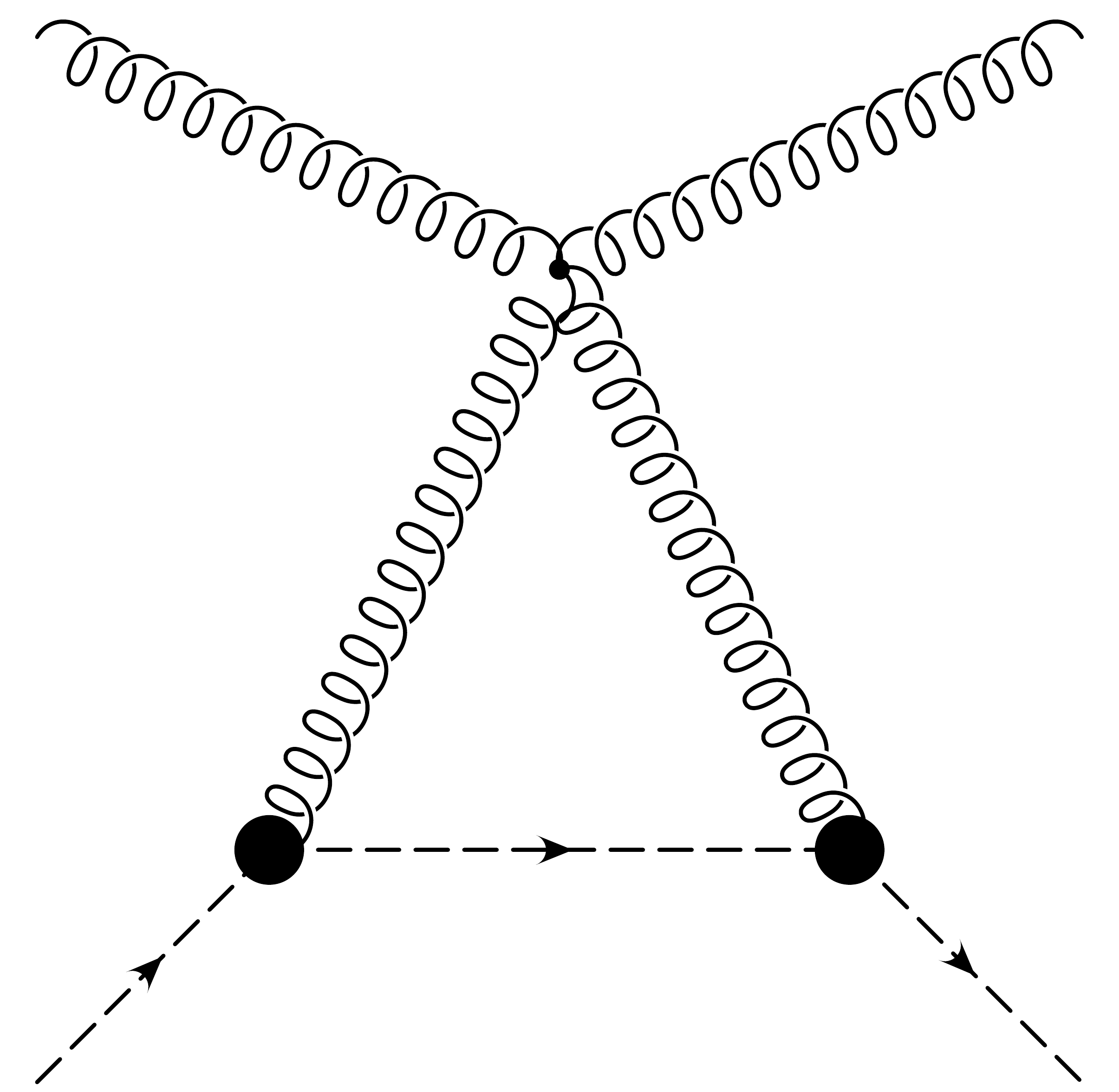}
\end{minipage}
\begin{minipage}{0.7\linewidth}
     \begin{center}
  %
\end{center}
\end{minipage}

\subsection{Matter sector}

\subsubsection{2-scalar-2-ghost vertex DSE}

\begin{minipage}{0.2\linewidth}
     \includegraphics[width=1.0\linewidth]{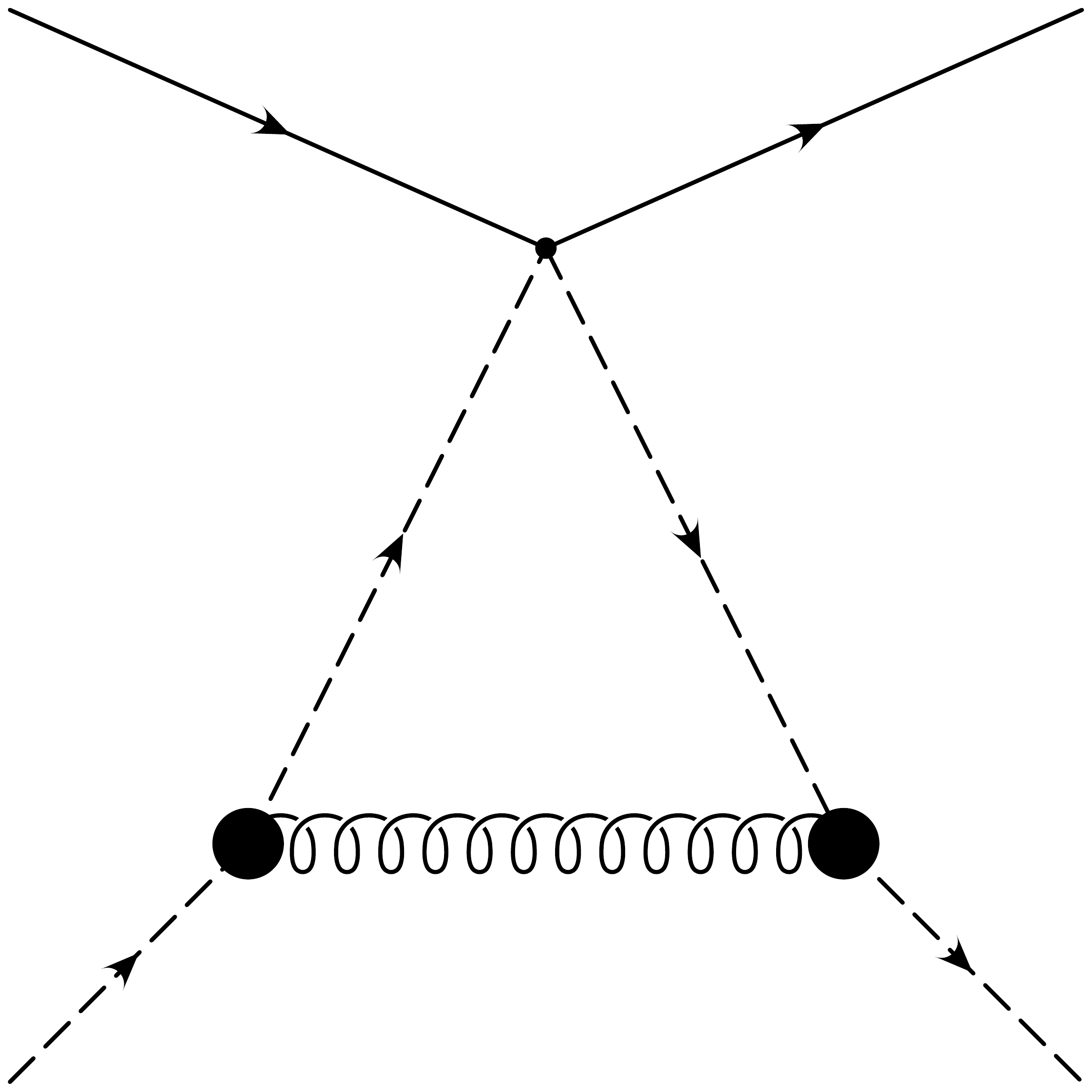}
\end{minipage}
\begin{minipage}{0.7\linewidth}
     \begin{center}
  %
\end{center}
\end{minipage}

\no\begin{minipage}{0.2\linewidth}
     \includegraphics[width=1.0\linewidth]{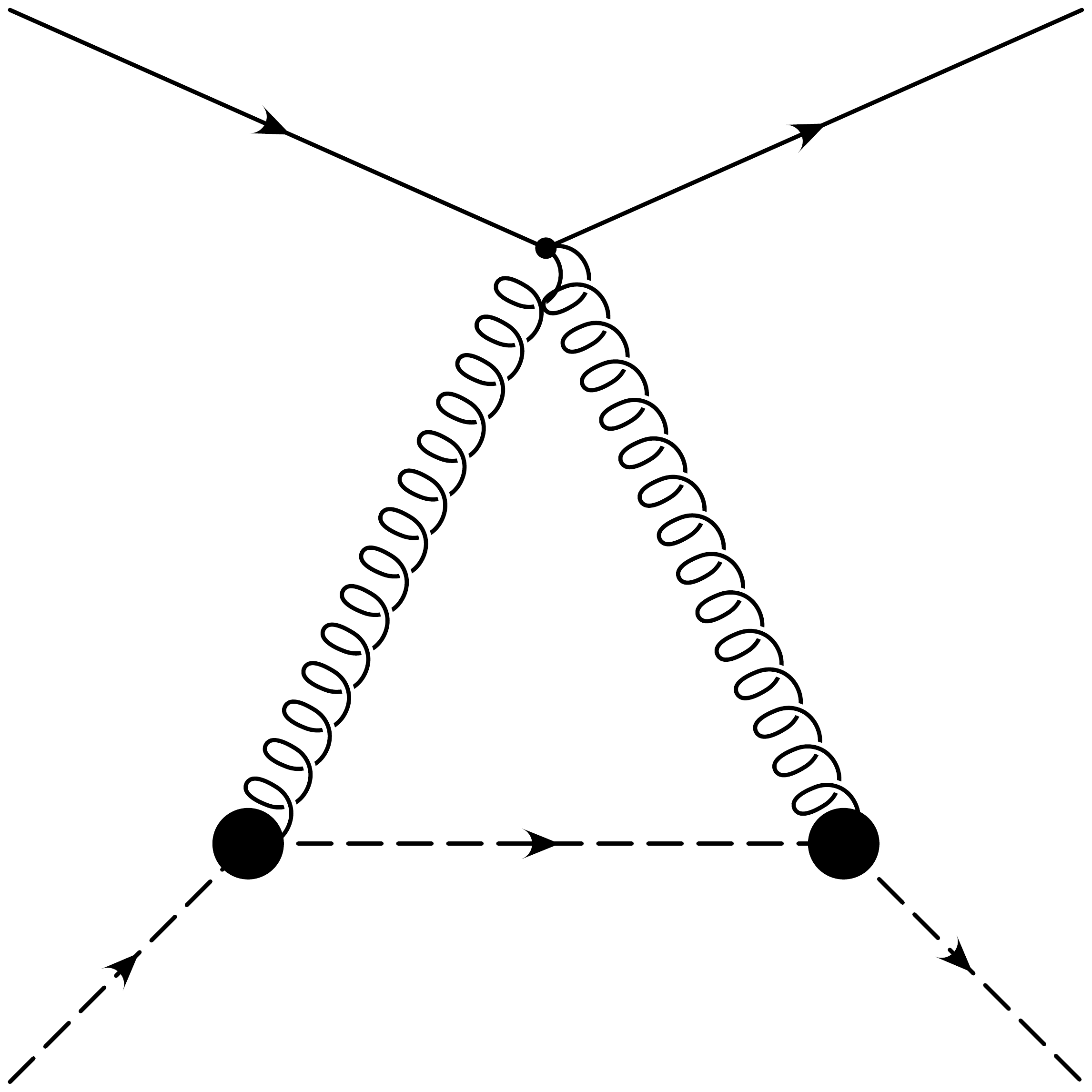}
\end{minipage}
\begin{minipage}{0.7\linewidth}
     \begin{center}
  %
\end{center}
\end{minipage}

\no\begin{minipage}{0.2\linewidth}
     \includegraphics[width=1.0\linewidth]{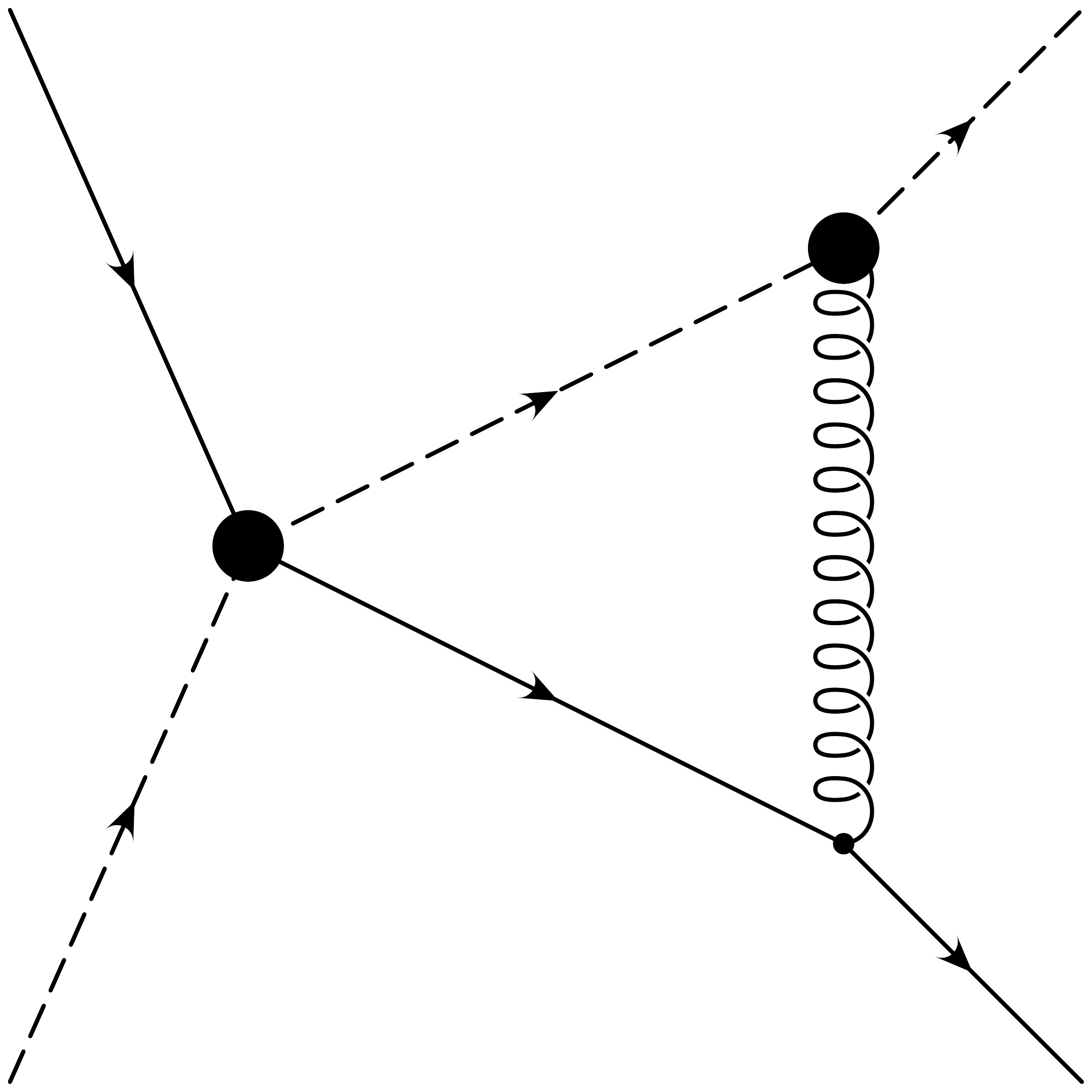}
\end{minipage}
\begin{minipage}{0.7\linewidth}
     \begin{center}
  %
\end{center}
\end{minipage}

\no\begin{minipage}{0.2\linewidth}
     \includegraphics[width=1.0\linewidth]{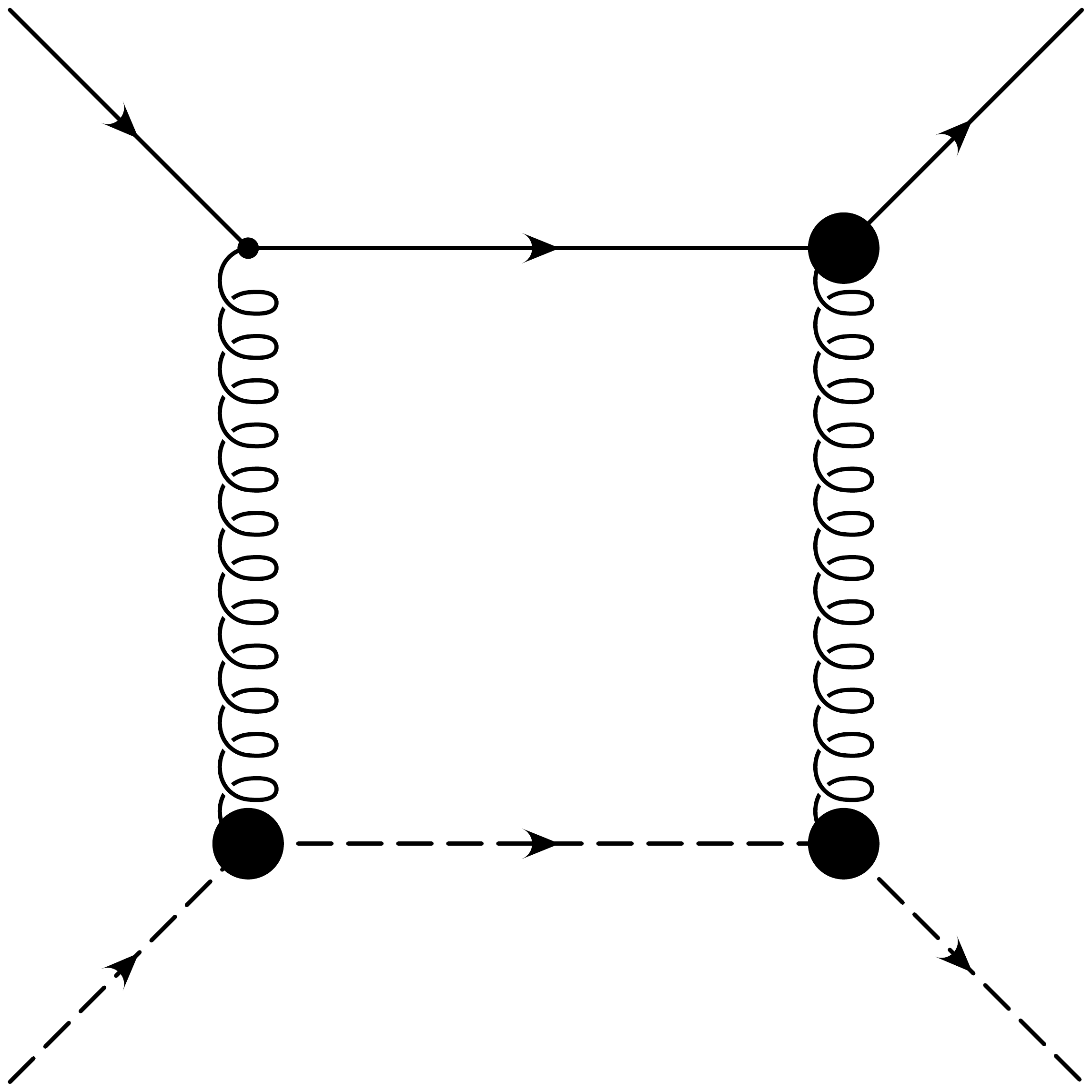}
\end{minipage}
\begin{minipage}{0.7\linewidth}
     \begin{center}
  %
\end{center}
\end{minipage}

\bibliographystyle{bibstyle}
\bibliography{bib}


\end{document}